\documentclass[a4paper,11pt]{article}
\usepackage{jheppub}
\usepackage[utf8]{inputenc}
\usepackage{amsmath,amssymb,slashed,braket}
\usepackage{dsfont}
\usepackage{amsfonts}
\usepackage{bm,bbm}
\usepackage[dvipsnames,table]{xcolor}
\usepackage[normalem]{ulem}
\usepackage{hyperref}
\hypersetup{colorlinks,citecolor= blue,linkcolor= blue, urlcolor=blue}
\usepackage{array}
\usepackage{verbatim}
\usepackage{epsfig}
\usepackage{multirow, booktabs}
\usepackage{makecell}
\usepackage{hhline}
\usepackage{bbold}
\usepackage[version=4]{mhchem}
\usepackage{cleveref}
\usepackage{orcidlink}
\usepackage{subcaption}
\usepackage{graphicx}
\usepackage{caption}
\usepackage{placeins} 

\allowdisplaybreaks[4]
\usepackage{tikz} 
\usepackage{tkz-euclide}
\usetikzlibrary{backgrounds} 
\usetikzlibrary{decorations.pathmorphing}
\usetikzlibrary{arrows.meta}
\tikzset{
mystyle/.style={line width=1, baseline, scale=0.6, every node/.style={scale=1}},
v/.style={decorate, draw, decoration={snake, segment length=2.mm, amplitude=0.5mm}},
f/.style={draw, decoration={markings,mark=at position #1 with {\arrow[]{Latex[length=1.5mm,width=1.5mm]}}},
    postaction={decorate},node contents=#1},
f/.default=.6,
fb/.style={draw,decoration={markings,mark=at position #1 with {\arrowreversed[]{Latex[length=1.5mm,width=1.5mm]}}},
    postaction={decorate},node contents=#1},
fb/.default=.6,
s/.style={dashed,draw, decoration={markings,mark=at position #1 with {\arrow[]{Latex[length=1.5mm,width=1.5mm]}}},
    postaction={decorate},node contents=#1},
s/.default=.6,    
sb/.style={dashed,draw,decoration={markings,mark=at position #1 with {\arrowreversed[]{Latex[length=1.5mm,width=1.5mm]}}},
    postaction={decorate},node contents=#1},
sb/.default=.4,
snar/.style={dashed,draw,line width =1.25pt},
cross/.style={cross out, draw=black, minimum size=2*(#1-\pgflinewidth), inner sep=0pt, outer sep=0pt}, 
photon/.style={decorate, draw, decoration={snake, segment length=2mm, amplitude=0.3mm}, line width=1pt},
         }

\newcommand{\calO}{\mathcal{O}}
\newcommand{\C}{ {\tt C} }

\newcommand{\tL}{ {\tt L} }
\newcommand{\tR}{ {\tt R} }
\newcommand{\tN}{ {\tt N} }
\newcommand{\N}{ {\tt N} }
\newcommand{\nn}{ {\nonumber} }

\title{Baryon-number-violating nucleon decays into a dark photon particle}
\author[a,b,c]{Jin-Han Liang\,\orcidlink{0000-0002-6141-216X},}
\emailAdd{jinhanliang@wyu.edu.cn}
\author[a,b]{Yi Liao\,\orcidlink{0000-0002-1009-5483},}
\emailAdd{liaoy@m.scnu.edu.cn}
\author[a,b]{Xiao-Dong Ma\,\orcidlink{0000-0001-7207-7793}}
\emailAdd{maxid@scnu.edu.cn}
\author[a,b]{and Xiang Zhao\,\orcidlink{0009-0008-6024-7722}}
\emailAdd{zhaox@m.scnu.edu.cn}
\affiliation[a]{State Key Laboratory of Nuclear Physics and
Technology, Institute of Quantum Matter, \\
South China Normal
University, \\
Guangzhou 510006, China}
\affiliation[b]{Guangdong Basic Research Center of Excellence for
Structure and Fundamental Interactions of Matter, 
Guangdong Provincial Key Laboratory of Nuclear Science, \\
Guangzhou 510006, China}
\affiliation[c]{School of Applied Physics and Materials, Wuyi University,
Jiangmen 529020, China}
\abstract{
Baryon-number-violating (BNV) nucleon decays into a 
light new particle represent an exciting yet experimentally unexplored frontier.
In this work, we systematically study nucleon decays into a dark photon using a low-energy effective field theory extended with a dark photon $X$, referred to as $X$LEFT. We first construct a complete set of leading-order BNV $X$LEFT operators and then perform a systematic matching onto the chiral perturbation theory for operators involving light $u,d,s$ quarks that dominantly contribute to nucleon decays.
Within the chiral framework, we derive general expressions for the decay widths of both two- and three-body nucleon decays and analyze the momentum distributions in the latter. 
Finally, we thoroughly reinterpret the existing experimental data on conventional two-body modes (into a lepton and a meson) to set lower bounds on partial lifetimes of the corresponding three-body modes involving an additional dark photon.
These bounds allow us to further set stringent constraints on the $X$LEFT operators and other correlated decay modes.
Our results provide a toolkit for future experimental and theoretical studies of these exotic nucleon decays. 
}


\makeatletter
\gdef\@fpheader{}
\makeatother
\begin{document} 

\maketitle
\setcounter{page}{2}

\section{Introduction}

Exotic nucleon decays featuring a light new particle in the final state provide new insights to test baryon number violation and explore new physics beyond the standard model (SM). 
Motivated by ongoing and future large-fiducial-mass neutrino experiments including JUNO~\cite{JUNO:2015zny}, Hyper-Kamiokande~\cite{Hyper-Kamiokande:2018ofw}, DUNE~\cite{DUNE:2020ypp}, and Theia~\cite{Theia:2019non}, these decay modes have recently attracted considerable theoretical interest~\cite{Davoudiasl:2013pda,Davoudiasl:2014gfa,Heeck:2020nbq,Fridell:2023tpb,Domingo:2024qoj,Li:2024liy, Li:2025slp, Fan:2025xhi, Liao:2025vlj,Ma:2025mjy,Heeck:2025uwh,Helo:2025kgx,Adolf:2026khc}.
The underlying new physics driving such decays could shed light on long-standing conundrums in particle physics and cosmology, such as the nature of dark matter~\cite{Fornal:2020poq}, matter-antimatter asymmetry in the Universe~\cite{Davoudiasl:2013pda,Elor:2018twp}, and even the smallness of neutrino masses. 
Moreover, such decays may resolve the neutron lifetime discrepancy observed between beam and bottle experiments via neutron dark decays~\cite{Fornal:2018eol,Liang:2023yta}.

By restricting the light particle's spin up to one, these nucleon decays can be classified into four categories in terms of the new particle: $\tN\to l+\varphi(+M)$ with a spin-0 scalar $\varphi$~\cite{Ma:2025mjy}, $\tN\to l+a(+M)$ with a spin-0 axion-like particle (ALP) $a$~\cite{Li:2024liy,Fan:2025xhi}, $\tN\to N+M$ with a spin-1/2 sterile neutrino $N$~\cite{Davoudiasl:2014gfa,Li:2025slp}, and $\tN\to l+X(+M)$ with a spin-1 dark photon $X$~\cite{Liao:2025vlj,Adolf:2026khc}. Here,
the nucleon $\tN$ denotes either proton $p$ or neutron $n$, $l$ represents a generic SM charged lepton $\ell^\pm =e^\pm, \mu^\pm$ or a (anti)neutrino $\nu(\bar\nu)$, and $M=K,\pi,\eta$ is a generic octet pseudoscalar meson. 
Going beyond a single light particle, nucleon decays involving two or three new light particles were also considered in the literature~\cite{Helo:2025kgx}. 

Since nucleon decays typically occur at an $\calO(\rm GeV)$ energy scale, effective field theory (EFT) emerges as the most convenient and model-independent tool for their systematic study. 
While the leading-order BNV EFT operators and their resulting nucleon decay widths have been well established for scalar and fermionic new particles ($\varphi, a, N$) in recent works~\cite{Ma:2025mjy,Li:2024liy,Fan:2025xhi,Li:2025slp},
a comparable formalism for the dark photon case remains largely unexplored.
In light of both the growing interest in dark photon searches and increasing sensitivity of neutrino experiments, 
developing a comprehensive EFT formalism for nucleon decays into a dark photon has become crucial. 
Such a formalism would not only facilitate dark photon searches through nucleon decays but could also reveal potential  
dark-photon-participated BNV interactions in future studies.
 
In this work, we fill this gap by systematically examining nucleon decays into a dark photon within the low-energy EFT extended with a dark photon $X$, named $X$LEFT. First,  
we construct a complete set of leading-order BNV $X$LEFT operators at dimensions seven and eight, corresponding to the dark photon being parametrized by a four-potential and a field strength tensor, respectively.  
These operators consist of a dark photon, a lepton, and three light quark fields. 
Focusing on operators involving the light $u,d,s$ quarks, which contribute to nucleon decays at leading order, 
we classify them according to irreducible representations (irreps) under the approximate QCD chiral symmetry $G_\chi\equiv\rm\,SU(3)_\tL\otimes SU(3)_\tR$.
To handle the nucleon matrix elements systematically, we employ chiral perturbation theory by matching these $X$LEFT interactions onto the BNV chiral framework developed in~\cite{Claudson:1981gh,Liao:2025vlj}, where quark fields are traded for octet baryons and pseudoscalar mesons.
Within this framework, we derive general expressions for the decay widths of both two- and three-body nucleon decays involving the dark photon. The resulting decay widths are only functions of $X$LEFT Wilson coefficients (WCs) and hadronic low-energy constants (LECs). 
To discriminate between different operator structures in future experimental searches, we further analyze the momentum distributions in the three-body decays. 

Besides the Super-Kamiokande (Super-K) search for the two-body decays $p\to e^+\phi$ and $p\to \mu^+\phi$ with a massless boson $\phi$~\cite{Super-Kamiokande:2015pys},
no dedicated experiments have probed these exotic decay modes.
However, we demonstrate that existing data on conventional two-body decays $\tN \to l M$ can be reanalyzed to constrain the corresponding three-body mode $\tN\to l M X$. 
Given the invisible nature of the dark photon in our consideration, the experimental signatures for both processes are nearly identical after accounting for nuclear effects and detector resolution. Adopting this approach, we reinterpret the Super-K data for $p\to \ell^+K^0$~\cite{Super-Kamiokande:2005lev,Super-Kamiokande:2022egr}, 
$n\to\ell^+\pi^-$~\cite{Super-Kamiokande:2017gev}, 
$n\to\nu(\bar\nu)\pi^0$~\cite{Super-Kamiokande:2025lxa},
and $p\to\nu(\bar\nu) K^+$~\cite{Super-Kamiokande:2014otb} 
to derive partial lifetime (i.e., the inverse decay width) bounds for their corresponding three-body modes with an additional massive $X$. 
By combining these results with existing limits on $p\to \ell^+ X$ from 
Ref.\,~\cite{Ma:2025mjy}, 
which extends the experimental bounds for the massless $\phi$ in~\cite{Super-Kamiokande:2015pys} to the full kinematically allowed range,
we impose stringent constraints on the effective scales of the relevant $X$LEFT operators across a wide range of dark photon masses. 
Finally, leveraging these effective scale bounds, we also establish partial lifetime limits for other correlated decay modes governed by the same $X$LEFT operators. 

The remainder of this paper is organized as follows. 
In \cref{sec:XLEFT}, we construct the leading-order BNV $X$LEFT operators involving a light dark photon below the electroweak scale.
In \cref{sec:nucleon_decay_EFT}, we reorganize the $X$LEFT operators responsible for nucleon decays into chiral irreps and perform chiral matching to derive the corresponding interaction vertices involving the dark photon, octet baryons, and pseudoscalar mesons. 
Section \ref{sec:decay_width} presents our calculations of the decay widths for two- and three-body nucleon decays, along with an analysis of their momentum distributions.  
In \cref{sec:constraint}, we reanalyze the existing Super-K data to establish lower partial lifetime bounds for eight representative three-body decay modes and derive constraints on the associated effective scales and correlated processes.  
Our conclusion is given in \cref{sec:Conclusion}. 
Additionally, \cref{app:Gamma_exp} collects general decay width expressions for two- and three-body nucleon decays involving a massless dark photon.

\section{BNV dark photon interactions in $X$LEFT}
\label{sec:XLEFT}

We work in the low-energy EFT extended with a dark photon field $X$, named $X$LEFT. The relevant SM degrees of freedom include five light quarks $u,d,s,c,b$, three charged leptons $e,\mu,\tau$, and three neutrinos $\nu_e,\nu_\mu,\nu_\tau$. 
To construct the relevant BNV operators involving a dark photon, we consider two different parametrizations of the dark photon field: a vector four-potential $X_\mu$ (case A) and a field strength tensor $X_{\mu\nu}$ (case B). 
Both cases can be realized in UV-complete models in a manner similar to the $W$ and $Z$ bosons in the SM. 
We construct the operator basis using the left- and right-handed chiral fields of the SM fermions in the mass eigenstates. Consequently, operators involving the Levi-Civita tensor $\epsilon_{\mu\nu\rho\sigma}$ are redundant and will be omitted from the subsequent discussion.

{\bf Case A:}
In this case the leading-order BNV interactions appear at dimension 7 and take the form of four general Lorentz-invariant structures
\begin{subequations}
\begin{align}
&X_\mu(\overline{l_\chi}\gamma^\mu q_{\chi_1}^\alpha)(\overline{q_{\chi_2}^{\beta\C}}q_{\chi_3}^\gamma)\epsilon_{\alpha\beta\gamma},
\quad\quad\quad
X_\mu(\overline{l_\chi} q_{\chi_1}^\alpha)(\overline{q_{\chi_2}^{\beta\C}} \gamma^\mu q_{\chi_3}^\gamma)\epsilon_{\alpha\beta\gamma},
\label{eq:Vlqqqvector}\\
& X_\mu(\overline{l_\chi} \gamma_\nu q_{\chi_1}^\alpha)(\overline{q_{\chi_2}^{\beta\C}} \sigma^{\mu\nu} q_{\chi_3}^\gamma)\epsilon_{\alpha\beta\gamma},
\quad~~
X_\mu(\overline{l_\chi}\sigma^{\mu\nu} q_{\chi_1}^\alpha)(\overline{q_{\chi_2}^{\beta\C}} \gamma_\nu q_{\chi_3}^\gamma)\epsilon_{\alpha\beta\gamma},
\label{eq:Vlqqqtensor}
\end{align}
\end{subequations}
where $l_\chi =\ell_{\tL,\tR},\ell_{\tL,\tR}^\C, \nu_\tL,\nu_\tL^\C$, 
$q_{\chi_{i}}$ represents the $i$-th chiral quark field with chirality $\chi_i =\tL{~\rm or~}\tR$, and $\alpha,\beta,\gamma$ are contracted color indices. 
For the two structures involving a tensor current in \cref{eq:Vlqqqtensor},
they can be converted into the first two involving a vector current by using the following Fierz transformations~\cite{Liao:2019gex}
\begin{subequations}
\begin{align}
(\overline{\psi_1}\sigma^{\mu\nu}P_\pm\psi_2)(\overline{\psi_3}\gamma_\nu P_\pm \psi_4)    
=\,& i (\overline{\psi_1} P_\pm\psi_2)(\overline{\psi_3}\gamma^\mu P_\pm \psi_4)   
+ 2 i (\overline{\psi_1} P_\pm \psi_4)(\overline{\psi_3}\gamma^\mu P_\pm\psi_2),  
\\
(\overline{\psi_1}\sigma^{\mu\nu}P_\pm\psi_2)(\overline{\psi_3}\gamma_\nu P_\mp \psi_4)    
=\,& - i (\overline{\psi_1} P_\pm\psi_2)(\overline{\psi_3}\gamma^\mu P_\mp \psi_4)   
-2 i (\overline{\psi_1}\gamma^\mu P_\mp \psi_4)(\overline{\psi_3}P_\pm\psi_2),  
\end{align}
\end{subequations}
where $P_-\equiv P_\tL$ and $P_+\equiv P_\tR$ are chirality projectors. 
Thus, we can parameterize all relevant operators using the first two structures in \cref{eq:Vlqqqvector}. 
Considering all possible field configurations, the final operator basis is summarized in \cref{tab:XLEFTopedim7}. 
It is evident that, without considering fermion flavors, there are 12 operators in the $\Delta(B-L)=0$ sector and 8 operators in the $\Delta(B+L)=0$ sector. 
Restricting to the light $u,d,s$ quarks, there are $36n_{\ell}$ operators corresponding to $n_\ell$ flavors of charged leptons, and $32n_\nu$ operators associated with $n_\nu$ neutrino flavors. 
When the dark photon field $X_\mu$ is replaced by the derivative of an ALP field $a$, i.e., $X_\mu \to \partial_\mu a$, the operator basis coincides exactly with the BNV operator basis in the $a$LEFT framework used in~\cite{Fan:2025xhi}.  

\begin{table}[ht]
\center
\resizebox{0.95\linewidth}{!}{
\renewcommand{\arraystretch}{1.}
\begin{tabular}{|c|c|c|c|c|}
\hline   
\multicolumn{5}{|c|}{Case A}
\\\hline
& Notation & Operator & Chiral Irrep. & 
\# of operators
\\\hline
\multirow{12}*{\rotatebox[origin=c]{90}{
$\Delta(B-L)=0$ }}
&$ \calO_{X\ell\rm uud}^{\tt VL,SL} $  
& $   X_\mu (\overline{\ell_\tR^\C}\gamma^\mu u_\tL^\alpha) (\overline{u_\tL^{\beta\C}}d_\tL^\gamma) \epsilon_{\alpha\beta\gamma}$ 
& ${\color{purple}\pmb{8}_\tL \otimes  \pmb{1}_\tR}$
& $n_\ell n_u^2 n_d \,[2n_\ell]$
\\
& $ \calO_{X\ell\rm uud}^{\tt SL,VR} $
& $   X_\mu  (\overline{\ell_\tL^\C} u_\tL^\alpha) (\overline{u_\tL^{\beta \C}}\gamma^\mu  d_\tR^\gamma) \epsilon_{\alpha\beta\gamma}$
& ${\color{Green}\pmb{6}_\tL \otimes \pmb{3}_\tR}$
& $n_\ell n_u^2 n_d \,[2n_\ell]$
\\
& $\calO_{X\ell\rm udu}^{\tt SL,VR}$
&$~ X_\mu  (\overline{\ell_\tL^\C} u_\tL^\alpha) (\overline{d_\tL^{\beta \C}}\gamma^\mu  u_\tR^\gamma) \epsilon_{\alpha\beta\gamma}~$
& $~{\color{Green}\pmb{6}_\tL \otimes \pmb{3}_\tR} \oplus 
{\color{Purple}\bar{\pmb{3}}_\tL \otimes \pmb{3}_\tR}~$
&  $n_\ell n_u^2 n_d \,[2n_\ell]$
\\
& $\calO_{X\ell\rm duu}^{\tt SL,VR} $
&$  X_\mu (\overline{\ell_\tL^\C} d_\tL^\alpha) (\overline{u_\tL^{\beta \C}}\gamma^\mu  u_\tR^\gamma) \epsilon_{\alpha\beta\gamma}$
& ${\color{Green}\pmb{6}_\tL \otimes \pmb{3}_\tR} \oplus 
{\color{Purple}\bar{\pmb{3}}_\tL \otimes \pmb{3}_\tR}$
& $n_\ell n_u^2 n_d \,[2n_\ell]$
\\
& $\calO_{X\ell\rm uud}^{\tt VR,SR}$
&$  X_\mu (\overline{\ell_\tL^\C} \gamma^\mu u_\tR^\alpha) (\overline{u_\tR^{\beta \C}} d_\tR^\gamma) \epsilon_{\alpha\beta\gamma}$
& ${\color{purple}\pmb{1}_\tL \otimes  \pmb{8}_\tR}$ 
& $n_\ell n_u^2 n_d \,[2n_\ell]$
\\
& $\calO_{X\ell\rm uud}^{\tt SR,VL}$
& $  X_\mu  (\overline{\ell_\tR^\C} u_\tR^\alpha) (\overline{u_\tR^{\beta \C}} \gamma^\mu  d_\tL^\gamma) \epsilon_{\alpha\beta\gamma}$
& ${\color{Green}\pmb{3}_\tL \otimes \pmb{6}_\tR}$
& $n_\ell n_u^2 n_d \,[2n_\ell]$
\\
& $ \calO_{X\ell\rm udu}^{\tt SR,VL}$ 
&$  X_\mu (\overline{\ell_\tR^\C} u_\tR^\alpha) (\overline{d_\tR^{\beta \C}} \gamma^\mu  u_\tL^\gamma) \epsilon_{\alpha\beta\gamma}$
& ${\color{Green}\pmb{3}_\tL \otimes \pmb{6}_\tR} \oplus 
{\color{Purple} \pmb{3}_\tL \otimes \bar{\pmb{3}}_\tR}$
& $n_\ell n_u^2 n_d \,[2n_\ell]$
\\
& $ \calO_{X\ell\rm duu}^{\tt SR,VL} $
& $  X_\mu  (\overline{\ell_\tR^\C} d_\tR^\alpha) (\overline{u_\tR^{\beta \C}} \gamma^\mu  u_\tL^\gamma) \epsilon_{\alpha\beta\gamma}$
& ${\color{Green}\pmb{3}_\tL \otimes \pmb{6}_\tR} \oplus 
{\color{Purple} \pmb{3}_\tL \otimes \bar{\pmb{3}}_\tR}$
& $n_\ell n_u^2 n_d \,[2n_\ell]$
\\\cline{2-5}
&$ \calO_{X \nu\rm ddu}^{\tt VR,SR}$
&$  X_\mu  (\overline{\nu_\tL^\C} \gamma^\mu d_\tR^\alpha) (\overline{d_\tR^{\beta \C}} u_\tR^\gamma) \epsilon_{\alpha\beta\gamma}$
& ${\color{purple}\pmb{1}_\tL \otimes  \pmb{8}_\tR}$ 
& $n_\nu n_u n_d^2 \,[4n_\nu]$
\\
& $\calO_{X \nu\rm ddu}^{\tt SL,VR} $
&$  X_\mu  (\overline{\nu_\tL^\C} d_\tL^\alpha) (\overline{d_\tL^{\beta \C}} \gamma^\mu  u_\tR^\gamma) \epsilon_{\alpha\beta\gamma}$
& ${\color{Green}\pmb{6}_\tL \otimes \pmb{3}_\tR} \oplus 
{\color{Purple}\bar{\pmb{3}}_\tL \otimes \pmb{3}_\tR}$
& $n_\nu n_u n_d^2 \,[4n_\nu]$
\\
& $\calO_{X \nu\rm dud}^{\tt SL,VR} $
& $  X_\mu  (\overline{\nu_\tL^\C} d_\tL^\alpha) (\overline{u_\tL^{\beta \C}} \gamma^\mu  d_\tR^\gamma) \epsilon_{\alpha\beta\gamma}$
& ${\color{Green}\pmb{6}_\tL \otimes \pmb{3}_\tR} \oplus 
{\color{Purple}\bar{\pmb{3}}_\tL \otimes \pmb{3}_\tR}$
& $n_\nu n_u n_d^2 \,[4n_\nu]$
\\
& $ \calO_{X \nu\rm udd}^{\tt SL,VR} $
&$  X_\mu  (\overline{\nu_\tL^\C} u_\tL^\alpha) (\overline{d_\tL^{\beta \C}} \gamma^\mu  d_\tR^\gamma) \epsilon_{\alpha\beta\gamma}$
& ${\color{Green}\pmb{6}_\tL \otimes \pmb{3}_\tR} \oplus 
{\color{Purple}\bar{\pmb{3}}_\tL \otimes \pmb{3}_\tR}$
& $n_\nu n_u n_d^2 \,[4n_\nu]$
\\\hline
& $\calO_{X \bar\nu\rm ddu}^{\tt VL,SL} $ 
& $  X_\mu  (\overline{\nu_\tL} \gamma^\mu d_\tL^\alpha) (\overline{d_\tL^{\beta \C}} u_\tL^\gamma) \epsilon_{\alpha\beta\gamma}$
& ${\color{purple}\pmb{8}_\tL \otimes  \pmb{1}_\tR}$
& $n_\nu n_u n_d^2 \,[4n_\nu]$
\\
& $\calO_{X \bar\nu\rm ddu}^{\tt SR,VL} $
&$  X_\mu  (\overline{\nu_\tL} d_\tR^\alpha) (\overline{d_\tR^{\beta \C}} \gamma^\mu  u_\tL^\gamma) \epsilon_{\alpha\beta\gamma}$
& ${\color{Green}\pmb{3}_\tL \otimes \pmb{6}_\tR} \oplus 
{\color{Purple} \pmb{3}_\tL \otimes \bar{\pmb{3}}_\tR}$
& $n_\nu n_u n_d^2 \,[4n_\nu]$
\\
&$\calO_{X \bar\nu\rm dud}^{\tt SR,VL} $
&$ X_\mu  (\overline{\nu_\tL} d_\tR^\alpha) (\overline{u_\tR^{\beta \C}} \gamma^\mu  d_\tL^\gamma) \epsilon_{\alpha\beta\gamma}$
& ${\color{Green}\pmb{3}_\tL \otimes \pmb{6}_\tR} \oplus 
{\color{Purple} \pmb{3}_\tL \otimes \bar{\pmb{3}}_\tR}$
& $n_\nu n_u n_d^2 \,[4n_\nu]$
\\
&$\calO_{X \bar\nu\rm udd}^{\tt SR,VL} $
& $ X_\mu  (\overline{\nu_\tL} u_\tR^\alpha) (\overline{d_\tR^{\beta \C}} \gamma^\mu  d_\tL^\gamma) \epsilon_{\alpha\beta\gamma}$
& ${\color{Green}\pmb{3}_\tL \otimes \pmb{6}_\tR} \oplus 
{\color{Purple} \pmb{3}_\tL \otimes \bar{\pmb{3}}_\tR}$
& $n_\nu n_u n_d^2 \,[4n_\nu]$
\\\cline{2-5}
& $\calO_{X\bar\ell\rm ddd}^{\tt VR,SR} $
& $ X_\mu (\overline{\ell_\tR} \gamma^\mu d_\tR^\alpha) (\overline{d_\tR^{\beta \C}} d_\tR^\gamma) \epsilon_{\alpha\beta\gamma}$
& ${\color{purple}\pmb{1}_\tL \otimes  \pmb{8}_\tR}$ 
& ${1\over 3} n_\ell n_d(n_d^2-1) \,[2n_\ell]$
\\
&$ \calO_{X\bar\ell\rm ddd}^{\tt SR,VL} $
&$ X_\mu (\overline{\ell_\tL} d_\tR^\alpha) (\overline{d_\tR^{\beta \C}} \gamma^\mu  d_\tL^\gamma) \epsilon_{\alpha\beta\gamma}$
& ${\color{Green}\pmb{3}_\tL \otimes \pmb{6}_\tR} \oplus 
{\color{Purple} \pmb{3}_\tL \otimes \bar{\pmb{3}}_\tR}$
& $n_\ell n_d^3 \,[8n_\ell]$
\\
& $\calO_{X\bar\ell\rm ddd}^{\tt VL,SL}$
&$ X_\mu (\overline{\ell_\tL} \gamma^\mu d_\tL^\alpha) (\overline{d_\tL^{\beta \C}} d_\tL^\gamma) \epsilon_{\alpha\beta\gamma}$
& ${\color{purple}\pmb{8}_\tL \otimes  \pmb{1}_\tR}$ 
&~~~~${1\over 3} n_\ell n_d(n_d^2-1)\,[2n_\ell]$~~~~
\\
~\multirow{-8}*{\rotatebox[origin=c]{90}{
$\Delta(B+L)=0$ } }~ 
&~~~~$ \calO_{X\bar\ell\rm ddd}^{\tt SL,VR} $~~~~ 
&~~~~$ X_\mu (\overline{\ell_\tR} d_\tL^\alpha) (\overline{d_\tL^{\beta \C}} \gamma^\mu  d_\tR^\gamma) \epsilon_{\alpha\beta\gamma}$~~~~
& ~~~~${\color{Green}\pmb{6}_\tL \otimes \pmb{3}_\tR} \oplus 
{\color{Purple}\bar{\pmb{3}}_\tL \otimes \pmb{3}_\tR}$~~~~
&  $n_\ell n_d^3 \,[8n_\ell]$
\\\hline
\end{tabular}}
\caption{The dimension-7 (dim-7) $X$LEFT BNV operators involving a dark photon, parametrized in terms of its field potential. $\alpha,\beta,\gamma$ are color indices while the flavor indices are omitted for simplicity.  
The fourth column indicates chiral irreducible representations of the operators involving the light $u,d,s$ quarks.
The last column counts the number of independent operators with general $n_\ell$ charged leptons, $n_\nu$ neutrinos, and $n_u$ up-type and $n_d$ down-type quarks; the number in square bracket represents the case with only $u,d,s$ quarks ($n_u=1$ and $n_d=2$). }
\label{tab:XLEFTopedim7}
\end{table}

\begin{table}[t]
\center
\resizebox{0.95\linewidth}{!}{
\renewcommand{\arraystretch}{1.}
\begin{tabular}{|c|c|c|c|c|}
\hline  
\multicolumn{5}{|c|}{Case B}
\\\hline
& ~Notation~ & Operator &~Chiral Irrep.~ & 
\# of operators 
\\\hline
\multirow{15}*{\rotatebox[origin=c]{90}{
$\Delta(B-L)=0$ }}
&$ \tilde\calO_{X\ell\rm uud}^{\tt TL,SL} $  
& $ X_{\mu\nu} (\overline{\ell_\tL^\C}\sigma^{\mu\nu} u_\tL^\alpha) (\overline{u_\tL^{\beta\C}}d_\tL^\gamma) \epsilon_{\alpha\beta\gamma}$ 
& ${\color{purple}\pmb{8}_\tL \otimes  \pmb{1}_\tR}$ 
& $n_\ell n_u^2 n_d \,[2n_\ell]$
\\
& $ \tilde\calO_{X\ell\rm duu}^{\tt SL,TL} $
& $ X_{\mu\nu} (\overline{\ell_\tL^\C}\{d_\tL^\alpha) (\overline{u_\tL^{\beta\C}}\sigma^{\mu\nu} u_\tL^\gamma\}) \epsilon_{\alpha\beta\gamma}$ 
& ${\color{Blue}\pmb{10}_\tL \otimes \pmb{1}_\tR}$
& $\frac{1}{2}n_\ell n_u(n_u+1) n_d \,[2n_\ell]$
\\
& $ \tilde\calO_{X\ell\rm duu}^{\tt VR,VL} $
&$ X_{\mu\nu}  (\overline{\ell_\tL^\C}\gamma^\mu d_\tR^\alpha) (\overline{u_\tR^{\beta \C}}\gamma^\nu  u_\tL^\gamma) \epsilon_{\alpha\beta\gamma}$ 
& ${\color{Green}\pmb{3}_\tL \otimes \pmb{6}_\tR}\oplus {\color{Purple}\pmb{3}_\tL \otimes \bar{\pmb{3}}_\tR}$
& $n_\ell n_u^2 n_d \,[2n_\ell]$
\\
& $\tilde\calO_{X\ell\rm udu}^{\tt VR,VL} $
&$ X_{\mu\nu}  (\overline{\ell_\tL^\C}\gamma^\mu u_\tR^\alpha) (\overline{d_\tR^{\beta \C}}\gamma^\nu  u_\tL^\gamma) \epsilon_{\alpha\beta\gamma}$
& ${\color{Green}\pmb{3}_\tL \otimes \pmb{6}_\tR}\oplus {\color{Purple}\pmb{3}_\tL \otimes \bar{\pmb{3}}_\tR}$
& $n_\ell n_u^2 n_d \,[2n_\ell]$
\\
& $\tilde\calO_{X\ell\rm uud}^{\tt VR,VL} $
&$ X_{\mu\nu}  (\overline{\ell_\tL^\C}\gamma^\mu u_\tR^\alpha) (\overline{u_\tR^{\beta \C}}\gamma^\nu  d_\tL^\gamma) \epsilon_{\alpha\beta\gamma}$
& ${\color{Green}\pmb{3}_\tL \otimes \pmb{6}_\tR}$
& $n_\ell n_u^2 n_d \,[2n_\ell]$
\\
&$ \tilde\calO_{X\ell\rm uud}^{\tt TR,SR} $  
& $ X_{\mu\nu} (\overline{\ell_\tR^\C}\sigma^{\mu\nu} u_\tR^\alpha) (\overline{u_\tR^{\beta\C}}d_\tR^\gamma) \epsilon_{\alpha\beta\gamma}$ 
& ${\color{purple}\pmb{1}_\tL \otimes  \pmb{8}_\tR}$ 
& $n_\ell n_u^2 n_d \,[2n_\ell]$
\\
& $ \tilde\calO_{X\ell\rm duu}^{\tt SR,TR} $
& $ X_{\mu\nu} (\overline{\ell_\tR^\C}\{d_\tR^\alpha) (\overline{u_\tR^{\beta\C}}\sigma^{\mu\nu} u_\tR^\gamma\}) \epsilon_{\alpha\beta\gamma}$ 
& ${\color{Blue}\pmb{1}_\tL \otimes \pmb{10}_\tR}$
& $\frac{1}{2}n_\ell n_u(n_u+1) n_d \,[2n_\ell]$
\\
& $ \tilde\calO_{X\ell\rm duu}^{\tt VL,VR} $
&$ X_{\mu\nu}  (\overline{\ell_\tR^\C}\gamma^\mu d_\tL^\alpha) (\overline{u_\tL^{\beta \C}}\gamma^\nu  u_\tR^\gamma) \epsilon_{\alpha\beta\gamma}$
& ${\color{Green}\pmb{6}_\tL \otimes \pmb{3}_\tR}\oplus {\color{Purple}\bar{\pmb{3}}_\tL \otimes \pmb{3}_\tR}$
& $n_\ell n_u^2 n_d \,[2n_\ell]$
\\
& $\tilde\calO_{X\ell\rm udu}^{\tt VL,VR} $
&$ X_{\mu\nu}  (\overline{\ell_\tR^\C}\gamma^\mu u_\tL^\alpha) (\overline{d_\tL^{\beta \C}}\gamma^\nu  u_\tR^\gamma) \epsilon_{\alpha\beta\gamma}$
& ${\color{Green}\pmb{6}_\tL \otimes \pmb{3}_\tR}\oplus {\color{Purple}\bar{\pmb{3}}_\tL \otimes \pmb{3}_\tR}$
& $n_\ell n_u^2 n_d \,[2n_\ell]$
\\
& $\tilde\calO_{X\ell\rm uud}^{\tt VL,VR} $
&$ X_{\mu\nu}  (\overline{\ell_\tR^\C}\gamma^\mu u_\tL^\alpha) (\overline{u_\tL^{\beta \C}}\gamma^\nu  d_\tR^\gamma) \epsilon_{\alpha\beta\gamma}$
& ${\color{Green}\pmb{6}_\tL \otimes \pmb{3}_\tR}$
& $n_\ell n_u^2 n_d \,[2n_\ell]$
\\\cline{2-5}
&$ \tilde\calO_{X \nu\rm ddu}^{\tt TL,SL} $  
& $ X_{\mu\nu} (\overline{\nu_\tL^\C}\sigma^{\mu\nu} d_\tL^\alpha) (\overline{d_\tL^{\beta\C}}u_\tL^\gamma) \epsilon_{\alpha\beta\gamma}$ 
& ${\color{purple}\pmb{8}_\tL \otimes  \pmb{1}_\tR}$ 
& $n_\nu n_u n_d^2 \,[4n_\nu]$
\\
& $ \tilde\calO_{X \nu\rm udd}^{\tt SL,TL} $
& $ X_{\mu\nu} (\overline{\nu_\tL^\C}\{u_\tL^\alpha) (\overline{d_\tL^{\beta\C}}\sigma^{\mu\nu} d_\tL^\gamma\}) \epsilon_{\alpha\beta\gamma}$ 
& ${\color{Blue}\pmb{10}_\tL \otimes \pmb{1}_\tR}$
& $\frac{1}{2}n_\nu n_u n_d(n_d+1) \,[3n_\nu]$
\\
& $ \tilde\calO_{X \nu\rm dud}^{\tt VR,VL} $
&$ X_{\mu\nu}  (\overline{\nu_\tL^\C}\gamma^\mu d_\tR^\alpha) (\overline{u_\tR^{\beta \C}}\gamma^\nu  d_\tL^\gamma) \epsilon_{\alpha\beta\gamma}$
& ${\color{Green}\pmb{3}_\tL \otimes \pmb{6}_\tR}\oplus {\color{Purple}\pmb{3}_\tL \otimes \bar{\pmb{3}}_\tR}$
& $n_\nu n_u n_d^2 \,[4n_\nu]$
\\
& $\tilde\calO_{X \nu\rm udd}^{\tt VR,VL} $
&$ X_{\mu\nu}  (\overline{\nu_\tL^\C}\gamma^\mu u_\tR^\alpha) (\overline{d_\tR^{\beta \C}}\gamma^\nu  d_\tL^\gamma) \epsilon_{\alpha\beta\gamma}$
& ${\color{Green}\pmb{3}_\tL \otimes \pmb{6}_\tR}\oplus {\color{Purple}\pmb{3}_\tL \otimes \bar{\pmb{3}}_\tR}$
& $n_\nu n_u n_d^2 \,[4n_\nu]$
\\
& $\tilde\calO_{X \nu\rm ddu}^{\tt VR,VL} $
&$ X_{\mu\nu}  (\overline{\nu_\tL^\C}\gamma^\mu d_\tR^\alpha) (\overline{d_\tR^{\beta \C}}\gamma^\nu  u_\tL^\gamma) \epsilon_{\alpha\beta\gamma}$
& ${\color{Green}\pmb{3}_\tL \otimes \pmb{6}_\tR}\oplus {\color{Purple}\pmb{3}_\tL \otimes \bar{\pmb{3}}_\tR}$
& $n_\nu n_u n_d^2\,[4n_\nu]$
\\\hline
&$ \tilde\calO_{X \bar\nu\rm ddu}^{\tt TR,SR} $  
& $ X_{\mu\nu} (\overline{\nu_\tL}\sigma^{\mu\nu} d_\tR^\alpha) (\overline{d_\tR^{\beta\C}}u_\tR^\gamma) \epsilon_{\alpha\beta\gamma}$ 
& ${\color{purple}\pmb{1}_\tL \otimes  \pmb{8}_\tR}$ 
& $n_\nu n_u n_d^2 \,[4n_\nu]$
\\
& $ \tilde\calO_{X \bar\nu\rm udd}^{\tt SR,TR} $
& $ X_{\mu\nu} (\overline{\nu_\tL}\{u_\tR^\alpha) (\overline{d_\tR^{\beta\C}}\sigma^{\mu\nu} d_\tR^\gamma\}) \epsilon_{\alpha\beta\gamma}$ 
& ${\color{Blue}\pmb{1}_\tL \otimes \pmb{10}_\tR}$
& $\frac{1}{2}n_\nu n_u n_d(n_d+1) \,[3n_\nu]$
\\
& $ \tilde\calO_{X \bar\nu\rm dud}^{\tt VL,VR} $
&$ X_{\mu\nu}  (\overline{\nu_\tL}\gamma^\mu d_\tL^\alpha) (\overline{u_\tL^{\beta \C}}\gamma^\nu  d_\tR^\gamma) \epsilon_{\alpha\beta\gamma}$ 
& ${\color{Green}\pmb{6}_\tL \otimes \pmb{3}_\tR}\oplus {\color{Purple}\bar{\pmb{3}}_\tL \otimes \pmb{3}_\tR}$
& $n_\nu n_u n_d^2 \,[4n_\nu]$
\\
& $\tilde\calO_{X \bar\nu\rm udd}^{\tt VL,VR} $
&$ X_{\mu\nu}  (\overline{\nu_\tL}\gamma^\mu u_\tL^\alpha) (\overline{d_\tL^{\beta \C}}\gamma^\nu  d_\tR^\gamma) \epsilon_{\alpha\beta\gamma}$
& ${\color{Green}\pmb{6}_\tL \otimes \pmb{3}_\tR}\oplus {\color{Purple}\bar{\pmb{3}}_\tL \otimes \pmb{3}_\tR}$
& $n_\nu n_u n_d^2 \,[4n_\nu]$
\\
& $\tilde\calO_{X \bar\nu\rm ddu}^{\tt VL,VR} $
&$ X_{\mu\nu}  (\overline{\nu_\tL}\gamma^\mu d_\tL^\alpha) (\overline{d_\tL^{\beta \C}}\gamma^\nu  u_\tR^\gamma) \epsilon_{\alpha\beta\gamma}$
& ${\color{Green}\pmb{6}_\tL \otimes \pmb{3}_\tR}\oplus {\color{Purple}\bar{\pmb{3}}_\tL \otimes \pmb{3}_\tR}$
& $n_\nu n_u n_d^2\,[4n_\nu]$
\\\cline{2-5}
&$ \tilde\calO_{X\bar\ell\rm ddd}^{\tt TL,SL} $  
& $ X_{\mu\nu} (\overline{\ell_\tR}\sigma^{\mu\nu} d_\tL^\alpha) (\overline{d_\tL^{\beta\C}}d_\tL^\gamma) \epsilon_{\alpha\beta\gamma}$ 
& ${\color{purple}\pmb{8}_\tL \otimes  \pmb{1}_\tR}$ 
& $\frac{1}{3}n_\ell n_d(n_d^2-1) \,[2n_\ell]$
\\
& $ \tilde\calO_{X\bar\ell\rm ddd}^{\tt SL,TL} $
& $ X_{\mu\nu} (\overline{\ell_\tR}\{d_\tL^\alpha) (\overline{d_\tL^{\beta\C}}\sigma^{\mu\nu} d_\tL^\gamma\}) \epsilon_{\alpha\beta\gamma}$ 
& ${\color{Blue}\pmb{10}_\tL \otimes \pmb{1}_\tR}$
& $\frac{1}{6}n_\ell n_d(n_d^2+3 n_d +2) \,[4n_\ell]$
\\
& $\tilde\calO_{X\bar\ell\rm ddd}^{\tt VR,VL} $
&$ X_{\mu\nu}  (\overline{\ell_\tR}\gamma^\mu d_\tR^\alpha) (\overline{d_\tR^{\beta \C}}\gamma^\nu  d_\tL^\gamma) \epsilon_{\alpha\beta\gamma}$
& ${\color{Green}\pmb{3}_\tL \otimes \pmb{6}_\tR}\oplus {\color{Purple}\pmb{3}_\tL \otimes \bar{\pmb{3}}_\tR}$
& $ n_\ell n_d^3 \,[8n_\ell]$
\\
&$ \tilde\calO_{X\bar\ell\rm ddd}^{\tt TR,SR} $  
& $ X_{\mu\nu} (\overline{\ell_\tL}\sigma^{\mu\nu} d_\tR^\alpha) (\overline{d_\tR^{\beta\C}}d_\tR^\gamma) \epsilon_{\alpha\beta\gamma}$ 
& ${\color{purple}\pmb{1}_\tL \otimes  \pmb{8}_\tR}$ 
& $\frac{1}{3}n_\ell n_d(n_d^2-1) \,[2n_\ell]$
\\
& $ \tilde\calO_{X\bar\ell\rm ddd}^{\tt SR,TR} $
& $ X_{\mu\nu} (\overline{\ell_\tL}\{d_\tR^\alpha) (\overline{d_\tR^{\beta\C}}\sigma^{\mu\nu} d_\tR^\gamma\}) \epsilon_{\alpha\beta\gamma}$ 
& ${\color{Blue}\pmb{1}_\tL \otimes \pmb{10}_\tR}$
&~~~~$\frac{1}{6}n_\ell n_d(n_d^2+3 n_d +2) \,[4n_\ell]$~~~~
\\
~\multirow{-11}*{\rotatebox[origin=c]{90}{$\Delta(B+L)=0$ } }~ 
&~~~~$\tilde\calO_{X\bar\ell\rm ddd}^{\tt VL,VR} $~~~
&~~~~$X_{\mu\nu}  (\overline{\ell_\tL}\gamma^\mu d_\tL^\alpha) (\overline{d_\tL^{\beta \C}}\gamma^\nu  d_\tR^\gamma) \epsilon_{\alpha\beta\gamma}$~~~~
&~~~~${\color{Green}\pmb{6}_\tL \otimes \pmb{3}_\tR}\oplus {\color{Purple}\bar{\pmb{3}}_\tL \otimes \pmb{3}_\tR}$~~~~
& $n_\ell n_d^3 \,[8n_\ell]$
\\\hline
\end{tabular} }
\caption{The dim-8 $X$LEFT BNV operators involving a dark photon, parametrized by its field strength tensor. 
For the operators involving a tensor quark current, a totally-symmetric operation in the three like-chirality quark fields is performed, as indicated by the curly brackets $\{...\}$.}
\label{tab:XLEFTopedim8}
\end{table}

{\bf Case B:}
Next, we examine the operator basis involving a dark photon field strength tensor $X_{\mu\nu}$. 
The leading-order BNV operators arise at dimension 8 and there are also four general structures 
\begin{align}
& X_{\mu\nu} (\overline{l_\chi} \sigma^{\mu\nu} q_{\chi_1}^\alpha)(\overline{q_{\chi_2}^{\beta\C}} q_{\chi_3}^\gamma)\epsilon_{\alpha\beta\gamma},~
X_{\mu\nu} (\overline{l_\chi} q_{\chi_1}^\alpha)(\overline{q_{\chi_2}^{\beta\C}} \sigma^{\mu\nu} q_{\chi_3}^\gamma)\epsilon_{\alpha\beta\gamma},
\nn
\\
&X_{\mu\nu} (\overline{l_\chi}\gamma^\mu q_{\chi_1}^\alpha)(\overline{q_{\chi_2}^{\beta\C}}\gamma^\nu q_{\chi_3}^\gamma)\epsilon_{\alpha\beta\gamma},~ X_{\mu\nu} (\overline{l_\chi}\sigma^{\mu\rho}q_{\chi_1}^\alpha)(\overline{q_{\chi_2}^{\beta\C}} \sigma^{\nu}_{~\rho} q_{\chi_3}^\gamma)\epsilon_{\alpha\beta\gamma}\,(\times).
\label{eq:Xmunulqqq}
\end{align}
Because we work with chiral fermion fields, operators involving the dual field strength tensor, $\tilde X_{\mu\nu} = (1/2)\epsilon_{\mu\nu\rho\sigma}X^{\rho\sigma}$, are redundant and thus omitted.
They can be straightforwardly transformed into the forms shown in \cref{eq:Xmunulqqq} using the Fierz identities provided in Appendix A of~\cite{Liao:2019gex}.
Furthermore, the double-tensor structure with a '$\times$' in \cref{eq:Xmunulqqq} is also redundant and can be transformed into the other three  by applying the Fierz identities
\begin{align}
X_{\mu\nu} (\overline{\psi_1}\sigma^{\mu\rho}P_\pm \psi_2)(\overline{\psi_3}\sigma^{\nu}_{~\rho} P_\pm \psi_4) = &
i X_{\mu\nu} [ 
(\overline{\psi_1}\sigma^{\mu\nu} P_\pm \psi_4)
(\overline{\psi_3} P_\pm \psi_2)
-  (\overline{\psi_1} P_\pm\psi_4)
(\overline{\psi_3}\sigma^{\mu\nu} P_\pm \psi_2) ],
\nn
\\
X_{\mu\nu} (\overline{\psi_1} \sigma^{\mu\rho}P_\pm\psi_2)(\overline{\psi_3}\sigma^{\nu}_{~\rho} P_\mp \psi_4) = & 0.
\end{align} 
For the remaining three structures, we can further eliminate some of them in favor of others after considering the specific chirality configurations of the three quarks.
Evidently, there are only two chirality configurations: $q_{\tL} q_{\tL} q_{\tL}$
and $q_{\tL}q_{\tL} q_{\tR}$, plus their chiral partners with $\tL\leftrightarrow\tR$.
For operators with $q_{\tL} q_{\tL} q_{\tL}$ or $q_{\tR} q_{\tR} q_{\tR}$, the third structure involving double vector currents in \cref{eq:Xmunulqqq} vanishes due to chirality mismatch. 
For operators with $q_{\tL} q_{\tL} q_{\tR}$ or $q_{\tR} q_{\tR} q_{\tL}$, the first two structures involving a tensor current can be transformed into the third structure by using the following Fierz identities
\begin{align}
(\overline{\psi_1} \gamma^\mu \gamma^\nu P_\pm \psi_2) 
(\overline{\psi_3^\C} P_\mp \psi_4) =\,& 
-(\overline{\psi_1} \gamma^\mu P_\mp \psi_3) 
(\overline{\psi_2^\C}\gamma^\nu P_\mp \psi_4)
-(\overline{\psi_1} \gamma^\mu P_\mp \psi_4) 
(\overline{\psi_3^\C} \gamma^\nu P_\pm \psi_2),
\nn
\\
(\overline{\psi_1} P_\pm \psi_2) 
(\overline{\psi_3^\C}\gamma^\mu \gamma^\nu P_\mp \psi_4) =\,& 
-(\overline{\psi_1} \gamma^\mu P_\mp \psi_3) 
(\overline{\psi_2^\C}\gamma^\nu P_\mp \psi_4)
-(\overline{\psi_1} \gamma^\nu P_\mp \psi_4) 
(\overline{\psi_3^\C} \gamma^\mu P_\pm \psi_2).
\end{align}   

In summary, the final chosen independent structures are
\begin{align}
X_{\mu\nu} (\overline{l_\chi} \sigma^{\mu\nu} q_{\tL}^\alpha)(\overline{q_{\tL}^{\beta\C}} q_{\tL}^\gamma)\epsilon_{\alpha\beta\gamma},~
X_{\mu\nu} (\overline{l_\chi} q_{\tL}^\alpha)(\overline{q_{\tL}^{\beta\C}} \sigma^{\mu\nu} q_{\tL}^\gamma)\epsilon_{\alpha\beta\gamma};~
X_{\mu\nu} (\overline{l_\chi}\gamma^\mu q_{\tL}^\alpha)(\overline{q_{\tL}^{\beta\C}}\gamma^\nu q_{\tR}^\gamma)\epsilon_{\alpha\beta\gamma},
\end{align}
and their chirality partners with $\tL\leftrightarrow \tR$.
Taking into account the relevant field configurations allowed by $\rm SU(3)_c\times U(1)_{em}$ symmetries, the final basis is presented in \cref{tab:XLEFTopedim8}. 
As observed from the table, there are 15 flavor-blind operators in the $\Delta(B-L)=0$ sector and 11 operators in the $\Delta(B+L)=0$ sector.
Focusing on operators with the $u,d,s$ quarks, there are 
$48 n_{\ell}$ operators involving $n_\ell$ charged leptons and $38 n_{\nu}$ operators involving $n_\nu$ neutrinos.
Note that the operators containing a tensor quark current are constructed to be fully symmetric under the exchange of any two quarks. This is because any antisymmetric combination is redundant and can be Fierz-transformed into operators with a scalar quark current. For instance,
\begin{align}
&X_{\mu\nu}(\overline{l_\tR} 
q_{1\tL}^\alpha)(\overline{q_{2\tL}^{\beta\C} } \sigma^{\mu\nu} q_{3\tL}^\gamma) \epsilon_{\alpha \beta \gamma}
-q_{1\tL}\leftrightarrow q_{2\tL}
\nn\\ 
=\,&X_{\mu\nu}(\overline{l_\tR}\sigma^{\mu\nu}
q_{3\tL}^\alpha)(\overline{q_{2\tL}^{\beta\C} }  q_{1\tL}^\gamma) \epsilon_{\alpha \beta \gamma}
\nn\\
=\,&X_{\mu\nu}(\overline{l_\tR}\sigma^{\mu\nu}
q_{2\tL}^\alpha)(\overline{q_{3\tL}^{\beta\C} }  q_{1\tL}^\gamma)\epsilon_{\alpha \beta \gamma}
+X_{\mu\nu}(\overline{l_\tR}\sigma^{\mu\nu}
q_{1\tL}^\alpha)(\overline{q_{2\tL}^{\beta\C} }  q_{3\tL}^\gamma) \epsilon_{\alpha \beta \gamma}.
\end{align}

Denoting the WC of each $X$LEFT operator $\calO_i^j$ by $C_i^j$, the BNV $X$LEFT Lagrangian can be compactly written as follows, 
\begin{subequations}
\label{eq:XLEFTLag}
\begin{align}
{\cal L}_{q^3}^{\slashed{B},\rm (A)} =\,& 
\sum_{l,y,z,w}\big( 
C_{Xlyzw}^{{\tt VL,SL},x} \calO_{Xlyzw}^{{\tt VL,SL},x}
+ C_{Xlyzw}^{{\tt SL,VR},x} \calO_{Xlyzw}^{{\tt SL,VR},x}\big)
+\tL\leftrightarrow\tR,
\\
{\cal L}_{q^3}^{\slashed{B},\rm (B)} =\,& 
\sum_{l,y,z,w}\big( 
\tilde C_{Xlyzw}^{{\tt TL,SL},x} \tilde\calO_{Xlyzw}^{{\tt TL,SL},x}
+\tilde C_{Xlyzw}^{{\tt VR,VL},x} \tilde \calO_{Xlyzw}^{{\tt VR,VL},x}
+\tilde C_{Xlyzw}^{{\tt SL,TL},x} \tilde \calO_{Xlyzw}^{{\tt SL,TL},x}\big)
+\tL\leftrightarrow\tR,
\end{align}
\end{subequations}
where $x\in\{e,\mu,\tau\}$ denotes the lepton flavor and $y,w,z\in \{u,d,s,c,b\}$ represent the quark flavors. The summation is performed over all allowed field configurations specified in \cref{tab:XLEFTopedim7,tab:XLEFTopedim8}.
When operators exhibit flavor symmetries,
we retain only one flavor-specific representative operator in each case. 
For example, the operators $\calO_{X\bar\ell dds}^{\tt VR,SR}$ and $\calO_{X\bar\ell dsd}^{\tt VR,SR}$ are related by $\calO_{X\bar\ell dsd}^{\tt VR,SR}=-\calO_{X\bar\ell dds}^{\tt VR,SR}$,
consequently, we include only $\calO_{X\bar\ell dds}^{\tt VR,SR}$ in the Lagrangian.

The BNV $X$LEFT bases given in \cref{tab:XLEFTopedim7,tab:XLEFTopedim8} are invariant under $\rm SU(3)_c\times U(1)_{em}$ but not the full SM gauge symmetry $G_{\rm SM}=\rm SU(3)_c\times SU(2)_L\times U(1)_Y$. 
Above the electroweak scale, the appropriate EFT is the SMEFT extended to include the dark photon, referred to as $X$SMEFT. 
For the BNV $X$LEFT operators, their embedding into $X$SMEFT can be straightforwardly achieved by promoting the left-handed leptons and quarks to $\rm SU(2)_L$ doublets ($L$ and $Q$) and incorporating the Higgs doublet $H$ to ensure invariance under $\rm SU(2)_L\times U(1)_Y$.
For the case-A operators listed in \cref{tab:XLEFTopedim7}, the corresponding $X$SMEFT operators arise at dimension 8 [7] or at higher even [odd] dimensions in the $\Delta(B-L)=0$ [$\Delta(B+L)=0$] sector.
At dimension 7, there are two $\Delta(B+L)=0$ $X$SMEFT operators, $\epsilon_{\alpha\beta\gamma}X_\mu(\overline{L}d^\alpha)(\overline{d^{\beta\tt C}} \gamma^\mu Q^\gamma)$
and $\epsilon_{\alpha\beta\gamma}X_\mu(\overline{e}\gamma^\mu d^\alpha)(\overline{d^{\beta\tt C}} d^\gamma)$, where $L$ and $Q$ ($e$ and $d$) denote the SM lepton and quark doublets (singlets), respectively.
Below the electroweak scale, these operators match onto the three $X$LEFT operators: 
${\cal O}_{X\bar\nu\rm ddu}^{\tt SR,VL}$,
${\cal O}_{X\bar\ell\rm ddd}^{\tt SR,VL}$,
and ${\cal O}_{X\bar\ell\rm ddd}^{\tt VR,SR}$. 
The complete dim-8 $\Delta(B-L)=0$ $X$SMEFT basis is given in~\cite{Song:2023jqm};
see also Ref.~\cite{Adolf:2026khc} for explicit operator enumerations and possible UV-complete realizations.

Regarding the case-B operators in \cref{tab:XLEFTopedim8}, the corresponding $X$SMEFT operators arise at dimension 8 [9] and at higher even (odd) dimensions in the $\Delta(B-L)=0$ [$\Delta(B+L)=0$] sector. The leading dim-8 (dim-9) case-B $X$SMEFT operators can be obtained directly from the corresponding dim-8 (dim-9) BNV SMEFT operators in the $lq^3B$ ($\bar l q^3HB$) classes by replacing the hypercharge field-strength tensor $B_{\mu\nu}$ with $X_{\mu\nu}$.
For the relevant SMEFT operator classes,
see Table 15 of~\cite{Murphy:2020rsh} for dim-8 $lq^3B$ class and table 7 of~\cite{Liao:2020jmn} for dim-9 $\bar l q^3HB$ class.

It it worth noting that the dim-8 and dim-9 case-B $X$SMEFT operators have the same fermionic field contents
as the dim-6 and dim-7 BNV SMEFT operators, respectively. 
Any underlying new-physics scenario generating these $X$SMEFT operators would generically also induce the corresponding lower-dimensional SMEFT operators, in the absence of a symmetry forbidding them.
Consequently, the lower-dimensional contributions to conventional nucleon decays would typically dominate, making the exotic nucleon decays involving a dark photon, induced by these case-B operators, phenomenologically less interesting.

\section{Nucleon decays involving a dark photon in EFTs}
\label{sec:nucleon_decay_EFT}

\subsection{$X$LEFT operators contributing to nucleon decays}
\label{sec:nucleon_decay_LEFT}

The $X$LEFT operators that contribute to nucleon decays at leading order are those involving light $u,d,s$ quarks. 
Operators containing the $\tau$ lepton, charm quark, or bottom quark require additional SM weak interactions to mediate nucleon decays. Since their contributions are strongly suppressed, we neglect them in this work. 
For these light-quark operators, 
we classify their triple-quark components into irreps under the QCD chiral group $G_\chi$ for the $u,d,s$ quarks.
As these components are not acted upon by derivatives, they fall into four irreps of the chiral group and can be parametrized as follows~\cite{Liao:2025vlj}, 
\begin{subequations}
\label{eq:Nyzw}
\begin{align}
{\cal N}_{yzw}^{\tL\tL} & \equiv  q_{\tL, y}^\alpha (\overline{ q_{\tL, z}^{\beta \C} } q_{\tL, w}^\gamma )\epsilon_{\alpha \beta \gamma}\in \pmb{8}_\tL \otimes  \pmb{1}_\tR , 
\\
{\cal N}_{yzw}^{\tR\tL} & \equiv q_{\tR, y}^\alpha (\overline{ q_{\tL, z}^{\beta \C} } q_{\tL,w}^\gamma)\epsilon_{\alpha \beta \gamma} \in 
\bar{\pmb{3}}_\tL \otimes \pmb{3}_\tR , 
\\
{\cal N}_{yzw}^{\tL\tR,\mu} & \equiv q_{\tL,\{y}^\alpha (\overline{ q_{\tL, z\}}^{\beta \C} } \gamma^\mu q_{\tR,w}^\gamma)\epsilon_{\alpha \beta \gamma}
\in \pmb{6}_\tL \otimes \pmb{3}_\tR,
\\
{\cal N}_{yzw}^{\tL\tL,\mu\nu} & \equiv  q_{\tL, \{y}^\alpha (\overline{ q_{\tL, z}^{\beta \C} } \sigma^{\mu\nu} q_{\tL, w\} }^\gamma )\epsilon_{\alpha \beta \gamma}\in \pmb{10}_\tL \otimes  \pmb{1}_\tR, 
\end{align}
\end{subequations} 
plus their chirality partners with $\tL\leftrightarrow\tR$. 
Here, the curly brackets denote total symmetrization in the flavor indices, $A_{\{y}B_{z\}} \equiv (1/2)(A_{y} B_{z} + A_{z}B_{y})$ and 
$A_{\{y}B_z C_{w\}} \equiv (1/6)[ A_{y}B_{z} C_w + \mbox{5 perms of}~(y,z,w)]$. 
$\pmb{i}_\tL\otimes \pmb{j}_\tR$ represents the chiral irrep under $G_\chi$. For later chiral matching convenience, 
we follow~Refs.\,\cite{Fan:2024gzc,Liao:2025vlj} to organize the flavor components of the $\pmb{8}_{\tL} \otimes \pmb{1}_{\tR}$ and $\bar{\pmb{3}}_{\tL} \otimes \pmb{3}_{\tR}$ irreps in matrix form: 
\begin{align}
{\cal N}_{{\bf 8}_\tL\otimes {\bf 1}_\tR}
 =
\begin{pmatrix}
{\cal N}^{\tL\tL}_{uds}  &  {\cal N}^{\tL\tL}_{usu}  & {\cal N}^{\tL\tL}_{uud}  
\\[2pt]
{\cal N}^{\tL\tL}_{dds}  & {\cal N}^{\tL\tL}_{dsu} & {\cal N}^{\tL\tL}_{dud}  
\\[2pt]
{\cal N}^{\tL\tL}_{sds} & {\cal N}^{\tL\tL}_{ssu} & {\cal N}^{\tL\tL}_{sud}
\end{pmatrix}, \quad 
{\cal N}_{\bar{\pmb{3}}_\tL \otimes \pmb{3}_\tR }   = 
 \begin{pmatrix}
{\cal N}_{uds}^{\tR\tL}  
& {\cal N}_{usu}^{\tR\tL} 
& {\cal N}_{uud}^{\tR\tL} 
\\[2pt]
{\cal N}_{dds}^{\tR\tL}  
& {\cal N}_{dsu}^{\tR\tL} 
& {\cal N}_{dud}^{\tR\tL} 
\\[2pt] 
{\cal N}_{sds}^{\tR\tL}  
& {\cal N}_{ssu}^{\tR\tL} 
& {\cal N}_{sud}^{\tR\tL}
 \end{pmatrix},
\label{eq:3qpart}
\end{align}
and their chirality partners with $\tL\leftrightarrow\tR$.

In the fourth column of \cref{tab:XLEFTopedim7,tab:XLEFTopedim8},
we indicate chiral irreps of operators involving $u,d,s$ quarks.
As evident from the two tables, the operators group into three distinct classes based on their quark currents: scalar, vector, and tensor currents.  
Operators with a scalar quark current transform under the irreps $\pmb{8}_{\tL(\tR)}\otimes \pmb{1}_{\tR(\tL)}$, and they can be expressed as
$X_\mu\overline{l}\gamma^\mu {\cal N}_{yzw}^{\tL\tL(\tR\tR)}$ and $X_{\mu\nu}\overline{l}\sigma^{\mu\nu} {\cal N}_{yzw}^{\tL\tL(\tR\tR)}$ for cases A and B, respectively.
Operators with a tensor quark current in case B belong to the irreps $\pmb{10}_{\tL(\tR)}\otimes \pmb{1}_{\tR(\tL)}$ and take the form $X_{\mu\nu}\overline{l}{\cal N}_{yzw}^{\tL\tL(\tR\tR),\mu\nu}$.
In contrast, operators with a vector quark current 
generally do not form one irrep due to the positioning of the two like-chirality quark fields. 
However, they can be decomposed into antisymmetric and symmetric components which respectively belong to the $\bar{\pmb{3}}_{\tL(\tR)} \otimes \pmb{3}_{\tR(\tL)}$ and ${\pmb{6}}_{\tL(\tR)} \otimes \pmb{3}_{\tR(\tL)}$ irreps.  
As shown in~\cite{Liao:2025vlj}, 
a general vector-current operator $(\overline{\psi_\mu}q_{\tL, y}^\alpha) (\overline{ q_{\tL, z}^{\beta \C} } \gamma^\mu q_{\tR, w}^\gamma )\epsilon_{\alpha \beta \gamma}$ can be decomposed in terms of ${\cal N}_i$ in \cref{eq:Nyzw} as follows, 
\begin{align}
(\overline{\psi_\mu} q_{\tL,y}^\alpha) (\overline{ q_{\tL, z}^{\beta \C} } \gamma^\mu q_{\tR, w}^\gamma ) \epsilon_{\alpha \beta \gamma} 
= \frac{1}{2} \overline{\psi_\mu}\gamma^\mu 
{\cal N}_{wzy}^{\tR\tL} (\in \bar{\pmb{3}}_{\tL} \otimes \pmb{3}_{\tR})
+ \overline{\psi_\mu} \, {\cal N}_{yzw}^{\tL\tR, \mu} (\in \pmb{6}_{\tL} \otimes \pmb{3}_{\tR} ),
\label{eq:irrep_decom}
\end{align}
where we have employed Fierz transformations to combine the antisymmetric combinations in the flavors $y$ and $z$ into a single scalar quark current in ${\cal N}_{wzy}^{\tR\tL}$. 
Similarly, the chirality-flipped operators with $\tL\leftrightarrow\tR$ can be decomposed into $\pmb{3}_{\tL} \otimes \pmb{6}_{\tR}$ and $\pmb{3}_{\tL} \otimes \bar{\pmb{3}}_{\tR}$.

In terms of the triple-quark objects in \cref{eq:Nyzw}, the $\Delta B=1$ $X$LEFT interactions in \cref{eq:XLEFTLag} with $y,z,w=\{u,d,s\}$ quarks can be expressed as 
\begin{align}
{\cal L}_{q^3}^{\slashed{B}} = \,& 
{\rm Tr} \big[  
{\cal P}_{\pmb{8}_\tL \otimes \pmb{1}_\tR }
{\cal N}_{\pmb{8}_\tL \otimes \pmb{1}_\tR } 
+ {\cal P}_{ \pmb{1}_\tL \otimes \pmb{8}_\tR }  
{\cal N}_{  \pmb{1}_\tL \otimes \pmb{8}_\tR }
\big]  
+ {\rm Tr} \big[ 
{\cal P}_{\pmb{3}_\tL \otimes \bar{\pmb{3}}_\tR}
{\cal N}_{\bar{\pmb{3}}_\tL \otimes \pmb{3}_\tR} 
+ {\cal P}_{\bar{\pmb{3}}_\tL \otimes \pmb{3}_\tR}
{\cal N}_{\pmb{3}_\tL \otimes \bar{\pmb{3}}_\tR} \big] 
\nn 
\\
& + \big[ {\cal P}_{yzw}^{\tL\tR,\mu}{\cal N}_{yzw,\mu}^{\tL\tR}
+ {\cal P}_{yzw}^{\tR\tL,\mu}{\cal N}_{yzw,\mu}^{\tR\tL}
\big]
+ \big[ {\cal P}_{yzw}^{\tL\tL,\mu\nu}
{\cal N}_{yzw,\mu\nu}^{\tL\tL}
+ {\cal P}_{yzw}^{\tR\tR,\mu\nu}
{\cal N}_{yzw,\mu\nu}^{\tR\tR}
\big]
 +\text{H.c.},
\label{eq:q3LEFT1}
\end{align}
where the trace is over the three-flavor space and the repeated indices $y,z,w$ are summed over $u,d,s$. 
${\cal P}$s are the so-called spurion fields.
For a chiral irrep $X$LEFT operator $\calO=\overline{\psi}{\cal N}_i$, ${\cal P}_i=C \overline{\psi}$ denotes the product of its WC $C$ and the non-QCD component $\overline{\psi}$.
The spurion fields corresponding to ${\cal N}_{yzw}^{\tL\tL(\tR\tR)}$ and ${\cal N}_{yzw}^{\tR\tL(\tL\tR)}$ are organized into matrices:
\begin{align}
 {\cal P}_{\pmb{8}_\tL \otimes \pmb{1}_\tR} 
 = 
 \begin{pmatrix}
0   & {\cal P}_{dds}^{\tL\tL}  
& {\cal P}_{sds}^{ \tL\tL} 
\\
{\cal P}_{usu}^{\tL\tL}  
& {\cal P}_{dsu}^{\tL\tL} 
& {\cal P}_{ssu}^{\tL\tL} 
\\
{\cal P}_{uud}^{\tL\tL}  
& {\cal P}_{dud}^{\tL\tL} 
& {\cal P}_{sud}^{\tL\tL}
 \end{pmatrix}, 
\,\,\,
 {\cal P}_{\pmb{3}_\tL \otimes \bar{\pmb{3}}_\tR }
 = 
 \begin{pmatrix}
{\cal P}_{uds}^{\tR\tL}  
&  {\cal P}_{dds}^{\tR\tL} 
& {\cal P}_{sds}^{\tR\tL} 
\\
{\cal P}_{usu}^{\tR\tL}  
&  {\cal P}_{dsu}^{\tR\tL} 
& {\cal P}_{ssu}^{\tR\tL} 
\\
{\cal P}_{uud}^{\tR\tL}  
&  {\cal P}_{dud}^{\tR\tL} 
& {\cal P}_{sud}^{\tR\tL}
\end{pmatrix},  
\end{align}
and the chirality partners with $\tL \leftrightarrow \tR$.
Note that the condition $\textrm{Tr}{\cal N}_{{\bf 8}_\tL\otimes {\bf 1}_\tR}=0$ has been used to attribute the $(1,1)$ entry of ${\cal P}_{{\bf 8}_\tL\otimes {\bf 1}_\tR}$ to its $(2,2)$ and $(3,3)$ entries by treating ${\cal N}^{\tL\tL}_{uds}$ as redundant~\cite{Liao:2025sqt}.
The spurion fields accompanying the 
$\pmb{6}_{\tL(\tR)} \otimes \pmb{3}_{\tR(\tL)}$ irreps, 
${\cal N}_{yzw}^{\tL\tR(\tR\tL), \mu}$, 
are denoted by ${\cal P}_{yzw}^{\tL\tR(\tR\tL), \mu}$, 
while those associated with the 
$\pmb{10}_{\tL(\tR)} \otimes \pmb{1}_{\tR(\tL)}$ irreps, 
${\cal N}_{yzw}^{\tt LL(RR), \mu\nu}$, 
are denoted by ${\cal P}_{yzw}^{\tt LL(RR), \mu\nu}$.

\begin{table}[htbp]
\center
\resizebox{\linewidth}{!}{
\renewcommand{\arraystretch}{1.2}
\begin{tabular}{|c|l|l|l|l|l|l|}
\hline
\multirow{2}*{Irrep.} 
&\multirow{2}*{Spurions}  
&\multicolumn{2}{c|}{Expressions}   
&\multirow{2}*{Spurions} 
&\multicolumn{2}{c|}{Expressions}  
\\\hhline{~~--~--}
& &\multicolumn{1}{c|}{case A}
&\multicolumn{1}{c|}{case B}
& &\multicolumn{1}{c|}{case A}
&\multicolumn{1}{c|}{case B}
\\\hline
\multirow{8}*{\rotatebox[origin=c]{90}{
\color{purple}$\pmb{8}_{\tL(\tR)}\otimes \pmb{1}_{\tR(\tL)}$}}
&${\cal P}_{uud}^{\tL\tL}$ 
&$ C^{{\tt VL,SL},x}_{X\ell uud}
X_\mu \overline{\ell_{\tR,x}^\C} \gamma^\mu$
&$\tilde{C}^{{\tt TL,SL},x}_{X\ell uud} 
X_{\mu\nu} \overline{\ell_{\tL,x}^\C} \sigma^{\mu\nu}$
&${\cal P}_{uud}^{\tR\tR}$ 
&$C^{{\tt VR,SR},x}_{X\ell uud}
X_\mu \overline{\ell_{\tL,x}^\C} \gamma^\mu$
&$\tilde{C}^{{\tt TR,SR},x}_{X\ell uud} 
X_{\mu\nu} \overline{\ell_{\tR,x}^\C} \sigma^{\mu\nu}$
\\%
& ${\cal P}_{usu}^{\tL\tL}$ 
& $-C^{{\tt VL,SL},x}_{X\ell uus} 
X_\mu \overline{\ell_{\tR,x}^\C} \gamma^\mu$&$-\tilde{C}^{{\tt TL,SL},x}_{X  \ell uus} 
X_{\mu\nu} \overline{\ell_{\tL,x}^\C} \sigma^{\mu\nu}$
& ${\cal P}_{usu}^{\tR\tR}$ 
& $ - C^{{\tt VR,SR},x}_{X\ell uus} 
X_\mu \overline{\ell_{\tL,x}^\C} \gamma^\mu$
&$-\tilde{C}^{{\tt TR,SR},x}_{X\ell uus} 
X_{\mu\nu} \overline{\ell_{\tR,x}^\C} \sigma^{\mu\nu}$
\\%
& ${\cal P}_{dds}^{\tL\tL}$ 
&$C^{{\tt VL,SL},x}_{X\bar\ell dds} 
X_\mu \overline{\ell_{\tL,x}} \gamma^\mu\,(\dagger)$
&$\tilde{C}^{{\tt TL,SL},x}_{X\bar\ell dds}X_{\mu\nu} \overline{\ell_{\tR,x}} \sigma^{\mu\nu}\,(\dagger)$
&${\cal P}_{dds}^{\tR\tR}$ 
&$C^{{\tt VR,SR},x}_{X\bar\ell dds} 
X_\mu \overline{\ell_{\tR,x}} \gamma^\mu\,(\dagger)$
&$\tilde{C}^{{\tt TR,SR},x}_{X\bar\ell dds} X_{\mu\nu} \overline{\ell_{\tL,x}} \sigma^{\mu\nu}\,(\dagger)$
\\%
&${\cal P}_{dud}^{\tL\tL}$ 
&$-C^{{\tt VL,SL},x}_{X\bar\nu ddu} 
X_\mu \overline{\nu_{\tL,x}} \gamma^\mu\,(\dagger)$
&$-\tilde{C}^{{\tt TL,SL},x}_{X \nu ddu} 
X_{\mu\nu} \overline{\nu_{\tL,x}^\C} \sigma^{\mu\nu}$
&${\cal P}_{dud}^{\tR\tR}$ 
&$-C^{{\tt VR,SR},x}_{X   \nu ddu} 
X_\mu \overline{\nu_{\tL,x}^\C} \gamma^\mu$
&$-\tilde{C}^{{\tt TR,SR},x}_{X \bar\nu ddu} X_{\mu\nu} \overline{\nu_{\tL,x}} \sigma^{\mu\nu}\,(\dagger)$
\\%
& ${\cal P}_{dsu}^{\tL\tL}$ 
& $C^{{\tt VL,SL},x}_{X\bar\nu dsu} 
X_\mu \overline{\nu_{\tL,x}} \gamma^\mu\,(\dagger)$
&$\tilde{C}^{{\tt TL,SL},x}_{X \nu dsu} 
X_{\mu\nu} \overline{\nu_{\tL,x}^\C} \sigma^{\mu\nu}$
&${\cal P}_{dsu}^{\tR\tR}$ 
&$C^{{\tt VR,SR},x}_{X  \nu dsu} 
X_\mu \overline{\nu_{\tL,x}^\C} \gamma^\mu$
&$\tilde{C}^{{\tt TR,SR},x}_{X \bar\nu dsu} X_{\mu\nu} \overline{\nu_{\tL,x}} \sigma^{\mu\nu}\,(\dagger)$
\\%
&${\cal P}_{sud}^{\tL\tL}$ 
&$-C^{{\tt VL,SL},x}_{X\bar\nu sdu} 
X_\mu \overline{\nu_{\tL,x}} \gamma^\mu\,(\dagger)$
&$-\tilde{C}^{{\tt TL,SL},x}_{X \nu sdu} 
X_{\mu\nu} \overline{\nu_{\tL,x}^\C} \sigma^{\mu\nu}$
&${\cal P}_{sud}^{\tR\tR}$ 
&$-C^{{\tt VR,SR},x}_{X   \nu sdu} 
X_\mu \overline{\nu_{\tL,x}^\C} \gamma^\mu$
&$-\tilde{C}^{{\tt TR,SR},x}_{X \bar\nu sdu} X_{\mu\nu} \overline{\nu_{\tL,x}} \sigma^{\mu\nu}\,(\dagger)$
\\%
&${\cal P}_{ssu}^{\tL\tL}\,(\times)$ 
&$C^{{\tt VL,SL},x}_{X\bar\nu ssu} 
X_\mu \overline{\nu_{\tL,x}} \gamma^\mu\,(\dagger)$
&$\tilde{C}^{{\tt TL,SL},x}_{X \nu ssu} 
X_{\mu\nu} \overline{\nu_{\tL,x}^\C} \sigma^{\mu\nu}$
&${\cal P}_{ssu}^{\tR\tR}\,(\times)$ 
&$C^{{\tt VR,SR},x}_{X  \nu ssu} 
X_\mu \overline{\nu_{\tL,x}^\C} \gamma^\mu$
&$\tilde{C}^{{\tt TR,SR},x}_{X \bar\nu ssu} X_{\mu\nu} \overline{\nu_{\tL,x}} \sigma^{\mu\nu}\,(\dagger)$
\\%
& ${\cal P}_{sds}^{\tL\tL}\,(\times)$ 
& $C^{{\tt VL,SL},x}_{X\bar\ell sds} 
X_\mu \overline{\ell_{\tL,x}} \gamma^\mu\,(\dagger)$
&$\tilde{C}^{{\tt TL,SL},x}_{X\bar\ell sds} X_{\mu\nu} \overline{\ell_{\tR,x}} \sigma^{\mu\nu}\,(\dagger)$
& ${\cal P}_{sds}^{\tR\tR}\,(\times)$ 
& $ C^{{\tt VR,SR},x}_{X\bar\ell sds} 
X_\mu \overline{\ell_{\tR,x}} \gamma^\mu\,(\dagger)$
&$\tilde{C}^{{\tt TR,SR},x}_{X\bar\ell sds} X_{\mu\nu} \overline{\ell_{\tL,x}} \sigma^{\mu\nu}\,(\dagger)$
\\\hline
\multirow{9}*{\rotatebox[origin=c]{90}{
\color{Purple}$\pmb{3}_{\tL(\tR)}\otimes \bar{\pmb{3}}_{\tR(\tL)}$}} 
&${\cal P}_{uud}^{\tR\tL}$ 
&$-C^{{\tt SL,VR}-,x}_{X   \ell udu}
X_\mu \overline{\ell_{\tL,x}^\C} \gamma^\mu$
&$i\tilde{C}^{{\tt VL,VR}-,x}_{X  \ell udu} X_{\mu\nu} \overline{\ell_{\tR,x}^\C} \sigma^{\mu\nu}$
&${\cal P}_{uud}^{\tL\tR}$ 
&$-C^{{\tt SR,VL}-,x}_{X\ell udu}
X_\mu \overline{\ell_{\tR,x}^\C} \gamma^\mu$
&$i\tilde{C}^{{\tt VR,VL}-,x}_{X  \ell udu} X_{\mu\nu} \overline{\ell_{\tL,x}^\C} \sigma^{\mu\nu}$
\\%
&${\cal P}_{usu}^{\tR\tL}$ 
&$C^{{\tt SL,VR}-,x}_{X\ell usu} 
X_\mu \overline{\ell_{\tL,x}^\C} \gamma^\mu$
&$-i\tilde{C}^{{\tt VL,VR}-,x}_{X\ell usu} X_{\mu\nu} \overline{\ell_{\tR,x}^\C} \sigma^{\mu\nu}$
&${\cal P}_{usu}^{\tL\tR}$ 
&$C^{{\tt SR,VL}-,x}_{X\ell usu} 
X_\mu \overline{\ell_{\tR,x}^\C} \gamma^\mu$
&$-i\tilde{C}^{{\tt VR,VL}-,x}_{X\ell usu} X_{\mu\nu} \overline{\ell_{\tL,x}^\C} \sigma^{\mu\nu}$
\\%
&${\cal P}_{dds}^{\tR\tL}$ 
&$-C^{{\tt SL,VR}-,x}_{X\bar\ell dsd}
X_\mu \overline{\ell_{\tR,x}} \gamma^\mu\,(\dagger)$
&$i\tilde{C}^{{\tt VL,VR}-,x}_{X\bar\ell dsd} X_{\mu\nu} \overline{\ell_{\tL,x}} \sigma^{\mu\nu}\,(\dagger)$
&${\cal P}_{dds}^{\tL\tR}$ 
&$-C^{{\tt SR,VL}-,x}_{X\bar\ell dsd} 
X_\mu \overline{\ell_{\tL,x}}\gamma^\mu\,(\dagger)$
&$i\tilde{C}^{{\tt VR,VL}-,x}_{X\bar\ell dsd} X_{\mu\nu} \overline{\ell_{\tR,x}} \sigma^{\mu\nu}\,(\dagger)$
\\%
&${\cal P}_{dud}^{\tR\tL}$ 
&$-C^{{\tt SL,VR}-,x}_{X\nu udd} 
X_\mu \overline{\nu_{\tL,x}^\C} \gamma^\mu$
&$i\tilde{C}^{{\tt VL,VR}-,x}_{X  \bar\nu udd} X_{\mu\nu} \overline{\nu_{\tL,x}} \sigma^{\mu\nu}\,(\dagger)$
&${\cal P}_{dud}^{\tL\tR}$ 
&$-C^{{\tt SR,VL}-,x}_{X\bar\nu udd}
X_\mu \overline{\nu_{\tL,x}}\gamma^\mu\,(\dagger)$
&$i\tilde{C}^{{\tt VR,VL}-,x}_{X\nu udd} 
X_{\mu\nu} \overline{\nu_{\tL,x}^\C} \sigma^{\mu\nu}$
\\%
&${\cal P}_{uds}^{\tR\tL}$ 
&$-C^{{\tt SL,VR}-,x}_{X\nu dsu} 
X_\mu \overline{\nu_{\tL,x}^\C}\gamma^\mu$
&$i\tilde{C}^{{\tt VL,VR}-,x}_{X\bar\nu dsu}X_{\mu\nu} \overline{\nu_{\tL,x}} \sigma^{\mu\nu}\,(\dagger)$
&${\cal P}_{uds}^{\tL\tR}$ 
&$-C^{{\tt SR,VL}-,x}_{X   \nu dsu} 
X_\mu \overline{\nu_{\tL,x}} \gamma^\mu\,(\dagger)$
&$i\tilde{C}^{{\tt VR,VL}-,x}_{X \bar\nu dsu} X_{\mu\nu} \overline{\nu_{\tL,x}^\C} \sigma^{\mu\nu}$
\\%
&${\cal P}_{dsu}^{\tR\tL}$ 
&$C^{{\tt SL,VR}-,x}_{X  \nu usd}
X_\mu \overline{\nu_{\tL,x}^\C} \gamma^\mu$
&$-i\tilde{C}^{{\tt VL,VR}-,x}_{X\bar\nu usd} X_{\mu\nu} \overline{\nu_{\tL,x}} \sigma^{\mu\nu}\,(\dagger)$
&${\cal P}_{dsu}^{\tL\tR}$ 
&$C^{{\tt SR,VL}-,x}_{X  \bar\nu usd}
X_\mu \overline{\nu_{\tL,x}} \gamma^\mu\,(\dagger)$
&$-i\tilde{C}^{{\tt VR,VL}-,x}_{X \nu usd} 
X_{\mu\nu} \overline{\nu_{\tL,x}^\C} \sigma^{\mu\nu}$
\\%
&${\cal P}_{sud}^{\tR\tL}$ 
&$-C^{{\tt SL,VR}-,x}_{X  \nu uds}
X_\mu \overline{\nu_{\tL,x}^\C} \gamma^\mu$
&$i\tilde{C}^{{\tt VL,VR}-,x}_{X  \bar\nu uds} 
X_{\mu\nu} \overline{\nu_{\tL,x}} \sigma^{\mu\nu}\,(\dagger)$
&${\cal P}_{sud}^{\tL\tR}$ 
&$-C^{{\tt SR,VL}-,x}_{X\bar \nu uds}
X_\mu \overline{\nu_{\tL,x}} \gamma^\mu\,(\dagger)$
&$i\tilde{C}^{{\tt VR,VL}-,x}_{X  \nu uds} 
X_{\mu\nu} \overline{\nu_{\tL,x}^\C} \sigma^{\mu\nu}$
\\%
&${\cal P}_{ssu}^{\tR\tL}\,(\times)$ 
&$C^{{\tt SL,VR}-,x}_{X  \nu uss} 
X_\mu \overline{\nu_{\tL,x}^\C} \gamma^\mu$
&$-i\tilde{C}^{{\tt VL,VR}-,x}_{X\bar\nu uss} X_{\mu\nu} \overline{\nu_{\tL,x}} \sigma^{\mu\nu}\,(\dagger)$
&${\cal P}_{ssu}^{\tL\tR}\,(\times)$ 
&$C^{{\tt SR,VL}-,x}_{X\bar\nu uss}
X_\mu \overline{\nu_{\tL,x}} \gamma^\mu\,(\dagger)$
&$-i\tilde{C}^{{\tt VR,VL}-,x}_{X  \nu uss} 
X_{\mu\nu} \overline{\nu_{\tL,x}^\C}\sigma^{\mu\nu}$
\\%
&${\cal P}_{sds}^{\tR\tL}\,(\times)$ 
&$-C^{{\tt SL,VR}-,x}_{X\bar\ell dss} 
X_\mu \overline{\ell_{\tR,x}} \gamma^\mu\,(\dagger)$
&$i\tilde{C}^{{\tt VL,VR}-,x}_{X\bar\ell dss} X_{\mu\nu} \overline{\ell_{\tL,x}} \sigma^{\mu\nu}\,(\dagger)$
&${\cal P}_{sds}^{\tL\tR}\,(\times)$ 
&$- C^{{\tt SR,VL}-,x}_{X\bar\ell dss}
X_\mu \overline{\ell_{\tL,x}} \gamma^\mu\,(\dagger)$  
&$i\tilde{C}^{{\tt VR,VL}-,x}_{X\bar\ell dss} X_{\mu\nu} \overline{\ell_{\tR,x}} \sigma^{\mu\nu}\,(\dagger)$
\\\hline
\multirow{18}*{\rotatebox[origin=c]{90}{
\color{Green}$\pmb{6}_{\tL(\tR)}\otimes \pmb{3}_{\tR(\tL)}$}} 
& ${\cal P}_{uuu}^{\tL\tR,\mu}$ 
&  ---&  ---
& ${\cal P}_{uuu}^{\tR\tL,\mu}$ 
&  ---&  ---
\\
& ${\cal P}_{uud}^{\tL\tR,\mu}$ 
& $ C^{{\tt SL,VR},x}_{X  \ell uud}
X^\mu \overline{\ell_{\tL,x}^\C}$&$-\tilde{C}^{{\tt VL,VR},x}_{X  \ell uud} 
X_{\mu\nu} \overline{\ell_{\tR,x}^\C} \gamma^{\nu}$
& ${\cal P}_{uud}^{\tR\tL,\mu}$ 
& $ C^{{\tt SR,VL},x}_{X  \ell uud} 
X^\mu \overline{\ell_{\tR,x}^\C}$&$-\tilde{C}^{{\tt VR,VL},x}_{X  \ell uud} 
X_{\mu\nu} \overline{\ell_{\tL,x}^\C} \gamma^{\nu}$ 
\\%
& ${\cal P}_{udu}^{\tL\tR,\mu}$ 
& $ C^{{\tt SL,VR}+,x}_{X\ell udu}
X^\mu \overline{\ell_{\tL,x}^\C}$&$-\tilde{C}^{{\tt VL,VR}+,x}_{X  \ell udu} 
X_{\mu\nu} \overline{\ell_{\tR,x}^\C} \gamma^{\nu}$
& ${\cal P}_{udu}^{\tR\tL,\mu}$ 
& $C^{{\tt SR,VL}+,x}_{X\ell udu}
X^\mu \overline{\ell_{\tR,x}^\C}$&$-\tilde{C}^{{\tt VR,VL}+,x}_{X  \ell udu} 
X_{\mu\nu} \overline{\ell_{\tL,x}^\C} \gamma^{\nu}$
\\%
& ${\cal P}_{uus}^{\tL\tR,\mu}$ 
& $ C^{{\tt SL,VR},x}_{X  \ell uus}
X^\mu \overline{\ell_{\tL,x}^\C}$&$-\tilde{C}^{{\tt VL,VR},x}_{X  \ell uus} 
X_{\mu\nu} \overline{\ell_{\tR,x}^\C} \gamma^{\nu}$
& ${\cal P}_{uus}^{\tR\tL,\mu}$ 
& $ C^{{\tt SR,VL},x}_{X  \ell uus}
X^\mu \overline{\ell_{\tR,x}^\C}$&$-\tilde{C}^{{\tt VR,VL},x}_{X  \ell uus} 
X_{\mu\nu} \overline{\ell_{\tR,x}^\C} \gamma^{\nu}$
\\%
&${\cal P}_{usu}^{\tL\tR,\mu}$ 
&$C^{{\tt SL,VR}+,x}_{X\ell usu} 
X^\mu \overline{\ell_{\tL,x}^\C}$
&$-\tilde{C}^{{\tt VL,VR}+,x}_{X  \ell usu} X_{\mu\nu} \overline{\ell_{\tR,x}^\C} \gamma^{\nu}$
&${\cal P}_{usu}^{\tR\tL,\mu}$ 
&$C^{{\tt SR,VL}+,x}_{X   \ell usu}
X^\mu \overline{\ell_{\tR,x}^\C}$
&$-\tilde{C}^{{\tt VR,VL}+,x}_{X  \ell usu} X_{\mu\nu} \overline{\ell_{\tL,x}^\C} \gamma^{\nu}$
\\%
&${\cal P}_{ddd}^{\tL\tR,\mu}$ 
&$C^{{\tt SL,VR},x}_{X\bar\ell ddd} 
X^\mu\overline{\ell_{\tR,x}}\,(\dagger)$
&$-\tilde{C}^{{\tt VL,VR},x}_{X\bar\ell ddd} X_{\mu\nu} \overline{\ell_{\tL,x}} \gamma^{\nu}\,(\dagger)$
&${\cal P}_{ddd}^{\tR\tL,\mu}$ 
&$C^{{\tt SR,VL},x}_{X\bar\ell ddd} 
X^\mu \overline{\ell_{\tL,x}}\,(\dagger)$
&$-\tilde{C}^{{\tt VR,VL},x}_{X\bar\ell ddd} X_{\mu\nu} \overline{\ell_{\tR,x}} \gamma^{\nu}\,(\dagger)$
\\%
&${\cal P}_{dds}^{\tL\tR,\mu}$ 
&$C^{{\tt SL,VR},x}_{X\bar\ell dds} 
X^\mu\overline{\ell_{\tR,x}}\,(\dagger)$
&$-\tilde{C}^{{\tt VL,VR},x}_{X\bar\ell dds} X_{\mu\nu} \overline{\ell_{\tL,x}} \gamma^{\nu}\,(\dagger)$
&${\cal P}_{dds}^{\tR\tL,\mu}$ 
&$C^{{\tt SR,VL},x}_{X\bar\ell dds} 
X^\mu\overline{\ell_{\tL,x}}\,(\dagger)$
&$-\tilde{C}^{{\tt VR,VL},x}_{X\bar\ell dds} X_{\mu\nu} \overline{\ell_{\tR,x}} \gamma^{\nu}\,(\dagger)$
\\%
&${\cal P}_{dsd}^{\tL\tR,\mu}$ 
&$C^{{\tt SL,VR}+,x}_{X\bar\ell dsd}
X^\mu\overline{\ell_{\tR,x}}\,(\dagger)$
&$-\tilde{C}^{{\tt VL,VR}+,x}_{X\bar\ell dsd} X_{\mu\nu} \overline{\ell_{\tL,x}} \gamma^{\nu}\,(\dagger)$
&${\cal P}_{dsd}^{\tR\tL,\mu}$ 
&$C^{{\tt SR,VL}+,x}_{X\bar\ell dsd}
X^\mu \overline{\ell_{\tL,x}}\,(\dagger)$
&$-\tilde{C}^{{\tt VR,VL}+,x}_{X\bar\ell dsd} X_{\mu\nu} \overline{\ell_{\tR,x}} \gamma^{\nu}\,(\dagger)$
\\%
&${\cal P}_{ddu}^{\tL\tR,\mu}$ 
&$C^{{\tt SL,VR},x}_{X\nu ddu} 
X^\mu\overline{\nu_{\tL,x}^\C}$
&$-\tilde{C}^{{\tt VL,VR},x}_{X\bar\nu ddu} X_{\mu\nu} \overline{\nu_{\tL,x}} \gamma^{\nu}\,(\dagger)$
& ${\cal P}_{ddu}^{\tR\tL,\mu}$ 
&$C^{{\tt SR,VL},x}_{X\bar\nu ddu} 
X^\mu \overline{\nu_{\tL,x}}\,(\dagger)$
&$-\tilde{C}^{{\tt VR,VL},x}_{X\nu ddu} 
X_{\mu\nu} \overline{\nu_{\tL,x}^\C} \gamma^{\nu}$
\\%
&${\cal P}_{udd}^{\tL\tR,\mu}$ 
&$C^{{\tt SL,VR}+,x}_{X\nu udd}
X^\mu \overline{\nu_{\tL,x}^\C}$
&$-\tilde{C}^{{\tt VL,VR}+,x}_{X\bar\nu udd} X_{\mu\nu} \overline{\nu_{\tL,x}} \gamma^{\nu}\,(\dagger)$
&${\cal P}_{udd}^{\tR\tL,\mu}$ 
&$C^{{\tt SR,VL}+,x}_{X\bar\nu udd}X^\mu \overline{\nu_{\tL,x}}\,(\dagger)$
&$-\tilde{C}^{{\tt VR,VL}+,x}_{X\nu udd} 
X_{\mu\nu} \overline{\nu_{\tL,x}^\C} \gamma^{\nu}$
\\%
&${\cal P}_{dsu}^{\tL\tR,\mu}$ 
&$C^{{\tt SL,VR}+,x}_{X  \nu dsu}
X^\mu \overline{\nu_{\tL,x}^\C}$
&$-\tilde{C}^{{\tt VL,VR}+,x}_{X  \bar \nu dsu} X_{\mu\nu} \overline{\nu_{\tL,x}} \gamma^{\nu}\,(\dagger)$
&${\cal P}_{dsu}^{\tR\tL,\mu}$ 
&$C^{{\tt SR,VL}+,x}_{X\bar \nu dsu}
X^\mu \overline{\nu_{\tL,x}}\,(\dagger)$
&$-\tilde{C}^{{\tt VR,VL}+,x}_{X   \nu dsu} X_{\mu\nu} \overline{\nu_{\tL,x}^\C} \gamma^{\nu}$
\\%
&${\cal P}_{usd}^{\tL\tR,\mu}$ 
&$C^{{\tt SL,VR}+,x}_{X  \nu usd}
X^\mu \overline{\nu_{\tL,x}^\C}$
&$-\tilde{C}^{{\tt VL,VR}+,x}_{X\bar\nu usd} X_{\mu\nu} \overline{\nu_{\tL,x}} \gamma^{\nu}\,(\dagger)$
& ${\cal P}_{usd}^{\tR\tL,\mu}$ 
&$C^{{\tt SR,VL}+,x}_{X\bar\nu usd}
X^\mu \overline{\nu_{\tL,x}}\,(\dagger)$
&$-\tilde{C}^{{\tt VR,VL}+,x}_{X\nu usd} 
X_{\mu\nu} \overline{\nu_{\tL,x}^\C} \gamma^{\nu}$
\\%
&${\cal P}_{uds}^{\tL\tR,\mu}$ 
&$C^{{\tt SL,VR}+,x}_{X \nu uds}
X^\mu \overline{\nu_{\tL,x}^\C}$
&$-\tilde{C}^{{\tt VL,VR}+,x}_{X\bar\nu uds} X_{\mu\nu} \overline{\nu_{\tL,x}} \gamma^{\nu}\,(\dagger)$
&${\cal P}_{uds}^{\tR\tL,\mu}$ 
&$C^{{\tt SR,VL}+,x}_{X\bar\nu uds}
X^\mu \overline{\nu_{\tL,x}}\,(\dagger)$
&$-\tilde{C}^{{\tt VR,VL}+,x}_{X  \nu uds} 
X_{\mu\nu} \overline{\nu_{\tL,x}^\C} \gamma^{\nu}$
\\%
&${\cal P}_{uss}^{\tL\tR,\mu}\,(\times)$ 
&$C^{{\tt SL,VR}+,x}_{X\nu uss} 
X^\mu \overline{\nu_{\tL,x}^\C}$
&$-\tilde{C}^{{\tt VL,VR}+,x}_{X\bar\nu uss} X_{\mu\nu} \overline{\nu_{\tL,x}} \gamma^{\nu}\,(\dagger)$
&${\cal P}_{uss}^{\tR\tL,\mu}\,(\times)$ 
&$C^{{\tt SR,VL}+,x}_{X\bar\nu uss} 
X^\mu \overline{\nu_{\tL,x}}\,(\dagger)$
&$-\tilde{C}^{{\tt VR,VL}+,x}_{X\nu uss} 
X_{\mu\nu} \overline{\nu_{\tL,x}^\C} \gamma^{\nu}$
\\%
&${\cal P}_{ssu}^{\tL\tR,\mu}\,(\times)$ 
&$ C^{{\tt SL,VR},x}_{X \nu ssu} 
X^\mu \overline{\nu_{\tL,x}^\C}$
&$-\tilde{C}^{{\tt VL,VR},x}_{X  \bar \nu ssu} X_{\mu\nu} \overline{\nu_{\tL,x}} \gamma^{\nu}\,(\dagger)$
&${\cal P}_{ssu}^{\tR\tL,\mu}\,(\times)$ 
&$C^{{\tt SR,VL},x}_{X\bar\nu ssu} 
X^\mu \overline{\nu_{\tL,x}}\,(\dagger)$
&$-\tilde{C}^{{\tt VR,VL},x}_{X\nu ssu} 
X_{\mu\nu} \overline{\nu_{\tL,x}^\C} \gamma^{\nu}$
\\%
&${\cal P}_{dss}^{\tL\tR,\mu}\,(\times)$ 
&$C^{{\tt SL,VR}+,x}_{X\bar\ell dss}
X^\mu\overline{\ell_{\tR,x}}\,(\dagger)$
&$-\tilde{C}^{{\tt VL,VR}+,x}_{X\bar\ell dss} X_{\mu\nu} \overline{\ell_{\tL,x}} \gamma^{\nu}\,(\dagger)$
& ${\cal P}_{dss}^{\tR\tL,\mu}\,(\times)$ 
&$C^{{\tt SR,VL}+,x}_{X\bar\ell dss}
X^\mu \overline{\ell_{\tL,x}}\,(\dagger)$
&$-\tilde{C}^{{\tt VR,VL}+,x}_{X\bar\ell dss} X_{\mu\nu} \overline{\ell_{\tR,x}} \gamma^{\nu}\,(\dagger)$
\\%
&${\cal P}_{ssd}^{\tL\tR,\mu}\,(\times) $ 
&$ C^{{\tt SL,VR},x}_{X\bar\ell ssd} 
X^\mu \overline{\ell_{\tR,x}}\,(\dagger)$
&$-\tilde{C}^{{\tt VL,VR},x}_{X  \bar \ell ssd} X_{\mu\nu} \overline{\ell_{\tL,x}} \gamma^{\nu}\,(\dagger)$
&${\cal P}_{ssd}^{\tR\tL,\mu}\,(\times) $ 
&$C^{{\tt SR,VL},x}_{X\bar\ell ssd} 
X^\mu\overline{\ell_{\tL,x}}\,(\dagger)$
&$-\tilde{C}^{{\tt VR,VL},x}_{X  \bar \ell ssd} 
X_{\mu\nu} \overline{\ell_{\tR,x}} \gamma^{\nu}\,(\dagger)$
\\%
&${\cal P}_{sss}^{\tL\tR,\mu}\,(\times)$ 
&$ C^{{\tt SL,VR},x}_{X\bar\ell sss} 
X^\mu \overline{\ell_{\tR,x}}\,(\dagger)$
&$-\tilde{C}^{{\tt VL,VR},x}_{X\bar\ell sss} X_{\mu\nu} \overline{\ell_{\tL,x}} \gamma^{\nu}\,(\dagger)$
&${\cal P}_{sss}^{\tR\tL,\mu}\,(\times)$
&$C^{{\tt SR,VL},x}_{X\bar\ell sss} 
X^\mu \overline{\ell_{\tL,x}}\,(\dagger)$
&$-\tilde{C}^{{\tt VR,VL},x}_{X  \bar \ell sss} X_{\mu\nu} \overline{\ell_{\tR,x}} \gamma^{\nu}\,(\dagger)$
\\\hline%
\end{tabular} }
\caption{
Expressions of the spurion fields arising from the BNV XLEFT interactions involving $u,d,s$ quarks in \cref{tab:XLEFTopedim7,tab:XLEFTopedim8}.
The cells without or with a dagger correspond to the $\Delta(B-L)=0$ or $\Delta(B+L)=0$ interactions, respectively. 
The spurions (marked with a $\times$) that involve two or three strange quarks do not contribute to nucleon decay at leading order, and therefore are omitted from the calculation.  
The subscript $x$ stands for the lepton flavor.} 
\label{tab:spurion1}
\end{table}

The specific forms of the spurion fields derived from \cref{eq:XLEFTLag} and \cref{tab:XLEFTopedim7,tab:XLEFTopedim8} are summarized in \cref{tab:spurion1,tab:spurion2}. 
For the vector-quark-current operators in case A, we have employed the relation in \cref{eq:irrep_decom} and its chirality-flipped counterpart to decompose the relevant interactions into the irreps $\bar{\pmb{3}}_{\tL(\tR)} \otimes \pmb{3}_{\tR(\tL)}$ and ${\pmb{6}}_{\tL(\tR)} \otimes \pmb{3}_{\tR(\tL)}$. For example, the vector-quark-current Lagrangian terms in \cref{eq:XLEFTLag} reduce into
\begin{subequations}
\label{eq:irrep_reduction}
\begin{align}
&\big( C_{X l yzw}^{{\tt SL,VR},x}
\calO_{X l yzw}^{{\tt SL,VR},x}
+ C_{X l zyw}^{{\tt SL,VR},x}
\calO_{X l zyw}^{{\tt SL,VR},x} \big)_{y\neq z}
\nn\\
=\,&  
C_{X l yzw}^{{\tt SL,VR}-,x} 
\calO_{X l yzw}^{{\tt SL,VR}-,x} 
(\in \bar{\pmb{3}}_{\tL} \otimes \pmb{3}_{\tR})
+ C_{X l yzw}^{{\tt SL,VR}+,x} 
\calO_{X l yzw}^{{\tt SL,VR}+,x}
(\in {\pmb{6}}_{\tL} \otimes \pmb{3}_{\tR}),
\\
&
\big( \tilde C_{X l yzw}^{{\tt VL,VR},x}
\tilde \calO_{X l yzw}^{{\tt VL,VR},x} 
+\tilde C_{X l zyw}^{{\tt VL,VR},x}
\tilde \calO_{X l zyw}^{{\tt VL,VR},x} \big)_{y\neq z}
\nn\\
=\,& 
\tilde C_{X l yzw}^{{\tt VL,VR}-,x}
\tilde \calO_{X l yzw}^{{\tt VL,VR}-,x}
(\in \bar{\pmb{3}}_{\tL} \otimes \pmb{3}_{\tR})
+ \tilde C_{X l yzw}^{{\tt VL,VR}+,x}
\tilde \calO_{X l yzw}^{{\tt VL,VR}+,x}
(\in {\pmb{6}}_{\tL} \otimes \pmb{3}_{\tR}),
\end{align}
\end{subequations}
where
\begin{subequations}
\label{eq:irrep_reduction2}
\begin{align}
 C_{X l yzw}^{{\tt SL,VR}\pm,x} &\equiv 
\frac{1}{2} (C_{X l yzw}^{{\tt SL,VR},x} 
\pm C_{X l zyw}^{{\tt SL,VR},x} ), 
&
\tilde C_{X l yzw}^{{\tt VL,VR}\pm,x} &\equiv 
\frac{1}{2} (\tilde C_{X l yzw}^{{\tt VL,VR},x} 
\pm \tilde C_{X l zyw}^{{\tt VL,VR},x} ),
\\%
\calO_{X l yzw}^{{\tt SL,VR}-,x} &\equiv
-X_\mu \overline{l_x} \gamma^\mu 
{\cal N}_{wyz}^{\tR\tL}, 
&
\tilde \calO_{X l yzw}^{{\tt VL,VR}-,x} & \equiv 
i X_{\mu\nu}\overline{l_x}\sigma^{\mu\nu}
{\cal N}_{wyz}^{\tR\tL},
\\
\calO_{X l yzw}^{{\tt SL,VR}+,x} & \equiv 
X_\mu \overline{l_x}
({\cal N}_{yzw}^{\tL\tR, \mu}
+{\cal N}_{zyw}^{\tL\tR, \mu}),
&
\tilde \calO_{X l yzw}^{{\tt VL,VR}+,x} & \equiv 
-X_{\mu\nu}\overline{l_x}\gamma^\nu
({\cal N}_{yzw}^{\tL\tR, \mu}
+{\cal N}_{zyw}^{\tL\tR, \mu}).
\end{align}
\end{subequations}
Similarly, the decomposition of chirality-flipped terms can be obtained by exchanging $\tL\leftrightarrow\tR$ in the expressions above. 
From \cref{eq:irrep_reduction,eq:irrep_reduction2}, the corresponding spurion fields can be directly identified as the remaining components in each ${\cal N}_i$.  
Note that the spurion fields ${\cal P}_{yzw}^{\tL\tR(\tR\tL),\mu}$ inherit the same flavor symmetry in $y,z$ as their triple-quark partners ${\cal N}_{yzw}^{\tL\tR(\tR\tL), \mu}$, thus, only half of them are displayed in \cref{tab:spurion1}.
In addition, ${\cal P}_{yzw}^{\tL\tL(\tR\tR),\mu\nu}$
are totally symmetric in $y,z,w$ as ${\cal N}_{yzw}^{\tL\tL(\tR\tR), \mu\nu}$, so only 10 independent flavor combinations are shown in \cref{tab:spurion2}.   
For the spurion fields containing two or three strange quarks marked with a $\times$ in \cref{tab:spurion1,tab:spurion2}, 
we note that they cannot contribute to two- and three-body nucleon decays due to the mismatch of strangeness quantum number. Therefore, we will neglect them in the following.  

\begin{table}[t]
\centering
\renewcommand{\arraystretch}{1.1}
\resizebox{0.6\linewidth}{!}{
\begin{tabular}{|c|l|l|l|l|}
\hline
\multirow{2}*{Irrep.} 
&\multirow{2}*{~Spurions}  
&  \multicolumn{1}{c|}{Expressions}   
&\multirow{2}*{~Spurions} 
&  \multicolumn{1}{c|}{Expressions}  
\\\hhline{~~-~-}
& & \multicolumn{1}{c|}{case B} 
& & \multicolumn{1}{c|}{case B}
\\\hline
\multirow{10}*{\rotatebox[origin=c]{90}{
\color{Blue}$\pmb{10}_{\tL(\tR)}\otimes \pmb{1}_{\tR(\tL)}$}}
& ${\cal P}_{uuu}^{{\tL\tL},\mu\nu}$ &--- 
& ${\cal P}_{uuu}^{{\tR\tR},\mu\nu}$ &---
\\
&${\cal P}_{uud}^{{\tL\tL},\mu\nu}$ 
&$\frac{1}{3} \tilde{C}^{{\tt SL,TL},x}_{X\ell uud} 
X^{\mu\nu} \overline{\ell_{\tL,x}^\C} $
&${\cal P}_{uud}^{{\tR\tR},\mu\nu}$ 
&$\frac{1}{3}\tilde{C}^{{\tt SR,TR},x}_{X\ell uud} 
X^{\mu\nu} \overline{\ell_{\tR,x}^\C} $
\\
&${\cal P}_{uus}^{{\tL\tL},\mu\nu}$ 
&$\frac{1}{3}\tilde{C}^{{\tt SL,TL},x}_{X\ell uus} 
X^{\mu\nu} \overline{\ell_{\tL,x}^\C} $
&${\cal P}_{uus}^{{\tR\tR},\mu\nu}$ 
&$\frac{1}{3}\tilde{C}^{{\tt SR,TR},x}_{X\ell uus} 
X^{\mu\nu} \overline{\ell_{\tR,x}^\C} $
\\
&${\cal P}_{ddd}^{{\tL\tL},\mu\nu}$ 
&$\tilde{C}^{{\tt SL,TL},x}_{X\bar\ell ddd} X^{\mu\nu} \overline{\ell_{\tR,x}}\,(\dagger)$
&${\cal P}_{ddd}^{{\tR\tR},\mu\nu}$ 
&$\tilde{C}^{{\tt SR,TR},x}_{X\bar\ell ddd} 
X^{\mu\nu} \overline{\ell_{\tL,x}}\,(\dagger)$
\\
&${\cal P}_{dds}^{{\tL\tL},\mu\nu}$ 
&$\frac{1}{3}\tilde{C}^{{\tt SL,TL},x}_{X\bar\ell dds} X^{\mu\nu} \overline{\ell_{\tR,x}}\,(\dagger)$
&${\cal P}_{dds}^{{\tR\tR},\mu\nu}$ 
&$\frac{1}{3}\tilde{C}^{{\tt SR,TR},x}_{X\bar\ell dds} X^{\mu\nu} \overline{\ell_{\tL,x}}\,(\dagger)$
\\
&${\cal P}_{udd}^{{\tL\tL},\mu\nu}$ 
&$\frac{1}{3}\tilde{C}^{{\tt SL,TL},x}_{X \nu udd} 
X^{\mu\nu} \overline{\nu_{\tL,x}^\C} $
&${\cal P}_{udd}^{{\tR\tR},\mu\nu}$ 
&$\frac{1}{3}\tilde{C}^{{\tt SR,TR},x}_{X \bar\nu udd} X^{\mu\nu} \overline{\nu_{\tL,x}}\,(\dagger)$
\\
&${\cal P}_{uds}^{{\tL\tL},\mu\nu}$ 
&$\frac{1}{6}\tilde{C}^{{\tt SL,TL},x}_{X \nu uds} 
X^{\mu\nu} \overline{\nu_{\tL,x}^\C} $
&${\cal P}_{uds}^{{\tR\tR},\mu\nu}$ 
&$\frac{1}{6}\tilde{C}^{{\tt SR,TR},x}_{X \bar\nu uds} X^{\mu\nu} \overline{\nu_{\tL,x}}\,(\dagger)$
\\
&${\cal P}_{uss}^{{\tL\tL},\mu\nu}\,(\times)$ 
&$\frac{1}{3}\tilde{C}^{{\tt SL,TL},x}_{X \nu uss} X^{\mu\nu} \overline{\nu_{\tL,x}^\C}$
&${\cal P}_{uss}^{{\tR\tR},\mu\nu}\,(\times)$ 
&$\frac{1}{3}\tilde{C}^{{\tt SR,TR},x}_{X \bar\nu uss} X^{\mu\nu} \overline{\nu_{\tL,x}}\,(\dagger)$
\\
&${\cal P}_{dss}^{{\tL\tL},\mu\nu}\,(\times)$ 
&$\frac{1}{3}\tilde{C}^{{\tt SL,TL},x}_{X\bar\ell dss} X^{\mu\nu} \overline{\ell_{\tR,x}}\,(\dagger)$
&${\cal P}_{dss}^{{\tR\tR},\mu\nu}\,(\times)$ 
&$\frac{1}{3}\tilde{C}^{{\tt SR,TR},x}_{X\bar\ell dss} X^{\mu\nu} \overline{\ell_{\tL,x}}\,(\dagger)$
\\
&${\cal P}_{sss}^{{\tL\tL},\mu\nu}\,(\times)$ 
&$ \tilde{C}^{{\tt SL,TL},x}_{X\bar\ell sss} 
X^{\mu\nu} \overline{\ell_{\tR,x}}\,(\dagger)$
&${\cal P}_{sss}^{{\tR\tR},\mu\nu}\,(\times)$ 
&$\tilde{C}^{{\tt SR,TR},x}_{X\bar\ell sss} 
X^{\mu\nu} \overline{\ell_{\tL,x}}\,(\dagger)$\\
\hline%
\end{tabular}}
\caption{
Similar to \cref{tab:spurion1} but for the spurion fields associated with irreps $\pmb{10}_{\tL(\tR)}\otimes \pmb{1}_{\tR(\tL)}$ in case B.} 
\label{tab:spurion2}
\end{table}

\subsection{Matching onto ChPT}
\label{sec:nucleon_decay_ChPT}

For the general LEFT interactions in \cref{eq:q3LEFT1}, their leading-order chiral matching has been systematically realized in~\cite{Liao:2025vlj} for the case of pure pseudoscalar mesons and in~\cite{Liao:2025sqt} for those involving vector mesons. 
In this work, we focus only on two- and three-body nucleon decays involving at most a pseudoscalar meson, and thus utilize the chiral results in~\cite{Liao:2025vlj} for the subsequent analysis. 
We define the octet pseudoscalar field by 
$\Sigma(x) = \xi^2(x) = \exp[i\sqrt{2}\Pi(x)/F_0]$
and baryon field by $B(x)$, with
\begin{align}
\Pi =\,&   
\begin{pmatrix}
\frac{\pi^0}{\sqrt{2}}+\frac{\eta}{\sqrt{6}} & \pi^+ & K^+
\\
\pi^- & -\frac{\pi^0}{\sqrt{2}}+\frac{\eta}{\sqrt{6}} & K^0
\\
K^- & \bar{K}^0 & -\sqrt{\frac{2}{3}}\eta
\end{pmatrix},~
B=
\begin{pmatrix}
{\Sigma^{0}\over \sqrt{2}}+{\Lambda^0 \over \sqrt{6}}  & \Sigma^+ & p \\
\Sigma^- & -{\Sigma^{0} \over \sqrt{2}}+{\Lambda^0 \over \sqrt{6}} &  n \\ 
\Xi^- & \Xi^0 & - \sqrt{2\over 3}\Lambda^0
\end{pmatrix},
\end{align}
where $F_0=f_{\pi}/\sqrt{2}$ with the pion decay constant $f_{\pi}=130.41(20)~\rm MeV$~\cite{ParticleDataGroup:2024cfk}.

In terms of the spurion fields ${\cal P}_i$ in \cref{eq:q3LEFT1}, the leading-order chiral Lagrangian takes the form~\cite{Liao:2025vlj}:
\begin{align}
{\cal L}_{B}^{\slashed{B}} =\,&
c_1 {\rm Tr}\big[ 
{\cal P}_{  \bar{\pmb{3}}_\tL \otimes \pmb{3}_\tR} \xi B_\tL \xi -
{\cal P}_{\pmb{3}_\tL \otimes \bar{\pmb{3}}_\tR} \xi^\dagger B_\tR \xi^\dagger 
 \big] + c_2 {\rm Tr}\big[ 
{\cal P}_{\pmb{8}_\tL \otimes \pmb{1}_\tR}\xi B_\tL \xi^\dagger
- {\cal P}_{ \pmb{1}_\tL \otimes  \pmb{8}_\tR} \xi^\dagger B_\tR \xi
\big] \nn\\
& + {c_3 \over \Lambda_\chi} \big[ 
{\cal P}_{yzi}^{\tL\tR,\mu}
{\Gamma}_{\mu\nu}^{\tt L} 
(\xi i D^\nu B_\tL \xi)_{yj}
\Sigma_{zk} \epsilon_{ijk}
 - {\cal P}_{yzi}^{\tR\tL,\mu}
{\Gamma}_{\mu\nu}^{\tt R}
(\xi^\dagger i D^\nu B_\tR  \xi^\dagger)_{yj}
\Sigma^*_{kz} \epsilon_{ijk} \big]
\nn\\
&+{c_4 \over \Lambda_\chi^2} \big[ 
{\cal P}_{yzw}^{\tL\tL,\mu\nu}
\hat\Gamma_{\mu\nu \alpha\beta}^{\tt L} 
(\xi D^{\alpha} B_\tL \xi)_{yi}
\Sigma_{zj}(D^{\beta}\Sigma)_{wk} \epsilon_{ijk}
\nn
\\
& - {\cal P}_{yzw}^{\tR\tR,\mu\nu} 
\hat\Gamma_{\mu\nu \alpha\beta}^{\tt R}  
(\xi^\dagger D^{\alpha} B_\tR \xi^\dagger)_{yi}
\Sigma^*_{jz} (D^{\beta}\Sigma)^*_{kw} \epsilon_{ijk}\big]
+\text{H.c.},
\label{eq:chiB_next}
\end{align} 
where the indices $y, z$ and $i, j, k$ are summed over the three light quark flavors $u, d, s$.
The chiral baryon fields
$B_{\tL(\tR)} \equiv P_{\tL(\tR)} B$, and the Lorentz projectors associated with  
$\pmb{6}_{\tL(\tR)}\otimes \pmb{3}_{\tR(\tL)}$
and $\pmb{10}_{\tL(\tR)}\otimes \pmb{1}_{\tR(\tL)}$ irreps are defined as
\begin{align}
\label{eq:Lorentz_projectors}
{\Gamma}_{\mu\nu}^{\tL(\tR)} \equiv \Big(g_{\mu\nu} - {1\over 4} \gamma_\mu \gamma_\nu\Big)P_{\tL(\tR)}, \quad 
\hat{\Gamma}^{\tL(\tR)}_{\mu\nu\alpha\beta}\equiv\frac{1}{24}\left(2\left\{\sigma_{\mu\nu}, \sigma_{\alpha\beta}\right\}-[\sigma_{\mu\nu}, \sigma_{\alpha\beta}]\right)P_{\tL(\tR)}\;.
\end{align}
$c_{1,2,3,4}$ are the hadronic low-energy constants (LECs).
The lattice QCD calculations lead to  
$c_1=-0.01257(111)\,{\rm GeV}^3$ and $c_2=0.01269(107)\,{\rm GeV}^3$~\cite{Yoo:2021gql}. 
For $c_{3}$ and $c_4$, there is no available lattice result, and we use naive dimensional analysis (NDA) estimates provided in~\cite{Liao:2025vlj}. 

By expanding the pseudoscalar matrix in \cref{eq:chiB_next} to the zeroth order in the meson fields, we obtain the following general two-point spurion-baryon vertices 
\begin{align}
\label{eq:LPB}
{\cal L}_{{\cal P}B}
&\supset
\Big\{
\big[ c_1 {\cal P}_{uud}^{\tL\tR} + c_2 {\cal P}_{uud}^{\tL\tL} 
+ c_{3\chi} ( {\cal P}_{uud}^{\tL\tR,\mu} - {\cal P}_{udu}^{\tL\tR,\mu} ) 
i\tilde{\partial}_\mu \big]p_\tL
\nn\\
& + \big[ c_1 {\cal P}_{usu}^{\tL\tR} + c_2 {\cal P}_{usu}^{\tL\tL} 
- c_{3\chi} ( {\cal P}_{uus}^{\tL\tR,\mu} - {\cal P}_{usu}^{\tL\tR,\mu} ) i\tilde{\partial}_\mu \big] \Sigma^+_\tL 
\nn\\
& + \big[ c_1 {\cal P}_{dds}^{\tL\tR} + c_2 {\cal P}_{dds}^{\tL\tL} 
+ c_{3\chi} ({\cal P}_{dds}^{\tL\tR,\mu}-{\cal P}_{dsd}^{\tL\tR,\mu})
i\tilde{\partial}_\mu \big] 
\Sigma^-_\tL 
\nn\\
& + \big[ c_1 {\cal P}_{dud}^{\tL\tR} + c_2 {\cal P}_{dud}^{\tL\tL} 
- c_{3\chi} ({\cal P}_{ddu}^{\tL\tR,\mu} -{\cal P}_{udd}^{\tL\tR,\mu} ) 
i\tilde{\partial}_\mu \big] n_\tL
\\
&+ {1\over \sqrt{6} } \big[
  c_1 ( {\cal P}_{uds}^{\tL\tR} + {\cal P}_{dsu}^{\tL\tR}-2{\cal P}_{sud}^{\tL\tR})
+ c_2 ({\cal P}_{dsu}^{\tL\tL} - 2 {\cal P}_{sud}^{\tL\tL}) 
- 3 c_{3\chi} ({\cal P}_{usd}^{\tL\tR,\mu} - {\cal P}_{dsu}^{\tL\tR,\mu})
i \tilde{\partial}_\mu 
\big] \Lambda^0_\tL
\nn\\
& +{1\over \sqrt{2}} \big[ 
c_1 ( {\cal P}_{uds}^{\tL\tR} - {\cal P}_{dsu}^{\tL\tR})
- c_2 {\cal P}_{dsu}^{\tL\tL} 
+ c_{3\chi} ( 2 {\cal P}_{uds}^{\tL\tR,\mu} - {\cal P}_{usd}^{\tL\tR,\mu}- {\cal P}_{dsu}^{\tL\tR,\mu} ) 
i\tilde{\partial}_\mu \big] \Sigma^0_\tL 
\Big\} 
- \tL \leftrightarrow \tR, 
\nn
\end{align} 
where $c_{3\chi} \equiv c_3\Lambda_\chi^{-1}$ and
$\tilde{\partial}_\mu = \partial_\mu - \tfrac{1}{4}\gamma_\mu 
\slashed{\partial}$. 
The terms involving two or three $s$ quarks are irrelevant to nucleon decays and have been omitted. 
After incorporating the spurion fields from \cref{tab:spurion1} into the above terms, we obtain the three-point baryon-lepton-dark-photon vertices.

Similarly, expanding the pseudoscalar matrix in \cref{eq:chiB_next} to the linear order in the meson fields yields the following spurion-nucleon-meson vertices
\begin{align}
{\cal L}_{{\cal P}{\tN}M} 
&=\Big\{ 
 \frac{ \pi^+ }{\sqrt{2}}
\big[ c_1 {\cal P}_{uud}^{\tL\tR} 
+ c_2 {\cal P}_{uud}^{\tL\tL} 
+ c_{3\chi} ({\cal P}_{uud}^{\tL\tR,\mu}
-3{\cal P}_{udu}^{\tL\tR,\mu})
i\tilde{\partial}_\mu\big] n_\tL
+\sqrt{2} c_{4\chi}
{\cal P}_{uud}^{\tL\tL,\mu\nu }
\hat{\Gamma}^{\tL}_{\mu\nu\alpha\beta}
\partial^\alpha n_\tL \partial^\beta \pi^+   
\nn\\
&+ \frac{\pi^0}{2} \big[ 
c_1 {\cal P}_{uud}^{\tL\tR} 
+ c_2 {\cal P}_{uud}^{\tL\tL} 
+ c_{3\chi} ({\cal P}_{udu}^{\tL\tR,\mu} 
+ 3 {\cal P}_{uud}^{\tL\tR,\mu})
i\tilde{\partial}_\mu \big] p_\tL 
- 2 c_{4\chi} {\cal P}_{uud}^{\tL\tL,\mu\nu }
\hat{\Gamma}^{\tL}_{\mu\nu\alpha\beta}
\partial^\alpha p_\tL \partial^\beta \pi^0
\nn\\
&
- \frac{ \eta }{2\sqrt{3}} 
\big[ c_1 {\cal P}_{uud}^{\tL\tR} 
-3 c_2 {\cal P}_{uud}^{\tL\tL} 
+c_{3\chi} ({\cal P}_{udu}^{\tL\tR,\mu}
- {\cal P}_{uud}^{\tL\tR,\mu}) 
i \tilde{\partial}_\mu\big]p_\tL      
\nn\\
&
+ \frac{\bar{K}^0 }{\sqrt{2}} 
\big[ c_1 {\cal P}_{usu}^{\tL\tR} 
- c_2 {\cal P}_{usu}^{\tL\tL} 
- c_{3\chi}({\cal P}_{usu}^{\tL\tR,\mu} 
+ {\cal P}_{uus}^{\tL\tR,\mu}) 
i\tilde{\partial}_\mu \big] p_\tL
+ \sqrt{2}c_{4\chi} 
{\cal P}_{uus}^{\tL\tL,\mu\nu }
\hat{\Gamma}^{\tL}_{\mu\nu\alpha\beta}
\partial^\alpha p_\tL \partial^\beta \bar{K}^0
\nn\\
&
+ \frac{ K^- }{\sqrt{2}}\big[
c_1 {\cal P}_{dds}^{\tL\tR} 
- c_2 {\cal P}_{dds}^{\tL\tL} 
+ c_{3\chi}({\cal P}_{dds}^{\tL\tR,\mu} + {\cal P}_{dsd}^{\tL\tR,\mu})i\tilde{\partial}_\mu
\big]n_\tL    
-\sqrt{2}c_{4\chi} {\cal P}_{dds}^{\tL\tL,\mu\nu }
\hat{\Gamma}^{\tL}_{\mu\nu\alpha\beta}
\partial^\alpha n_\tL \partial^\beta {K}^-
\nn\\
&
+  \frac{\pi^- }{\sqrt{2}} \big[ 
c_1 {\cal P}_{dud}^{\tL\tR} 
+ c_2 {\cal P}_{dud}^{\tL\tL} 
- c_{3\chi} ({\cal P}_{ddu}^{\tL\tR,\mu}
-3{\cal P}_{udd}^{\tL\tR,\mu})
i\tilde{\partial}_\mu \big] p_\tL 
-\sqrt{2}c_{4\chi} {\cal P}_{udd}^{\tL\tL,\mu\nu}
\hat{\Gamma}^{\tL}_{\mu\nu\alpha\beta}
\partial^\alpha p_\tL \partial^\beta \pi^-   
\nn\\
&- \frac{\pi^0 }{2} \big[
c_1 {\cal P}_{dud}^{\tL\tR} 
+ c_2 {\cal P}_{dud}^{\tL\tL} 
- c_{3\chi} (3{\cal P}_{ddu}^{\tL\tR,\mu}
+ {\cal P}_{udd}^{\tL\tR,\mu})
i\tilde{\partial}_\mu \big] n_\tL
-2 c_{4\chi} {\cal P}_{udd}^{\tL\tL,\mu\nu} 
\hat{\Gamma}^{\tL}_{\mu\nu\alpha\beta} \partial^\alpha n_\tL \partial^\beta \pi^0
\nn\\
&
-  \frac{\eta }{2\sqrt{3}} \big[ 
c_1 {\cal P}_{dud}^{\tL\tR} 
-3 c_2 {\cal P}_{dud}^{\tL\tL} 
+ c_{3\chi} ({\cal P}_{ddu}^{\tL\tR,\mu}
- {\cal P}_{udd}^{\tL\tR,\mu}) 
i \tilde{\partial}_\mu \big] n_\tL
\nn\\
&
+ \frac{K^-}{\sqrt{2}} \big[ 
c_1 ( {\cal P}_{sud}^{\tL\tR} 
+ {\cal P}_{uds}^{\tL\tR}) 
+ c_2 {\cal P}_{sud}^{\tL\tL} 
- c_{3\chi} ({\cal P}_{dsu}^{\tL\tR,\mu}
-{\cal P}_{uds}^{\tL\tR,\mu}
-2{\cal P}_{usd}^{\tL\tR,\mu})
i \tilde{\partial}_\mu \big] p_\tL 
\nn\\
&
+ \frac{\bar{K}^0 }{\sqrt{2}}  
\big[ 
c_1 ( {\cal P}_{dsu}^{\tL\tR}
+ {\cal P}_{sud}^{\tL\tR} ) 
+ c_2 ( {\cal P}_{sud}^{\tL\tL} 
- {\cal P}_{dsu}^{\tL\tL} )
- c_{3\chi} (2{\cal P}_{dsu}^{\tL\tR,\mu} 
+ {\cal P}_{uds}^{\tL\tR,\mu} 
- {\cal P}_{usd}^{\tL\tR,\mu} ) 
i \tilde{\partial}_\mu  
\big]n_\tL 
\nn\\
&
-\sqrt{2}c_{4\chi} 
{\cal P}_{uds}^{\tL\tL,\mu\nu }
\hat{\Gamma}^{\tL}_{\mu\nu\alpha\beta}
\partial^\alpha p_\tL \partial^\beta {K}^-
+\sqrt{2} c_{4\chi} 
{\cal P}_{uds}^{\tL\tL,\mu\nu }
\hat{\Gamma}^{\tL}_{\mu\nu\alpha\beta}
\partial^\alpha n_\tL \partial^\beta \bar{K}^0
\nn\\
&+\sqrt{2}\pi^- c_{3\chi} 
{\cal P}_{ddd}^{\tL\tR,\mu} 
i \tilde{\partial}_\mu n_\tL
-\sqrt{2} c_{4\chi} {\cal P}_{ddd}^{\tL\tL,\mu\nu}
\hat{\Gamma}^{\tL}_{\mu\nu\alpha\beta}
\partial^\alpha n_\tL \partial^\beta \pi^- \Big\} {i \over F_0}
+ \tL \leftrightarrow \tR,
\label{eq:LPNM}    
\end{align}
where $c_{4\chi}\equiv c_4\Lambda_{\chi}^{-2}$.
The absence of $K^+\texttt{N}$ and $K^0\texttt{N}$ terms is because an anti-strange quark would be needed.
Again, replacing the spurion fields from \cref{tab:spurion1,tab:spurion2} into the above terms generates the local four-point vertices $XMl\tN$ contributing to three-body nucleon decays. 

For the three-body decay modes, 
besides the BNV interactions discussed above, 
the conventional baryon ChPT interactions due to the 
SM strong interactions are also relevant. Together with the BNV interactions in \cref{eq:LPB}, they can induce noncontact contributions. 
The leading-order Lagrangian is~\cite{Jenkins:1990jv,Bijnens:1985kj,Oller:2006yh}
\begin{align}
\label{L_Bchpt}
{\cal L}_{\tt ChPT}^B =\,& 
{\rm Tr}[\bar B (i \slashed{D} -M) B] 
+ \frac{D}{2} {\rm Tr}(\bar B \gamma^\mu \gamma_5\{u_\mu,B\}) 
+ \frac{F}{2} {\rm Tr}(\bar B \gamma^\mu \gamma_5 [u_\mu,B]).
\end{align}
In the absence of external sources, the axial-vector field $u_\mu$ is defined as 
$u_{\mu}=i(\xi\partial_{\mu}\xi^{\dagger}
-\xi^{\dagger}\partial_{\mu}\xi)$, 
and the covariant derivative of the baryon octet is given by $D_\mu B = \partial_\mu B + [\Gamma_\mu ,B]$, 
where $\Gamma_{\mu}=\tfrac{1}{2}
(\xi\partial_{\mu}\xi^{\dagger}
+\xi^{\dagger}\partial_{\mu}\xi)$. 
For the axial couplings we use the 
lattice results $D=0.730(11)$ and $F=0.447^{+0.006}_{-0.007}$~\cite{Bali:2022qja}. 
Expanding the $\xi$ matrix in 
\cref{L_Bchpt} to linear order in the pseudoscalar fields yields the following relevant interaction terms 
involving a nucleon, a meson, and a conjugate baryon field:
\begin{align}
{\cal L}_{\bar{B}\texttt{N}M} & \supset 
\frac{D-F}{2F_0} 
\big[ 
\overline{\Sigma^0} \gamma^\mu \gamma_5 p \partial_\mu K^-
- \overline{\Sigma^0} \gamma^\mu \gamma_5 n \partial_\mu \bar K^0 
+ \sqrt{2} (\overline{\Sigma^+}\gamma^\mu \gamma_5 p \partial_\mu \bar K^0 
+ \overline{\Sigma^-} \gamma^\mu \gamma_5 n \partial_\mu K^- ) 
\big]
\nn  \\ 
& 
+ \frac{3F-D}{2\sqrt{3}F_0} 
(\overline{p} \gamma^\mu \gamma_5 p\partial_\mu \eta 
+ \overline{n} \gamma^\mu \gamma_5 n \partial_\mu \eta )
-\frac{D+3F}{2\sqrt{3}F_0} 
\big[ 
\overline{\Lambda^0} \gamma^\mu \gamma_5 p \partial_\mu K^- 
+ \overline{\Lambda^0} \gamma^\mu \gamma_5 n\partial_\mu \bar K^0
\big]
\nn  \\ 
&
+\frac{D+F}{2F_0} 
\big[
\overline{p} \gamma^\mu \gamma_5 p \partial_\mu \pi^0
- \overline{n} \gamma^\mu \gamma_5 n \partial_\mu \pi^0
+ \sqrt{2} (\overline{n} \gamma^\mu \gamma_5 p \partial_\mu \pi^- 
+ \overline{p} \gamma^\mu \gamma_5 n \partial_\mu \pi^+ ) \big]
\nn\\
& \equiv \sum_{\tN, B, M} \frac{ C_{\tN BM} }{ F_0 }
\overline{B}\,\gamma_\mu \gamma_5 \texttt{N}\,
\partial^\mu \bar M.
\label{eq:LBNM}
\end{align}
In the second step, we collectively express them in a general form, where the coefficients $C_{\tN BM}$ can be easily identified once the fields $(\tN,B,M)$ are specified. 

\section{Decay width calculation}
\label{sec:decay_width}

\begin{table}[t]
\center
\resizebox{\linewidth}{!}{
\renewcommand{\arraystretch}{1.1}
\begin{tabular}{|l|l l|l l|l l|l l|l l|l l l l| }
\hline
\multirow{4}*{\rotatebox[origin=c]{90}{Case A} }
&$\calO_{X\ell uud}^{{\tt VL,SL}}$ 
&$\calO_{X\ell uud}^{{\tt VR,SR},x}$%
&$\calO_{X\ell uus}^{{\tt VL,SL}}$
&$\calO_{X\ell uus}^{{\tt VR,SR}}$%
&$\calO_{X\bar{\ell} ddd}^{{\tt SL,VR}}$
&$\calO_{X\bar{\ell} ddd}^{{\tt SR,VL}}$%
&$\calO_{X\bar{\ell} dds}^{{\tt VL,SL}}$ 
&$\calO_{X\bar{\ell} dds}^{{\tt VR,SR}}$%
&$\calO_{X\bar{\nu} ddu}^{{\tt VL,SL}}$
&$\calO_{X\nu ddu}^{{\tt VR,SR}}$%
&$\calO_{X\bar\nu dsu}^{{\tt VL,SL}}$
&$\calO_{X\bar\nu sdu}^{{\tt VL,SL}}$
&$\calO_{X\nu dsu}^{{\tt VR,SR}}$
&$\calO_{X\nu sdu}^{{\tt VR,SR}}$
\\ 
&$\calO_{X\ell udu}^{{\tt SL,VR}-}$
&$\calO_{X\ell udu}^{{\tt SR,VL}-}$%
&$\calO_{X\ell usu}^{{\tt SL,VR}-}$
&$\calO_{X\ell usu}^{{\tt SR,VL}-}$%
& &%
&$\calO_{X\bar{\ell} dsd}^{{\tt SL,VR}-}$
&$\calO_{X\bar{\ell} dsd}^{{\tt SR,VL}-}$%
&$\calO_{X\bar\nu udd}^{{\tt SR,VL}-}$
&$\calO_{X\nu udd}^{{\tt SL,VR}-}$%
&$\calO_{X\bar\nu uds}^{{\tt SR,VL}-}$
&$\calO_{X\bar\nu usd}^{{\tt SR,VL}-}$~~
&$\calO_{X\nu uds}^{{\tt SL,VR}-}$
&$\calO_{X\nu usd}^{{\tt SL,VR}-}$ 
\\ 
&$\calO_{X\ell udu}^{{\tt SL,VR}+}$
&$\calO_{X\ell udu}^{{\tt SR,VL}+}$%
&$\calO_{X\ell usu}^{{\tt SL,VR}+}$
&$\calO_{X\ell usu}^{{\tt SR,VL}+}$%
& &%
&$\calO_{X\bar{\ell} dsd}^{{\tt SL,VR}+}$
&$\calO_{X\bar{\ell} dsd}^{{\tt SR,VL}+}$%
&$\calO_{X\bar\nu udd}^{{\tt SR,VL}+}$
&$\calO_{X\nu udd}^{{\tt SL,VR}+}$%
&$\calO_{X\bar\nu dsu}^{{\tt SR,VL}-}$
&$\calO_{X\bar\nu uds}^{{\tt SR,VL}+}$
&$\calO_{X\nu dsu}^{{\tt SL,VR}-}$
&$\calO_{X\nu uds}^{{\tt SL,VR}+}$
\\ 
&$\calO_{X\ell uud}^{{\tt SL,VR}}$
&$\calO_{X\ell uud}^{{\tt SR,VL}}$%
&$\calO_{X\ell uus}^{{\tt SL,VR}}$
&$\calO_{X\ell uus}^{{\tt SR,VL}}$%
& &%
&$\calO_{X\bar{\ell} dds}^{{\tt SL,VR}}$
&$\calO_{X\bar{\ell} dds}^{{\tt SR,VL}}$%
&$\calO_{X\bar{\nu} ddu}^{{\tt SR,VL}}$
&$\calO_{X\nu ddu}^{{\tt SL,VR}}$%
&$\calO_{X\bar\nu usd}^{{\tt SR,VL}+}$
&$\calO_{X\bar\nu dsu}^{{\tt SR,VL}+}$
&$\calO_{X\nu usd}^{{\tt SL,VR}+}$
&$\calO_{X\nu dsu}^{{\tt SL,VR}+}$
\\\hline
\multirow{4}*{\rotatebox[origin=c]{90}{Case B}}
&$\tilde{\calO}_{X\ell uud}^{{\tt TL,SL}}$ 
&$\tilde{\calO}_{X\ell uud}^{{\tt TR,SR}}$%
&$\tilde{\calO}_{X\ell uus}^{{\tt TL,SL}}$ 
&$\tilde{\calO}_{X\ell uus}^{{\tt TR,SR}}$%
&$\tilde{\calO}_{X\bar{\ell} ddd}^{{\tt VL,VR}}$
&$\tilde{\calO}_{X\bar{\ell} ddd}^{{\tt VR,VL}}$%
&$\tilde{\calO}_{X\bar{\ell} dds}^{{\tt TL,SL}}$
&$\tilde{\calO}_{X\bar{\ell} dds}^{{\tt TR,SR}}$%
&$\tilde{\calO}_{X\bar\nu ddu}^{{\tt TR,SR}}$
&$\tilde{\calO}_{X\nu ddu}^{{\tt TL,SL}}$%
&$\tilde{\calO}_{X\bar\nu dsu}^{{\tt TR,SR}}$
&$\tilde{\calO}_{X\bar\nu sdu}^{{\tt TR,SR}}$
&$\tilde{\calO}_{X\nu dsu}^{{\tt TL,SL}}$
&$\tilde{\calO}_{X\nu sdu}^{{\tt TL,SL}}$
\\ 
&$\tilde{\calO}_{X\ell udu}^{{\tt VL,VR}-}$
&$\tilde{\calO}_{X\ell udu}^{{\tt VR,VL}-}$%
&$\tilde{\calO}_{X\ell usu}^{{\tt VL,VR}-}$
&$\tilde{\calO}_{X\ell usu}^{{\tt VR,VL}-}$%
&$\tilde{\calO}_{X\bar{\ell} ddd}^{{\tt SL,TL}}$
&$\tilde{\calO}_{X\bar{\ell} ddd}^{{\tt SR,TR}}$%
&$\tilde{\calO}_{X\bar{\ell} dsd}^{{\tt VL,VR}-}$
&$\tilde{\calO}_{X\bar{\ell} dsd}^{{\tt VR,VL}-}$%
&$\tilde{\calO}_{X\bar{\nu} udd}^{{\tt VL,VR}-}$
&$\tilde{\calO}_{X\nu udd}^{{\tt VR,VL}-}$ %
&$\tilde{\calO}_{X\bar\nu uds}^{{\tt VL,VR}-}$
&$\tilde{\calO}_{X\bar\nu usd}^{{\tt VL,VR}-}$
&$\tilde{\calO}_{X\nu uds}^{{\tt VR,VL}-}$
&$\tilde{\calO}_{X\nu usd}^{{\tt VR,VL}-}$
\\ 
&$\tilde{\calO}_{X\ell udu}^{{\tt VL,VR}+}$
&$\tilde{\calO}_{X\ell udu}^{{\tt VR,VL}+}$%
&$\tilde{\calO}_{X\ell usu}^{{\tt VL,VR}+}$
&$\tilde{\calO}_{X\ell usu}^{{\tt VR,VL}+}$%
& &%
&$\tilde{\calO}_{X\bar{\ell} dsd}^{{\tt VL,VR}+}$
&$\tilde{\calO}_{X\bar{\ell} dsd}^{{\tt VR,VL}+}$%
&$\tilde{\calO}_{X\bar{\nu} udd}^{{\tt VL,VR}+}$
&$\tilde{\calO}_{X\nu udd}^{{\tt VR,VL}+}$%
&$\tilde{\calO}_{X\bar\nu dsu}^{{\tt VL,VR}-}$
&$\tilde{\calO}_{X\bar\nu uds}^{{\tt VL,VR}+}$
&$\tilde{\calO}_{X\nu dsu}^{{\tt VR,VL}-}$
&$\tilde{\calO}_{X\nu uds}^{{\tt VR,VL}+}$
\\ 
&$\tilde{\calO}_{X\ell uud}^{{\tt VL,VR}}$
&$\tilde{\calO}_{X\ell uud}^{{\tt VR,VL}}$%
&$\tilde{\calO}_{X\ell uus}^{{\tt VL,VR}}$
&$\tilde{\calO}_{X\ell uus}^{{\tt VR,VL}}$%
& &%
&$\tilde{\calO}_{X\bar{\ell} dds}^{{\tt VL,VR}}$
&$\tilde{\calO}_{X\bar{\ell} dds}^{{\tt VR,VL}}$%
&$\tilde{\calO}_{X\bar\nu ddu}^{{\tt VL,VR}}$
&$\tilde{\calO}_{X\nu ddu}^{{\tt VR,VL}}$%
&$\tilde{\calO}_{X\bar\nu usd}^{{\tt VL,VR}+}$
&$\tilde{\calO}_{X\bar\nu dsu}^{{\tt VL,VR}+}$
&$\tilde{\calO}_{X\nu usd}^{{\tt VR,VL}+}$
&$\tilde{\calO}_{X\nu dsu}^{{\tt VR,VL}+}$
\\
&$\tilde{\calO}_{X\ell uud}^{{\tt SL,TL}}$ 
&$\tilde{\calO}_{X\ell uud}^{{\tt SR,TR}}$
&$\tilde{\calO}_{X\ell uus}^{{\tt SL,TL}}$ 
&$\tilde{\calO}_{X\ell uus}^{{\tt SR,TR}}$
&
&%
&$\tilde{\calO}_{X\bar{\ell} dds}^{{\tt SL,TL}}$ 
&$\tilde{\calO}_{X\bar{\ell} dds}^{{\tt SR,TR}}$
&$\tilde{\calO}_{X\bar\nu udd}^{{\tt SR,TR}}$
&$\tilde{\calO}_{X\nu udd}^{{\tt SL,TL}}$%
&$\tilde{\calO}_{X\bar\nu uds}^{{\tt SR,TR}}$
&
&$\tilde{\calO}_{X\nu uds}^{{\tt SL,TL}}$
&
\\\hline
\multirow{8}*{\rotatebox[origin=c]{90}{Nucleon decay modes}}
&\multicolumn{2}{l|}{$p\to e^+ X$ 
(\checkmark)} 
&\multicolumn{2}{l|}{$p\to e^+ K^{0} X$  (\checkmark)}
&\multicolumn{2}{l|}{$n\to e^- \pi^+ X$ (\checkmark)}
&\multicolumn{2}{l|}{$n\to e^- K^+ X$  (\checkmark)}
&\multicolumn{2}{l|}{$n\to \hat\nu_x X$}
&\multicolumn{4}{l|}{$p\to \hat\nu_x K^{+} X$  (\checkmark)}
\\%
&\multicolumn{2}{l|}{$p\to \mu^+ X$
(\checkmark)} 
&\multicolumn{2}{l|}{$p\to \mu^+ K^{0} X$ (\checkmark)}
&\multicolumn{2}{l|}{$n\to \mu^- \pi^+ X$ (\checkmark)}
&\multicolumn{2}{l|}{$n\to \mu^- K^+ X$  (\checkmark)}
&\multicolumn{2}{l|}{$p\to \hat\nu_x \pi^{+} X$}
&\multicolumn{4}{l|}{$n\to \hat\nu_x K^0 X$} 
\\%
& \multicolumn{2}{l|}{$p\to e^+\pi^0 X $}
& & & & & &
& \multicolumn{2}{l|}{$n\to \hat\nu_x \pi^0 X$ (\checkmark)}
& & & & 
\\%
& \multicolumn{2}{l|}{$p\to \mu^+ \pi^0 X$}
& & & & & &
& \multicolumn{2}{l|}{$n\to \hat\nu_x \eta  X$} 
& & & & 
\\%
& \multicolumn{2}{l|}{$p\to e^+ \eta X$}
& & & & & & & & & & & &
\\%
& \multicolumn{2}{l|}{$p\to \mu^+ \eta X$} & & & & & & & & & & & &
\\%
& \multicolumn{2}{l|}{$n\to e^+ \pi^- X$}
& & & & & & & & & & & &
\\%
& \multicolumn{2}{l|}{$n\to \mu^+ \pi^- X$} & & & & & & & & & & & &
\\
\hline
\end{tabular} }
\caption{Summary of relevant nucleon decay modes involving a dark photon $X$ and their corresponding $X$LEFT operators in cases A and B that can induce them. 
The operators in irreps $\bar{\pmb{3}}_{\tL(\tR)}\otimes\pmb{3}_{\tR(\tL)}$ and $\pmb{6}_{\tL(\tR)}\otimes\pmb{3}_{\tR(\tL)}$ follow the convention in \cref{eq:irrep_reduction2}.} 
\label{tab:ope_process}
\end{table}

In this work, we focus on the most prominent two- and three-body nucleon decays induced at leading order by the insertion of $X$LEFT interactions. These processes, along with the corresponding $X$LEFT operators, are summarized in \cref{tab:ope_process}. They include 3 two-body channels $p\to e^+ X$, $p\to \mu^+ X$, and $n\to \hat\nu X$, and 17 three-body modes that contain an additional $\pi$, $\eta$, or $K$ meson.
As neutrinos and antineutrinos are invisible to the detectors, we collectively denote them by $\hat\nu=\nu,\bar\nu$.
Note that the processes are grouped into six categories according to the quark field configurations of the relevant operators. 
The $uud$-type operators give rise to 6 proton decay modes $p\to (e^+,\mu^+) X, (e^+,\mu^+)(\pi^0,\eta) X$, and 2 neutron decays $n\to (e^+,\mu^+)\pi^- X$. 
The $uus$-type operators can only induce three-body decays $p\to (e^+,\mu^+)K^0 X$ due to charge and strangeness conservation. 
The $ddd$-type and $dds$-type operators induce the three-body $\Delta(B+L)=0$ processes $n\to (e^-,\mu^-)\pi^+X$ and $n\to (e^-,\mu^-)K^+ X$, respectively, with the former two processes uniquely changing isospin by $3/2$ units. 
The $udd$-type and $uds$-type operators contribute to the neutrino/antineutrino modes. 
In each field configuration, the operators are further divided into two columns, related to each other by a chirality flip $\tL\leftrightarrow\tR$ and $\nu_\tL\leftrightarrow \nu_\tL^\C$ when neutrinos are involved.

\begin{figure}[t]
\centering
\begin{subfigure}[b]{0.3\textwidth}
\centering
\begin{tikzpicture}[mystyle,scale=0.8]
\begin{scope}[shift={(1,1.5)}] 
\draw[f] (0, 0)node[left]{$\tN$} -- (1.5,0);
\draw[f] (1.5, 0) -- (3,0) node[right]{$l$};
\draw[photon, purple] (1.5,0) -- (2.5,1.2) node[right,yshift = 2 pt]{$X$};
\filldraw [cyan] (1.5,0) circle (3pt);
\end{scope}
\end{tikzpicture}
\caption{}
\label{fig:2body}
\end{subfigure}
\hspace{-1 cm} 
\begin{subfigure}[b]{0.65\textwidth}
\centering
\begin{tikzpicture}[mystyle,scale=0.8]
\begin{scope}[shift={(1,1.2)}] 
\draw[f] (0, 0)node[left]{$\texttt{N}$} -- (1.5,0);
\draw[f] (1.5, 0) -- (3,0) node[midway,yshift = 8 pt]{$B$};
\draw[snar, black] (1.5,0) -- (2.5,-1.2) node[right,yshift = 2 pt]{$M$};
\draw[f] (3.0, 0) -- (4.5,0) node[right]{$l$};
\draw[photon, purple] (3,0) -- (4,1.2) node[right,yshift = 2 pt]{$X$};
\filldraw [black] (1.38,-0.12) rectangle (1.62,0.12);
\filldraw [cyan] (3,0) circle (3pt);
\end{scope}
\end{tikzpicture}
\hspace{0.5cm} 
\begin{tikzpicture}[mystyle,scale=0.8]
\begin{scope}[shift={(1,1.2)}] 
\draw[f] (0, 0)node[left]{$\texttt{N}$} -- (1.5,0);
\draw[f] (1.5, 0) -- (3,0) node[right]{$l$};
\draw[photon, purple] (1.5,0) -- (2.5,1.2) node[right,yshift = 2 pt]{$X$};
\draw[snar, black] (1.5,0) -- (2.5,-1.2) node[right,yshift = 2 pt]{$M$};
\filldraw [cyan] (1.5,0) circle (3pt);
\end{scope}
\end{tikzpicture}
\caption{}
\label{fig:3body}
\end{subfigure}
\caption{Diagrams contributing to two- (a) and three-body (b) nucleon decays containing a dark photon. The cyan blob (black square) denotes a BNV (usual) chiral vertex.}
\label{fig:Feyndiagram}
\end{figure}
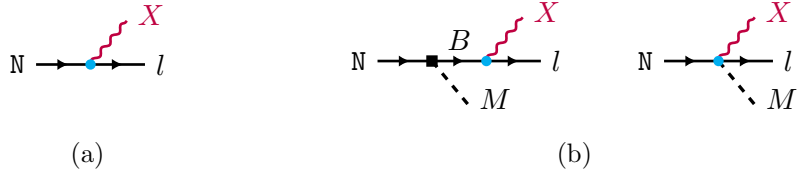

Figure \ref{fig:Feyndiagram} displays the leading-order Feynman diagrams contributing to two- and three-body nucleon decays within the ChPT framework.
The BNV three-point and four-point vertices are derived from \cref{eq:LPB} and \cref{eq:LPNM}, respectively, while the baryon-number-conserving (BNC) three-point vertices are given in \cref{eq:LBNM}.
Based on these vertices and diagrams, we derive general expressions for the squared amplitudes and decay widths in the following subsections.
These results serve as a bridge between theoretical predictions and the underlying models driving such decays. 
For the three-body modes, we also analyze the kinematic distributions of the final-state particles, which are crucial for probing the operator structures and extracting experimental constraints on the decay modes.

From \cref{eq:LPB} and \cref{eq:LPNM}, we find that operators in the irreps 
$\pmb{8}_{\tL(\tR)}\otimes \pmb{1}_{\tR(\tL)}$, 
$\pmb{6}_{\tL(\tR)}\otimes \pmb{3}_{\tR(\tL)}$, 
and $\bar{\pmb{3}}_{\tL(\tR)}\otimes \pmb{3}_{\tR(\tL)}$ 
can contribute to both two-body and three-body modes.  
In contrast, operators in the irreps 
$\pmb{10}_{\tL(\tR)}\otimes \pmb{1}_{\tR(\tL)}$ necessarily involve a pseudoscalar meson 
and thus only contribute to three-body modes ${\tt N}\to l MX$ through the contact diagram in \cref{fig:3body}. 

In case A of the dark photon parametrization, the longitudinal component of the polarization yields decay width contributions that contain inverse powers of dark photon mass. To prevent this superficial divergence in the $m_X\to 0$ limit, we introduce an operational factor of $m_X$ for each operator. This modification effectively promotes the operators to be dimension-8. Our subsequent calculations will explicitly incorporate this mass factor
into the relevant amplitudes.

\subsection{Two-body nucleon decays $\tN\to l+X$}

Since two-body hyperon decays share the same vertex structure as nucleon decays, we incorporate them in this derivation as well. 
For a general two-body baryon decay, $B \to l X$, the effective vertex $B\text{-}l\text{-}X$ can be systematically obtained by replacing the spurion fields ${\cal P}$ in \cref{eq:LPB} with the corresponding entries from \cref{tab:spurion1}. The relevant Lagrangian terms for each process $B\to l X$ in the two cases can then be concisely expressed as follows:
\begin{subequations}
\label{eq:LB2lX}
\begin{align}
{\cal L}_{\bar lBX}^{({\rm A})}  =\,& c_1 m_X X^\mu \,
\overline{l}\Big[ 
C^{1\tL}_{BlX} \gamma_\mu P_\tL
+C^{1\tR}_{BlX} \gamma_\mu P_\tR
+{C^{3\tL}_{BlX} \over \Lambda_\chi} i\tilde\partial_\mu P_\tL
+{ C^{3\tR}_{BlX} \over \Lambda_\chi } i\tilde\partial_\mu  P_\tR \Big] B,
\\
{\cal L}_{\bar lBX}^{({\rm B})}  =\,& 
c_1 X^{\mu\nu}\,\overline{l}\Big[ 
\tilde C^{1\tL}_{BlX} \sigma_{\mu\nu} P_\tL
+\tilde C^{1\tR}_{BlX} \sigma_{\mu\nu} P_\tR 
+{\tilde C^{3\tL}_{BlX} \over \Lambda_\chi}\gamma_\nu i\tilde\partial_\mu P_\tL
+{\tilde C^{3\tR}_{BlX} \over \Lambda_\chi}\gamma_\nu i\tilde\partial_\mu P_\tR
\Big] B,
\end{align}
\end{subequations}
where $m_X$ in case A is added by hand, as we elaborated above. 
The conjugate lepton field 
$\bar l = \overline{\ell^\C}$, $\overline{\ell}$, 
$\overline{\nu_\tL^\C}$, and $\overline{\nu_\tL}$
when the final state lepton $l =\ell^+$, $\ell^-$, $\bar\nu$, and $\nu$, respectively. 

The coefficients $C^{i}_{BlX}$ in \cref{eq:LB2lX} for each two-body process are summarized in \cref{tab:CB2lX}, 
where we have introduced the ratios 
$\kappa_2 \equiv c_2/c_1$ and $\kappa_3 \equiv c_3/c_1$ (and $\kappa_4 \equiv c_4/c_1$ in \cref{tab:CN2lMX_T}).
In particular, 
the coefficients $C^{1\tL(\tR)}_{BlX}$ and $C^{1\tt TL(\tt TR)}_{BlX}$ correspond to operators in the chiral irreps
$\pmb{8}_{\tL(\tR)} \otimes \pmb{1}_{\tR(\tL)}$ and 
$\bar{\pmb{3}}_{\tL(\tR)} \otimes \pmb{3}_{\tR(\tL)}$, 
while the coefficients $C^{3\tL(\tR)}_{BlX}$ and 
$C^{3\tt TL(\tt TR)}_{BlX}$ are instead associated with the irreps $\pmb{6}_{\tL(\tR)} \otimes \pmb{3}_{\tR(\tL)}$. 
These coefficients are also relevant to three-body nucleon decays via the noncontact diagram in \cref{fig:3body}. 

\begin{table}[t]
\centering
\renewcommand{\arraystretch}{1.4}
\resizebox{\linewidth}{!}{
\begin{tabular}{|l|l|l|l|l|}
\hline
\multirow{2}*{$B\to lX$}
& \multicolumn{2}{c|}{Case~A}
& \multicolumn{2}{c|}{Case~B}
\\\hhline{~----}
&  \multicolumn{1}{c|}{
$C^{1\tL}_{BlX}$ (upper) \& 
$C^{1\tR}_{BlX}$ (lower)}
& \multicolumn{1}{c|}{
$C^{3\tL}_{BlX}$ (upper) \& 
$C^{3\tR}_{BlX}$ (lower)}
&  \multicolumn{1}{c|}{
${\tilde C}^{1 \tL}_{BlX}$ (upper) \& 
${\tilde C}^{1 \tR}_{BlX}$ (lower)}
& \multicolumn{1}{c|}{
${\tilde C}^{3 \tL}_{BlX}$ (upper) \& 
${\tilde C}^{3 \tR}_{BlX}$ (lower)}
\\\hline\hline
\multirow{2}*{$p \to \ell_x^+ X$}
& $-C^{{\tt SR,VL}-,x}_{X\ell udu}
+\kappa_2 {C}^{{\tt VL,SL},x}_{X\ell uud}$ 
& $- \kappa_3 \big( C^{{\tt SL,VR}+,x}_{X\ell udu} 
-C^{{\tt SL,VR},x}_{X\ell uud} \big) $
& $i\tilde{C}^{{\tt VR,VL}-,x}_{X   \ell udu}
+ \kappa_2 \tilde{C}^{{\tt TL,SL},x}_{X  \ell uud}$ 
& $ \kappa_3 \big( \tilde{C}^{{\tt VL,VR}+,x}_{X   \ell udu} 
-\tilde{C}^{{\tt VL,VR},x}_{X   \ell uud} \big) $
\\\hhline{~----}
& $ C^{{\tt SL,VR}-,x}_{X\ell udu}
- \kappa_2 {C}^{{\tt VR,SR},x}_{X\ell uud}$ 
& $  \kappa_3 \big( C^{{\tt SR,VL}+,x}_{X\ell udu} 
-C^{{\tt SR,VL},x}_{X\ell uud} \big) $
& $-i\tilde{C}^{{\tt VL,VR}-,x}_{X   \ell udu}
- \kappa_2 \tilde{C}^{{\tt TR,SR},x}_{X  \ell uud}$ 
& $  -\kappa_3 \big( \tilde{C}^{{\tt VR,VL}+,x}_{X\ell udu} 
-\tilde{C}^{{\tt VR,VL},x}_{X\ell uud} \big) $
\\\hline\hline
\multirow{2}*{$\Sigma^+ \to \ell_x^+ X$}
& $  C^{{\tt SR,VL}-,x}_{X   \ell usu} 
- \kappa_2 {C}^{{\tt VL,SL},x}_{X  \ell uus}$
& $ \kappa_3 \big( C^{{\tt SL,VR}+,x}_{X  \ell usu}
- C^{{\tt SL,VR},x}_{X  \ell uus} \big) $
& $  -i\tilde{C}^{{\tt VR,VL}-,x}_{X   \ell usu} 
- \kappa_2 \tilde{C}^{{\tt TL,SL},x}_{X  \ell uus}$
& $ -\kappa_3 \big( \tilde{C}^{{\tt VL,VR}+,x}_{X  \ell usu}
- \tilde{C}^{{\tt VL,VR},x}_{X  \ell uus} \big) $
\\\hhline{~----}
& $ 
- C^{{\tt SL,VR}-,x}_{X\ell usu}
+ \kappa_2 {C}^{{\tt VR,SR},x}_{X\ell uus}$
& $ - \kappa_3 \big( C^{{\tt SR,VL}+,x}_{X\ell usu}
- C^{{\tt SR,VL},x}_{X\ell uus}\big) $
& $  i\tilde{C}^{{\tt VL,VR}-,x}_{X   \ell usu} 
+ \kappa_2 \tilde{C}^{{\tt TR,SR},x}_{X  \ell uus}$
& $  \kappa_3 \big( \tilde{C}^{{\tt VR,VL}+,x}_{X\ell usu}
- \tilde{C}^{{\tt VR,VL},x}_{X\ell uus}\big) $
\\\hline\hline
&$ -C^{{\tt SR,VL}-,x}_{X\bar\ell dsd} 
+ \kappa_2 {C}^{{\tt VL,SL},x}_{X\bar\ell dds} $
&$-\kappa_3 \big(C^{{\tt SL,VR}+,x}_{X\bar\ell dsd}
 - C^{{\tt SL,VR},x}_{X\bar\ell dds} \big) $
&$i\tilde{C}^{{\tt VR,VL}-,x}_{X\bar\ell dsd} 
+ \kappa_2 \tilde{C}^{{\tt TL,SL},x}_{X\bar\ell dds} $
&$ \kappa_3\big(\tilde{C}^{{\tt VL,VR}+,x}_{X\bar \ell dsd}
 - \tilde{C}^{{\tt VL,VR},x}_{X\bar\ell dds} \big) $
\\\hhline{~----}
\multirow{-2}*{$\Sigma^-\to \ell_x^- X$}
&$C^{{\tt SL,VR}-,x}_{X\bar\ell dsd}
- \kappa_2 {C}^{{\tt VR,SR},x}_{X\bar\ell dds} $
&$ \kappa_3 \big(
C^{{\tt SR,VL}+,x}_{X\bar\ell dsd} 
- C^{{\tt SR,VL},x}_{X\bar\ell dds} \big) $
&$-i\tilde{C}^{{\tt VL,VR}-,x}_{X\bar\ell dsd}
- \kappa_2 \tilde{C}^{{\tt TR,SR},x}_{X\bar\ell dds} $
&$-\kappa_3 \big(
\tilde{C}^{{\tt VR,VL}+,x}_{X\bar\ell dsd} 
- \tilde{C}^{{\tt VR,VL},x}_{X\bar\ell dds} \big) $
\\\hline\hline
\multirow{2}*{$n\to \hat\nu_x X$}
&$- C^{{\tt SR,VL}-,x}_{X\bar \nu udd} 
-\kappa_2 {C}^{{\tt VL,SL},x}_{X\bar \nu ddu}$
& $ \kappa_3\big( C^{{\tt SL,VR}+,x}_{X \nu udd}
- C^{{\tt SL,VR},x}_{X\nu ddu} \big)$
& $i\tilde{C}^{{\tt VR,VL}-,x}_{X \nu udd} 
-\kappa_2 \tilde{C}^{{\tt TL,SL},x}_{X \nu ddu}$
&$ -\kappa_3\big( \tilde{C}^{{\tt VL,VR}+,x}_{X \bar\nu udd}
- \tilde{C}^{{\tt VL,VR},x}_{X\bar\nu ddu} \big)$
\\\hhline{~----}
& $ C^{{\tt SL,VR}-,x}_{X \nu udd}
+ \kappa_2 {C}^{{\tt VR,SR},x}_{X\nu ddu} $
&$ -\kappa_3 \big( 
C^{{\tt SR,VL}+,x}_{X\bar \nu udd}
-C^{{\tt SR,VL},x}_{X\bar\nu ddu}
\big)$ 
&$ -i\tilde{C}^{{\tt VL,VR}-,x}_{X\bar\nu udd}
+ \kappa_2 \tilde{C}^{{\tt TR,SR},x}_{X\bar\nu ddu} $
& $ \kappa_3 \big( 
\tilde{C}^{{\tt VR,VL}+,x}_{X \nu udd}
-\tilde{C}^{{\tt VR,VL},x}_{X\nu ddu}
\big)$ 
\\\hline\hline
\multirow{4}*{$\Lambda^0 \to \hat\nu_x X$}
&${1\over\sqrt{6}} \big[\big(
2C^{{\tt SR,VL}-,x}_{X  \bar \nu uds} 
+ C^{{\tt SR,VL}-,x}_{X  \bar \nu usd}-$
& $ - \sqrt{3 \over 2} \kappa_3 \big( 
C^{{\tt SL,VR}+,x}_{X  \nu usd}
- C^{{\tt SL,VR}+,x}_{X  \nu dsu}
 \big) $ 
& $- {1 \over \sqrt{6}}\big[ i\big(
2\tilde{C}^{{\tt VR,VL}-,x}_{X   \nu uds} 
+ \tilde{C}^{{\tt VR,VL}-,x}_{X   \nu usd}-$
&$  \sqrt{3 \over 2} \kappa_3 \big( 
\tilde{C}^{{\tt VL,VR}+,x}_{X  \bar\nu usd}
- \tilde{C}^{{\tt VL,VR}+,x}_{X  \bar\nu dsu}
 \big) $ 
\\
&$C^{{\tt SR,VL}-,x}_{X\bar\nu dsu} \big)
+\kappa_2\big( 
{C}^{{\tt VL,SL},x}_{X  \bar \nu dsu} 
+2 {C}^{{\tt VL,SL},x}_{X  \bar\nu sdu} \big)\big] $
&
&$\tilde{C}^{{\tt VR,VL}-,x}_{X  \nu dsu} \big)
-\kappa_2\big( 
\tilde{C}^{{\tt TL,SL},x}_{X   \nu dsu} 
+2 \tilde{C}^{{\tt TL,SL},x}_{X  \nu sdu} \big)\big] $
&
\\\hhline{~----}%
&$- {1\over \sqrt{6}} \big[\big( 
2C^{{\tt SL,VR}-,x}_{X\nu uds}
+ C^{{\tt SL,VR}-,x}_{X\nu usd}-$
&$ \sqrt{3\over 2} \kappa_3 \big( 
C^{{\tt SR,VL}+,x}_{X\bar\nu usd}
-C^{{\tt SR,VL}+,x}_{X\bar\nu dsu} \big) $
&$ {1\over \sqrt{6}} \big[i\big( 
2\tilde{C}^{{\tt VL,VR}-,x}_{X\bar\nu uds}
+ \tilde{C}^{{\tt VL,VR}-,x}_{X\bar\nu usd}- 
$
& $ -\sqrt{3\over 2} \kappa_3 \big( 
\tilde{C}^{{\tt VR,VL}+,x}_{X  \nu usd}
-\tilde{C}^{{\tt VR,VL}+,x}_{X  \nu dsu} \big) $
\\
&$C^{{\tt SL,VR}-,x}_{X\nu dsu} \big) 
+ \kappa_2 \big( 
{C}^{{\tt VR,SR},x}_{X  \nu dsu} 
+2 {C}^{{\tt VR,SR},x}_{X  \nu sdu}\big)\big] $
&
&$\tilde{C}^{{\tt VL,VR}-,x}_{X\bar\nu dsu} \big) 
- \kappa_2 \big( 
\tilde{C}^{{\tt TR,SR},x}_{X  \bar\nu dsu} 
+2 \tilde{C}^{{\tt TR,SR},x}_{X  \bar\nu sdu}\big)\big]$
&
\\\hline
\multirow{4}*{$\Sigma^0 \to \hat\nu_x X$}
&$ - {1 \over \sqrt{2}} \big( 
C^{{\tt SR,VL}-,x}_{X\bar \nu usd}
+ C^{{\tt SR,VL}-,x}_{X\bar\nu dsu} \big)$
&$ {\kappa_3 \over \sqrt{2}} \big( 
 2 C^{{\tt SL,VR}+,x}_{X   \nu uds}
-C^{{\tt SL,VR}+,x}_{X   \nu usd}$
&$ {i \over \sqrt{2}} \big( 
 \tilde{C}^{{\tt VR,VL}-,x}_{X \nu usd}
+ \tilde{C}^{{\tt VR,VL}-,x}_{X  \nu dsu} \big)$
&$ -{\kappa_3 \over \sqrt{2}} \big( 
2 \tilde{C}^{{\tt VL,VR}+,x}_{X  \bar \nu uds}
-\tilde{C}^{{\tt VL,VR}+,x}_{X  \bar \nu usd}
$
\\
& $- {\kappa_2 \over \sqrt{2}} 
C^{{\tt VL,SL},x}_{X  \bar\nu dsu}  $ 
& $- C^{{\tt SL,VR}+,x}_{X \nu dsu}  \big) $
& $- {\kappa_2 \over \sqrt{2}} 
\tilde{C}^{{\tt TL,SL},x}_{X  \nu dsu}  $
& $- \tilde{C}^{{\tt VL,VR}+,x}_{X \bar\nu dsu}  \big) $
\\\hhline{~----}
&$ {1\over \sqrt{2}}\big(
C^{{\tt SL,VR}-,x}_{X  \nu usd}
+C^{{\tt SL,VR}-,x}_{X\nu dsu}  \big)$
&$ - {\kappa_3\over \sqrt{2}} \big(
2 C^{{\tt SR,VL}+,x}_{X\bar \nu uds}
-C^{{\tt SR,VL}+,x}_{X\bar \nu usd}$
&${-i\over \sqrt{2}}\big(
\tilde{C}^{{\tt VL,VR}-,x}_{X  \bar\nu usd}
+\tilde{C}^{{\tt VL,VR}-,x}_{X\bar\nu dsu}  \big)$
&${\kappa_3\over \sqrt{2}} \big(
2 \tilde{C}^{{\tt VR,VL}+,x}_{X \nu uds}
-\tilde{C}^{{\tt VR,VL}+,x}_{X \nu usd}$
\\
&$ + {\kappa_2\over \sqrt{2}}
C^{{\tt VR,SR},x}_{X\nu dsu}$
& $-C^{{\tt SR,VL}+,x}_{X\bar \nu dsu} \big)$
& $ + {\kappa_2\over \sqrt{2}}
\tilde{C}^{{\tt TR,SR},x}_{X\bar\nu dsu}$
& $-\tilde{C}^{{\tt VR,VL}+,x}_{X \nu dsu} \big)$
\\\hline
\end{tabular} }
\caption{The explicit expressions of the coefficients $C_{B lX}^{1,3\tL(\tR)}$ in case A and $\tilde C_{BlX}^{1,3\tL(\tR)}$ in case B for each two-body transition mode.}
\label{tab:CB2lX}
\end{table}

From the effective Lagrangian in \cref{eq:LB2lX}, the amplitude for the process $B(p_B)\to l(p_l)X(p_X)$ can be directly obtained for both cases A and B,   
\begin{subequations}
\begin{align}
{\cal M}_{B\to l X}^{\rm (A)}
=\,& c_1 m_X  \epsilon_{X,\mu}^* 
\overline{u_l} \Big[\gamma^\mu 
(C_{BlX }^{1\tL}P_{\tL}+C_{BlX }^{1\tR}P_{\tR})
\nn\\
& +\Lambda_\chi^{-1}\big(p_B^\mu-\frac{1}{4}\gamma^\mu\slashed{p}_B\big)(C_{BlX }^{3\tL}P_{\tL}+C_{BlX }^{3\tR}P_{\tR}) \Big] u_B,
\label{eq:amp2body_caseA}
\\%
{\cal M}_{B\to l X}^{\rm (B)}
=\,& ic_1 \big(p_{X,\mu}\epsilon_{X,\nu}^*-p_{X,\nu}\epsilon_{X,\mu}^*\big)
\overline{u_l} \Big[\sigma^{\mu\nu} 
(\tilde C_{BlX}^{1\tL}P_{\tL}
+\tilde C_{BlX}^{1\tR}P_{\tR})
\nn\\
&+\Lambda_\chi^{-1}\gamma^\nu(p_B^\mu-\frac{1}{4}\gamma^\mu\slashed{p}_B)
(\tilde C_{BlX}^{3\tL}P_{\tL}
+\tilde C_{BlX}^{3\tR}P_{\tR}) \Big] u_B.
\label{eq:amp2body_caseB}
\end{align}
\end{subequations}
Defining the mass ratios $x_l \equiv m_l^2/m_B^2$ and $x_X\equiv m_X^2/m_B^2$, and introducing $\rho_+\equiv 1+x_l-x_X$, 
the spin-summed and -averaged matrix element squared takes the following compact form: 
\begin{subequations}
\begin{align}
\overline{\big|{\cal M}_{B\to lX}^{\rm (A)}\big|^2}
=\,&\frac{c_1^2 m_B^4}{16}\Big\{ 
8 \big[ \lambda(1,x_l,x_X) + 3 x_X\rho_+\big]
|C_{B lX}^{1\tL}|^2
\nn\\
& +\frac{m_B^2}{2\Lambda_\chi^2}  
\big[ (4\rho_+-3)\lambda(1,x_l,x_X) 
+ 3x_X \rho_+\big]
|C_{B lX}^{3\tL}|^2
\nn\\
& + \frac{8 m_B}{\Lambda_\chi}
\sqrt{x_l}\big[ \lambda(1,x_l,x_X) +3 x_X \big]
\Re({C_{B lX}^{1\tL}C_{B lX}^{{3\tL}*}})
\nn\\
& + \frac{4 m_B}{\Lambda_\chi}
\big[ \lambda(1,x_l,x_X) -3 x_X \rho_+ \big]
\Re({C_{B lX}^{1\tL}C_{B lX}^{{3\tR}*}})
-48\sqrt{x_l}x_X\Re({C_{B lX}^{1\tL}C_{B lX}^{{1\tR}*}})
\nn\\
&+\frac{m_B^2}{\Lambda_\chi^2}
\sqrt{x_l} \big[ 2\lambda(1,x_l,x_X) - 3 x_X \big]
\Re({C_{B lX}^{3\tL}C_{B lX}^{3\tR*}})\Big\}+\tL\leftrightarrow\tR,
\label{eq:Amp_2body_caseA}
\\%
\overline{\big|{\cal M}_{B\to lX}^{\rm (B)}\big|^2}
=\,& c_1^2m_B^4\Big\{ 
\big[ 4\lambda(1,x_l,x_X)+6x_X\rho_+\big] 
|{\tilde C}_{BlX}^{1\tL}|^2
\nn\\
&+\frac{m_B^2}{8\Lambda_\chi^2}
\big[ 2(\rho_+ - 1)\lambda(1,x_l,x_X)+3x_X\rho_+\big]
|{\tilde C}_{BlX}^{3\tL}|^2
\nn\\
&-\frac{2m_B}{\Lambda_\chi}\sqrt{x_l} \big[\lambda(1,x_l,x_X)+3x_X \big]
\Im({\tilde C}_{BlX}^{1\tL} {\tilde C}_{BlX}^{3\tL*})
+\frac{3m_B}{\Lambda_\chi} x_X \rho_+
\Im({\tilde C}_{BlX}^{1\tL} {\tilde C}_{BlX}^{3\tR*})
\nn\\
&-12 \sqrt{x_l} x_X 
\Re({\tilde C}_{BlX}^{1\tL} {\tilde C}_{BlX}^{1\tR*})
-\frac{3m_B^2}{4\Lambda_\chi^2}\sqrt{x_l} x_X 
\Re({\tilde C}_{BlX}^{3\tL}{\tilde C}_{BlX}^{3\tR*}) 
\Big\}+\tL\leftrightarrow\tR,
\label{eq:Amp_2body_caseB}
\end{align}
\end{subequations}
where $\lambda(x,y,z)\equiv x^2+y^2+z^2-2xy-2yz-2zx$ is the triangle function. For notational simplicity, the real and imaginary parts of a complex quantity $z$ are denoted by $\Re(z)$ and $\Im(z)$, respectively. 
In the joint limit of $m_l\to 0$ and $m_X \to 0$, we see that the terms associated with $\tilde C_{BlX}^{3\tL(\tR)}$ vanish.

Finally, the decay width is given by
\begin{align}
\Gamma_{B \to l X} =\,& \frac{ \overline{|{\cal M}_{B\to lX}|^2} }{ 16\pi m_B } \lambda^{1/2}(1,x_l, x_X).
\end{align}
Using the formalism above, the decay widths for $B\to lX$ can be expressed in terms of the XLEFT WCs for a given dark photon mass. The expressions for the 3 two-body nucleon decay modes with a massless dark photon are collected in \cref{eq:Ga_N2lXA} and \cref{eq:Ga_N2lXB} of \cref{app:Gamma_exp}, corresponding 
to the operators in cases A and B, respectively.
For instance, for the $\pmb{8}_\tL\otimes \pmb{1}_\tR$ irrep operators $\calO_{X\ell uud}^{{\tt VL,SL},e}$ and
$\tilde\calO_{X\ell uud}^{{\tt TL,SL},e}$, their contributions to $p\to e^+ X$ are 
\begin{subequations}
\begin{align}
\Gamma_{p\to e^+ X}^{\rm (A)-1} =\,&  \frac{1}{1.3\cdot 10^{-6} ({\rm GeV})^9 |C_{X\ell uud}^{{\tt VL,SL},e}|^2}   
=(7.9\cdot 10^{32}\,{\rm yr})\left(\frac{\Lambda_{\rm eff}}{2.17\cdot10^7{\rm GeV}}\right)^8,
\\
\Gamma_{p\to e^+ X}^{\rm (B)-1} =\,& \frac{1}{1.1\cdot 10^{-5} ({\rm GeV})^9 |\tilde C_{X\ell uud}^{{\tt TL,SL},e}|^2}
=(7.9\cdot 10^{32}\,{\rm yr})\left(\frac{\Lambda_{\rm eff}}{2.83\cdot10^7{\rm GeV}}\right)^8,
\end{align}
\end{subequations}
where the number in parentheses corresponds to the current experimental limit on the decay~\cite{ParticleDataGroup:2024cfk}, and 
the effective scale is defined as $\Lambda_{\rm eff} = |C_i|^{-1/4}$.

\subsection{Three-body nucleon decays $\texttt{N}\to l+M+X$}

\begin{table}[t]
\center
\resizebox{\linewidth}{!}{
\renewcommand{\arraystretch}{1.1}
\begin{tabular}{|l| l | l | c | c| }
\hline
\multirow{2}*{$\texttt{N}\to lMX$}
&\multicolumn{2}{c|}{case A}
&\multicolumn{2}{c|}{Noncontact diagram}
\\\hhline{~----}
&\multicolumn{1}{c|}{$C^{1\tL}_{\tN lMX}$ (upper) and $C^{1\tR}_{\tN lMX}$ (lower)}
&\multicolumn{1}{c|}{$C^{3\tL}_{\tN lMX}$ (upper) and $C^{3\tR}_{\tN lMX}$ (lower)} 
& $C_{\tN B M}$
& $C_{BlX}^{1,3\tL(\tR)}$
\\\hline\hline
\multirow{2}*{$p\to\ell_x^+ \pi^0 X$}
&$-{1\over 2}\big( 
C^{{\tt SR,VL}-,x}_{X\ell udu}
-\kappa_2 C^{{\tt VL,SL},x}_{X\ell uud}\big)$ 
&${\kappa_3 \over 2}\big( 
C^{{\tt SL,VR}+,x}_{X\ell udu} 
+3C^{{\tt SL,VR},x}_{X\ell uud}\big)$
& \multirow{2}*{$C_{pp\pi^0}$}
& \multirow{6}*{$C_{p\ell^+ X}^{1,3\tL(\tR)}$}
\\\hhline{~--}
&$-{1 \over 2}\big( 
C^{{\tt SL,VR}-,x}_{X\ell udu} 
-\kappa_2 C^{{\tt VR,SR},x}_{X\ell uud}\big)$
&${\kappa_3 \over 2}\big( 
C^{{\tt SR,VL}+,x}_{X\ell udu}
+3C^{{\tt SR,VL},x}_{X\ell uud}\big)$
& &
\\\hhline{----}
\multirow{2}*{$p\to\ell_x^+ \eta X$}
&${1\over 2\sqrt{3}}\big(
C^{{\tt SR,VL}-,x}_{X\ell udu}
+3\kappa_2C^{{\tt VL,SL},x}_{X\ell uud}\big)$ 
&$-{\kappa_3 \over 2\sqrt{3}}\big( 
C^{{\tt SL,VR}+,x}_{X\ell udu} 
-C^{{\tt SL,VR},x}_{X\ell uud}\big)$
& \multirow{2}*{$C_{pp\eta}$}
&
\\\hhline{~--}
&${1\over 2\sqrt{3}}\big( 
C^{{\tt SL,VR}-,x}_{X\ell udu}
+ 3\kappa_2 C^{{\tt VR,SR},x}_{X\ell uud}\big)$ 
&$-{\kappa_3 \over 2\sqrt{3}}\big( 
C^{{\tt SR,VL}+,x}_{X\ell udu}
-C^{{\tt SR,VL},x}_{X\ell uud}\big)$
& &
\\\hhline{----}
\multirow{2}*{$n\to\ell_x^+ \pi^- X$}
&$-{ 1\over \sqrt{2}}\big( 
C^{{\tt SR,VL}-,x}_{X\ell udu} 
-\kappa_2 C^{{\tt VL,SL},x}_{X\ell uud}\big)$
&$-{\kappa_3 \over\sqrt{2}}\big(
3C^{{\tt SL,VR}+,x}_{X\ell udu} 
-C^{{\tt SL,VR},x}_{X\ell uud}\big)$
& \multirow{2}*{$C_{np\pi^-}$}
&
\\\hhline{~--}
&$- {1\over\sqrt{2}}\big( 
C^{{\tt SL,VR}-,x}_{X\ell udu} 
-\kappa_2 C^{{\tt VR,SR},x}_{X\ell uud}\big)$
&$-{\kappa_3\over\sqrt{2}}\big(
3C^{{\tt SR,VL}+,x}_{X\ell udu} 
-C^{{\tt SR,VL},x}_{X\ell uud}\big)$
& &
\\\hline\hline
\multirow{2}*{$p\to\ell_x^+ K^0 X$}
&${1\over\sqrt{2}}\big( 
C^{{\tt SR,VL}-,x}_{X\ell usu}
+\kappa_2 C^{{\tt VL,SL},x}_{X\ell uus}\big)$ 
&$-{\kappa_3\over\sqrt{2}}\big(
C^{{\tt SL,VR}+,x}_{X\ell usu} 
+C^{{\tt SL,VR},x}_{X\ell uus}\big)$
& \multirow{2}*{$C_{p\Sigma^+ K^0}$}
& \multirow{2}*{$C_{\Sigma^+\ell^+ X}^{1,3\tL(\tR)}$}
\\\hhline{~--}
&${1\over \sqrt{2}}\big( 
C^{{\tt SL,VR}-,x}_{X\ell usu}
+\kappa_2 C^{{\tt VR,SR},x}_{X\ell uus}\big)$
&$-{\kappa_3\over\sqrt{2}}\big(
C^{{\tt SR,VL}+,x}_{X\ell usu}
+C^{{\tt SR,VL},x}_{X\ell uus}\big)$
& &
\\\hline\hline
\multirow{2}*{$n\to\ell_x^-\pi^+ X$}
&---
&$\sqrt{2}\kappa_3 C^{{\tt SL,VR},x}_{X\bar\ell ddd}$
& \multirow{2}*{---}
& \multirow{2}*{---}
\\\hhline{~--}
&---
&$\sqrt{2}\kappa_3 C^{{\tt SR,VL},x}_{X\bar\ell ddd}$
& &
\\\hline\hline
\multirow{2}*{$n\to\ell_x^- K^+ X$} 
&$-{1\over \sqrt{2}}\big(
C^{{\tt SR,VL}-,x}_{X\bar\ell dsd}
+\kappa_2 C^{{\tt VL,SL},x}_{X\bar\ell dds}\big)$ 
&${\kappa_3 \over \sqrt{2}}\big( 
C^{{\tt SL,VR}+,x}_{X\bar\ell dsd}
+C^{{\tt SL,VR},x}_{X\bar\ell dds}\big)$
& \multirow{2}*{$C_{n\Sigma^- K^+}$}
& \multirow{2}*{$C_{\Sigma^-\ell^- X}^{1,3\tL(\tR)}$}
\\\hhline{~--}
&$-{1\over\sqrt{2}}\big(
C^{{\tt SL,VR}-,x}_{X\bar\ell dsd} 
+\kappa_2 C^{{\tt VR,SR},x}_{X\bar\ell dds}\big)$
&${\kappa_3\over\sqrt{2}}\big(
C^{{\tt SR,VL}+,x}_{X\bar\ell dsd}
+C^{{\tt SR,VL},x}_{X\bar\ell dds}\big)$
& & 
\\\hline\hline
\multirow{2}*{$p\to\hat\nu_x\pi^+ X$}
&$-{1\over \sqrt{2}}\big( 
C^{{\tt SR,VL}-,x}_{X\bar\nu udd} 
+\kappa_2 C^{{\tt VL,SL},x}_{X\bar\nu ddu}\big)$ 
&${\kappa_3 \over \sqrt{2}}\big(
3 C^{{\tt SL,VR}+,x}_{X\nu udd}
- C^{{\tt SL,VR},x}_{X\nu ddu}\big)$
& \multirow{2}*{$C_{pn\pi^+}$}
& \multirow{6}*{$C_{n\hat\nu X}^{1,3\tL(\tR)}$}
\\\hhline{~--}
&$-{1\over\sqrt{2}}\big( 
C^{{\tt SL,VR}-,x}_{X\nu udd} 
+\kappa_2C^{{\tt VR,SR},x}_{X\nu ddu}\big)$
&${\kappa_3\over\sqrt{2}}\big( 
3C^{{\tt SR,VL}+,x}_{X\bar\nu udd} 
-C^{{\tt SR,VL},x}_{X\bar\nu ddu}\big)$
& &
\\\hhline{----}
\multirow{2}*{$n\to\hat\nu_x\pi^0 X$}
&${1\over 2}\big( 
C^{{\tt SR,VL}-,x}_{X\bar\nu udd} 
+\kappa_2 C^{{\tt VL,SL},x}_{X\bar\nu ddu}\big)$
&${\kappa_3 \over 2}\big( 
C^{{\tt SL,VR}+,x}_{X\nu udd}
+3C^{{\tt SL,VR},x}_{X\nu ddu}\big)$
& \multirow{2}*{$C_{nn\pi^0}$}
&
\\\hhline{~--}
&${1\over 2}\big( 
C^{{\tt SL,VR}-,x}_{X\nu udd} 
+\kappa_2 C^{{\tt VR,SR},x}_{X\nu ddu}\big)$
&${\kappa_3 \over 2}\big(
C^{{\tt SR,VL}+,x}_{X\bar\nu udd}
+3C^{{\tt SR,VL},x}_{X\bar\nu ddu}\big)$
& & 
\\\hhline{----}
\multirow{2}*{$n\to\hat\nu_x \eta X$}
&${1 \over 2\sqrt{3}}\big( 
C^{{\tt SR,VL}-,x}_{X\bar\nu udd} 
-3\kappa_2 C^{{\tt VL,SL},x}_{X\bar\nu ddu}\big)$
&${\kappa_3 \over 2\sqrt{3}}\big( 
C^{{\tt SL,VR}+,x}_{X\nu udd}
- C^{{\tt SL,VR},x}_{X\nu ddu}\big)$
& \multirow{2}*{$C_{nn\eta}$}
&
\\\hhline{~--}
&${1\over 2\sqrt{3}}\big( 
C^{{\tt SL,VR}-,x}_{X\nu udd}
-3\kappa_2 C^{{\tt VR,SR},x}_{X\nu ddu}\big)$
&${\kappa_3\over 2\sqrt{3}}\big( 
C^{{\tt SR,VL}+,x}_{X\bar\nu udd}
-C^{{\tt SR,VL},x}_{X\bar\nu ddu}\big)$
& &
\\\hline\hline
\multirow{2}*{$p\to\hat\nu_x K^+ X$}
&$-{1\over \sqrt{2}}\big( 
C^{{\tt SR,VL}-,x}_{X\bar\nu uds}
+C^{{\tt SR,VL}-,x}_{X\bar\nu dsu} 
+\kappa_2 C^{{\tt VL,SL},x}_{X\bar\nu sdu}\big)$
&${\kappa_3\over\sqrt{2}}\big( 
C^{{\tt SL,VR}+,x}_{X\nu uds}
+2C^{{\tt SL,VR}+,x}_{X\nu usd}
-C^{{\tt SL,VR}+,x}_{X\nu dsu}\big)$
& \multirow{2}*{$C_{p\Lambda^0 K^+}$}
& \multirow{2}*{$C_{\Lambda^0\hat\nu X}^{1,3\tL(\tR)}$}
\\\hhline{~--}
&$-{1\over \sqrt{2}}\big( 
C^{{\tt SL,VR}-,x}_{X\nu uds}
+C^{{\tt SL,VR}-,x}_{X\nu dsu}
+\kappa_2 C^{{\tt VR,SR},x}_{X\nu sdu}\big)$
&${\kappa_3 \over \sqrt{2}}\big( 
C^{{\tt SR,VL}+,x}_{X\bar\nu uds}
+2C^{{\tt SR,VL}+,x}_{X\bar\nu usd}
-C^{{\tt SR,VL}+,x}_{X\bar\nu dsu}\big)$
& &
\\\hhline{---}
\multirow{2}*{$n\to\hat\nu_x K^0 X$}
&$-{1\over\sqrt{2}}\big( 
C^{{\tt SR,VL}-,x}_{X\bar\nu uds} 
-C^{{\tt SR,VL}-,x}_{X\bar\nu usd}\big) 
-{\kappa_2 \over\sqrt{2}}\big( 
C^{{\tt VL,SL},x}_{X\bar\nu dsu}
+C^{{\tt VL,SL},x}_{X\bar\nu sdu}\big)$
&$-{\kappa_3 \over\sqrt{2}}\big( 
C^{{\tt SL,VR}+,x}_{X\nu uds} 
-C^{{\tt SL,VR}+,x}_{X\nu usd}
+2C^{{\tt SL,VR}+,x}_{X\nu dsu}\big)$
& \multirow{2}*{$C_{n\Sigma^0 K^0}$}
& \multirow{2}*{$C_{\Sigma^0\hat\nu X}^{1,3\tL(\tR)}$}
\\\hhline{~--}
&$-{1 \over \sqrt{2}}\big( 
C^{{\tt SL,VR}-,x}_{X\nu uds} 
-C^{{\tt SL,VR}-,x}_{X\nu usd}\big)
-{\kappa_2\over\sqrt{2}}\big(
C^{{\tt VR,SR},x}_{X\nu dsu}
+C^{{\tt VR,SR},x}_{X\nu sdu} \big)$
&$-{\kappa_3\over\sqrt{2}}\big( 
C^{{\tt SR,VL}+,x}_{X\bar\nu uds}
-C^{{\tt SR,VL}+,x}_{X\bar\nu usd}
+2C^{{\tt SR,VL}+,x}_{X\bar\nu dsu}\big)$
& &
\\\hline
\end{tabular}}
\caption{The explicit expressions of the coefficients $C_{\tN lMX}^{1,3\tL(\tR)}$ governing the contact contributions to each three-body nucleon decay in case A.}
\label{tab:CN2lMX}
\end{table}

We begin with an
analysis of the contact contribution represented by the second diagram of \cref{fig:3body}. 
The local BNV vertices involve a nucleon, a meson, a lepton, and a dark photon.
By taking the spurion fields listed in 
\cref{tab:spurion1} into \cref{eq:LPNM}, 
we identify four (six) structures for the local interactions governing the general three-body nucleon decay $\tN\to l X M$ in case A (B). These can be parametrized as follows: 
{\small
\begin{subequations}
\label{eq:LN2lMX}
\begin{align}
{\cal L}_{\bar l \tN M X}^{({\rm A})} =\,&
\frac{ic_1}{F_0} m_X {X^\mu}
\bar{M} \overline{l}\Big[ 
C^{1\tL}_{\tN lMX}\, \gamma_\mu P_{\tL}
+C^{1\tR}_{\tN lMX}\,\gamma_\mu P_{\tR} 
+{ C^{3\tL}_{\tN lMX} \over \Lambda_\chi}  i\tilde\partial_\mu P_{\tL}
+{ C^{3\tR}_{\tN lMX} \over \Lambda_\chi}  i\tilde\partial_\mu P_{\tR}
\Big]\tN\;,
\label{eq:LN2lMXa}
\\
{\cal L}_{\bar l \tN M X}^{({\rm B})} =\,&
\frac{ic_1}{F_0} X^{\mu\nu} \Big\{ 
\bar{M} \overline{l}
\Big[ {\tilde C}^{1\tL}_{\tN lMX}\sigma_{\mu\nu}P_{\tL}+{\tilde C}^{1\tR}_{\tN lMX}\sigma_{\mu\nu}P_{\tR}
+{ {\tilde C}^{3\tL}_{\tN lMX} \over \Lambda_\chi} \gamma_\nu i\tilde\partial_\mu P_{\tL}
+{ {\tilde C}^{3 \tR}_{\tN lMX} \over \Lambda_\chi}  \gamma_\nu i\tilde\partial_\mu P_{\tR}
\Big] \tN
\nn\\
&+\Lambda_\chi^{-2} \overline{l} \big(
{\tilde C}^{4\tL}_{\tN lMX} 
\hat{\Gamma}^{\tL}_{\mu\nu\alpha\beta} 
+{\tilde C}^{4 \tR}_{\tN lMX}
\hat{\Gamma}^{\tR}_{\mu\nu\alpha\beta} \big)
\partial^\alpha\tN\, \partial^\beta\bar{M}\Big\}.
\label{eq:LN2lMXb}
\end{align}
\end{subequations}}%
From \cref{tab:spurion1} and \cref{eq:LPNM},
we can extract explicit expressions for the coefficients 
$C_{\tN lMX}^{1,3\tL(\tR)}$ 
and ${\tilde C}^{1,3,4\tL(\tR)}_{\tN lMX}$ corresponding to each three-body nucleon decay process. 
These results are systematically summarized in \cref{tab:CN2lMX} (case A) and \cref{tab:CN2lMX_T} (case B).
Here, the coefficients 
$C^{1\tL(\tR)}_{\tN lMX}$ and 
${\tilde C}^{1\tL(\tR)}_{\tN lMX}$,
$C^{3\tL(\tR)}_{\tN lMX}$ and 
$\tilde C^{3\tL(\tR)}_{\tN lMX}$,
and $\tilde C^{4\tL(\tR)}_{\tN lMX}$
are associated with the operators in the irreps 
$\pmb{8}_{\tL(\tR)} \otimes \pmb{1}_{\tR(\tL)}$ and 
$\bar{\pmb{3}}_{\tL(\tR)} \otimes \pmb{3}_{\tR(\tL)}$, 
$\pmb{6}_{\tL(\tR)} \otimes \pmb{3}_{\tR(\tL)}$, 
and $\pmb{10}_{\tL(\tR)} \otimes \pmb{1}_{\tR(\tL)}$,
respectively.
As shown in the tables, the $ddd$-type operators belong exclusively to the irreps $\pmb{6}_{\tL(\tR)} \otimes \pmb{3}_{\tR(\tL)}$ and $\pmb{10}_{\tL(\tR)} \otimes \pmb{1}_{\tR(\tL)}$. They contribute uniquely to the $\Delta I=3/2$ transition processes $n\to (e^-,\mu^-)\pi^+X$.

\begin{table}[t]
\center
\resizebox{\linewidth}{!}{
\renewcommand{\arraystretch}{1.2}
\begin{tabular}{|l| l | l | l |}
\hline
\multirow{2}*{$\texttt{N}\to lMX$}
&\multicolumn{3}{c|}{Case~B}
\\\hhline{~---}
&  \multicolumn{1}{c|}{
${\tilde C}^{1 \tL}_{\tN lMX}$ (upper) and 
${\tilde C}^{1 \tR}_{\tN lMX}$ (lower)}
& \multicolumn{1}{c|}{
${\tilde C}^{3 \tL}_{\tN lMX}$ (upper) and 
${\tilde C}^{3 \tR}_{\tN lMX}$ (lower)} 
& \multicolumn{1}{c|}{
${\tilde C}^{4 \tL}_{\tN lMX}$ and 
${\tilde C}^{4 \tR}_{\tN lMX}$ } 
\\\hline\hline
\multirow{2}*{$p\to\ell_x^+ \pi^0 X$}
&$-{1\over 2}\big( 
-i\tilde{C}^{{\tt VR,VL}-,x}_{X\ell udu}
-\kappa_2 \tilde{C}^{{\tt TL,SL},x}_{X\ell uud}\big)$
&$-{\kappa_3 \over 2}\big( 
\tilde{C}^{{\tt VL,VR}+,x}_{X\ell udu} 
+3 \tilde{C}^{{\tt VL,VR},x}_{X\ell uud}\big)$
&$-\frac{2}{3}\kappa_4
\tilde{C}^{{\tt SL,TL},x}_{X\ell uud} $
\\\hhline{~---}
&$- {1 \over 2}\big( 
-i\tilde{C}^{{\tt VL,VR}-,x}_{X\ell udu} 
-\kappa_2 \tilde{C}^{{\tt TR,SR},x}_{X\ell uud}\big)$
&$-{\kappa_3 \over 2}\big( 
\tilde{C}^{{\tt VR,VL}+,x}_{X\ell udu}
+ 3 \tilde{C}^{{\tt VR,VL},x}_{X\ell uud}\big)$
& $-\frac{2}{3}\kappa_4
\tilde{C}^{{\tt SR,TR},x}_{X\ell uud}$
\\\hline
\multirow{2}*{$p\to\ell_x^+ \eta X$}
&${1\over 2\sqrt{3}} \big(
-i\tilde{C}^{{\tt VR,VL}-,x}_{X\ell udu}
+3\kappa_2 \tilde{C}^{{\tt TL,SL},x}_{X\ell uud}\big)$ 
&${\kappa_3 \over 2\sqrt{3}} \big( 
\tilde{C}^{{\tt VL,VR}+,x}_{X\ell udu} 
-\tilde{C}^{{\tt VL,VR},x}_{X\ell uud}\big)$
& ---
\\\hhline{~---}
&${1\over 2\sqrt{3}} \big(
-i\tilde{C}^{{\tt VL,VR}-,x}_{X\ell udu}
+3\kappa_2 \tilde{C}^{{\tt TR,SR},x}_{X\ell uud}\big)$ 
&${\kappa_3 \over 2\sqrt{3}}\big( 
\tilde{C}^{{\tt VR,VL}+,x}_{X\ell udu}
-\tilde{C}^{{\tt VR,VL},x}_{X\ell uud}\big)$
& ---
\\\hline
\multirow{2}*{$n\to\ell_x^+ \pi^- X$}
&$-{ 1\over \sqrt{2}}\big( 
-i\tilde{C}^{{\tt VR,VL}-,x}_{X\ell udu} 
-\kappa_2 \tilde{C}^{{\tt TL,SL},x}_{X\ell uud}\big)$
&${\kappa_3 \over \sqrt{2}}\big( 
3\tilde{C}^{{\tt VL,VR}+,x}_{X\ell udu} 
-\tilde{C}^{{\tt VL,VR},x}_{X\ell uud}\big)$
&$\frac{\sqrt{2}}{3}\kappa_4
\tilde{C}^{{\tt SL,TL},x}_{X\ell uud}$
\\\hhline{~---}
&$-{ 1\over \sqrt{2}}\big( 
-i\tilde{C}^{{\tt VL,VR}-,x}_{X\ell udu} 
-\kappa_2 \tilde{C}^{{\tt TR,SR},x}_{X\ell uud}\big)$
&${\kappa_3 \over \sqrt{2}}\big(
3\tilde{C}^{{\tt VR,VL}+,x}_{X\ell udu} 
-\tilde{C}^{{\tt VR,VL},x}_{X\ell uud}\big)$
&$\frac{\sqrt{2}}{3}\kappa_4
\tilde{C}^{{\tt SR,TR},x}_{X\ell uud} $
\\\hline\hline
\multirow{2}*{$p\to\ell_x^+ K^0 X$}
&${1\over \sqrt{2}}\big( 
-i\tilde{C}^{{\tt VR,VL}-,x}_{X\ell usu}
+\kappa_2 \tilde{C}^{{\tt TL,SL},x}_{X\ell uus}\big)$  
&${\kappa_3 \over \sqrt{2}}\big(
\tilde{C}^{{\tt VL,VR}+,x}_{X\ell usu} 
+\tilde{C}^{{\tt VL,VR},x}_{X\ell uus}\big)$
& $\frac{\sqrt{2}}{3}\kappa_4
\tilde{C}^{{\tt SL,TL},x}_{X   \ell uus} $
\\\hhline{~---}
&${1\over \sqrt{2}}\big( 
-i\tilde{C}^{{\tt VL,VR}-,x}_{X\ell usu}
+\kappa_2\tilde{C}^{{\tt TR,SR},x}_{X\ell uus}\big)$  
&${\kappa_3 \over \sqrt{2}}\big(
\tilde{C}^{{\tt VR,VL}+,x}_{X\ell usu}
+ \tilde{C}^{{\tt VR,VL},x}_{X\ell uus}\big)$
&$\frac{\sqrt{2}}{3}\kappa_4
\tilde{C}^{{\tt SR,TR},x}_{X\ell uus}$
\\\hline\hline
\multirow{2}*{$n\to\ell_x^-\pi^+ X$}
&---
&$-\sqrt{2}\kappa_3 
\tilde{C}^{{\tt VL,VR},x}_{X\bar\ell ddd}$
&$-\sqrt{2}\kappa_4
\tilde{C}^{{\tt SL,TL},x}_{X\bar\ell ddd}$
\\\hhline{~---}
&---
&$-\sqrt{2}\kappa_3 
\tilde{C}^{{\tt VR,VL},x}_{X\bar\ell ddd}$
&$-\sqrt{2}\kappa_4
\tilde{C}^{{\tt SR,TR},x}_{X\bar\ell ddd}$
\\\hline\hline
\multirow{2}*{$n\to\ell_x^- K^+ X$} 
&$- {1\over \sqrt{2}}\big(
-i\tilde{C}^{{\tt VR,VL}-,x}_{X\bar\ell dsd}
+\kappa_2\tilde{C}^{{\tt TL,SL},x}_{X\bar\ell dds}\big)$  
&$-{\kappa_3 \over \sqrt{2}} \big( 
\tilde{C}^{{\tt VL,VR}+,x}_{X\bar\ell dsd}
+\tilde{C}^{{\tt VL,VR},x}_{X\bar\ell dds}\big)$
&$-\frac{\sqrt{2}}{3}\kappa_4
\tilde{C}^{{\tt SL,TL},x}_{X\bar\ell dds}$
\\\hhline{~---}
&$-{1 \over \sqrt{2}}\big(
-i\tilde{C}^{{\tt VL,VR}-,x}_{X\bar\ell dsd} 
+\kappa_2 \tilde{C}^{{\tt TR,SR},x}_{X\bar\ell dds}\big)$
&$-{\kappa_3 \over \sqrt{2}}\big( 
\tilde{C}^{{\tt VR,VL}+,x}_{X\bar\ell dsd}
+ \tilde{C}^{{\tt VR,VL},x}_{X\bar\ell dds}\big)$
&$-\frac{\sqrt{2}}{3}\kappa_4
\tilde{C}^{{\tt SR,TR},x}_{X\bar\ell dds}$
\\\hline\hline
\multirow{2}*{$p\to\hat\nu_x\pi^+ X$}
&$- {1\over \sqrt{2}}\big(
-i\tilde{C}^{{\tt VR,VL}-,x}_{X\nu udd} 
+ \kappa_2 \tilde{C}^{{\tt TL,SL},x}_{X\nu ddu}\big)$ 
&$-{\kappa_3 \over \sqrt{2}}\big( 
3\tilde{C}^{{\tt VL,VR}+,x}_{X\bar\nu udd}
-\tilde{C}^{{\tt VL,VR},x}_{X\bar\nu ddu}\big)$
&$-\frac{\sqrt{2}}{3}\kappa_4
\tilde{C}^{{\tt SL,TL},x}_{X\nu udd}$
\\\hhline{~---}
&$-{1 \over \sqrt{2}}\big( 
-i\tilde{C}^{{\tt VL,VR}-,x}_{X\bar\nu udd} 
+\kappa_2\tilde{C}^{{\tt TR,SR},x}_{X\bar\nu ddu}\big)$
&$-{\kappa_3 \over \sqrt{2}}\big( 
3\tilde{C}^{{\tt VR,VL}+,x}_{X\nu udd} 
-\tilde{C}^{{\tt VR,VL},x}_{X\nu ddu}\big)$
&$-\frac{\sqrt{2}}{3}\kappa_4
\tilde{C}^{{\tt SR,TR},x}_{X\bar\nu udd}$
\\\hline
\multirow{2}*{$n\to\hat\nu_x \pi^0 X$}
&$ {1\over 2}\big( 
-i\tilde{C}^{{\tt VR,VL}-,x}_{X\nu udd} 
+\kappa_2\tilde{C}^{{\tt TL,SL},x}_{X\nu ddu}\big)$
&$-{\kappa_3 \over 2}\big( 
\tilde{C}^{{\tt VL,VR}+,x}_{X\bar\nu udd}
+3\tilde{C}^{{\tt VL,VR},x}_{X\bar\nu ddu}\big)$
&$-\frac{2}{3}\kappa_4
\tilde{C}^{{\tt SL,TL},x}_{X\nu udd}$
\\\hhline{~---}
&${1\over 2}\big( 
-i\tilde{C}^{{\tt VL,VR}-,x}_{X\bar\nu udd} 
+\kappa_2\tilde{C}^{{\tt TR,SR},x}_{X\bar\nu ddu}\big)$
&$-{\kappa_3 \over 2}\big(
\tilde{C}^{{\tt VR,VL}+,x}_{X\nu udd}
+3\tilde{C}^{{\tt VR,VL},x}_{X\nu ddu}\big)$
&$-\frac{2}{3}\kappa_4
\tilde{C}^{{\tt SR,TR},x}_{X\bar\nu udd}$
\\\hline
\multirow{2}*{$n\to\hat\nu_x \eta X$}
&${1 \over 2\sqrt{3}}\big( 
-i\tilde{C}^{{\tt VR,VL}-,x}_{X\nu udd} 
-3\kappa_2\tilde{C}^{{\tt TL,SL},x}_{X\nu ddu}\big)$
&$-{\kappa_3 \over 2\sqrt{3}}\big( 
\tilde{C}^{{\tt VL,VR}+,x}_{X\bar\nu udd}
-\tilde{C}^{{\tt VL,VR},x}_{X\bar\nu ddu}\big)$
&---
\\\hhline{~---}
&$ {1\over 2\sqrt{3}}\big( 
-i\tilde{C}^{{\tt VL,VR}-,x}_{X\bar\nu udd}
-3\kappa_2 \tilde{C}^{{\tt TR,SR},x}_{X\bar\nu ddu}\big)$
&$-{\kappa_3\over 2\sqrt{3}}\big( 
\tilde{C}^{{\tt VR,VL}+,x}_{X\nu udd}
-\tilde{C}^{{\tt VR,VL},x}_{X\nu ddu}\big)$
&---
\\\hline\hline
\multirow{2}*{$p\to\hat\nu_x K^+ X$}
&$-{1\over \sqrt{2}}\big( 
-i\tilde{C}^{{\tt VR,VL}-,x}_{X\nu uds}
-i\tilde{C}^{{\tt VR,VL}-,x}_{X\nu dsu} 
+\kappa_2\tilde{C}^{{\tt TL,SL},x}_{X\nu sdu}\big)$
&$-{\kappa_3 \over\sqrt{2}}\big( 
\tilde{C}^{{\tt VL,VR}+,x}_{X\bar\nu uds}
+2\tilde{C}^{{\tt VL,VR}+,x}_{X\bar\nu usd}
-\tilde{C}^{{\tt VL,VR}+,x}_{X\bar\nu dsu}\big)$
&$-\frac{\sqrt{2}}{6}\kappa_4
\tilde{C}^{{\tt SL,TL},x}_{X\nu uds}$
\\\hhline{~---}
&$-{1\over \sqrt{2}}\big( 
-i\tilde{C}^{{\tt VL,VR}-,x}_{X\bar\nu uds}
-i\tilde{C}^{{\tt VL,VR}-,x}_{X\bar\nu dsu}
+\kappa_2\tilde{C}^{{\tt TR,SR},x}_{X\bar\nu sdu}\big)$
&$-{\kappa_3 \over \sqrt{2}}\big( 
\tilde{C}^{{\tt VR,VL}+,x}_{X\nu uds}
+2\tilde{C}^{{\tt VR,VL}+,x}_{X\nu usd}
-\tilde{C}^{{\tt VR,VL}+,x}_{X\nu dsu}\big)$
&$-\frac{\sqrt{2}}{6}\kappa_4
\tilde{C}^{{\tt SR,TR},x}_{X\bar\nu uds}$
\\\hline
\multirow{2}*{$n\to\hat\nu_x K^0 X$}
&$ {i\over \sqrt{2}}\big( 
\tilde{C}^{{\tt VR,VL}-,x}_{X\nu uds} 
- \tilde{C}^{{\tt VR,VL}-,x}_{X\nu usd}\big)
-{\kappa_2 \over \sqrt{2}}\big(
\tilde{C}^{{\tt TL,SL},x}_{X\nu dsu}
+\tilde{C}^{{\tt TL,SL},x}_{X\nu sdu}\big)$
&${\kappa_3 \over \sqrt{2}}\big( 
\tilde{C}^{{\tt VL,VR}+,x}_{X\bar\nu uds} 
-\tilde{C}^{{\tt VL,VR}+,x}_{X\bar\nu usd}
+2\tilde{C}^{{\tt VL,VR}+,x}_{X\bar\nu dsu}\big)$
&$\frac{\sqrt{2}}{6}\kappa_4
\tilde{C}^{{\tt SL,TL},x}_{X\nu uds} $
\\\hhline{~---}
&${i \over \sqrt{2}}\big( 
\tilde{C}^{{\tt VL,VR}-,x}_{X\bar\nu uds} 
-\tilde{C}^{{\tt VL,VR}-,x}_{X\bar\nu usd}\big)
-{\kappa_2 \over \sqrt{2}}\big(
\tilde{C}^{{\tt TR,SR},x}_{X\bar\nu dsu}
+ \tilde{C}^{{\tt TR,SR},x}_{X\bar\nu sdu}\big)$
&$ {\kappa_3 \over \sqrt{2}}\big( 
\tilde{C}^{{\tt VR,VL}+,x}_{X\nu uds}
-\tilde{C}^{{\tt VR,VL}+,x}_{X\nu usd}
+2\tilde{C}^{{\tt VR,VL}+,x}_{X\nu dsu}\big)$
&$\frac{\sqrt{2}}{6}\kappa_4
\tilde{C}^{{\tt SR,TR},x}_{X\bar\nu uds}$
\\\hline
\end{tabular} }
\caption{The explicit expressions of the coefficients $\tilde C_{\tN lMX}^{1,3,4\tL(\tR)}$ governing the contact contributions to each three-body nucleon decay in case B. }
\label{tab:CN2lMX_T}
\end{table}

Now we turn to the noncontact contribution in \cref{fig:3body}. 
For each nucleon decay mode $\tN\to l X M$, the intermediate baryon state $B$ can be readily identified by combining the BNV interactions in \cref{eq:LB2lX} with the BNC terms in \cref{eq:LBNM}. 
In the last two columns of \cref{tab:CN2lMX}, we list the corresponding $C_{\tN BM}$ and $C_{BlX}^{1,3\tL(\tR)}$ for each process. There is only one intermediate baryon state for all processes, except for $\tN \to \hat\nu K X$, whose noncontact contributions can be mediated by both $\Lambda^0$ and $\Sigma^0$. 
For the noncontact contributions in case B, $C_{BlX}^{1,3\tL(\tR)}$s in the last column should be replaced by $\tilde C_{BlX}^{1,3\tL(\tR)}$s.

Based on the general Lagrangian terms in \cref{eq:LB2lX,eq:LBNM} (noncontact diagram) and \cref{eq:LN2lMX} (contact diagram), the amplitude for each three-body nucleon decay mode $\tN(p_\tN)\to l(p_l) \,X(p_X)\,M(p_M)$ can be expressed in the following form: 
{\small
\begin{subequations}
\begin{align}
{\cal M}_{\texttt{N}\to l M X}^{\rm (A)}
=\,& \frac{ic_1}{F_0}m_X\overline{u_l}\Big\{
\gamma^\mu(C_{\tN lMX}^{1\tL}P_\tL+C_{\tN lMX}^{1\tR}P_\tR)
+\Lambda_\chi^{-1}
(p_\tN^\mu-\frac{1}{4}\gamma^\mu\slashed{p}_\tN)
(C_{\tN lMX}^{3\tL} P_\tL + C_{\tN lMX}^{3\tR}P_\tR)
\nn\\
& +\frac{C_{\tN BM}}{m_B^2-q^2} \Big[
\Lambda_\chi^{-1}
(q^\mu-\frac{1}{4}\gamma^\mu\slashed{q}) 
(C_{BlX}^{3\tL} P_\tL+C_{BlX}^{3\tR} P_\tR)
+\gamma^\mu(C_{BlX}^{1\tL} P_\tL+C_{BlX}^{1\tR}P_\tR)\Big]
\nn\\
&
\times (\slashed{q}+m_B)(m_{\tN}+\slashed{q})
(P_\tL-P_\tR) \Big\} u_\tN \epsilon_{X,\mu}^*\;,
\label{eq:amp3body_caseA}
\\%
{\cal M}_{\texttt{N}\to l M X}^{\rm (B)}
=\,& -\frac{c_1}{F_0}\overline{u_l}\Big\{
\sigma^{\mu\nu} 
({\tilde C}_{\tN lMX}^{1\tL} P_\tL
+{\tilde C}_{\tN lMX}^{1\tR} P_\tR)
+\Lambda_\chi^{-1} \gamma^\nu
(p_\tN^\mu-\frac{1}{4}\gamma^\mu\slashed{p}_\tN)
({\tilde C}_{\tN lMX}^{3\tL} P_\tL
+{\tilde C}_{\tN lMX}^{3\tR} P_\tR)
\nn\\
&+ \Lambda_\chi^{-2} 
({\tilde C}_{\N lMX}^{4\tL}
\hat{\Gamma}_{\mu\nu\alpha\beta}^\tL
+{\tilde C}_{\N lMX}^{4\tR}
\hat{\Gamma}_{\mu\nu\alpha\beta}^\tR)
p_\tN^\alpha p_M^\beta
\nn\\
&+\frac{C_{\tN BM}}{m_B^2-q^2}\Big[
\Lambda_\chi^{-1} \gamma^\nu 
(q^\mu-\frac{1}{4}\gamma^\mu\slashed{q}) 
({\tilde C}_{BlX}^{3\tL} P_\tL
+{\tilde C}_{BlX}^{3\tR} P_\tR)
+\sigma^{\mu\nu} 
({\tilde C}_{BlX}^{1\tL} P_\tL
+{\tilde C}_{BlX}^{1\tR} P_\tR)\Big]
\nn\\
&\times(\slashed{q}+m_B)(m_{\tN}+\slashed{q})
(P_\tL-P_\tR)
\Big\} u_\tN 
(p_{X,\mu}\epsilon_{X,\nu}^* 
-p_{X,\nu}\epsilon_{X,\mu}^*),
\label{eq:amp3body_caseB}
\end{align}
\end{subequations}}%
where $m_B$ and $q=p_\tN-p_M$ denote the mass and momentum of the intermediate baryon state $B$ in the noncontact diagram, respectively.
For each process, the coefficients $C_i$ can be substituted using the expressions in \cref{tab:CB2lX,tab:CN2lMX,tab:CN2lMX_T} and \cref{eq:LBNM}. It is important to emphasize that for the decay $\tN\to \hat\nu K X$, the contributions from the noncontact diagram are obtained by summing over the two intermediate states $\Lambda^0$ and $\Sigma^0$. 

To proceed with the calculation, we employ the {\tt FeynCalc} package~\cite{Shtabovenko:2020gxv} to compute the matrix element squared.
The resulting expressions are particularly cumbersome in both cases, and therefore we omit them from the main text. The complete calculations are provided in the auxiliary {\tt Mathematica} notebook.  
Defining the invariant mass squares $s\equiv q^2=(p_l+p_X)^2$ and $t=(p_M+p_X)^2$, the decay width is given by
\begin{align}
\Gamma_{\texttt{N}\to lMX}
= \frac{1}{256\pi^3 m_{\tN}^3}  
\int d s \int_{t_-}^{t_+} d t \;
\overline{|{\cal M}_{\texttt{N}\to lMX}|^2},
\end{align}
where the kinematic limits are 
\begin{align}
& (m_l + m_X)^2 \leq s \leq (m_{\tN} - m_M)^2, 
\nn
\\
& t_\pm = (E_2^* + E_3^*)^2 -
\Big(\sqrt{E_2^{*2} - m_X^2} 
\mp \sqrt{E_3^{*2} - m_M^2}\Big)^2, 
\nn
\\
& E_2^* = \frac{s - m_l^2 + m_X^2}{2\sqrt{s}},\qquad
E_3^* = \frac{m_{\tN}^2 - s - m_M^2}{2\sqrt{s}}.
\end{align}
Using the central values of particle masses and hadronic LECs, we perform full phase-space integration for a given dark photon mass and express the decay widths in terms of the $X$LEFT WCs. For the massless dark photon case, the complete numerical results for all relevant three-body nucleon decay modes are provided in \cref{eq:Ga_N2lXMA} and \cref{eq:Ga_N2lXMB} of \cref{app:Gamma_exp}, corresponding to operators in case A and case B, respectively.

\subsection{Momentum distribution}
\label{subsub:P_dis}

Experimentally, event numbers are typically binned according to kinematic variables of the final state particles, such as momentum, missing energy, or invariant mass. These distributions encode key information about the underlying dynamics.
If a signal is observed, they can help discriminate between operator structures and determine the mass of the invisible particle. 
To illustrate this, we analyze the momentum spectra of visible SM particles in three-body nucleon decays induced by individual operators. 
For each decay mode, we examine the normalized differential width with respect to the momentum of the charged lepton and/or the octet meson,  
\begin{align}
\frac{d\tilde \Gamma}{d |\pmb{p}_\ell|} \equiv \frac{1}{\Gamma} \frac{d\Gamma}{d |\pmb{p}_\ell|}, \quad 
\frac{d\tilde \Gamma}{d |\pmb{p}_M|} \equiv \frac{1}{\Gamma} \frac{d\Gamma}{d |\pmb{p}_M|}, 
\end{align}
where $\pmb{p}_\ell$ and $\pmb{p}_M$ are the three-momenta of the final-state charged lepton and the octet meson, respectively. As both the neutrino and dark photon are invisible to the detector, their spectra are omitted in this analysis. 

We concentrate on proton and neutron decays $p\to \ell^+(\pi^0,\eta)X$ and $n\to \ell^+ \pi^- X$ that are induced by $X\ell uud$-type operators listed in the second column of \cref{tab:ope_process}, where the charged lepton $\ell$ can be either $e$ or $\mu$.
We take the single-operator-dominance analysis by working with only one operator at a time. 
Since the hadronic counterparts of the $X$LEFT operators are constructed to be parity-invariant due to the QCD matching nature, 
this implies that a pair of operators with interchange of fermion chiralities yields identical kinematic distributions and decay widths.
Thus, we restrict ourselves to the 9 operators on the left side of the second column in  \cref{tab:ope_process}, 
including 4 operators ($\calO_{X\ell uud}^{{\tt VL,SL},x}$, 
$\calO_{X\ell udu}^{{\tt SL,VR}-,x}$, $\calO_{X\ell udu}^{{\tt SL,VR}+,x}$,
$\calO_{X\ell uud}^{{\tt SL,VR},x}$) in case A and 5 operators 
($\tilde \calO_{X\ell uud}^{{\tt TL,SL},x}$, $\tilde \calO_{X\ell udu}^{{\tt VL,VR}-,x}$, $\tilde \calO_{X\ell udu}^{{\tt VL,VR}+,x}$, $\tilde \calO_{X\ell uud}^{{\tt VL,VR},x}$, $\tilde \calO_{X\ell uud}^{{\tt SL,TL},x}$) in case B. 
Among these, 
$\{\calO_{X\ell uud}^{{\tt VL,SL},x}, 
\tilde \calO_{X\ell uud}^{{\tt TL,SL},x}\}\in\pmb{8}_\tL\otimes \pmb{1}_\tR$ 
and $\{\calO_{X\ell udu}^{{\tt SL,VR}-,x},
\tilde \calO_{X\ell udu}^{{\tt VL,VR}-,x}\}\in \bar{\pmb{3}}_\tL\otimes \pmb{3}_\tR$, 
and all of them exclusively induce an isospin change of $\Delta I=1/2$ unit.
The 4 operators $\calO_{X\ell udu}^{{\tt SL,VR}+,x}$, 
$\calO_{X\ell uud}^{{\tt SL,VR},x}$, 
$\tilde \calO_{X\ell udu}^{{\tt VL,VR}+,x}$, 
and  $\tilde \calO_{X\ell uud}^{{\tt VL,VR},x}$ belong to the irrep $\pmb{6}_\tL\otimes \pmb{3}_\tR$ and contain both $\Delta I=1/2$ and $\Delta I=3/2$ isospin components. The last operator $\tilde \calO_{X\ell uud}^{{\tt SL,TL},x}\in \pmb{10}_\tL\otimes \pmb{1}_\tR$ changes isospin by $\Delta I=3/2$ units. 
Below, we will utilize their isospin property to understand the features in the distributions. 
To analyze the dark photon mass dependence, we consider three benchmark masses in each case: 
$m_X=(0,\,0.3,\,0.5)\,\text{GeV}$ for the 
$\pi$ final state and $m_X=(0,\,0.1,\,0.2)\,\text{GeV}$ for the $\eta$ final state. 

\begin{figure}[t]  
\centering
\includegraphics[width=0.325\linewidth]{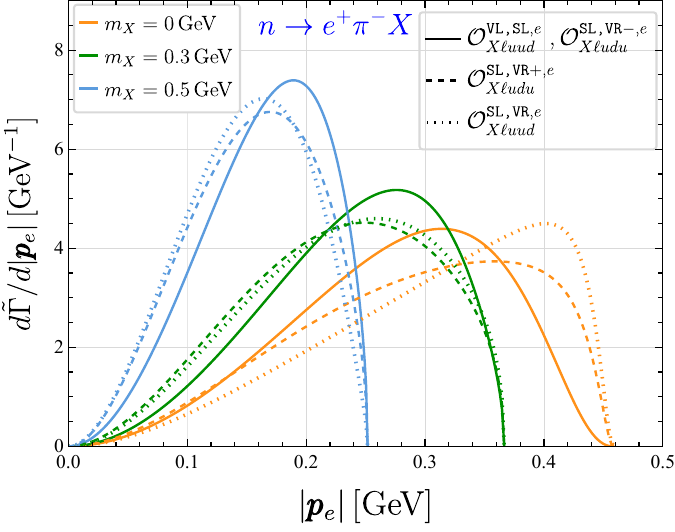}
\includegraphics[width=0.325\linewidth]{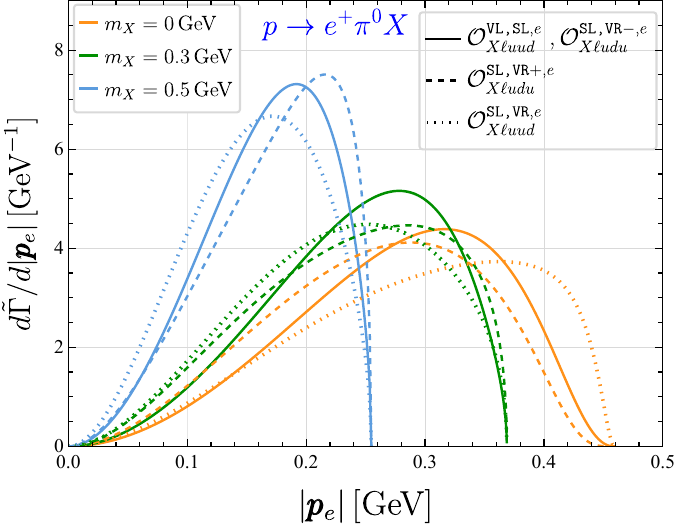}
\includegraphics[width=0.325\linewidth]{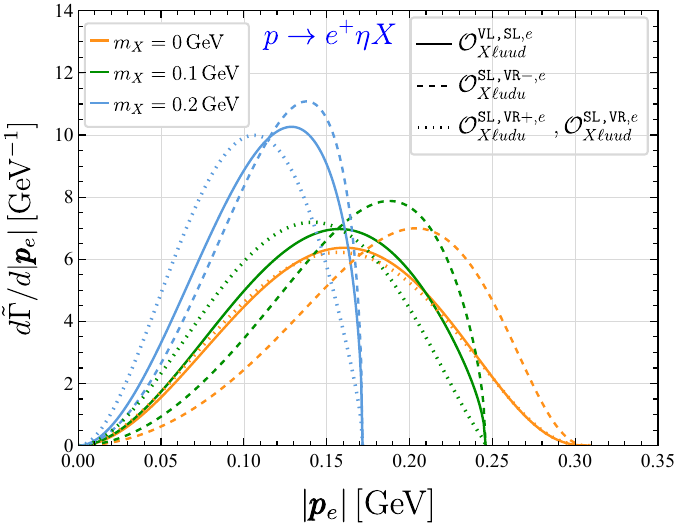}
\\
\vspace{0.1em}
\includegraphics[width=0.325\linewidth]{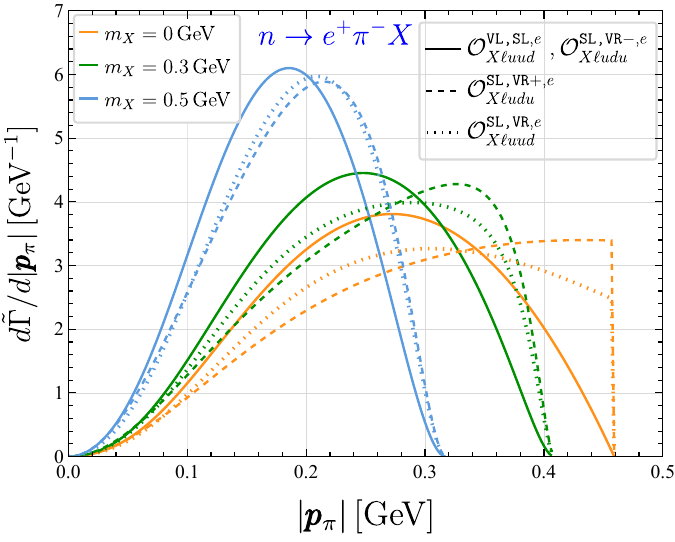}
\includegraphics[width=0.325\linewidth]{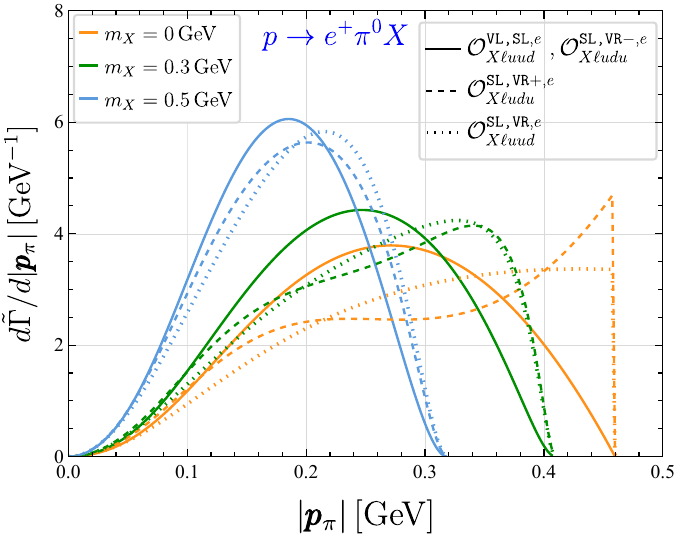}
\includegraphics[width=0.325\linewidth]{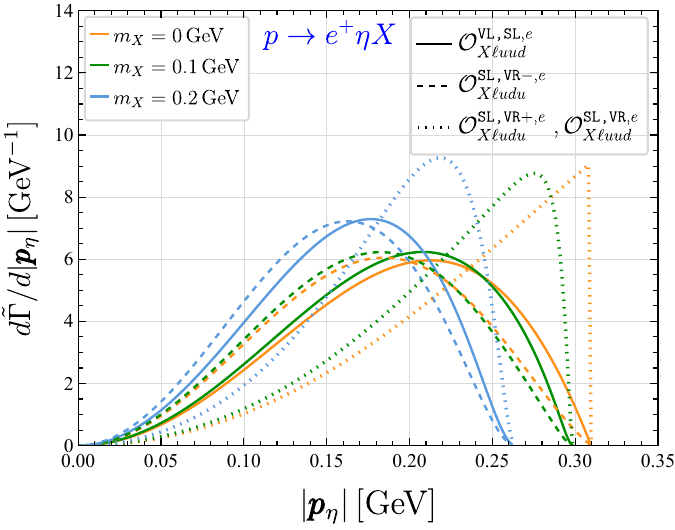}
\\
\vspace{0.2em}
\includegraphics[width=0.325\linewidth]{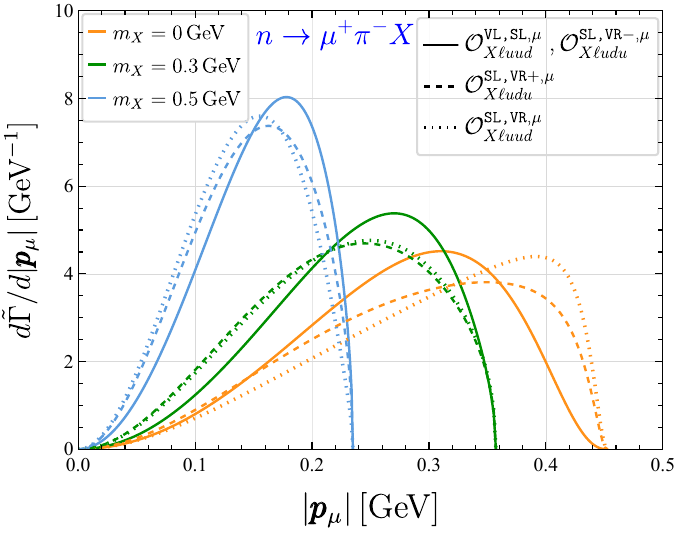}
\includegraphics[width=0.325\linewidth]{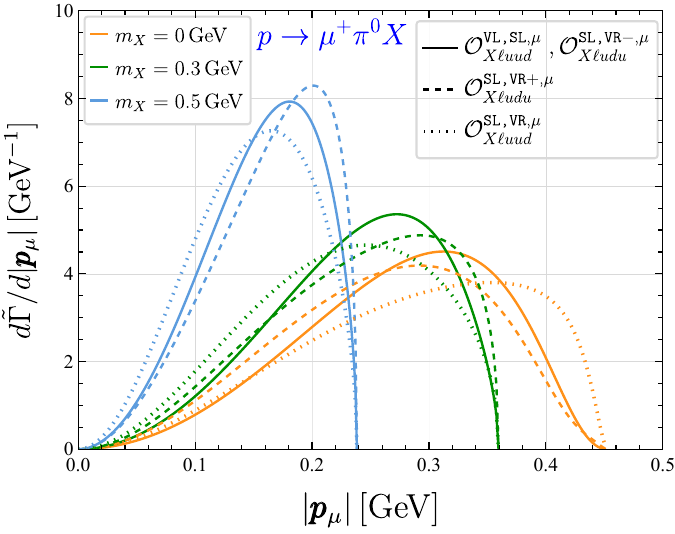}
\includegraphics[width=0.325\linewidth]{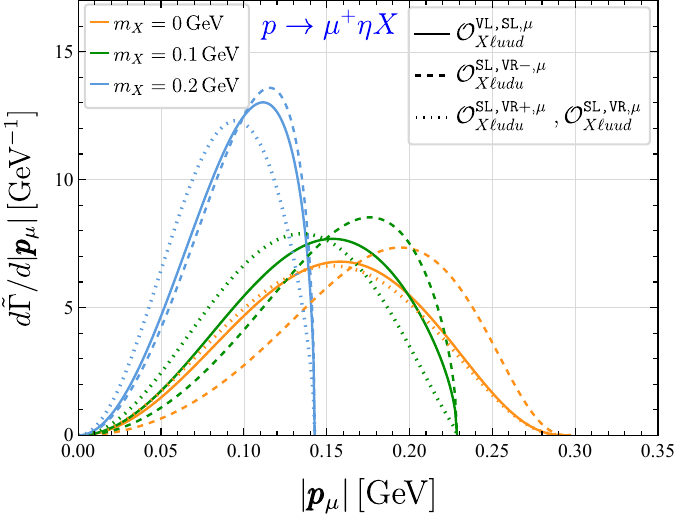}
\\
\vspace{0.1em}
\includegraphics[width=0.325\linewidth]{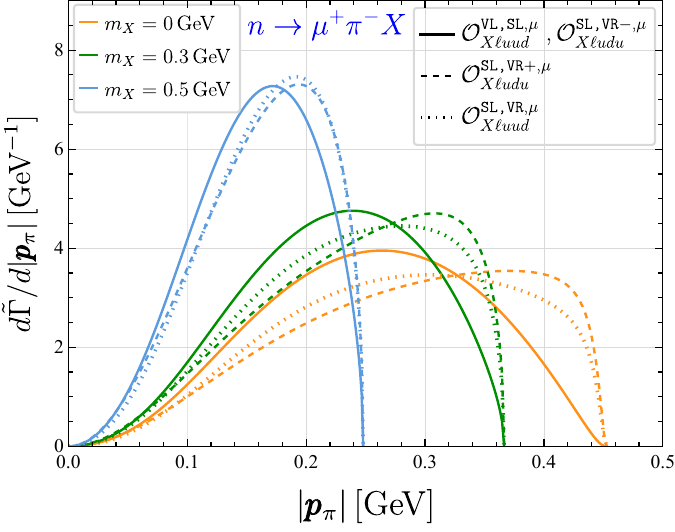}
\includegraphics[width=0.325\linewidth]{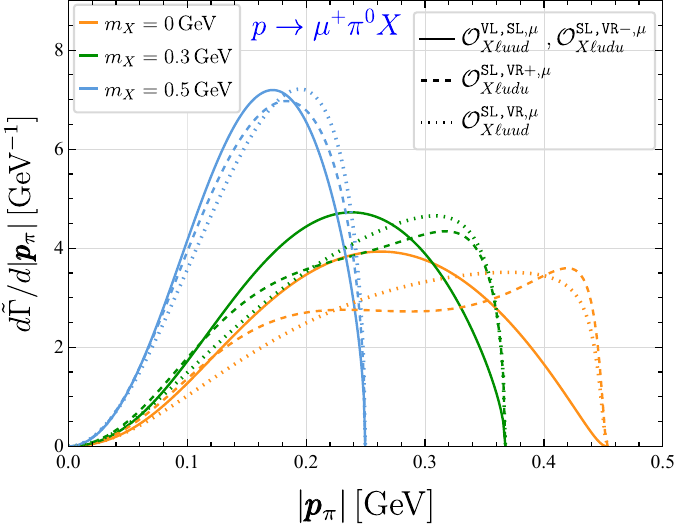}
\includegraphics[width=0.325\linewidth]{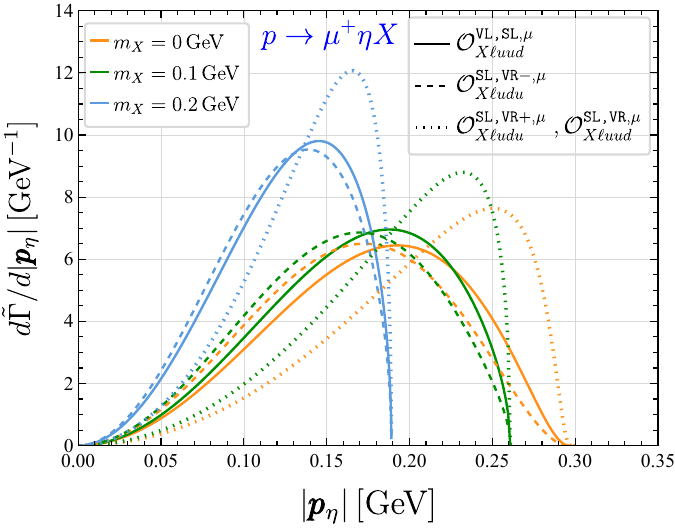}
\caption{
The normalized momentum distributions for  $n \to \ell^+ \pi^- X$ (left column) and $p \to \ell^+ (\pi^0,\eta) X$ (middle and right columns) from insertion of various operators in case A.
} 
\label{fig:3body_dis_A}
\end{figure}

Figure \ref{fig:3body_dis_A} displays the normalized decay width spectra as a function of the momenta of the charged leptons (first and third rows) and pseudoscalar mesons (second and fourth rows) for the decay modes 
$n \to \ell^+ \pi^- X$ (left panels) and 
$p \to \ell^+ (\pi^0,\eta) X$ (middle and right panels) that are induced by the four $X\ell uud$-type operators in case A. 
For the processes $n\to \ell^+\pi^-X$ and $p\to \ell^+\pi^0X$ ($\ell=e,\mu$), the operators $\calO_{X\ell uud}^{{\tt VL,SL},x} $ and $ \calO_{X\ell udu}^{{\tt SL,VR}-,x}$ produce identical charged lepton and pion momentum distributions at each dark photon mass benchmark point.
This equality arises because their WCs (including their chirality partners) always appear in the linear combinations, $C_{X\ell udu}^{{\tt SR,VL}-,x} -\kappa_2 C_{X\ell uud}^{{\tt VL,SL},x}$ and $C_{X\ell udu}^{{\tt SL,VR}-,x} -\kappa_2 C_{X\ell uud}^{{\tt VR,SR},x}$, 
in the BNV vertices contributing to these processes. 
Furthermore, the fact that these operators have a definite isospin ($\Delta I = 1/2$), together with the Wigner-Eckart theorem, explains the nearly identical distributions observed between the two decay modes.
In contrast, the two operators $\calO_{X\ell udu}^{{\tt SL,VR}+,x}$ and $\calO_{X\ell uud}^{{\tt SL,VR},x}$ in the irrep $\pmb{6}_{\tL} \otimes \pmb{3}_{\tR}$  
yield noticeably different $|\pmb{p}_\ell|$ and  $|\pmb{p}_\pi|$ distributions between the two 
decay channels, owing to the involvement of both $\Delta I=1/2$ and $\Delta I=3/2$ isospin components.
However, for the process $p\to \ell^+ \eta X$, the two operators in the irrep $\pmb{6}_{\tL} \otimes \pmb{3}_{\tR}$ yield identical distributions in each case. 
This occurs because only the $\Delta I=1/2$ components of these operators contribute, and these components are the same up to a minus sign difference. This similarity also explains why both operators make equal contributions to the decay widths of $\Delta I=1/2$ nucleon decay processes summarized in \cref{app:Gamma_exp}.

Since both the electron and muon are much lighter than the nucleon, the momentum distributions are largely similar between the $e^+$ and $\mu^+$ modes. 
However, they differ in the higher momentum region where the mass difference between the electron and the muon becomes pronounced. 
From the plots in \cref{fig:3body_dis_A}, it is evident that different operators generally yield distinct distribution behavior when both $\pi$ and $\eta$ modes are considered together, particularly in the positions of their peaks. 
These characteristic signatures provide a valuable tool for distinguishing between operator structures in future experimental searches.
Moreover, the endpoints of these distributions exhibit high sensitivity to the dark photon mass, making them a promising way to determine its value.

\begin{figure}[t]  
\centering
\includegraphics[width=0.325\linewidth]{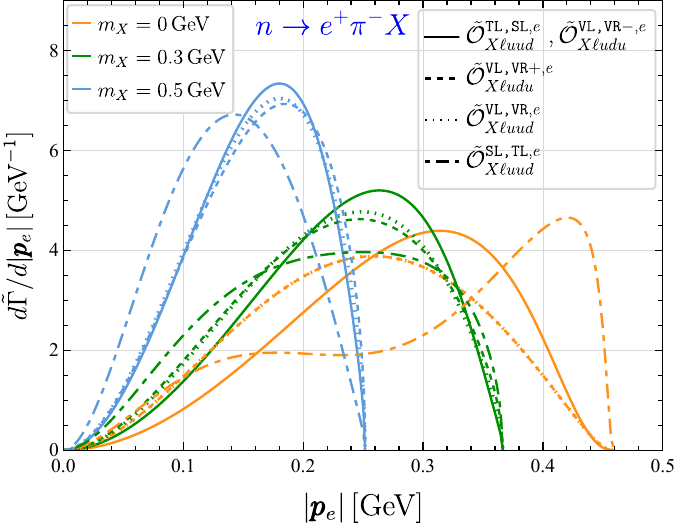}
\includegraphics[width=0.325\linewidth]{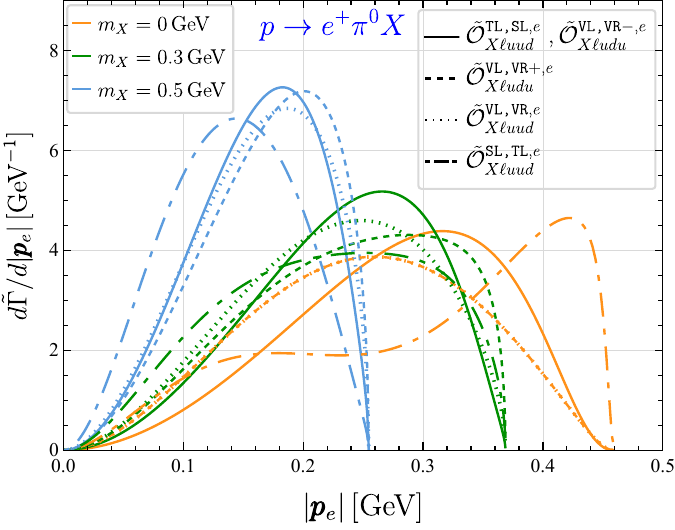}
\includegraphics[width=0.325\linewidth]{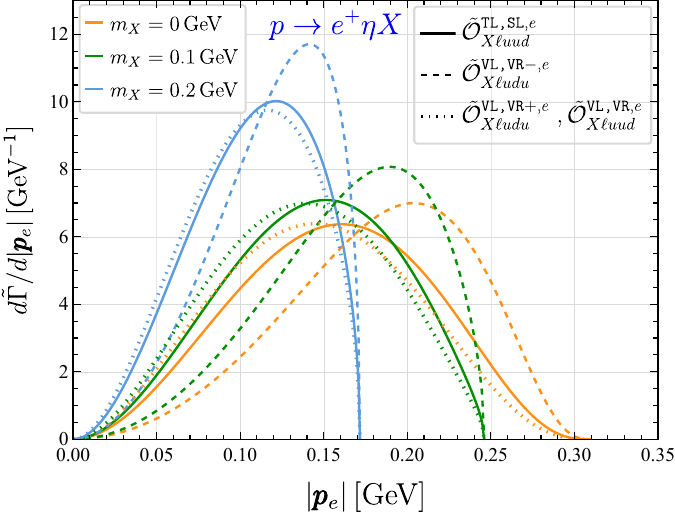}
\\
\vspace{0.1em}
\includegraphics[width=0.325\linewidth]{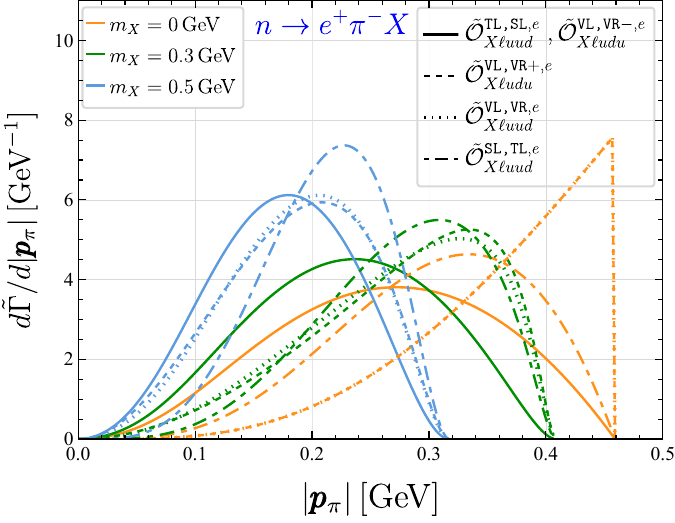}
\includegraphics[width=0.325\linewidth]{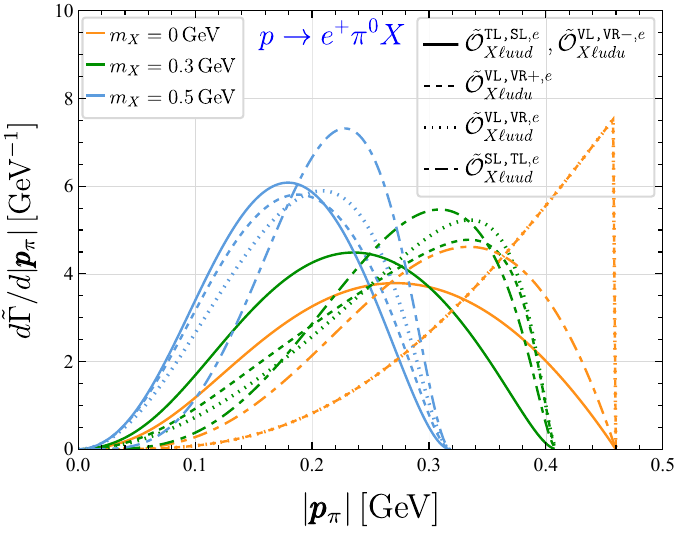}
\includegraphics[width=0.325\linewidth]{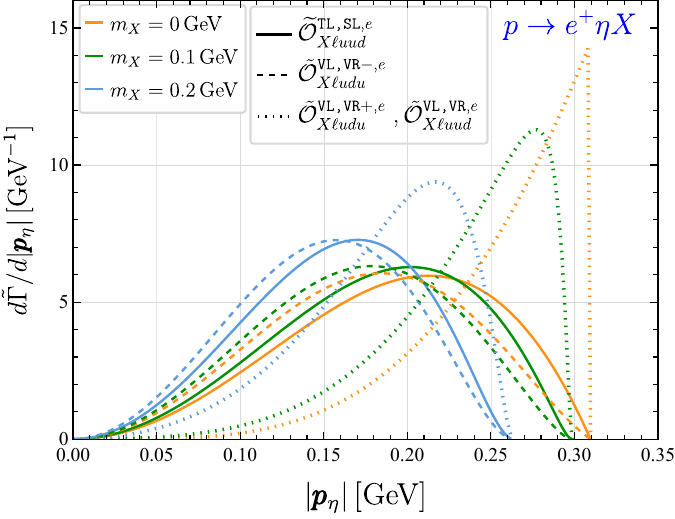}
\\
\vspace{0.2em}
\includegraphics[width=0.325\linewidth]{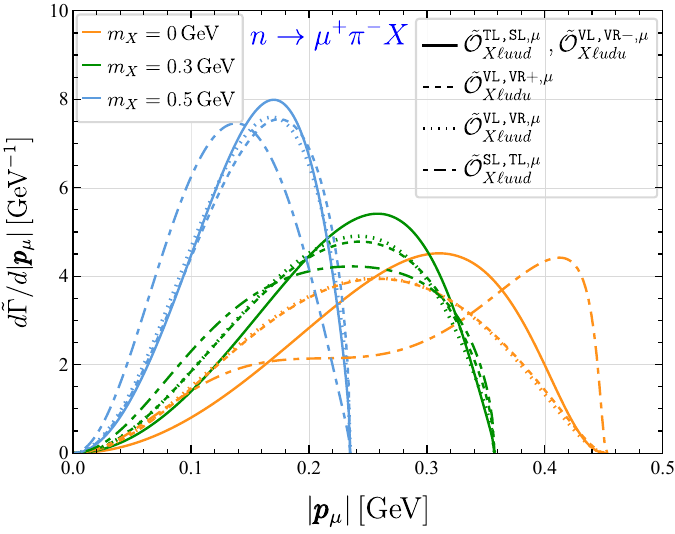}
\includegraphics[width=0.325\linewidth]{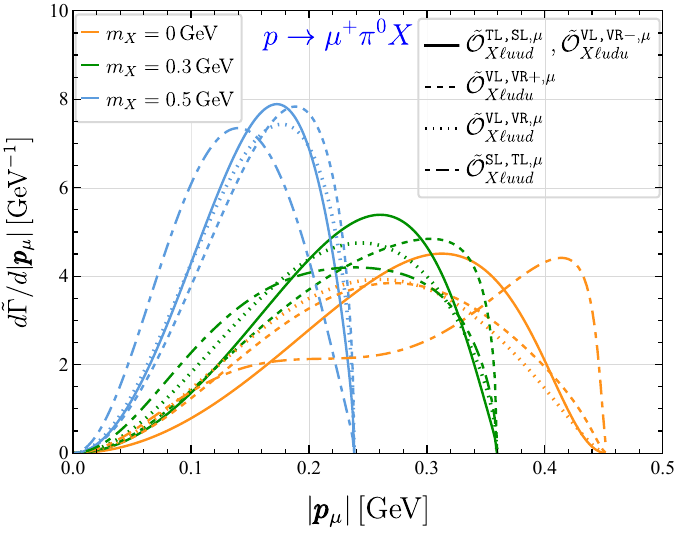}
\includegraphics[width=0.325\linewidth]{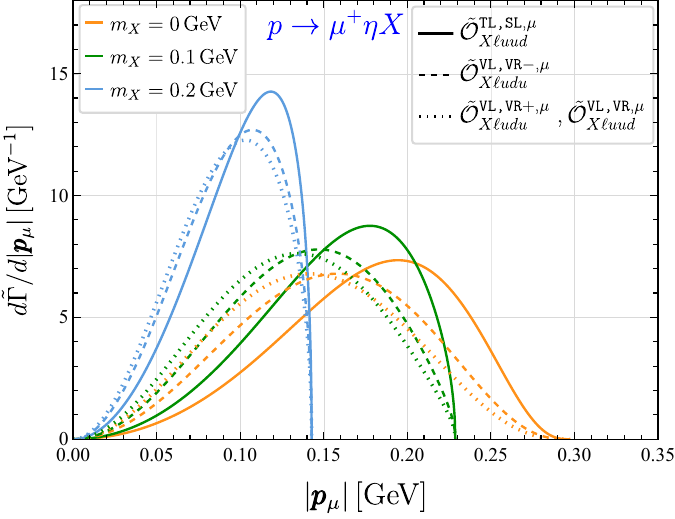}
\\
\vspace{0.1em}
\includegraphics[width=0.325\linewidth]{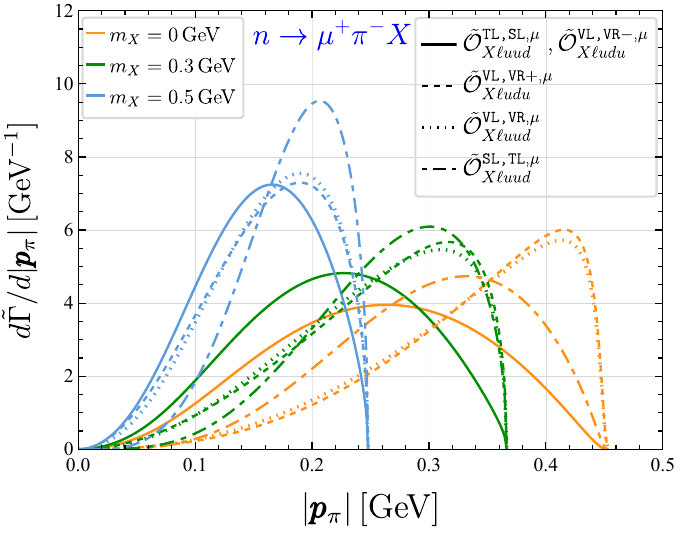}
\includegraphics[width=0.325\linewidth]{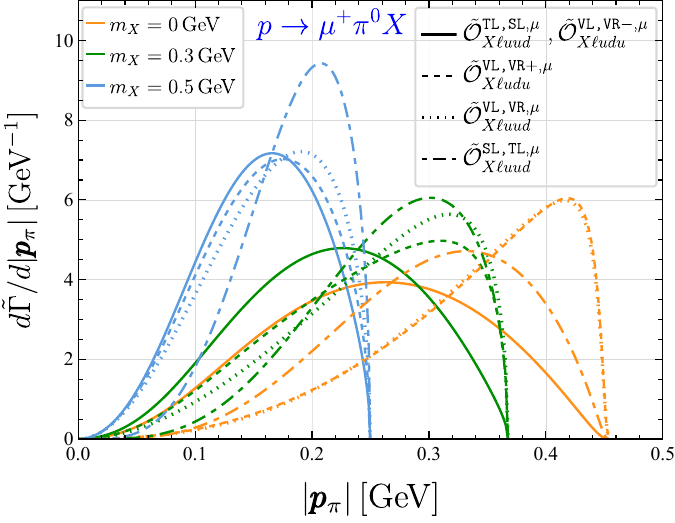}
\includegraphics[width=0.325\linewidth]{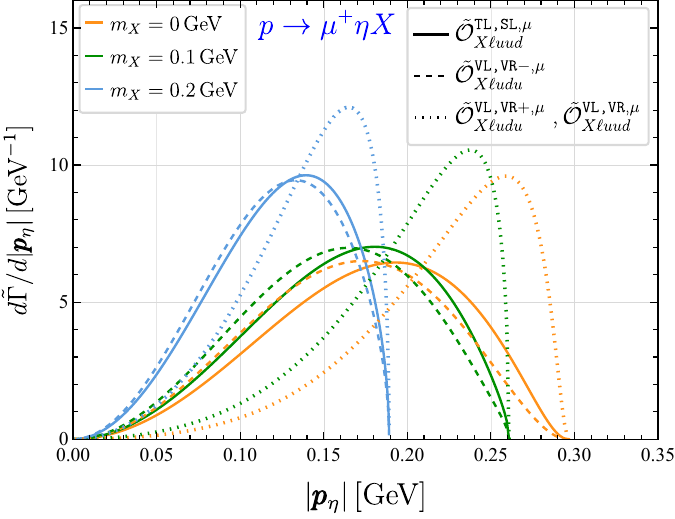}
\caption{
Similar to \cref{fig:3body_dis_A} but for case B.} 
\label{fig:3body_dis_B}
\end{figure}

Figure \ref{fig:3body_dis_B} shows similar distributions for the case where the dark photon is parametrized by a field strength tensor. 
With the exception of the additional operator $\tilde\calO^{{\tt SL,TL},x}_{X\ell uud}$ in the chiral irrep $\pmb{10}_{\tL}\otimes 
\pmb{1}_{\tR}$, the remaining four operators share the same chiral properties as their counterparts in case A. Consequently, the momentum distributions for each process from the insertion of these operators can be analyzed in a similar way to those in case A. 
Therefore, we focus our discussion only on the operator $\tilde\calO^{{\tt SL,TL},x}_{X\ell uud}$. 
Since this operator has a definite isospin component of $\Delta I = 3/2$, it does not contribute to the $\Delta I = 1/2$ process $p\to \ell^+\eta X$. 
In addition, this isospin structure also leads to identical distributions for the two decay modes $n\to \ell^+ \pi^- X$ and $p\to \ell^+ \pi^0 X$ in each scenario.
Noticeably, the distribution in $|\pmb{p}_\ell|$ from this operator peaks in the higher momentum region when $m_X$ decreases. 

For the remaining three-body modes involving a kaon or a (anti)neutrino in the final state, 
similar momentum distributions can be studied.
First, the $|\pmb{p}_\pi|$-distributions in the process $p\to \hat\nu\pi^+ X$ closely follow those of $n\to e^+\pi^- X$ for the corresponding operators under the simultaneous interchanges of $u\leftrightarrow d$ and $e\leftrightarrow \hat\nu$. 
This happens because
the nucleon-to-pion matrix elements are related by interchanging $u \leftrightarrow d$ and because both neutrino and electron masses are negligible. 
A slightly larger difference may occur for the $\pmb{6}_{\tL(\tR)} \otimes \pmb{3}_{\tR(\tL)}$ operators because the amplitudes of the two processes differ by a relative sign originating from the two isospin components.
In the same vein, such similarity in distributions is also anticipated between $p \to e^+ (\pi^0,\eta) X$ and $n \to \hat\nu (\pi^0,\eta) X$, $p \to \hat\nu K^+ X$ and $n \to \hat\nu K^0 X$,
as well as $p \to \ell^+ K^0 X$ and $n \to \ell^- K^+ X$ for the corresponding operators with the simultaneous interchanges of $u\leftrightarrow d$ and the involved lepton fields. 

\section{Numerical constraints and implications}
\label{sec:constraint}

Having established the general formalism for the decay widths, we now study constraints on these exotic decay modes and the associated $X$LEFT operators.
As noted in the introduction, there are currently no dedicated experiments studying these decays (except for $p\to \ell^+\phi$ with a massless $\phi$ in~\cite{Super-Kamiokande:2015pys}).
However, due to the similarity in the experimental signatures between the conventional two-body modes $\tN\to l M$ and the new three-body modes $\tN\to l M X$, the existing experimental data from two-body searches can be reinterpreted to constrain the three-body modes. 
To simplify the analysis, we select one representative three-body mode from each field configuration (with the exception of $X\ell uud$) to compute partial lifetime bounds, as indicated by checkmarks in \cref{tab:ope_process}.
For the two-body decays $p\to \ell^+ X$ in the $X\ell uud$ class, we employ directly the recasting limits on $p\to \ell^+\varphi$ involving a scalar $\varphi$ presented in~\cite{Ma:2025mjy},
since the kinematics of a two-body decay is independent of the specific interaction structures. 

In \cref{subsec:recast_bounds}, we first present our reanalysis of existing Super-K data to establish constraints on the selected three-body modes. After deriving partial lifetime bounds on these modes, we translate them into bounds on the effective scales of all relevant $X$LEFT operators in \cref{subsec:Lambda_bounds}.
As illustrated in \cref{tab:ope_process}, operators with each field configuration of $X\ell uud$, $X\hat\nu udd$, and $X\hat\nu uds$ induce multiple decay processes. Utilizing the operator limits obtained in \cref{subsec:Lambda_bounds}, we can derive new partial lifetime bounds for the remaining decay modes not considered in \cref{subsec:recast_bounds}, as detailed in \cref{subsec:converted_bounds}.  

\subsection{Reinterpretation of existing Super-K data}
\label{subsec:recast_bounds}

We exclusively employ the Super-K datasets to place constraints on the five classes of three-body decays: $p\to \ell^+ K^0 X$,
$n\to \ell^- \pi^+ X$, $n\to \ell^- K^+ X$, $n\to \nu \pi^0 X$, and $p\to \nu K^+ X$.
These modes provide relatively stringent constraints on the corresponding operators listed in \cref{tab:ope_process}. 
Since Super-K is a water Cherenkov detector, 
our simulation incorporates both initial-state nuclear effects and final-state intranuclear interactions relevant to oxygen nuclei. 
Before carrying out the analysis of each decay mode, 
we first discuss these nuclear effects together with the detector resolutions.    

\textbf{Initial-state nuclear effects:} In water Cherenkov detectors, proton decay can occur through either free protons in hydrogen or bound protons in oxygen nuclei, with a relative abundance of $1:4$. 
In contrast, all neutrons are bound in oxygen nuclei.
In our simulation, free protons are taken to be at rest, 
whereas bound nucleons are modeled by including both nuclear Fermi motion and binding-energy effects.
According to the nuclear shell model, 
each nucleon in an oxygen nucleus is assigned to the s- or p-shell with relative probabilities $1:3$~\cite{Mayer_nuclear_state}.
We follow Super-K analysis and use the shell-dependent Fermi-momentum distributions extracted from electron--$^{12}{\rm C}$ scattering data in~\cite{Nakamura:1976mb}.
For the nucleon binding energy ($E_b$), 
we adopt Gaussian distributions with mean and standard deviation
$(\mu,\sigma)=(39.0,10.2)\,{\rm MeV}$ for the s-shell and
$(15.5,3.82)\,{\rm MeV}$ for the p-shell.
The binding energy reduces the available mass of the decaying nucleon $\tN$ to an effective value $m_{\tN}^{\rm eff}=m_{\tN}-E_b$. 
Consequently, when calculating the bound nucleon decays, we replace $m_{\tN}$ appearing in the squared amplitude for free nucleon decays with the effective mass $m_{\tN}^{\rm eff}$. 

\textbf{Final-state intranuclear interactions:} Mesons produced in bound nucleon decays may undergo
hadronic interactions with the daughter nucleus before escaping, including
scattering, absorption, charge exchange, and secondary hadron
production~\cite{Super-Kamiokande:2022egr}. These intranuclear processes can
alter both momentum distributions and the particle identities of the primary mesons.

For the final-state $K^+$ mesons, the Super-K analysis identifies them through their decay products after the kaons come to rest.
Consequently, modifications to the $K^+$ momentum distribution are negligible for event selection. 
On the other hand, the charge-exchange reaction $K^+ n\to K^0 p$ causes $K^+$ to be misidentified as a $K^0$. 
However, only about $1\,\%$ of $K^+$ undergo this conversion~\cite{Super-Kamiokande:2005lev}, and its effect can therefore be neglected in a good approximation. 

For $\pi^0$ and $K_S^0$, 
they decay promptly into visible particles after escaping the nucleus. 
Their survival probability and momentum distributions are therefore important ingredients in the simulation.
Since detailed momentum distributions for the interacting component of the mesons are not publicly available, we retain only the non-interacting component to derive conservative bounds. 
According to the Super-K simulations~\cite{Super-Kamiokande:2013rwg,Super-Kamiokande:2022egr,Super-Kamiokande:2025lxa},
the non-interacting fractions are about $50\,\%$ and $80\,\%$ for  $\pi^0$ and $K^0$, respectively.  

\textbf{Detector resolutions:} 
The Super-K detector reconstructs both energies and directions of the Cherenkov rings produced by the charged particles and photons. 
For energy measurement, we use the parametrized energy resolutions from~\cite{Super-Kamiokande:2005mbp}
\begin{align}
\sigma_{\rm sh}=\Big(0.6+\frac{2.6}{\sqrt{p_{\rm sh}/{\rm GeV}}}\Big)\% ,
\qquad
\sigma_{\rm nsh}=\Big(1.7+\frac{0.7}{\sqrt{p_{\rm nsh}/{\rm GeV}}}\Big)\% ,
\end{align}
for showering and non-showering rings, respectively, where $p_{\rm sh}$ and
$p_{\rm nsh}$ denote the corresponding particle momenta. 
The angular resolutions are taken to be 
$3.0^\circ$ and $1.8^\circ$ for showering and non-showering rings~\cite{Super-Kamiokande:2005mbp}, respectively.

\subsubsection{$p\to \ell^+ K^0 X$\,($X\ell uus$)}
\label{subsubsec:p2lK0X}

We constrain $p\to e^+ K^0 X$ and $p\to \mu^+ K^0 X$ by reinterpreting the
Super-K searches for the corresponding two-body modes $p\to e^+K^0$~\cite{Super-Kamiokande:2005lev} and
$p\to \mu^+K^0$~\cite{Super-Kamiokande:2022egr},
respectively.
Since $K^0$ is reconstructed through its decay products, 
we focus on the decay mode $K_S^0\to 2\pi^0$. This mode provides a  
clean experimental signature via the subsequent decay $\pi^0\to\gamma\gamma$ and yields the strongest constraints. 
Consequently, after accounting for the $K^0$ decay, the final state consists of one charged lepton and four photons.

The event selection criteria used in the two-body decay searches~\cite{Super-Kamiokande:2022egr,Super-Kamiokande:2005lev} can be divided into
non-kinematic (A-1,2,3,4,9 in~\cite{Super-Kamiokande:2022egr}) and kinematic (A-5,6,7,8 in~\cite{Super-Kamiokande:2022egr}) cuts. 
The non-kinematic cuts are mainly based on
the Cherenkov-ring topology, and 
their overall selection efficiencies are estimated to be 
$0.10$ for the $e^+K^0$ mode and $0.12$ for the $\mu^+K^0$ mode from the simulation results provided in~\cite{Super-Kamiokande:2022egr,Super-Kamiokande:2005lev}. 
In deriving limits on the corresponding three-body modes, we assume similar selection efficiencies.

For the kinematic cuts, we reevaluate their efficiencies for the three-body modes,
as their visible kinematic distributions differ from those of the two-body modes and depend on the underlying operator structures. For the $p\to\mu^+K^0 X$ mode, we
adopt the same muon-momentum and kaon-mass selection criteria as those used in the Super-K analysis~\cite{Super-Kamiokande:2022egr},
\begin{align}
150\,{\rm MeV}<|\pmb{p}_{\mu}|<400\,{\rm MeV},\qquad
400\,{\rm MeV}<M_{K^0}<600\,{\rm MeV},
\label{eq:cut1}
\end{align}
where $K^0$ is reconstructed from the four photons. 
The lower cut on muon momentum is slightly above its Cherenkov threshold (120\,MeV).
By contrast, no analogous cuts can be imposed for the $p\to e^+K^0 X$ mode, 
because the electron cannot be reliably distinguished from the four photons, as they produce similar showering Cherenkov rings.

\begin{figure}[t]
\centering
\includegraphics[width=0.442\linewidth]{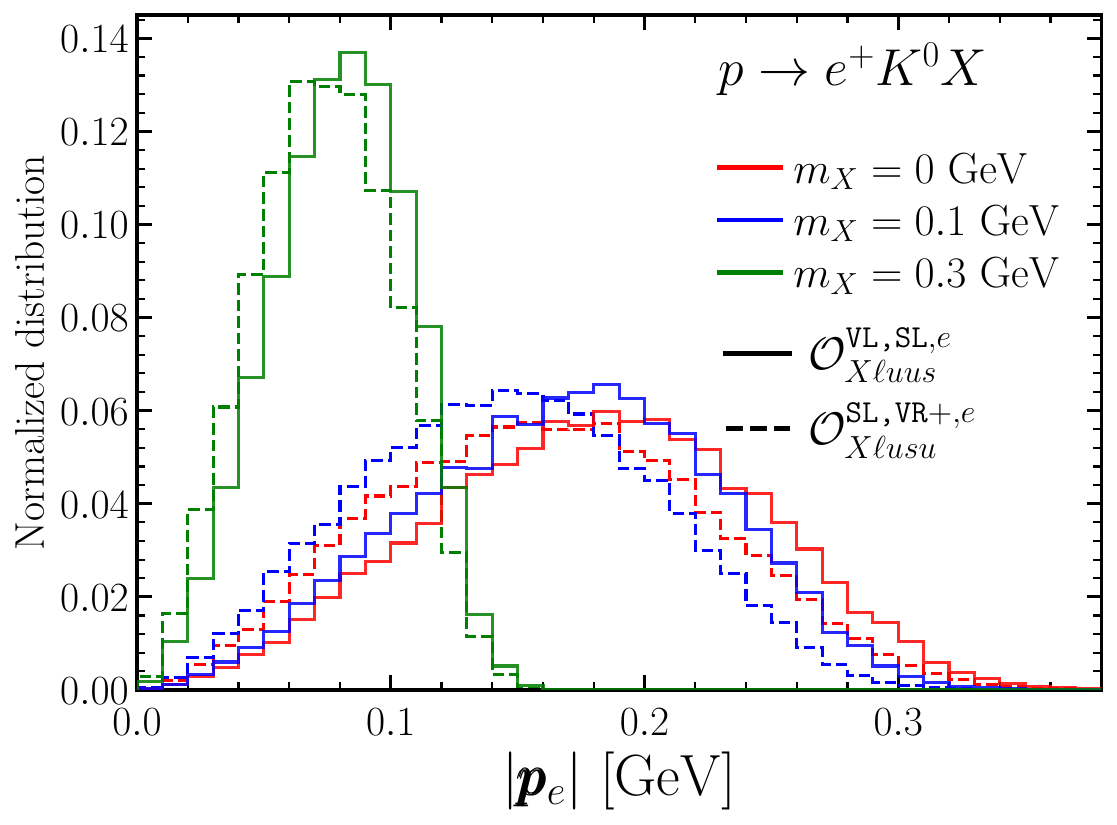}
\includegraphics[width=0.45\linewidth]{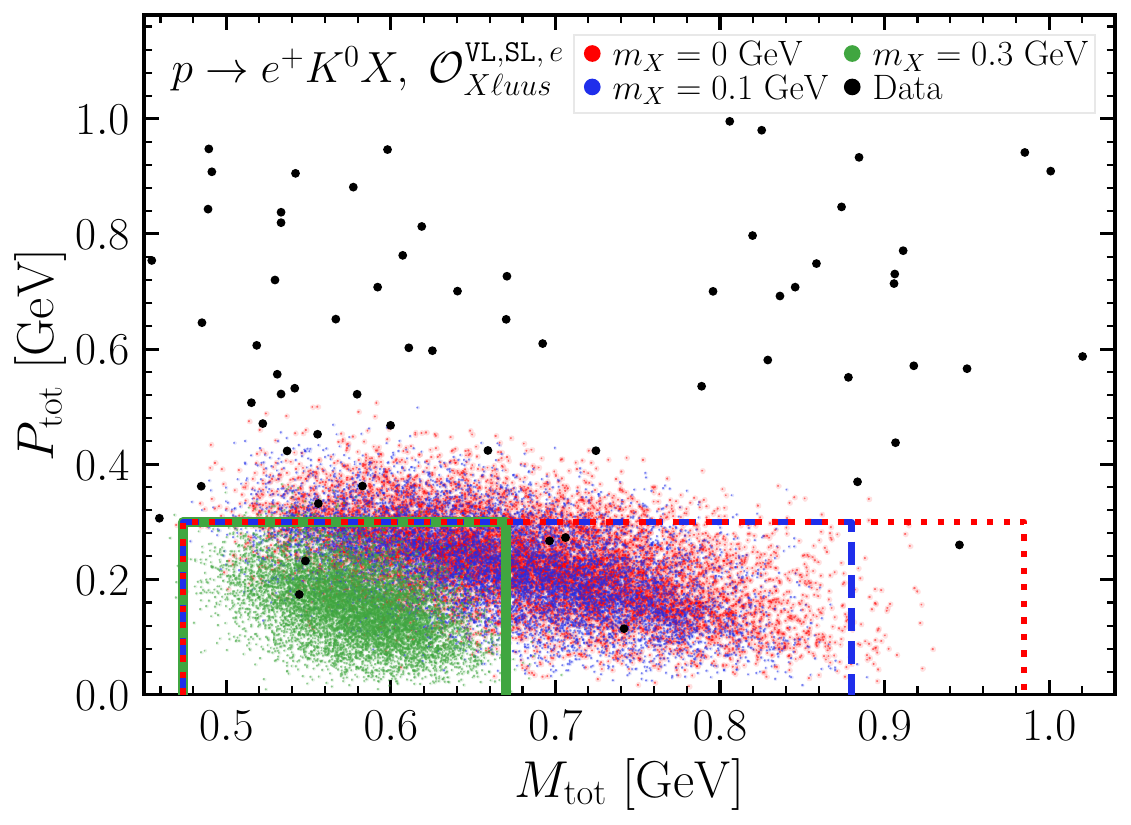}
\\ 
\includegraphics[width=0.442\linewidth]{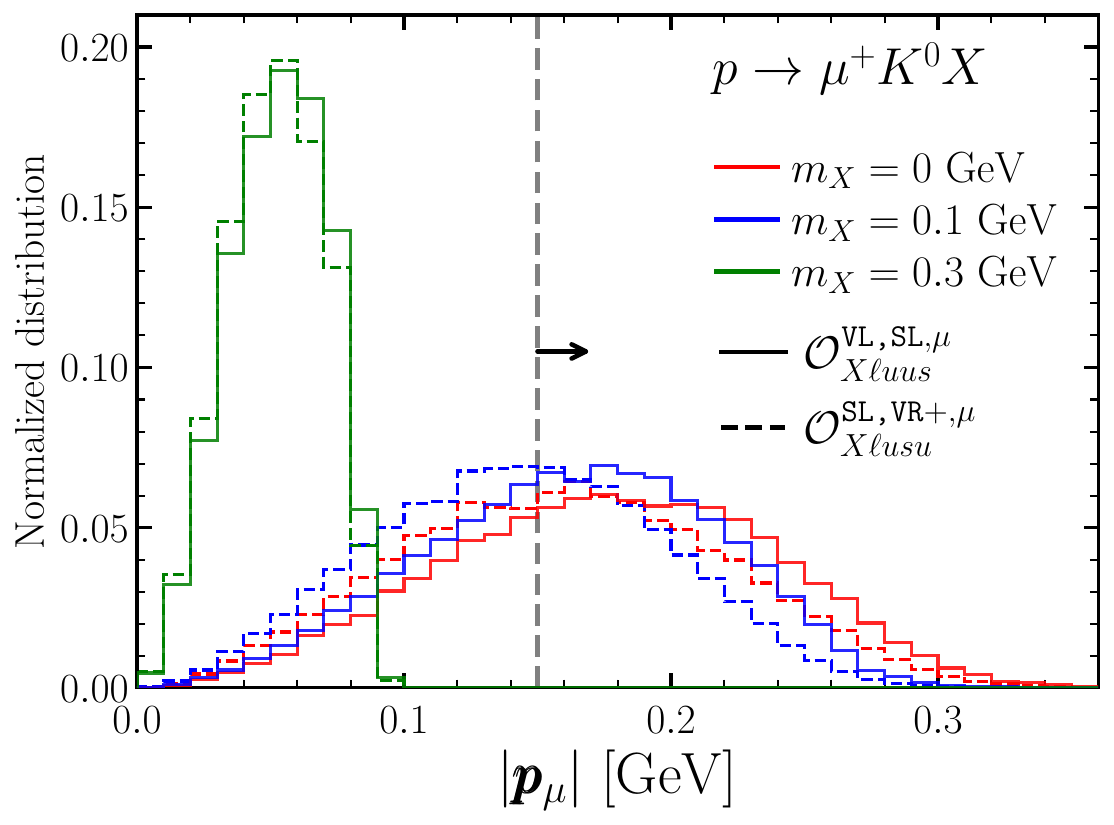}
\includegraphics[width=0.45\linewidth]{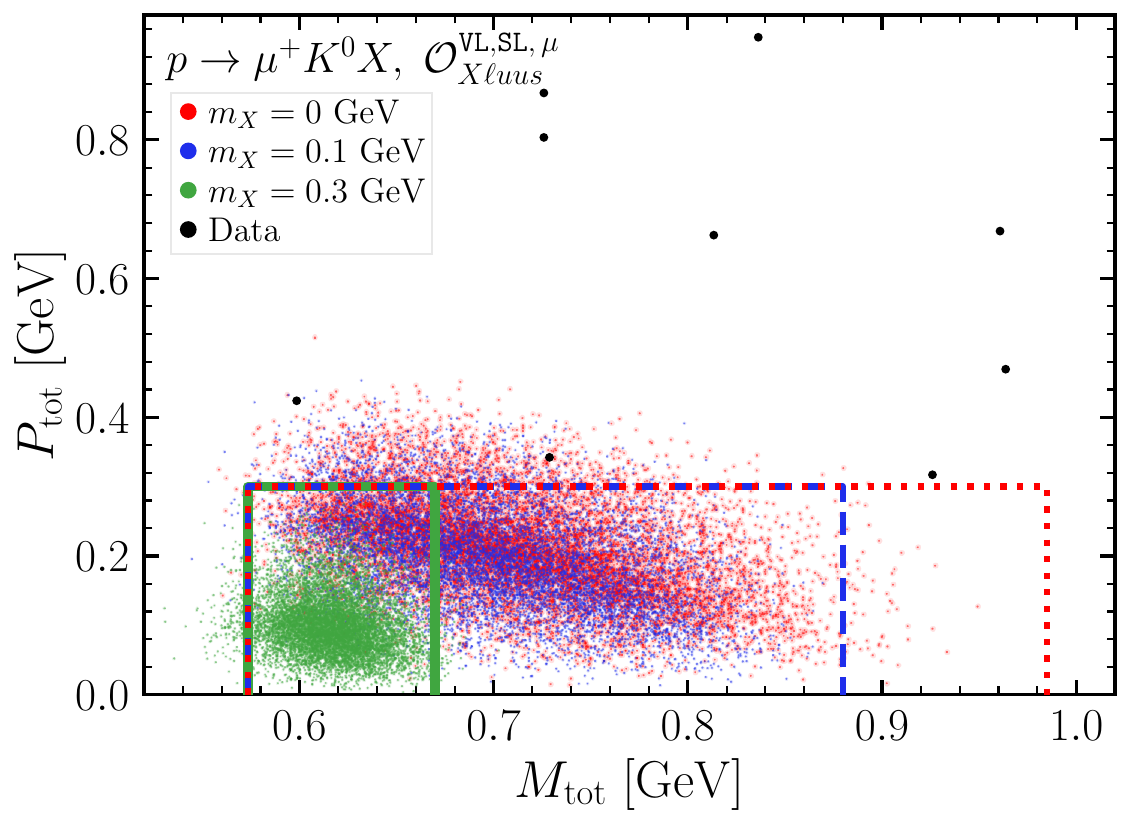}
\caption{Kinematic distributions for $p\to e^+K^0X$ (upper) and $p\to \mu^+K^0X$ (lower) at detector level. The left panels show
the normalized $e^+$ and $\mu^+$ momentum spectra, while the right panels show their $M_{\rm tot}\text{--}P_{\rm tot}$ distributions.}
\label{fig:fig_pmu}
\end{figure}

For each decay mode, we define a total invariant mass $M_{\rm tot}$ and a total momentum $P_{\rm tot}$ from all its visible final-state particles, respectively.
In \cref{fig:fig_pmu}, we show the lepton momentum spectra (left panels) and $M_{\rm tot}$--$P_{\rm tot}$ distributions (right panels) for the two three-body modes $p\to e^+K^0X$ (upper panels) and $p\to \mu^+K^0X$ (lower panels), considering three mass points $m_X=0$, $0.1$, and $0.3~\mathrm{GeV}$.
For the lepton momentum spectra shown in the left panels, 
the solid and dashed histograms correspond to contributions from two representative operators
$\calO_{X\ell uus}^{{\tt VL,SL},x}$ and
$\calO_{X\ell uus}^{{\tt SL,VR}+,x}$, respectively. 
As $m_X$ increases, the invisible dark photon carries away a larger fraction of
the available energy, which causes the charged-lepton spectra to shift toward lower values.
The spectra are also sensitive to the operator structures: the spectra from the operator
$\calO_{X\ell uus}^{{\tt VL,SL},x}$ 
concentrate on a larger momentum region than those of $\calO_{X\ell uus}^{{\tt SL,VR}+,x}$.
This difference is particularly important for the muon channel because of the imposed momentum cut (the gray vertical line), which implies that there are more surviving events from the harder spectra. 

For the $M_{\rm tot}$--$P_{\rm tot}$ distributions shown in the right panels, 
we take the operator $\calO_{X\ell uus}^{{\tt VL,SL},x}$ as an example and generate $10^4$
unweighted events for each benchmark mass.
For comparison, we also show the observed data points as black dots. These events are primarily produced by atmospheric neutrinos, with the vast majority located in the region $P_{\rm tot}>300\,\rm{MeV}$.
As $m_X$ increases, the signal distributions become increasingly concentrated at smaller values of both $M_{\rm tot}$ and $P_{\rm tot}$.
Based on these behaviors, we define the signal region as
\begin{align}
0.95\,(m_\ell+m_{K^0}) < M_{\rm tot} < 1.05\,(m_p-m_X)
~\&~
P_{\rm tot}<300\,{\rm MeV}.
\label{eq:cut2}
\end{align}
The resulting signal regions for the three benchmark masses are indicated by the colored rectangular boxes in \cref{fig:fig_pmu}. 

To derive a lower limit on the partial lifetime $\tau \equiv \Gamma^{-1}$ for each three-body decay mode, we use a Poisson likelihood function for the observed event count in the signal region. 
For a given value of $\tau$, the expected number of events is
$\mu(\tau,b)=\lambda\varepsilon/\tau+b$, where $\lambda$ denotes the experimental exposure.  
The final signal efficiency $\varepsilon$ is obtained as the product of the individual efficiencies for the non-kinematic and kinematic selection requirements described in \cref{eq:cut1,eq:cut2}.
The expected background yield $b$ is treated as a nuisance parameter and is modeled by a Gaussian distribution $P(b)$ with mean $b_0$ and standard deviation $\sigma_b$. 
The likelihood function is then obtained by marginalizing the Poisson distribution over the background uncertainty,
\begin{align}
P(\tau\,|\,n)
\propto
\frac{1}{n!}\int_0^\infty db\,
e^{-\mu(\tau,b)}\mu(\tau,b)^n\,
P(b),
\end{align}
where $n$ is the number of events observed.
The integration over $b$ is restricted to the physical region $b>0$. The 90\% C.L. lower limit is then determined by
\begin{align}
\frac{\int_{\tau_{90}}^\infty d\tau\, P(\tau\,|\,n)}
{\int_0^\infty d\tau\, P(\tau\,|\,n)} = 0.9 .
\label{eq:90CL_bound}
\end{align}

\begin{figure}[t]
\centering
\includegraphics[width=0.44\linewidth]{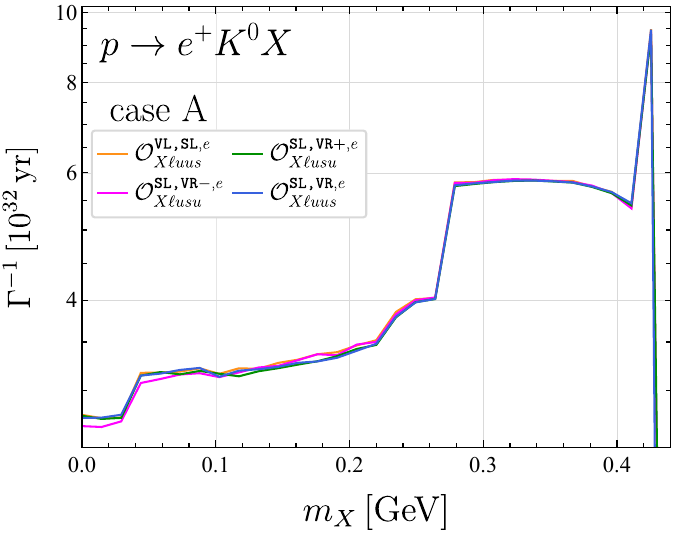}
\includegraphics[width=0.44\linewidth]{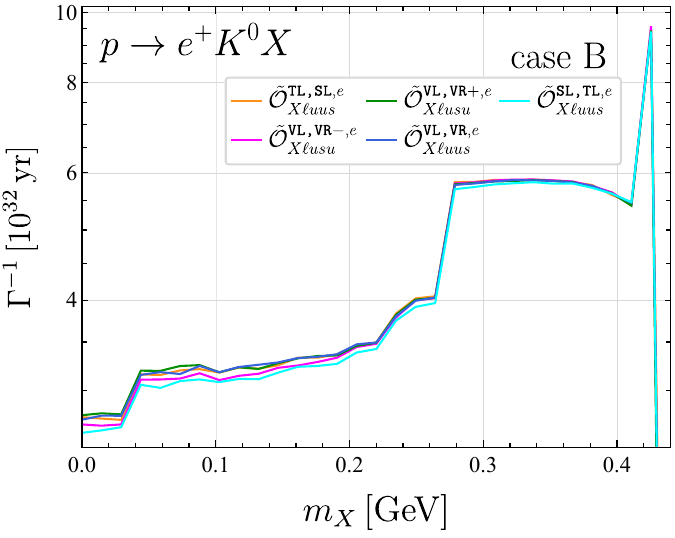}
\\
\includegraphics[width=0.44\linewidth]{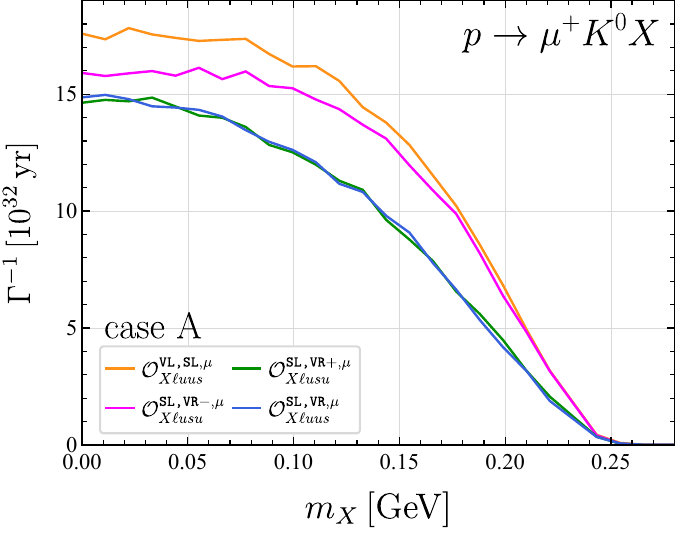}
\includegraphics[width=0.44\linewidth]{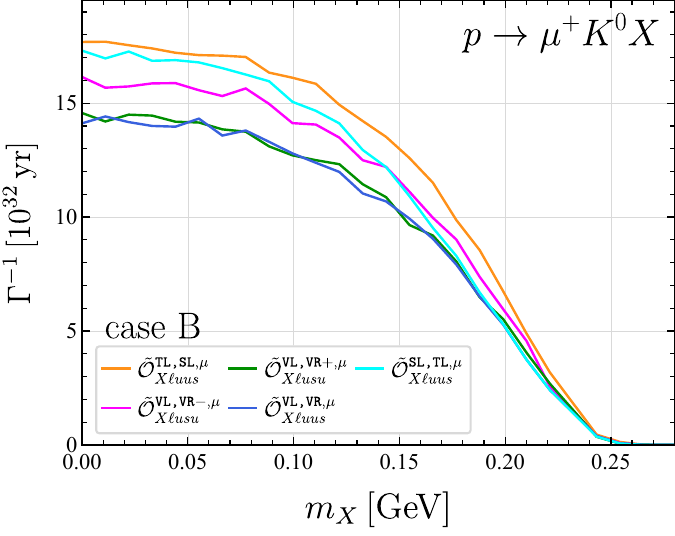}
\caption{Recasting constraints on the partial lifetimes of $p\to e^+ K^0 X$ (upper) and $p\to \mu^+ K^0 X$ (lower).}
\label{fig:InGamma_p2lK0X}
\end{figure}

For the $p\to e^+K^0X$ and $p\to \mu^+K^0X$ modes, we use exposures of 92\,kton$\cdot$yr from the SK-I dataset and 199\,kton$\cdot$yr from the SK-IV dataset, respectively. 
The observed event count $n$ is obtained from the right panels of fig.\,11 in~\cite{Super-Kamiokande:2005lev}
and fig.\,7 in~\cite{Super-Kamiokande:2022egr}. 
The expected background $b_0$ is determined by counting the simulated background events in the corresponding middle panels after rescaling them to the same experimental exposures.
For both parameters $n$ and $b_0$, they are restricted to the signal region defined in \cref{eq:cut2}. 
The background uncertainty $\sigma_b$ is taken to be proportional to the expected background yield, with $\sigma_b = 0.62\,(0.33)\, b_0$ for the $e(\mu)$-mode~\cite{Super-Kamiokande:2005lev,Super-Kamiokande:2022egr}. 

The resulting 90\,\% C.L. lower limits on partial lifetimes for the two modes $p\to e^+K^0X$ and $p\to \mu^+K^0X$ as a function of $m_X$ are shown in
\cref{fig:InGamma_p2lK0X}.
The left (right) panels correspond to the $X\ell uus$-class operators in case A (case B).
The markedly different $m_X$ dependence of the limits in the two modes can be understood from the interplay
between the kinematic distributions in the $M_{\rm tot}\text{--}P_{\rm tot}$ plane and
the $\mu^+$ momentum cut, as illustrated in \cref{fig:fig_pmu}. 
In the $e^+$ channel, the signal becomes increasingly concentrated within the signal region as $m_X$ increases, leading to a higher signal-to-background ratio and consequently stronger limits. 
The step-like behavior of the constraints arises from
the discrete change in the number of data points falling within the signal region as $m_X$ varies. 
In contrast, in the $\mu^+$ channel, the 
muon momentum cut removes an increasing fraction of low-momentum $\mu^+$s as $m_X$ increases, thereby reducing the signal efficiency and weakening the limits. 
This momentum cut also makes the limits in the $\mu^+$ channel more sensitive to the underlying operator structure, since it preserves the differences in the muon momentum spectra after the event selection, as discussed above. 
Overall, the resulting lower limits on the partial lifetimes are approximately $(2-6)\cdot 10^{32}\,\rm yr$ for $p\to e^+ K^0 X$ and $(1-19)\cdot 10^{32}\,\rm yr$ for $p\to \mu^+ K^0 X$.

\subsubsection{$n\to \ell^-\pi^+ X$\,($X\bar\ell ddd$)}
\label{subsubsec:n2lpipX}

For the decay modes $n\to e^-\pi^+ X$ and $n\to \mu^-\pi^+ X$, there are no dedicated Super-K searches for the corresponding two-body modes $n\to e^-\pi^+$ and $n\to \mu^-\pi^+$. We therefore derive constraints by reinterpreting the Super-K analysis of the charge-symmetric channels $n\to e^+\pi^-$ and $n\to \mu^+\pi^-$ reported in~\cite{Super-Kamiokande:2017gev}, corresponding to a total exposure of $0.316~{\rm Mton}\cdot{\rm yr}$. Although each pair of charge-symmetric processes corresponds to different patterns of baryon and lepton number violation, they produce nearly identical prompt Cherenkov signatures in the detector. 
In particular, the visible final state consists of a charged lepton and a charged pion, leading to one showering and one non-showering Cherenkov ring in the electron channel, and two non-showering rings in the muon channel.

After requiring two reconstructed Cherenkov rings, the observed numbers of events are $2400$ and
$570$ in the $e$ and $\mu$ channels, respectively, with expected atmospheric neutrino backgrounds of $2200$ and $600$~\cite{Super-Kamiokande:2017gev}. 
The corresponding signal efficiencies, including the effects of intranuclear pion interactions and Cherenkov-ring reconstruction, are $0.204$ for $n\to e^+\pi^-$ and $0.228$ for $n\to  \mu^+\pi^-$.
Since $\pi^+$ and $\pi^-$ exhibit comparable intranuclear propagation in the oxygen nucleus~\cite{Saunders:1996ic}, we adopt the same signal efficiencies for the charged-conjugated decay modes in our analysis.

In our recast, we consider only the prompt two-ring topology to derive the constraints.
We require all visible charged particles to be above
their Cherenkov thresholds in water,
corresponding to momentum thresholds of 
$p_{\rm th}\simeq0.58$, $120$, and $159\,{\rm MeV}$ for $e^\pm$, $\mu^\pm$, and $\pi^\pm$, respectively. 
We do not impose the Michel-electron requirement or subsequent kinematic
selections adopted in~\cite{Super-Kamiokande:2017gev}. The former is charge dependent: a stopped $\pi^+$ typically decays via $\pi^+\to\mu^+\to e^+$, whereas a stopped $\pi^-$ is often
captured by nuclei before decaying~\cite{Numao:2006br}. The latter are optimized for two-body 
decay $n\to\ell^+\pi^-$ and are therefore not directly applicable to the three-body process $n\to\ell^-\pi^+X$. 

\begin{figure}[t]
\centering
\includegraphics[width=0.45\linewidth]{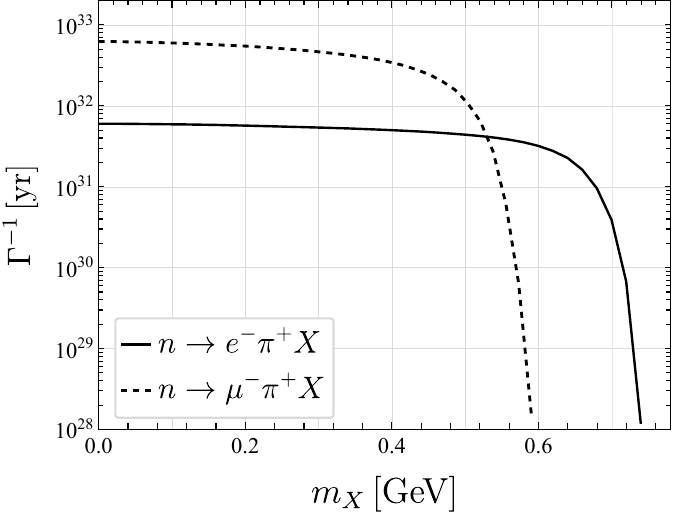}
\includegraphics[width=0.45\linewidth] {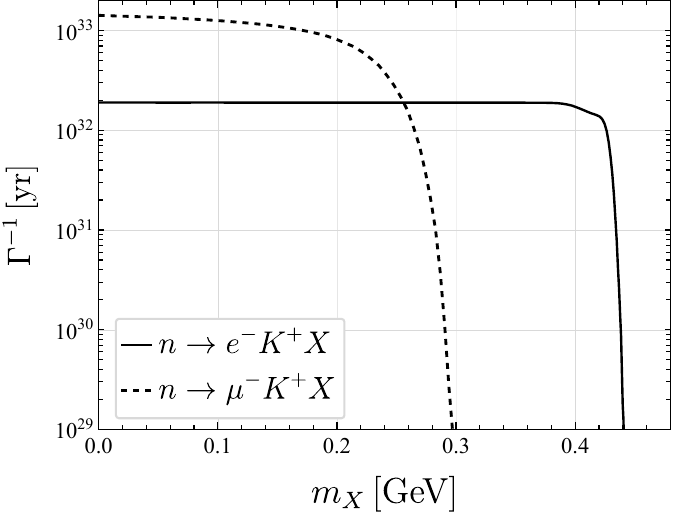}
\caption{Recasting constraints on the partial lifetimes of
$n\to \ell^- \pi^+ X$ (left) and $n\to \ell^- K^+ X$ (right).}
\label{fig:InGamma_n2lMX}
\end{figure}

All analysis steps, in addition to the omission of kinematic selections, are the same as those used for $p\to \ell^+ K^0 X$ in \cref{subsubsec:p2lK0X}. 
The resulting 90\,\% C.L. partial lifetime limits are
shown in the left panel of \cref{fig:InGamma_n2lMX}. Since this recast relies only on the prompt
two-ring topology and does not apply additional kinematic cuts, the limits 
exhibit minimal dependence on the underlying operator structure. Thus, a single curve is shown for each mode.
These bounds are conservative. A dedicated analysis with channel-specific selection criteria could more effectively
suppress the atmospheric-neutrino backgrounds, further improving the sensitivity.
Across a broad range of $m_X$, the partial lifetime bounds are approximately $6\cdot 10^{31}\,\rm yr$ for the electron mode and $5\cdot 10^{32}\,\rm yr$ for the muon mode. The roughly one-order-of-magnitude stronger limit for the muon mode arises from its lower expected background events. 

\subsubsection{$n\to \ell^- K^+ X$\,($X\bar\ell dds$)}
\label{subsubsec:n2lKpX}

Similarly to the treatment of $n\to\ell^- \pi^+ X$ in \cref{subsubsec:n2lpipX}, we  constrain $n\to \ell^-K^+X$ by reinterpreting the Super-K data for $n\to \ell^+\pi^-$ in~\cite{Super-Kamiokande:2017gev}.
This is justified because the produced $K^+$ is identified through its decay products, as discussed in \cref{subsubsec:p2lK0X}.
Its dominant decay mode, $K^+\to\mu^+\nu_\mu$, produces a daughter muon that forms a non-showering Cherenkov ring. 
Consequently, the process $n\to e^-K^+X$ yields one showering ring and one non-showering ring, while $n\to \mu^-K^+X$ yields two non-showering rings. 
These Cherenkov ring topologies are
identical to those in $n\to e^-\pi^+$ and $n\to \mu^-\pi^+$, respectively.
We therefore use the same two-ring event samples, 
including the observed and expected event numbers.

We require the primary charged lepton to be above its Cherenkov threshold. 
The secondary muon from the kaon decay automatically satisfies its Cherenkov threshold because of the two-body decay kinematics. 
Since the intranuclear interaction of
$K^+$ in oxygen is negligible, no additional nuclear suppression factor is
needed. 
We adopt the ring-identification efficiencies from~\cite{Super-Kamiokande:2017gev}: $94.0\,\%$ for showering rings and
$76.4\,\%$ ($91.7\,\%$) for non-showering rings in SK-I--III (SK-IV).
Incorporating the branching ratio $\mathcal{B}(K^+\to\mu^+\nu_\mu)=0.636$ into the overall efficiency, we obtain signal efficiencies of 0.457 (0.548) for $n\to e^- K^+X$ and 0.371 (0.535) for $n\to \mu^- K^+X$ in SK-I--III (SK-IV), respectively. 

The resulting 90\,\% C.L. lower limits on the partial lifetimes are shown in the right panel of
\cref{fig:InGamma_n2lMX}. As in the pion case, our recast limits depend only on the
two-ring topology and are therefore insensitive to the underlying operator structure.
The bounds exhibit behaviors similar to those of $n\to (e^-,\mu^-)\pi^+ X$ shown in the left panel, due to the similarity of the analysis.
The slightly stronger constraints on the kaon modes than on the corresponding pion modes with the same charged lepton arise from their higher signal efficiencies. 
Our final results indicate that 
$\Gamma^{-1}(n\to e^- K^+ X)\gtrsim 2\cdot 10^{32}\,\rm yr$ for $m_X \lesssim 0.4 \,\rm GeV$,
and $\Gamma^{-1}(n\to \mu^- K^+ X)\gtrsim 10^{33}\,\rm yr$ for $m_X \lesssim 0.2 \,\rm GeV$.

\subsubsection{$n\to \nu \pi^0 X$\,($X\bar\nu udd$)}

We constrain $n\to \nu \pi^0 X$ by reinterpreting the latest
Super-K search for $n\to \nu\pi^0$ in~\cite{Super-Kamiokande:2025lxa}, which is based on SK-I--V data with a total exposure of $0.484~{\rm Mton\cdot yr}$.
We adopt a signal efficiency of $0.32$ for the non-kinematic selections and
impose the kinematic requirements for the reconstructed di-photon as in the experimental paper:
\begin{align}
85\,{\rm MeV}<M_{\pi^0}<185\,{\rm MeV},\qquad
0<|\pmb{p}_{\pi^0}|<1000\,{\rm MeV}.
\end{align}
Since neutrinos cannot be detected, the experimental data for $n\to \nu \pi^0$ is binned solely in pion momentum (fig.\,4 in~\cite{Super-Kamiokande:2025lxa}), unlike the previous cases which used the $M_{\rm tot}$-$P_{\rm tot}$ plane. 
In this case, we follow Ref.\,\cite{Super-Kamiokande:2025lxa} and use a simpler $\chi^2$-statistics method to derive the bound. 
It is defined by
\begin{align}
\chi^2(\tau) = 2
\sum_i
\Big[ \mu_i(\tau)-n_i+n_i\ln\frac{n_i}{\mu_i(\tau)} \Big],
\quad
\mu_i(\tau)=s_i(\tau)+b_i ,
\label{eq:chi2_poisson_bkg}
\end{align}
where $i$ runs over all $|\pmb{p}_{\pi^0}|$-bins.  
The expected signal, background, and observed event counts in the $i$th bin are denoted by $s_i(\tau)$, $b_i$, and $n_i$, respectively,
where the latter two are extracted directly from fig.\,4 of the paper. 
In the above formula, systematic uncertainties are neglected due to insufficient information from the experimental paper. 
The final 90\% C.L. lower bound on the partial lifetime is defined by
\begin{align}
\chi^2(\tau_{90})-\chi^2(\tau\to\infty)=2.71.
\end{align}

\begin{figure}[t]
\centering
\includegraphics[width=0.425\linewidth]{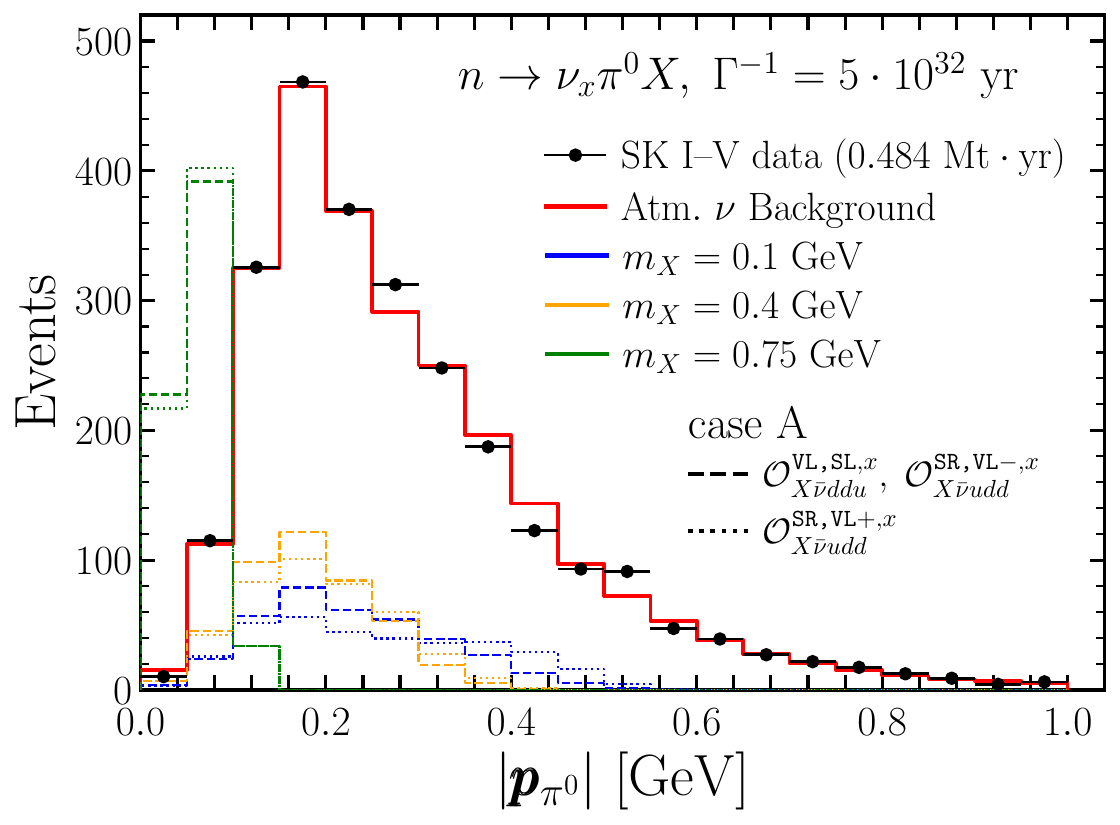}
\includegraphics[width=0.425\linewidth]{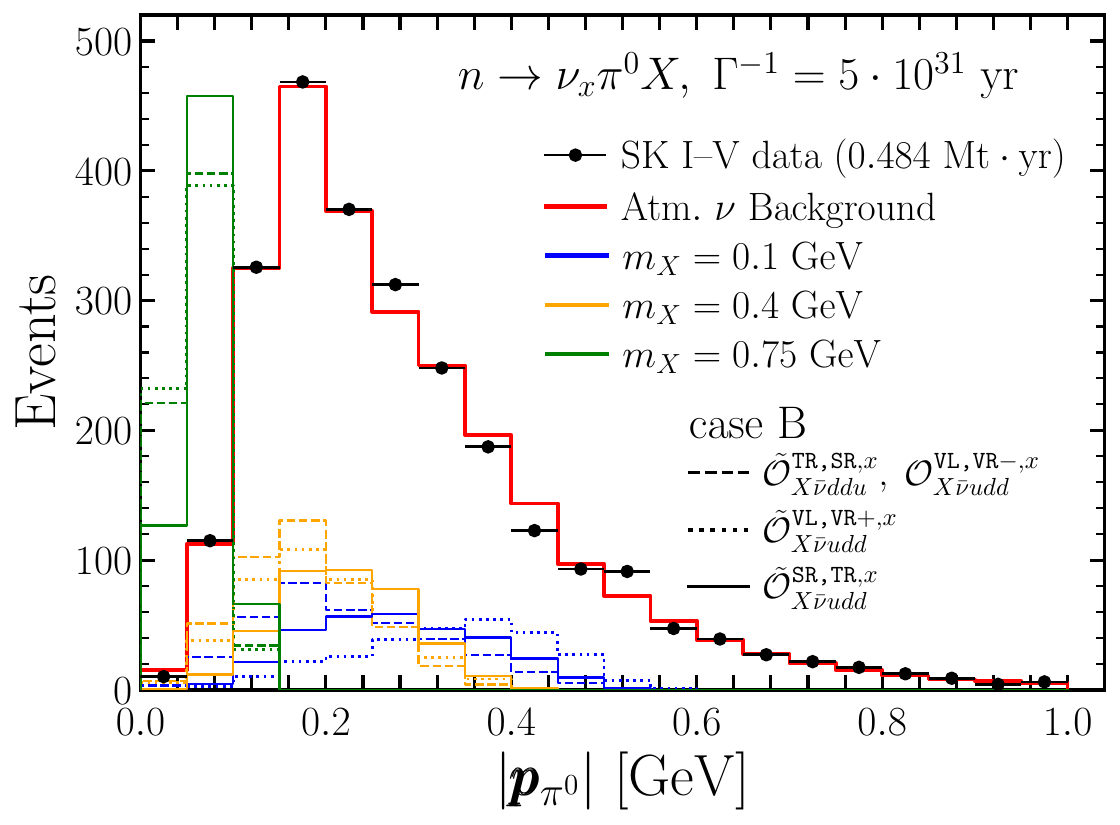}
\includegraphics[width=0.425\linewidth]{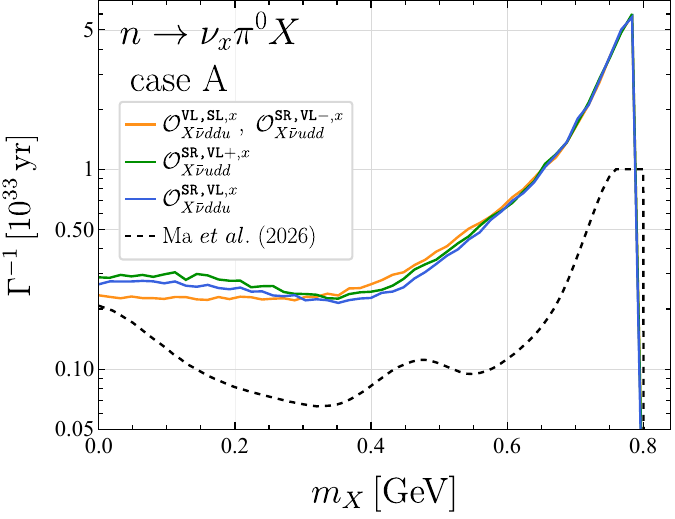}
\includegraphics[width=0.425\linewidth]{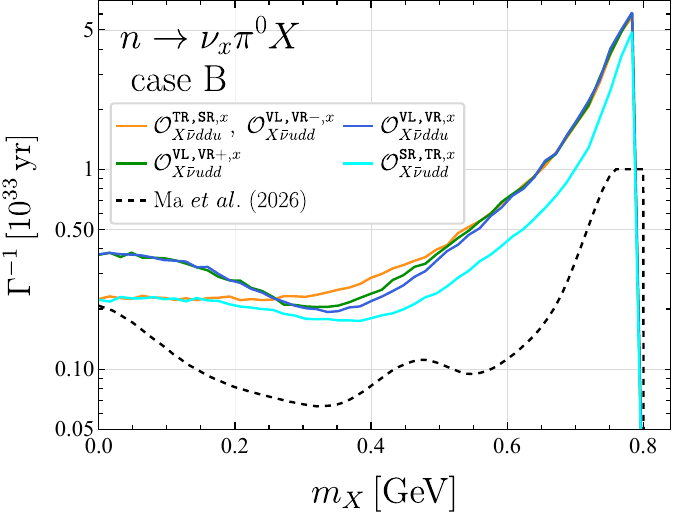}
\caption{Event distributions and recasting bound on the partial lifetime of $n\to \nu \pi^0 X$. }
\label{fig:InGamma_n2vpi0X}
\end{figure}

The top two panels of~\cref{fig:InGamma_n2vpi0X} show the binned pion-momentum spectra for several representative operators in case A (left) and B (right) and three benchmark masses, $m_X=0.1$, $0.4$, and $0.75\,\rm GeV$,  after incorporating the experimental effects.
The observed data and simulated background event counts are also shown.
Each signal spectrum for the $n\to \nu\pi^0 X$ decay is evaluated assuming a benchmark partial lifetime $\Gamma^{-1}=10^{32}\,\rm yr$.
As in the previous channels, increasing $m_X$ shifts the pion spectrum toward lower momenta while making it more concentrated.
Meanwhile, the relatively low background in the low-momentum region enhances the sensitivity, leading to progressively stronger lower limit on the partial lifetime as $m_X$ approaches 0.784\,GeV, as illustrated in the bottom panels.

The final results are shown as the colored solid curves in the bottom two panels, corresponding to the two dark photon cases. 
Generally, the limits derived from different operators tend to converge for $m_X\gtrsim0.5\,\rm GeV$ because the signal spectra become increasingly similar as $m_X$ increases, as exemplified in the top panels. 
A notable exception is the operator $\tilde \calO_{X\bar\nu udd}^{{\tt SR,TR},x}$, which consistently yields stronger limits across the entire kinematically allowed range. This behavior can be understood by examining the top right panel in which the signal spectra exhibit a distinctly different shape compared to those of the other operators. 
Additionally, since the pion momentum distributions for $n\to \nu\pi^0 X$ are similar to those for $p\to e^+ \pi^0 X$ shown in \cref{fig:3body_dis_A,fig:3body_dis_B}, due to the $u\leftrightarrow d$ relationship and negligible electron mass, these constraints can also be interpreted in terms of the normalized distributions.

Beyond the peak, the limit deteriorates rapidly as $m_X$ approaches the kinematic endpoint, $m_n-m_{\pi^0}\simeq0.805~\mathrm{GeV}$. 
In this region, the decay is kinematically allowed only for the high-mass tail of the effective bound-neutron mass distribution, leading to a rapid suppression of the signal rate. 
Overall, the lower limit on the partial lifetime lies in the range of $\calO(10^{32\text{--}34}\,\rm yr)$.  
For comparison, we also include the recast bound on $n\to \nu\pi^0 \varphi$ from~\cite{Ma:2025mjy} (dashed curve), which was derived using the earlier Super-K data reported in~\cite{Super-Kamiokande:2013rwg}. 
As observed, the new bounds are stronger by a factor of a few across the entire mass range. This improvement comes from a more comprehensive treatment of nuclear effects and the use of the full SK-I--V data in the latest Super-K analysis~\cite{Super-Kamiokande:2025lxa}.

\subsubsection{$p\to \nu K^+ X$\,($X\bar\nu uds$)}

We constrain $p\to \nu K^+X$ by reinterpreting the Super-K search for $p\to \nu K^+$~\cite{Super-Kamiokande:2014otb}.
This analysis is based on a data set with a total exposure of $0.26\,\rm {Mton\cdot yr}$ and sets a 90\,\% C.L. lower limit on the partial lifetime, 
$\Gamma^{-1}_{\tt SK}(p\to\nu K^+) > 5.9\cdot10^{33}\,\rm yr$.
The signal is identified through the decay products of a stopped $K^+$,
since approximately 89\,\% of the produced kaons come to rest in water before decaying~\cite{Super-Kamiokande:2014otb}.
For the three-body decay $p\to\nu K^+X$, the additional dark photon modifies the initial $K^+$ momentum spectrum but does not affect the signature of the stopped-$K^+$ component. 
We therefore conservatively apply the Super-K bound for $p\to\nu K^+$ directly to $p\to\nu K^+X$, yielding $\Gamma^{-1}(p\to\nu K^+X)>5.9\cdot10^{33}\,\rm yr$ at 90\,\% C.L..

\subsection{Constraints on effective scales associated with $X$LEFT operators}
\label{subsec:Lambda_bounds}

\begin{figure}[t]
\centering
\includegraphics[width=0.325\linewidth]{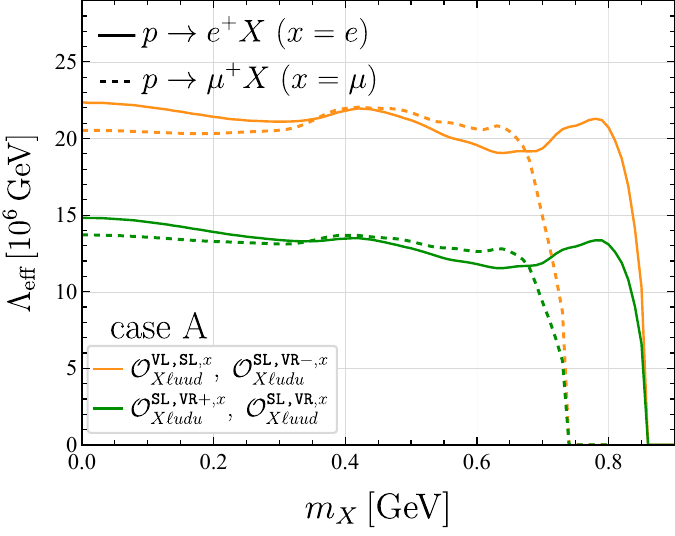}
\includegraphics[width=0.325\linewidth]{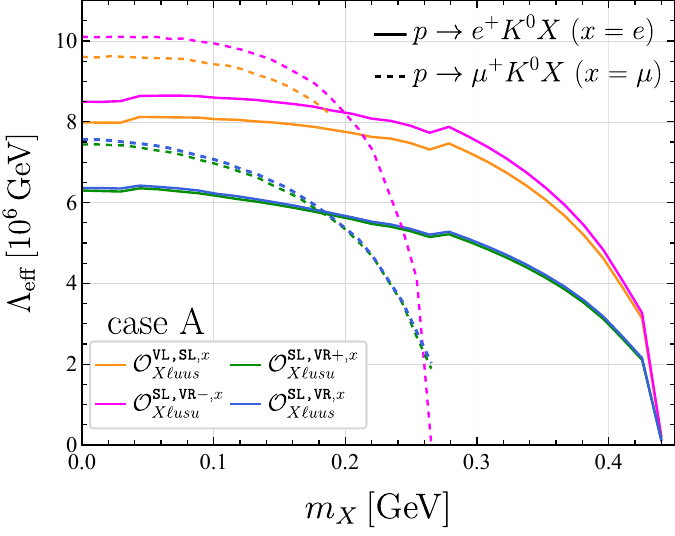}
\includegraphics[width=0.325\linewidth]{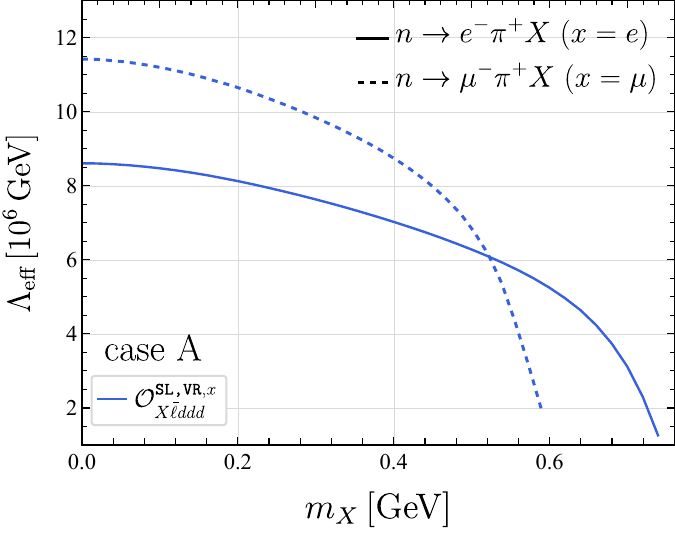}
\\
\includegraphics[width=0.325\linewidth]{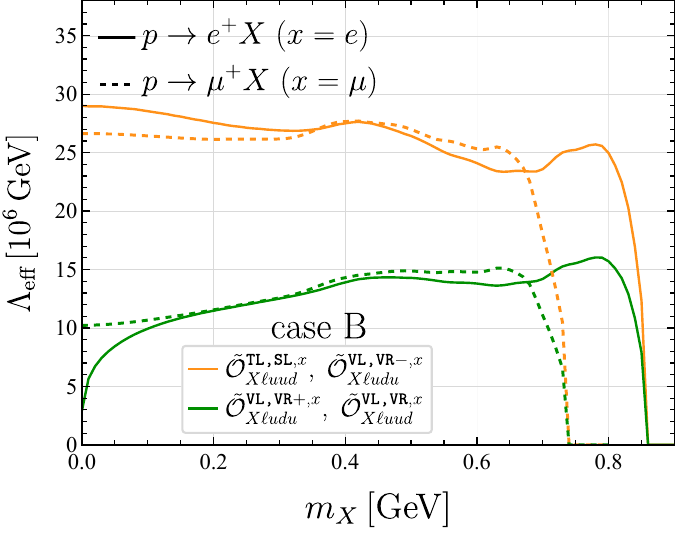}
\includegraphics[width=0.325\linewidth]{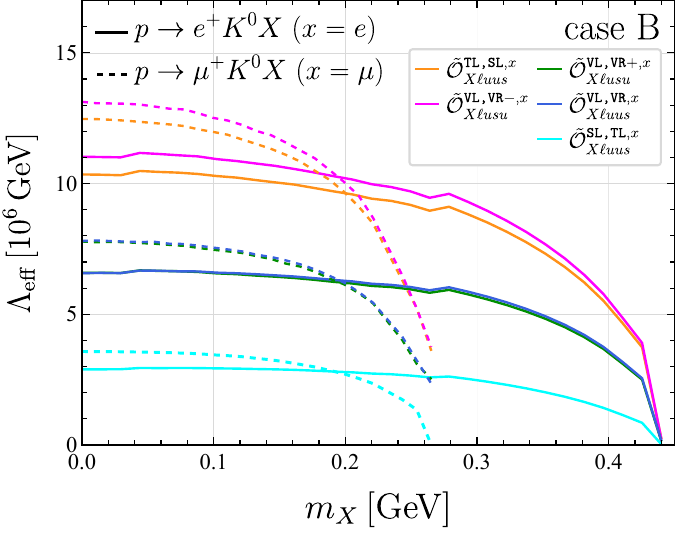}
\includegraphics[width=0.325\linewidth]{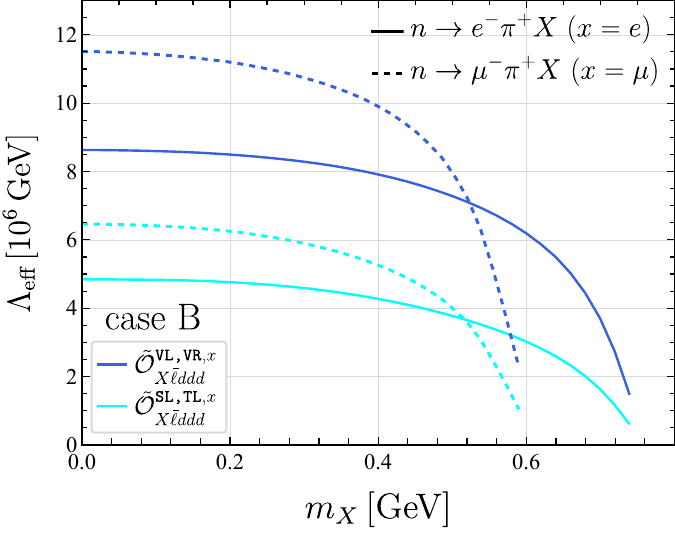}
\\%
\vspace{1em}
\includegraphics[width=0.325\linewidth]{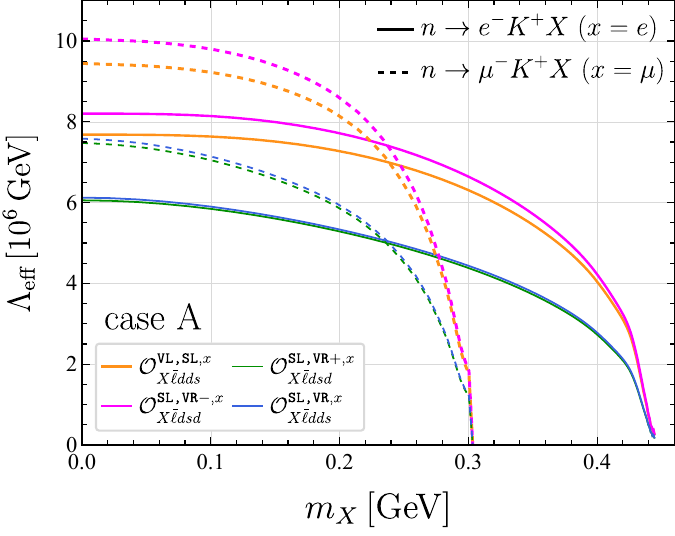}
\includegraphics[width=0.325\linewidth]{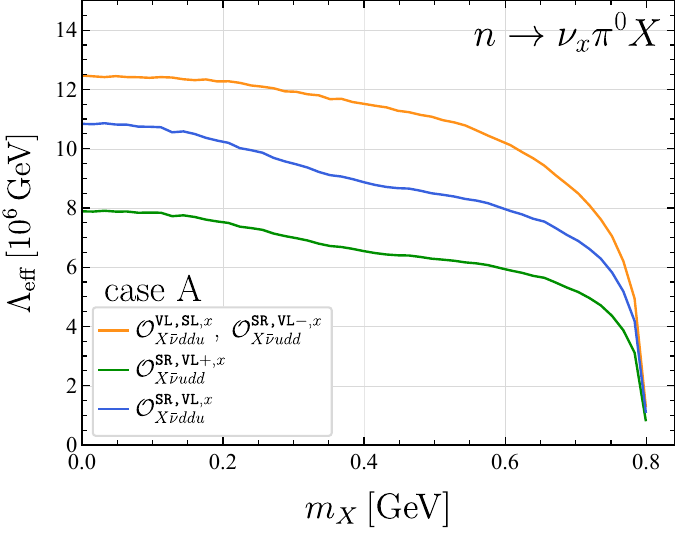}
\includegraphics[width=0.325\linewidth]{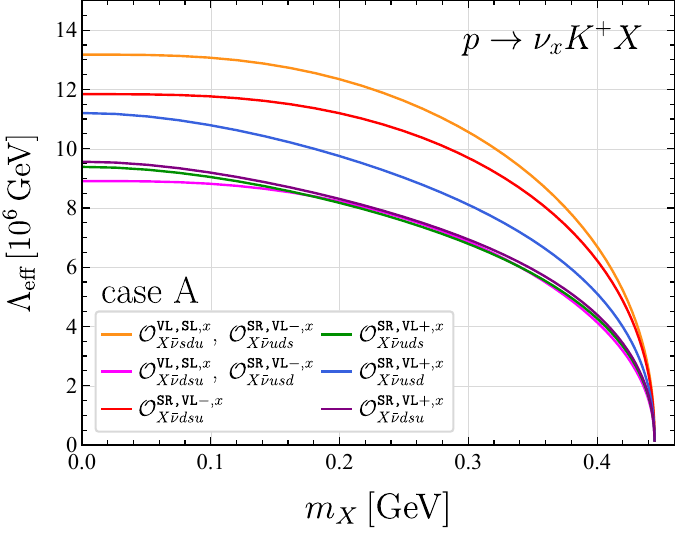}
\\
\includegraphics[width=0.325\linewidth]{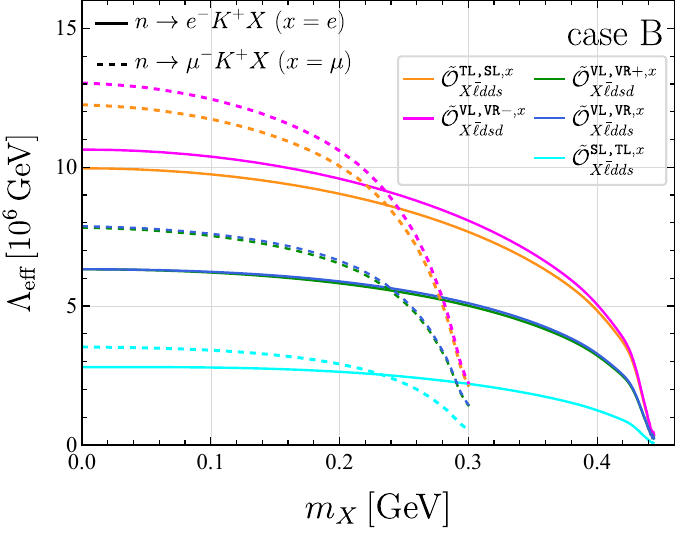}
\includegraphics[width=0.325\linewidth]{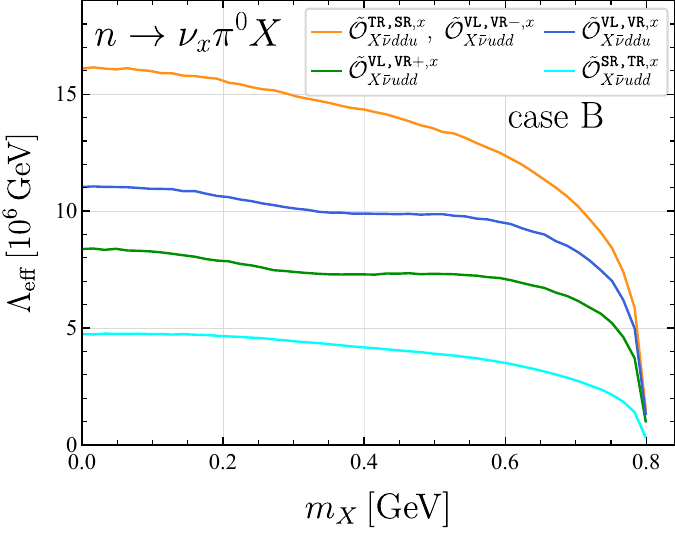}
\includegraphics[width=0.325\linewidth]{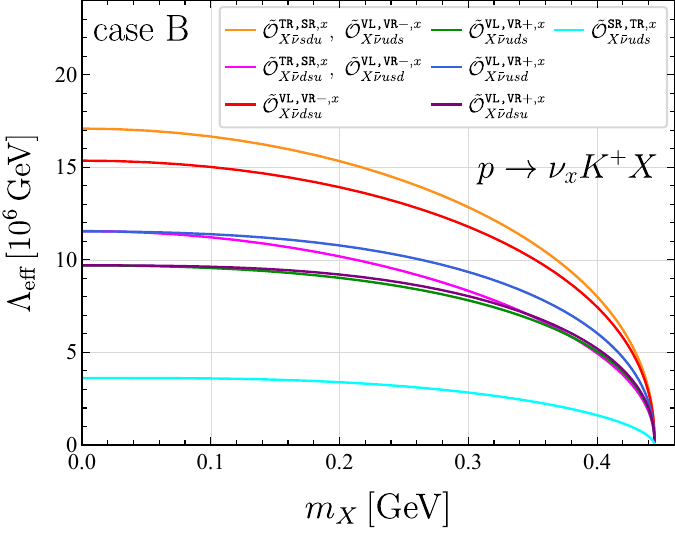}
\caption{Constraints on the effective scales associated with various $X$LEFT operators.}
\label{fig:Lambda_eff_limits}
\end{figure}

We now translate the partial lifetime limits for the three-body decay modes derived in \cref{subsec:recast_bounds}, together with those for the two-body modes $p\to (e^+,\mu^+) X$ obtained in~\cite{Ma:2025mjy},
into constraints on the effective scales of the relevant $X$LEFT operators. 
The effective scale associated with each operator is defined through its WC as $\Lambda_{\rm eff}\equiv |C_i|^{-1/4}$. 
For the case-A operators, the mass dimension of the WCs is reduced by one because an explicit mass factor is included into the decay amplitudes by hand. 
Since the operator $\tilde \calO_{X\ell uud}^{\tt SL,TL}$ in the irrep $\pmb{10}_\tL\otimes \pmb{1}_\tR$ does not contribute to the two-body decay $p\to \ell^+ X$, we exclude it from the present analysis.
In addition, we adopt the NDA estimates for $c_{3,4}$ in~\cite{Liao:2025vlj} when deriving the results. 

For the 25 case-A operators and 30 case-B operators (without counting lepton flavors and excluding $\tilde \calO_{X\ell uud}^{\tt SL,TL}$) listed in the left columns of the six operator classes in \cref{tab:ope_process}, 
the resulting lower bounds on the effective scales as a function of the dark photon mass are presented in \cref{fig:Lambda_eff_limits}. 
The other half chirality-flipped operators are subject to identical bounds and are therefore omitted.  
The top two rows of \cref{fig:Lambda_eff_limits} (the upper row for case A and the lower row for case B) display the results for the $X\ell uud$ (left), 
$X\ell uus$ (middle), and $X\bar\ell ddd$ (right) operator classes.
The bottom two rows show the corresponding results for the  operator classes of $X\bar\ell dds$ (left), 
$X\bar\nu udd$ (middle), and $X\bar\nu uds$ (right) respectively, 
following exactly the ordering in \cref{tab:ope_process}.
For the reader's convenience, each plot also indicates the decay modes used to derive the corresponding bounds.

We now briefly comment on these results. 
First, for all of the operators considered, the effective scales are constrained to be of $\calO(10^{6\text{--}7}\,\rm GeV)$ over a broad range of dark photon masses, demonstrating that nucleon decays provide a sensitive probe of such new physics scenarios. 
Second, although the operators in the irrep $\pmb{6}_{\tL(\tR)}\otimes \pmb{3}_{\tR(\tL)}$ have the same leading-order chiral power counting as those in $\pmb{8}_{\tL(\tR)}\otimes \pmb{1}_{\tR(\tL)}$ and $\bar{\pmb{3}}_{\tL(\tR)}\otimes \pmb{3}_{\tR(\tL)}$, the former generally receive weaker constraints on the effective scales (or equivalently the WCs) for each field configuration. 
This difference originates from the special Lorentz structure of the former's chiral counterparts and the potential underestimate of its corresponding LEC $c_3$ based on the NDA.  
Third, the case-B operators in the irrep $\pmb{10}_{\tL(\tR)}\otimes \pmb{1}_{\tR(\tL)}$ are subject to the weakest constraints (cyan and red dotted curves), reflecting their higher chiral order. 
Finally, the two operators $\tilde\calO_{X\ell udu}^{{\tt VL,VR}+,e}$ and $\tilde\calO_{X\ell uud}^{{\tt VL,VR},e}$ in the irrep $\pmb{6}_{\tL}\otimes \pmb{3}_{\tR}$ exhibit weaker constraints as $m_X\to0$. This behavior occurs because the corresponding squared amplitude is proportional to $2(x_\ell-x_X)\lambda(1,x_\ell,x_X)+3x_X\rho_+$ (see the second line in \cref{eq:Amp_2body_caseB}), which is strongly suppressed when both the charged-lepton and dark-photon masses are sufficiently small.  
In summary, we obtain stringent constraints on 55 lepton-flavor-blind $X$LEFT operators over a wide range of dark photon masses. 

\subsection{Implications for other decay modes}
\label{subsec:converted_bounds}

From the obtained bounds on all relevant WCs of $X$LEFT operators, we examine their implications for other decay processes listed in \cref{tab:ope_process}, i.e., those without a checkmark. 
Inspecting the table, only three groups of operators contribute to more than one nucleon decay channel. These include the operator groups that are characterized by the field configurations: 
$X\ell uud$, $X\hat\nu udd$, $X\hat\nu uds$. 
Based on the decay width results in \cref{sec:decay_width} and the effective scale bounds shown in \cref{fig:Lambda_eff_limits}, 
we can predict lower bounds on the partial lifetimes of the decay modes not covered in \cref{subsec:recast_bounds}.
Under the single-operator-dominance assumption, the partial widths of distinct decay modes induced by the same operator are proportional to each other. This leads to the resulting relative lifetime limits that are essentially independent of the LECs values $c_{1,2,3,4}$.      
We now analyze the three classes one by one. 

\begin{figure}[t]
\centering
\includegraphics[width=0.325\linewidth]{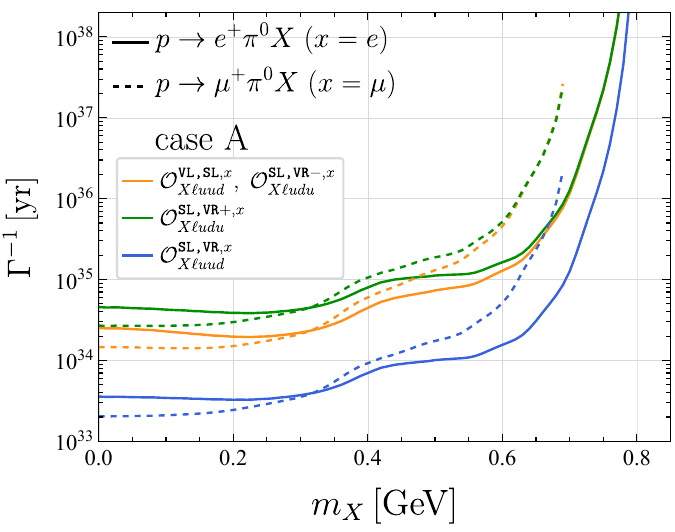}
\includegraphics[width=0.325\linewidth]{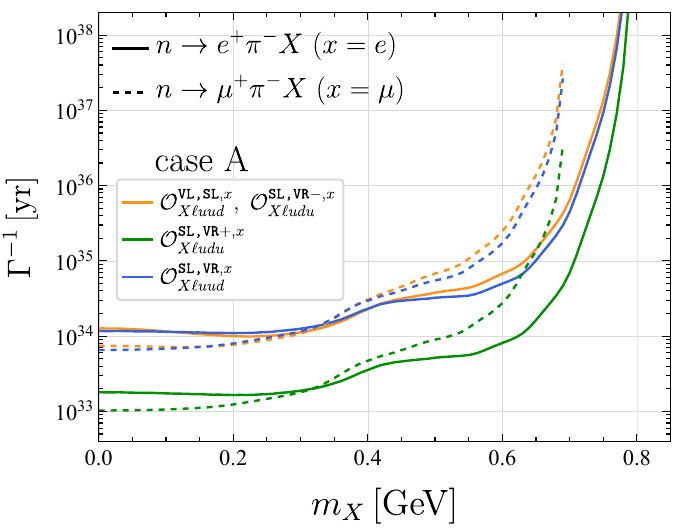}
\includegraphics[width=0.325\linewidth]{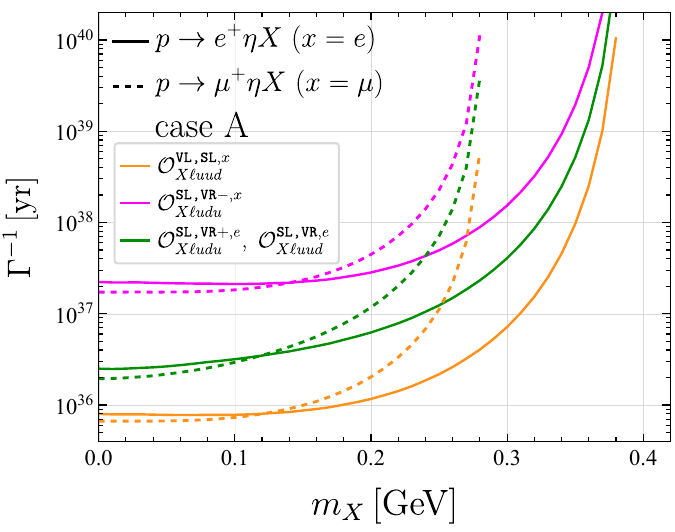}
\\
\includegraphics[width=0.325\linewidth]{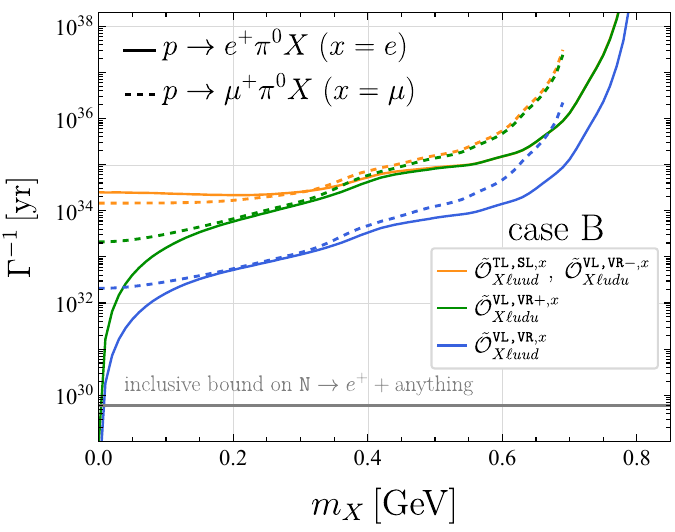}
\includegraphics[width=0.325\linewidth]{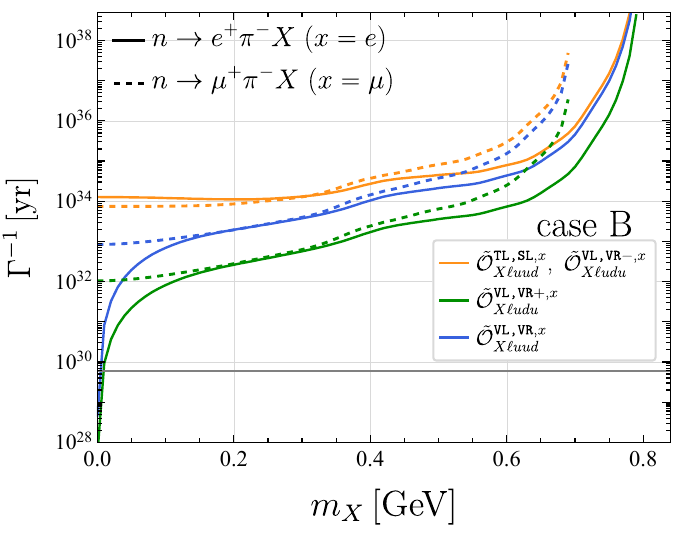}
\includegraphics[width=0.325\linewidth]{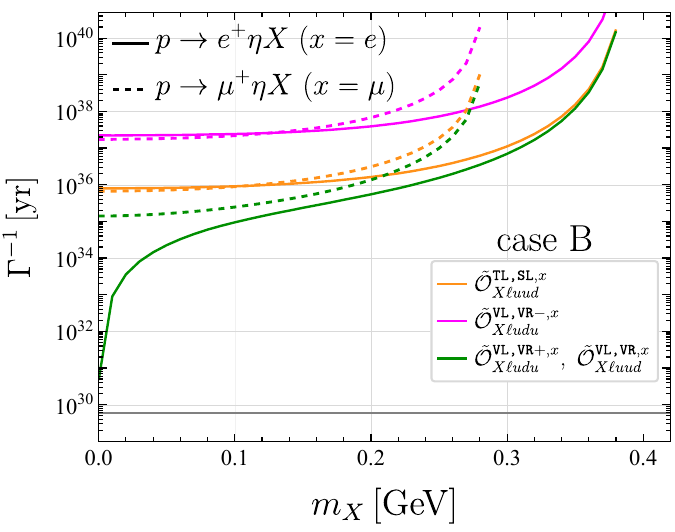}
\caption{Derived partial lifetime bounds on $p\to \ell^+(\pi^0,\eta)X$ and $n\to \ell^+\pi^- X$ based on the WC limits from $p\to \ell^+ X$ in \cref{fig:Lambda_eff_limits}. }
\label{fig:InGamma_bound_uud}
\end{figure}

\pmb{Processes in $X\ell uud$-class}: 
In addition to the two-body decay $p\to \ell^+ X$ ($\ell=e,\mu$), the operators in this class also induce three three-body decays: $p\to \ell^+ (\pi^0,\eta) X$ and $n\to \ell^+ \pi^- X$. 
The partial lifetime bounds obtained for these decays are 
shown in \cref{fig:InGamma_bound_uud}.
The upper and lower panels correspond to operators in cases A and B, respectively. 
Solid curves denote the results for $e^+$-related modes, while dashed curves represent $\mu^+$-related modes.
The bounds increase monotonically as $m_X$ runs from zero to its kinematic endpoint, reflecting the fact that the three-body phase space decreases more rapidly than that of the two-body modes with increasing $m_X$.  
For the operators in the chiral irreps $\pmb{8}_{\tL(\tR)}\otimes \pmb{1}_{\tR(\tL)}$ and  
$\bar{\pmb{3}}_{\tL(\tR)}\otimes \pmb{3}_{\tR(\tL)}$ (orange and magenta curves), 
the bounds derived on the decay modes $p\to \ell^+\pi^0 X$ and $n\to \ell^+\pi^- X$ [$p\to \ell^+\eta X$] generally exceed $\calO(10^{34}\,\rm yr)$ [$\calO(10^{36}\,\rm yr)$].
The remaining four operators in the irrep $\pmb{6}_{\tL(\tR)}\otimes \pmb{3}_{\tR(\tL)}$
(green and blue curves) yield similarly stringent bounds, with variations of up to $\calO(100)$ depending on the mode and mass $m_X$. 

For the operators $\tilde{\calO}^{{\tt VL,VR}+,e}_{X\ell udu}$ and $\tilde{\calO}^{{\tt VL,VR},e}_{X\ell uud}$ in case B, the derived bounds in the $e^+$-modes weaken significantly as $m_X$ goes to zero.
This effect arises from the diminishing constraints on $\Lambda_{\rm eff}$ (see the corresponding panel in \cref{fig:Lambda_eff_limits}), which follows from the suppression of the amplitude squared in the joint limit $(m_X, m_l)\to 0$. 
In such scenarios, the inclusive bounds from the $\tN\to e^+$+anything search conducted in~\cite{Learned:1979gp} provide complementary constraints, as indicated by the gray solid lines. 
Lastly, the degenerate behavior observed for the pair of operators in each decay mode is a direct consequence of their shared isospin properties, as discussed in \cref{subsub:P_dis}.

\begin{figure}[t]
\centering
\includegraphics[width=0.325\linewidth]{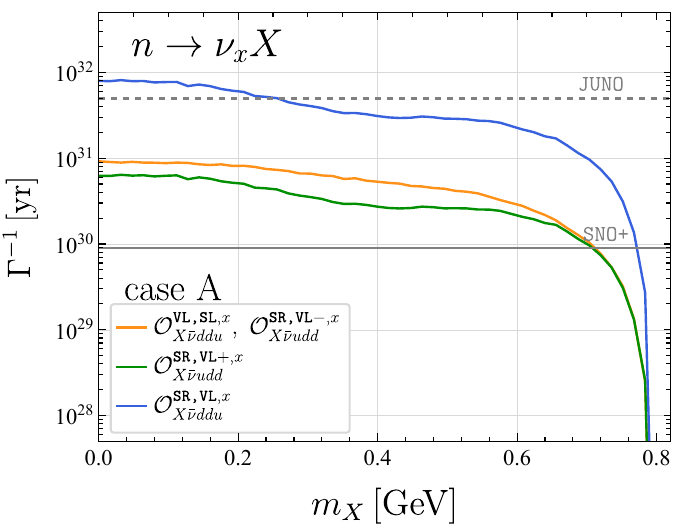}
\includegraphics[width=0.325\linewidth]{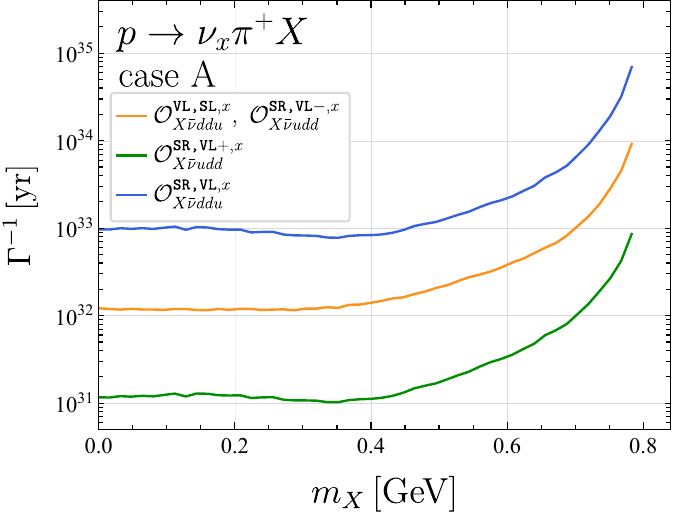}
\includegraphics[width=0.325\linewidth]{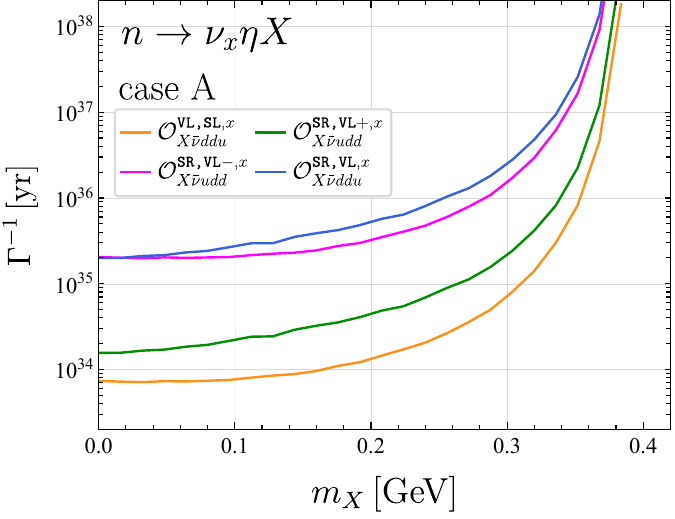}
\\
\includegraphics[width=0.325\linewidth]{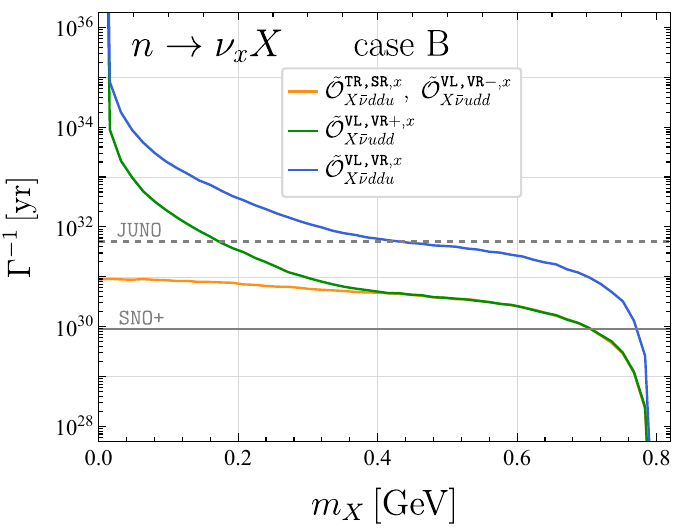}
\includegraphics[width=0.325\linewidth]{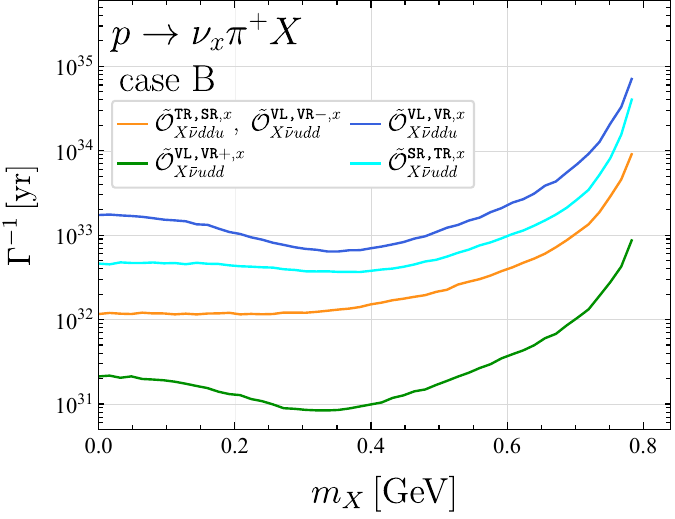}
\includegraphics[width=0.325\linewidth]{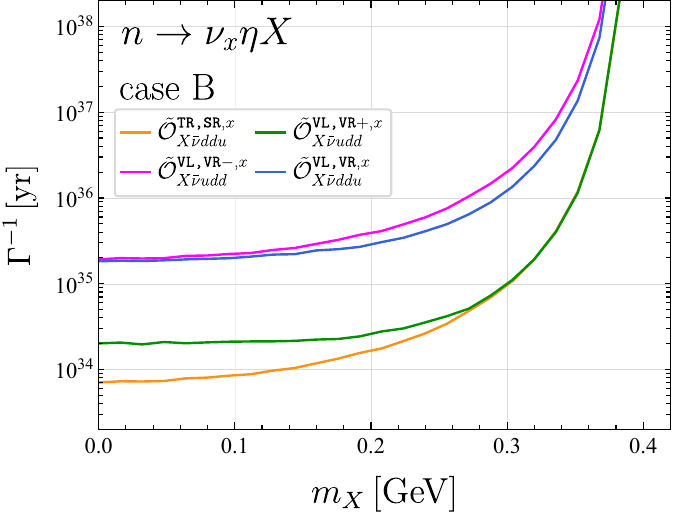}
\caption{Derived partial lifetime bounds on $n\to \nu_x X$, $p\to \nu_x \pi^+ X$, and $n\to \nu_x\eta X$ based on the WC limits from $n\to \nu_x \pi^0 X$ in \cref{fig:Lambda_eff_limits}.  }
\label{fig:InGamma_bound_udd}
\end{figure}

\pmb{Processes in $X\hat\nu udd$-class}: 
Similar to the $X\ell uud$ class, operators in this class induce four correlated decay modes. The quark fields and their associated hadronic states in the two classes are related by the interchange of $u$ and $d$ quarks. 
Based on the $\Lambda_{\rm eff}$ limits derived from $n\to \nu \pi^0 X$, the corresponding bounds on the two-body decay $n\to \nu X$ and the three-body modes $p\to \nu \pi^+ X$ and $n\to \nu\eta X$ are presented in the left, middle, and right panels of \cref{fig:InGamma_bound_udd}, respectively.

For the two three-body decays, the partial lifetime bounds become stronger as $m_X$ increases because their available phase space decreases more quickly than that of $n\to \nu \pi^0 X$, owing to the larger meson masses, $m_{\eta}>  m_{\pi^+} > m_{\pi^0}$. 
For $p\to \nu \pi^+ X$, the derived lower limits range from $\calO(10^{31}\,\rm yr)$ to $\calO(10^{34}\,\rm yr)$ for $m_X \lesssim 0.7\,\rm GeV$. 
In contrast, for $n\to \nu \eta X$, the limits reach $\calO(10^{34\text{--}37}\,\rm yr)$ for $m_X\lesssim 0.35\,\rm GeV$, due to the much larger mass of the eta meson. 

For the two-body neutron invisible decay,
$n\to \nu X$, the bounds decrease monotonically as $m_X$ increases.
This behavior, opposite to that of the two three-body decays, is a consequence of the phase space enhancement in $n\to \nu X$. 
As a result, the constraints become significantly weaker for $m_X\gtrsim0.7\,\rm GeV$, as shown in the plots.
On the other hand, the increasingly stringent bounds related to the two case-B operators as $m_X \to 0$ originate from a similar  suppression in the matrix element squared as those for the $e^+$-related modes discussed in \cref{fig:InGamma_bound_uud}.  

Experimentally, the neutron invisible decay  can be searched for directly through nuclear deexcitation accompanied by gamma-ray emission. 
The SNO+ experiment has established a 90\,\% C.L. lower limit,
$\Gamma^{-1}_{\tt SNO+}(n\to {\rm inv}) > 9.0\cdot 10^{29}\,{\rm yr}$~\cite{SNO:2022trz}.
This limit is expected to be further improved by the JUNO experiment, with a sensitivity of $5\cdot 10^{31}\,\rm yr$ after two years of data taking~\cite{JUNO:2024pur}.
These existing and projected limits are also shown in the two left panels to illustrate  their complementarity with the translated bounds in probing the allowed parameter space. 

\begin{figure}[t]
\centering
\includegraphics[width=0.45\linewidth]{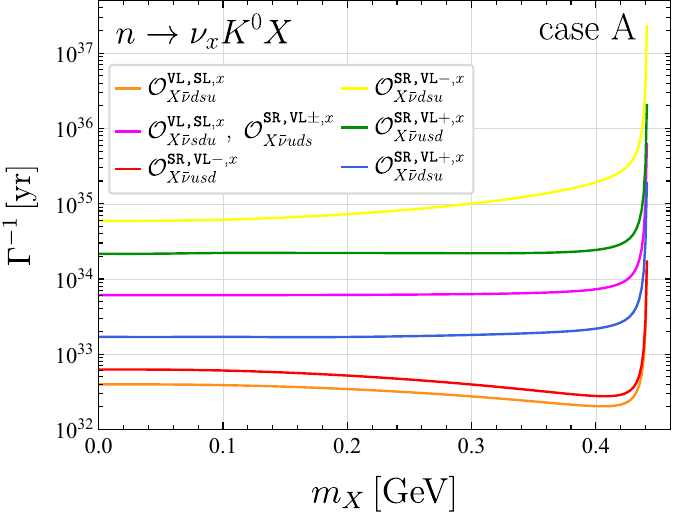}
\includegraphics[width=0.45\linewidth]{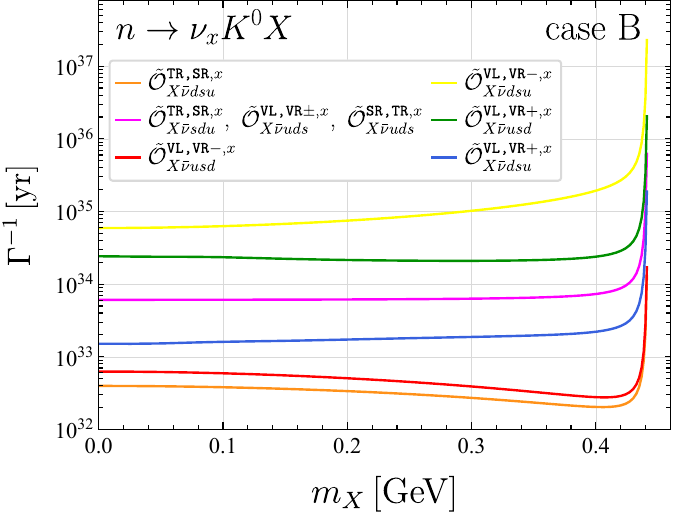}
\caption{Derived partial lifetime bounds on $n\to \nu_x K^0 X$ based on the WC limits from $p\to \nu_x K^+ X$ in \cref{fig:Lambda_eff_limits}.  }
\label{fig:InGamma_bound_uds}
\end{figure}

\pmb{Processes in $X\hat\nu uds$-class}: 
Using the effective scale limits for the $X\bar\nu uds$-class operators derived from the $p\to \nu K^+ X$ decay in
\cref{fig:Lambda_eff_limits}, we obtain the corresponding partial lifetime bounds for the $(u\leftrightarrow d)$-related mode $n\to \nu K^0 X$, presented in \cref{fig:InGamma_bound_uds}.
These bounds range from $\calO(10^{32}\,\rm yr)$ to $\calO(10^{35}\,\rm yr)$ across different operators, comparable to the input constraint of $5.9\cdot 10^{33}\,\rm yr$ for $p\to \nu K^+ X$.
For masses above $0.42\,\rm GeV$, the bounds quickly strengthen significantly as phase space suppression becomes more severe for the neutral kaon channel compared to the charged mode, due to the mass difference $m_{K^0}-m_{K^+}\sim 4\,\rm MeV$. 

We observe that in case A, the three operators $\calO_{X\bar\nu sdu}^{{\tt VL,SL},x}, \calO_{X\bar\nu uds}^{{\tt SR,VL}\pm,x}$ exhibit degenerate limits, while in case B, the four operators $\tilde\calO_{X\bar\nu sdu}^{{\tt TR,SR},x}, \tilde\calO_{X\bar\nu uds}^{{\tt VL,VR}\pm,x}, \tilde\calO_{X\bar\nu uds}^{{\tt SR,TR},x}$ display the same behavior.  
This degeneracy can be  naturally understood from isospin analysis.
Since both the initial nucleon and final kaon states belong to the isospin $I=1/2$ representation, 
and each of the seven operators mentioned above corresponds to a specific isospin change (either $\Delta I=0$ or $\Delta I=1$).
Consequently, 
they produce identical hadronic matrix elements for both decay modes, leading to the same bounds. 
The remaining operators involve both isospin components, but typically with distinct weighting factors. As a result, the matrix elements of these operators differ between the two modes, leading to different bounds.

Before closing this section, 
we note that liquid-scintillator detectors such as JUNO can provide complementary sensitivity to these decay modes beyond that of water-Cherenkov detectors. 
In particular, for channels with a final-state $K^+$, JUNO can reconstruct the kinetic energy of the primary $K^+$ and separate it from the energy deposited by its decay products using  their characteristic time topology~\cite{JUNO:2022qgr}. This capability, together with the absence of a Cherenkov-threshold limitation, makes JUNO particularly sensitive to $K^+$-associated modes, especially in the large-$m_X$ region where visible final-state particles become soft~\cite{Heeck:2025uwh}. 
A dedicated analysis based on the JUNO experiment is left for future work.

\section{Summary}
\label{sec:Conclusion}

In this work, we have carried out a systematic study of BNV two- and three-body nucleon decays involving an invisible dark photon within the framework of $X$LEFT and ChPT.
We first constructed a complete set of leading-order BNV $X$LEFT operators for two different scenarios: one where the dark photon field is parametrized as a four-potential $X_\mu$ (case A), and the other where it is described by a field-strength tensor $X_{\mu\nu}$ (case B). 
Without accounting for fermion flavors, there are in total 20 operators in case A and 26 operators in case B. 

We then concentrated on operators involving the light $u,d,s$ quarks relevant to nucleon decays and performed a chiral decomposition under the QCD chiral group. By identifying the corresponding spurion fields that enter into the recently developed chiral Lagrangian for the BNV sector, we obtained the hadron-level counterparts of the $X$LEFT interactions.  Within the chiral framework, we derived general expressions for the decay widths in terms of $X$LEFT WCs and hadronic LECs.  
From the matrix element squared, we further analyzed the kinematic distributions of the charged leptons and mesons in three-body decays. Our results show that these distributions can help distinguish between different operator structures and determine the dark photon mass in future experimental searches.

Based on our established formalism, we extended our analysis to place constraints on these exotic decay modes and their associated $X$LEFT operators. 
Due to the similarity in experimental signatures between the three-body modes listed in \cref{tab:ope_process} and their conventional two-body counterparts without the $X$ particle, 
we reanalyzed the Super-K data for the processes
$p\to \ell^+K^0$, 
$n\to\ell^+\pi^-$, 
$n\to\hat\nu\pi^0$,
and $p\to\hat\nu K^+$, 
to derive first constraints on the corresponding three-body modes with an additional $X$ in the final state across a broad range of dark photon masses. 
Our results show that the partial lifetime bounds for $p\to \ell^+ K^0 X$ and $n\to \ell^- (\pi^+, K^+)X$ can reach up to $\calO(10^{33})\,\rm yr$ when $m_X$ approaches 0.
From these results, we derived stringent limits on the effective scales of the dim-8 $X$LEFT operators that are around $10^6$ to $10^7\,\rm GeV$.
These WC constraints are further applied to predict lower lifetime limits for the other decay modes, including 
$p\to \ell^+(\pi^0,\eta)X$, 
$n\to \ell^+\pi^-X$; 
$n\to \hat\nu X$,
$p\to \hat\nu \pi^+ X$,
$n\to \hat\nu \eta X$;
and $n\to \hat\nu K^0  X$.
The projected limits range from $\calO(10^{30\text{--}40}\,\rm yr)$, with the exact value depending on the specific mode, the operator structure, and the dark photon mass.

Finally, our formalism provides a bridge for further theoretical investigation of these decay modes, especially by enabling more efficient calculations from a UV-complete perspective. For a given model, one can integrate out the heavy particles, derive the matching results for the WCs, and directly incorporate them into the decay width expressions. 
At the same time, our established partial lifetime limits for these exotic decay modes offer valuable guidance for future experimental searches for them.
The bounds we obtained can help the experimentalists identify the most promising channels for discovery.

\acknowledgments
This work was supported 
by the Grants 
No.\,NSFC-12305110, 
No.\,NSFC-12247151, 
and No.\,NSFC-12035008.  

\appendix
\section{Complete width expressions for nucleon decays in $X$LEFT}
\label{app:Gamma_exp}

In this Appendix, we present the full decay widths for 20 two- and three-body nucleon decays involving a massless dark photon. These results are explicitly expressed in terms of the WCs of the $X$LEFT operators.
Using central values for particle masses and hadronic LECs, we numerically evaluate the three-body phase-space integrals. 
Since the LECs $c_3$ and $c_4$ remain undetermined, the results maintain an explicit dependence on them through the ratios $\kappa_3\equiv c_3/c_1$ and $\kappa_4\equiv c_4/c_1$. 

For decay modes involving a neutrino or an antineutrino, we show only the results for modes with a neutrino final state. The expressions for the corresponding antineutrino modes can be obtained from those of neutrino cases by performing a joint exchange of $\tL\leftrightarrow\tR$ and $\nu\leftrightarrow\bar\nu$ in the WC labels.
Although these expressions are derived in the limit of $m_X\to 0$, they remain a good approximation for a massive dark photon as long as $m_X \ll m_\mu$.  

\subsection{Case A}
\label{app:Gamma_exp_A}
In this subsection, we summarize the width results in the case where the dark photon is described by a four-potential $X_\mu$. 
Note that a mass factor $m_X$ is manually incorporated into each operator here, resulting in the WCs with dimension $\rm [energy]^{-4}$. 
For the two-body decay processes, we obtain
{\small
\begin{subequations}	
\label{eq:Ga_N2lXA}
\begin{align}
{\Gamma_{p\to e^+ X} \over (0.1\rm GeV)^9} =\,&   
1300  |C_{X\ell uud}^{{\tt VL,SL},e}|^2 
+ 1300  |C_{X\ell udu}^{{\tt SL,VR}-,e}|^2
+ 50 \kappa_3^2 \big( |C_{X\ell udu}^{{\tt SL,VR}+,e}|^2
+ |C_{X\ell uud}^{{\tt SL,VR},e}|^2  \big) 
\nn\\
&+ 2600 \Re\big( C_{X\ell udu}^{{\tt SL,VR}-,e} C_{X\ell uud}^{{\tt  VR,SR},e*}  \big)
+510 \kappa_3 \Re\big( (C_{X\ell uud}^{{\tt SL,VR},e} 
- C_{X\ell udu}^{{\tt SL,VR}+,e})
C_{X\ell uud}^{{\tt VR,SR},e*}\big)
\nn\\
&+ 510 \kappa_3 \Re\big( (C_{X\ell uud}^{{\tt SL,VR},e}-C_{X\ell udu}^{{\tt SL,VR}+,e} )
C_{X\ell udu}^{{\tt SL,VR}-,e*}  \big)
-99 \kappa_3^2 \Re\big( C_{X\ell udu}^{{\tt SL,VR}+,e} C_{X\ell uud}^{{\tt  SL,VR},e*}\big)
\nn\\
&+ 0.6 \kappa_3 \Re\big( (C_{X\ell udu}^{{\tt SL,VR}+,e}-C_{X\ell uud}^{{\tt SL,VR},e})
C_{X\ell uud}^{{\tt VL,SL},e*}  \big)
+ 0.2 \kappa_3^2 \Re\big( C_{X\ell udu}^{{\tt SL,VR}+,e} 
C_{X\ell uud}^{{\tt SR,VL},e*}\big) 
\nn\\
&-0.5\kappa_3 \Re\big( (C_{X\ell uud}^{{\tt SL,VR},e} 
-C_{X\ell udu}^{{\tt SL,VR}+,e} )
C_{X\ell udu}^{{\tt SR,VL}-,e*} \big)
- 0.1 \kappa_3^2 \Re\big( C_{X\ell udu}^{{\tt SL,VR}+,e} 
C_{X\ell udu}^{{\tt SR,VL }+,e*}\big)
\nn\\
&-0.1\kappa_3^2 \Re\big(C_{X\ell uud}^{{\tt SL,VR},e}
C_{X\ell uud}^{{\tt SR,VL},e*}\big) +\tL\leftrightarrow \tR,
\\%
{\Gamma_{p\to \mu^+ X} \over (0.1\rm GeV)^9} =\,&   
1300 |C_{X\ell uud}^{{\tt VL,SL},\mu}|^2
+ 1200 |C_{X\ell udu}^{{\tt SL,VR}-,\mu}|^2
+ 50 \kappa_3^2 \big( |C_{X\ell udu}^{{\tt SL,VR}+,\mu}|^2
+ |C_{X\ell uud}^{{\tt SL,VR},\mu}|^2\big) 
\nn\\
&+ 2500 \Re\big( C_{X\ell udu}^{{\tt SL,VR}-,\mu} C_{X\ell uud}^{{\tt  VR,SR},\mu*}  \big)
-490 \kappa_3 \Re\big( (C_{X\ell udu}^{{\tt SL,VR}+,\mu} 
-C_{X\ell uud}^{{\tt SL,VR},\mu})
C_{X\ell uud}^{{\tt VR,SR},\mu*}\big)
\nn\\
&-490 \kappa_3 \Re\big( (C_{X\ell udu}^{{\tt SL,VR}+,\mu} 
-C_{X\ell uud}^{{\tt SL,VR},\mu} )
C_{X\ell udu}^{{\tt SL,VR}-,\mu*}  \big)
-100 \kappa_3^2 \Re\big( C_{X\ell udu}^{{\tt SL,VR}+,\mu} C_{X\ell uud}^{{\tt  SL,VR},\mu*}\big)
\nn\\
&- 110 \kappa_3 \Re\big( (C_{X\ell uud}^{{\tt SL,VR},\mu} 
-C_{X\ell udu}^{{\tt SL,VR}+,\mu} )
C_{X\ell uud}^{{\tt VL,SL},\mu*}\big)
+ 43 \kappa_3^2 \Re\big( C_{X\ell udu}^{{\tt SL,VR}+,\mu} 
C_{X\ell uud}^{{\tt SR,VL},\mu*}\big)    
\nn\\
&- 110 \kappa_3 \Re\big( (C_{X\ell uud}^{{\tt SL,VR},\mu} 
-C_{X\ell udu}^{{\tt SL,VR}+,\mu} )
C_{X\ell udu}^{{\tt SR,VL}-,\mu*}\big)
- 22 \kappa_3^2 \Re\big( C_{X\ell udu}^{{\tt SL,VR}+,\mu} 
C_{X\ell udu}^{{\tt SR,VL}+,\mu*}\big)
\nn\\
&-22\kappa_3^2 \Re\big(C_{X\ell uud}^{{\tt SL,VR},\mu}
C_{X\ell uud}^{{\tt SR,VL},\mu*}\big)+\tL\leftrightarrow \tR ,
\\%
{\Gamma_{n\to\nu_x X} \over (0.1\rm GeV)^9} =\,&
1300 |C_{X\bar\nu ddu}^{{\tt VL,SL},x}|^2
+ 1300 |C_{X\bar\nu udd}^{{\tt SR,VL}-,x}|^2
+ 50 \kappa_3^2 \big( |C_{X\bar\nu ddu}^{{\tt SR,VL},x}|^2
+ |C_{X\bar\nu udd}^{{\tt SR,VL}+,x}|^2 \big)
\nn\\
&- 2600 \Re\big( C_{X\bar\nu udd}^{{\tt SR,VL}-,x}
C_{X\bar\nu ddu}^{{\tt VL,SL},x*}\big)
- 520 \kappa_3 \Re\big( (C_{X\bar\nu udd}^{{\tt SR,VL}+,x} 
-C_{X\bar\nu ddu}^{{\tt SR,VL},x})
C_{X\bar\nu ddu}^{{\tt VL,SL},x*}\big)
\nn\\
&- 510 \kappa_3 \Re\big( (C_{X\bar\nu ddu}^{{\tt SR,VL},x}
-C_{X\bar\nu udd}^{{\tt SR,VL}+,x})
C_{X\bar\nu udd}^{{\tt SR,VL}-,x*}\big)
-100\kappa_3^2 \Re\big( C_{X\bar\nu udd}^{{\tt SR,VL}+,x}
C_{X\bar\nu ddu}^{{\tt SR,VL},x*}\big).
\end{align}
\end{subequations}}\normalsize
The subscript $x=e,~\mu,~\tau$ denotes any of the three neutrino flavors. 
For the three-body processes:
{\small
\begin{subequations}
\label{eq:Ga_N2lXMA}
\begin{align}
{\Gamma_{p\to e^+ \pi^0 X} \over (0.1\rm GeV)^9} =\,&
52 |C_{X\ell uud}^{{\tt VL,SL},e}|^2 
+51 |C_{X\ell udu}^{{\tt SL,VR}-,e}|^2
+14\kappa_3^2 |C_{X\ell uud}^{{\tt SL,VR},e}|^2 
+\kappa_3^2 |C_{X\ell udu}^{{\tt SL,VR}+,e}|^2
\nn\\
&+102\Re\big(C_{X\ell udu}^{{\tt SL,VR}-,e} 
C_{X\ell uud}^{{\tt VR,SR},e*} \big)
-14\kappa_3 \Re\big( (C_{X\ell uud}^{{\tt VR,SR},e} 
+C_{X\ell udu}^{{\tt SL,VR}-,e})
C_{X\ell uud}^{{\tt SL,VR},e*} \big)
\nn\\
&-9.6\kappa_3 \Re\big( (C_{X\ell uud}^{{\tt VR,SR},e} 
+C_{X\ell udu}^{{\tt SL,VR}-,e} )
C_{X\ell udu}^{{\tt SL,VR}+,e*} \big)
+6.3\kappa_3^2 \Re\big(C_{X\ell udu}^{{\tt SL,VR}+,e} 
C_{X\ell uud}^{{\tt SL,VR},e*} \big)
\nn\\
&-0.1\kappa_3 \Re\big( (C_{X\ell uud}^{{\tt VL,SL},e} 
+C_{X\ell udu}^{{\tt SR,VL}-,e})
C_{X\ell uud}^{{\tt SL,VR},e*} \big)
+0.03\kappa_3^2 \Re\big(C_{X\ell uud}^{{\tt SL,VR},e} 
C_{X\ell uud}^{{\tt SR,VL},e*} \big)
\nn\\
&-0.02\kappa_3 \Re\big( (C_{X\ell uud}^{{\tt VR,SR},e} 
+C_{X\ell udu}^{{\tt SL,VR}-,e} )
C_{X\ell udu}^{{\tt SR,VL}+,e*} \big)
+0.02\kappa_3^2 \Re\big(C_{X\ell udu}^{{\tt SL,VR}+,e} 
C_{X\ell uud}^{{\tt SR,VL},e*} \big)
\nn\\
&+0.003\kappa_3^2 \Re\big(C_{X\ell udu}^{{\tt SL,VR}+,e} 
C_{X\ell udu}^{{\tt SR,VL}+,e*} \big)
+ \tL \leftrightarrow \tR,
\\%
{\Gamma_{p\to \mu^+ \pi^0 X} \over (0.1\rm GeV)^9} =\,&
45 |C_{X\ell uud}^{{\tt VL,SL},\mu}|^2
+44 |C_{X\ell udu}^{{\tt SL,VR}-,\mu}|^2 
+13\kappa_3^2 |C_{X\ell uud}^{{\tt SL,VR},\mu}|^2 
+\kappa_3^2 |C_{X\ell udu}^{{\tt SL,VR}+,\mu}|^2
\nn\\
&+89\Re\big(C_{X\ell udu}^{{\tt SL,VR}-,\mu} 
C_{X\ell uud}^{{\tt VR,SR},\mu*} \big)
-16\kappa_3 \Re\big( (C_{X\ell uud}^{{\tt VR,SR},\mu} 
+C_{X\ell udu}^{{\tt SL,VR}-,\mu})
C_{X\ell uud}^{{\tt SR,VL},\mu*} \big)
\nn\\
&-13\kappa_3 \Re\big( (C_{X\ell uud}^{{\tt VR,SR},\mu} 
+C_{X\ell udu}^{{\tt SL,VR}-,\mu})
C_{X\ell uud}^{{\tt SL,VR},\mu*} \big)
+5.7\kappa_3^2 \Re\big(C_{X\ell udu}^{{\tt SL,VR}+,\mu} 
C_{X\ell uud}^{{\tt SL,VR},\mu*} \big)
\nn\\
&-8.6\kappa_3 \Re\big( (C_{X\ell uud}^{{\tt VR,SR},\mu} 
+C_{X\ell udu}^{{\tt SL,VR}-,\mu})
C_{X\ell udu}^{{\tt SL,VR}+,\mu*} \big)
+4.6\kappa_3^2 \Re\big(C_{X\ell uud}^{{\tt SL,VR},\mu} 
C_{X\ell uud}^{{\tt SR,VL},\mu*} \big)
\nn\\
&-2.6\kappa_3 \Re\big( (C_{X\ell uud}^{{\tt VR,SR},\mu} 
+C_{X\ell udu}^{{\tt SL,VR}-,\mu} )
C_{X\ell udu}^{{\tt SR,VL}+,\mu*} \big)
+3.6\kappa_3^2 \Re\big(C_{X\ell udu}^{{\tt SL,VR}+,\mu} 
C_{X\ell uud}^{{\tt SR,VL},\mu*} \big)
\nn\\
&+0.45\kappa_3^2 \Re\big(C_{X\ell udu}^{{\tt SL,VR}+,\mu} 
C_{X\ell udu}^{{\tt SR,VL}+,\mu*} \big)
+\tL\leftrightarrow \tR,
\\%
{\Gamma_{p\to e^+ \eta X} \over (0.1\rm GeV)^9} =\,&
1.6 |C_{X\ell uud}^{{\tt VL,SL},e}|^2 
+0.06 |C_{X\ell udu}^{{\tt SL,VR}-,e}|^2
+0.02\kappa_3^2 \big(|C_{X\ell uud}^{{\tt SL,VR},e}|^2  
+|C_{X\ell udu}^{{\tt SL,VR}+,e}|^2\big)
\nn\\
&-0.6\Re\big(C_{X\ell udu}^{{\tt SL,VR}-,e} 
C_{X\ell uud}^{{\tt VR,SR},e*} \big)
+0.2\kappa_3 \Re\big( (C_{X\ell udu}^{{\tt SL,VR}+,e} 
- C_{X\ell uud}^{{\tt SL,VR},e})
C_{X\ell uud}^{{\tt VR,SR},e*} \big)
\nn\\
&+0.05\kappa_3 \Re\big( (C_{X\ell uud}^{{\tt SL,VR},e} 
- C_{X\ell udu}^{{\tt SL,VR}+,e})
C_{X\ell udu}^{{\tt SL,VR}-,e*} \big)
- 0.04\kappa_3^2 \Re\big(C_{X\ell udu}^{{\tt SL,VR}+,e} 
C_{X\ell uud}^{{\tt SL,VR},e*} \big)
\nn\\
&+10^{-3}\kappa_3 \Re\big(
(C_{X\ell udu}^{{\tt SR,VL}+,e} 
-C_{X\ell uud}^{{\tt SR,VL},e} )
C_{X\ell uud}^{{\tt VR,SR},e*} \big)
-10^{-4}\kappa_3^2 \Re\big(C_{X\ell udu}^{{\tt SL,VR}+,e} 
C_{X\ell uud}^{{\tt SR,VL},e*} \big)
\nn\\
&+2\cdot10^{-4}\kappa_3 \Re\big( 
(C_{X\ell uud}^{{\tt SR,VL},e} 
-C_{X\ell udu}^{{\tt SR,VL}+,e})
C_{X\ell udu}^{{\tt SL,VR}-,e*} \big) 
\nn\\ 
&+5\cdot10^{-5}\kappa_3^2 \Re\big(C_{X\ell udu}^{{\tt SL,VR}+,e} 
C_{X\ell udu}^{{\tt SR,VL}+,e*} 
+C_{X\ell uud}^{{\tt SL,VR},e} 
C_{X\ell uud}^{{\tt SR,VL},e*} \big) 
+\tL\leftrightarrow \tR,
\\%
{\Gamma_{p\to \mu^+ \eta X} \over (0.1\rm GeV)^9} =\,&
1.0 |C_{X\ell uud}^{{\tt VL,SL},\mu}|^2
+0.04 |C_{X\ell udu}^{{\tt SL,VR}-,\mu}|^2
+0.01\kappa_3^2 \big(|C_{X\ell uud}^{{\tt SL,VR},\mu}|^2  
+|C_{X\ell udu}^{{\tt SL,VR}+,\mu}|^2\big)
\nn\\
&-0.4 \Re\big(C_{X\ell udu}^{{\tt SL,VR}-,\mu} 
C_{X\ell uud}^{{\tt VR,SR},\mu*} \big) 
+0.1 \kappa_3 \Re\big( (C_{X\ell udu}^{{\tt SL,VR}+,\mu} 
-C_{X\ell uud}^{{\tt SL,VR},\mu})
C_{X\ell uud}^{{\tt VR,SR},\mu*} \big)
\nn\\
&+0.1\kappa_3 \Re\big( (C_{X\ell udu}^{{\tt SR,VL}+,\mu} 
-C_{X\ell uud}^{{\tt SR,VL},\mu})
C_{X\ell uud}^{{\tt VR,SR},\mu*} \big)
-0.03\kappa_3^2 \Re\big(C_{X\ell udu}^{{\tt SL,VR}+,\mu} 
C_{X\ell uud}^{{\tt  SL,VR},\mu*} \big)
\nn\\
&+0.03\kappa_3 \Re\big( (C_{X\ell uud}^{{\tt SL,VR},\mu} 
-C_{X\ell udu}^{{\tt SL,VR}+,\mu})
C_{X\ell udu}^{{\tt SL,VR}-,\mu*} \big)
-0.01\kappa_3^2 \Re\big(C_{X\ell udu}^{{\tt SL,VR}+,\mu} 
C_{X\ell uud}^{{\tt SR,VL},\mu*} \big)
\nn\\
&+0.02\kappa_3 \Re\big( (C_{X\ell uud}^{{\tt SR,VL},\mu} 
-C_{X\ell udu}^{{\tt SR,VL}+,\mu} )
C_{X\ell udu}^{{\tt SL,VR}-,\mu*} \big)
\nn\\ 
&+0.005\kappa_3^2 \Re\big(C_{X\ell udu}^{{\tt SL,VR}+,\mu} 
C_{X\ell udu}^{{\tt SR,VL}+,\mu*} 
+C_{X\ell uud}^{{\tt SL,VR},\mu} 
C_{X\ell uud}^{{\tt SR,VL},\mu*} \big)  
+ \tL \leftrightarrow \tR\;,
\\%
{\Gamma_{p\to e^+ K^0 X} \over (0.1\rm GeV)^9} =\,&
2.1 |C_{X\ell usu}^{{\tt SL,VR}-,e}|^2
+1.2|C_{X\ell uus}^{{\tt VL,SL},e}|^2
+0.2 \kappa_3^2 \big(|C_{X\ell uus}^{{\tt SL,VR},e}|^2 
+|C_{X\ell usu}^{{\tt SL,VR}+,e}|^2\big)
\nn\\
&-3.2 \Re\big( C_{X\ell usu}^{{\tt SL,VR}-,e} 
C_{X\ell uus}^{{\tt VR,SR},e*}\big)
-0.7\kappa_3 \Re\big( (C_{X\ell uus}^{{\tt SL,VR},e} 
+C_{X\ell usu}^{{\tt SL,VR}+,e})
C_{X\ell usu}^{{\tt SL,VR}-,e*} \big)
\nn\\
&+0.6\kappa_3 \Re\big( (C_{X\ell usu}^{{\tt SL,VR}+,e} 
+C_{X\ell uus}^{{\tt SL,VR},e})
C_{X\ell uus}^{{\tt VR,SR},e*} \big)
+0.4\kappa_3^2 \Re\big(C_{X\ell usu}^{{\tt SL,VR}+,e} 
C_{X\ell uus}^{{\tt  SL,VR},e*} \big)
\nn\\
&-0.004\kappa_3 \Re\big( (C_{X\ell uus}^{{\tt SR,VL},e} 
+C_{X\ell usu}^{{\tt SR,VL}+,e})
C_{X\ell usu}^{{\tt SL,VR}-,e*} \big)
+0.001\kappa_3^2 \Re\big(C_{X\ell usu}^{{\tt SL,VR}+,e} 
C_{X\ell uus}^{{\tt SR,VL},e*} \big) 
\nn\\
&+0.003\kappa_3 \Re\big( (C_{X\ell usu}^{{\tt SR,VL}+,e} 
+C_{X\ell uus}^{{\tt SR,VL},e})
C_{X\ell uus}^{{\tt VR,SR},e*} \big)
\nn\\
&+5\cdot10^{-4}\kappa_3^2 \Re\big(C_{X\ell usu}^{{\tt SL,VR}+,e} 
C_{X\ell usu}^{{\tt SR,VL}+,e*} 
+C_{X\ell uus}^{{\tt SL,VR},e} 
C_{X\ell uus}^{{\tt SR,VL},e*} \big)
+ \tL \leftrightarrow \tR,
\\%
{\Gamma_{p\to \mu^+ K^0 X} \over (0.1\rm GeV)^9} =\,&
1.4 |C_{X\ell usu}^{{\tt SL,VR}-,\mu}|^2
+0.9 |C_{X\ell uus}^{{\tt VL,SL},\mu}|^2
+0.1\kappa_3^2 \big(|C_{X\ell uus}^{{\tt SL,VR},\mu}|^2 
+|C_{X\ell usu}^{{\tt SL,VR}+,\mu}|^2\big)
\nn\\
&-2.2 \Re\big(C_{X\ell usu}^{{\tt SL,VR}-,\mu} 
C_{X\ell uus}^{{\tt VR,SR},\mu*} \big)
-0.5 \kappa_3 \Re\big( (C_{X\ell uus}^{{\tt SL,VR},\mu} 
+C_{X\ell usu}^{{\tt SL,VR}+,\mu})
C_{X\ell usu}^{{\tt SL,VR}-,\mu*} \big)
\nn\\
&-0.4\kappa_3 \Re\big( (C_{X\ell uus}^{{\tt SR,VL},\mu} 
+C_{X\ell usu}^{{\tt SR,VL}+,\mu})
C_{X\ell usu}^{{\tt SL,VR}-,\mu*}\big)
+0.3\kappa_3^2 \Re\big(C_{X\ell usu}^{{\tt SL,VR}+,\mu} 
C_{X\ell uus}^{{\tt SL,VR},\mu*}\big)
\nn\\
&+0.4\kappa_3 \Re\big( (C_{X\ell usu}^{{\tt SL,VR}+,\mu} 
+C_{X\ell uus}^{{\tt SL,VR},\mu})
C_{X\ell uus}^{{\tt VR,SR},\mu*} \big)
+0.1\kappa_3^2 \Re\big(C_{X\ell usu}^{{\tt SL,VR}+,\mu} 
C_{X\ell uus}^{{\tt SR,VL},\mu*} \big) 
\nn\\
&+0.3\kappa_3 \Re\big( (C_{X\ell usu}^{{\tt SR,VL}+,\mu} 
+C_{X\ell uus}^{{\tt SR,VL},\mu})
C_{X\ell uus}^{{\tt VR,SR},\mu*} \big)  
\nn\\
&+0.05\kappa_3^2 \Re\big(C_{X\ell usu}^{{\tt SL,VR}+,\mu} 
C_{X\ell usu}^{{\tt SR,VL}+,\mu*} 
+C_{X\ell uus}^{{\tt SL,VR},\mu} 
C_{X\ell uus}^{{\tt SR,VL},\mu*} \big)
+ \tL \leftrightarrow \tR,
\\%
{\Gamma_{n\to e^+\pi^- X} \over (0.1\rm GeV)^9} =\,&
102|C_{X\ell uud}^{{\tt VL,SL},e}|^2
+100|C_{X\ell udu}^{{\tt SL,VR}-,e}|^2
+27\kappa_3^2 |C_{X\ell udu}^{{\tt SL,VR}+,e}|^2
+4.2\kappa_3^2 |C_{X\ell uud}^{{\tt SL,VR},e}|^2
\nn\\
&+202\Re\big(C_{X\ell udu}^{{\tt SL,VR}-,e} 
C_{X\ell uud}^{{\tt VR,SR},e*} \big)
+29\kappa_3 \Re\big( (C_{X\ell uud}^{{\tt VR,SR},e} 
+C_{X\ell udu}^{{\tt SL,VR}-,e})
C_{X\ell udu}^{{\tt SL,VR}+,e*} \big)
\nn\\
&-21\kappa_3^2 \Re\big(C_{X\ell udu}^{{\tt SL,VR}+,e} 
C_{X\ell uud}^{{\tt SL,VR},e*} \big)
- 5\kappa_3 \Re\big( (C_{X\ell uud}^{{\tt VR,SR},e} 
+C_{X\ell udu}^{{\tt SL,VR}-,e})
C_{X\ell uud}^{{\tt SL,VR},e*} \big)
\nn\\
&+ 0.2\kappa_3 \Re\big( (C_{X\ell uud}^{{\tt VR,SR},e} 
+C_{X\ell udu}^{{\tt SL,VR}-,e})
C_{X\ell udu}^{{\tt SR,VL}+,e*} \big)
-0.03\kappa_3^2 \Re\big(C_{X\ell udu}^{{\tt SL,VR}+,e} 
C_{X\ell uud}^{{\tt SR,VL},e*} \big)
\nn\\
&- 0.08\kappa_3 \Re\big( (C_{X\ell uud}^{{\tt VL,SL},e} 
+C_{X\ell udu}^{{\tt SR,VL}-,e})
C_{X\ell uud}^{{\tt SL,VR},e*} \big)
+0.05\kappa_3^2 \Re\big( C_{X\ell udu}^{{\tt SL,VR}+,e} 
C_{X\ell udu}^{{\tt SR,VL}+,e*} \big)
\nn\\
&+0.005\kappa_3^2 \Re\big(C_{X\ell uud}^{{\tt SL,VR},e} 
C_{X\ell uud}^{{\tt SR,VL},e*}\big) 
+ \tL \leftrightarrow \tR,
\\%
{\Gamma_{n\to\mu^+\pi^- X} \over (0.1\rm GeV)^9} =\,&
89 |C_{X\ell uud}^{{\tt VL,SL},\mu}|^2
+87 |C_{X\ell udu}^{{\tt SL,VR}-,\mu}|^2
+25\kappa_3^2 |C_{X\ell udu}^{{\tt SL,VR}+,\mu}|^2
+4\kappa_3^2 |C_{X\ell uud}^{{\tt SL,VR},\mu}|^2 
\nn\\
&+176\Re\big(C_{X\ell udu}^{{\tt SL,VR}-,\mu} 
C_{X\ell uud}^{{\tt VR,SR},\mu*} \big)
+ 31\kappa_3 \Re\big( (C_{X\ell uud}^{{\tt VR,SR},\mu} 
+C_{X\ell udu}^{{\tt SL,VR}-,\mu})
C_{X\ell udu}^{{\tt SR,VL}+,\mu*}
\nn\\
&+26\kappa_3 \Re\big( (C_{X\ell uud}^{{\tt VR,SR},\mu} 
+C_{X\ell udu}^{{\tt SL,VR}-,\mu})
C_{X\ell udu}^{{\tt SL,VR}+,\mu*}\big)
-20\kappa_3^2 \Re\big(C_{X\ell udu}^{{\tt SL,VR}+,\mu} 
C_{X\ell uud}^{{\tt  SL,VR},\mu*}\big)
\nn\\
&-13\kappa_3 \Re\big( (C_{X\ell uud}^{{\tt VL,SL},\mu} 
+C_{X\ell udu}^{{\tt SR,VL}-,\mu} )
C_{X\ell uud}^{{\tt SL,VR},\mu*}  \big)
-5.5\kappa_3^2 \Re\big(C_{X\ell udu}^{{\tt SL,VR}+,\mu} 
C_{X\ell uud}^{{\tt SR,VL},\mu*}  \big)
\nn\\
&+9\kappa_3^2 \Re\big(C_{X\ell udu}^{{\tt SL,VR}+,\mu} 
C_{X\ell udu}^{{\tt SR,VL}+,\mu*}\big)
-4.4\kappa_3 \Re\big( (C_{X\ell uud}^{{\tt VR,SR},\mu} 
+C_{X\ell udu}^{{\tt SL,VR}-,\mu})
C_{X\ell uud}^{{\tt SL,VR},\mu*} \big)
\nn\\
&+0.7\kappa_3^2 \Re\big(C_{X\ell uud}^{{\tt SL,VR},\mu} 
C_{X\ell uud}^{{\tt SR,VL},\mu*}\big)
+ \tL\leftrightarrow\tR,
\\%
{\Gamma_{p\to\nu _x \pi^+ X} \over (0.1\rm GeV)^9} =\,&
101 |C_{X\bar\nu ddu}^{{\tt VL,SL},x}|^2 
+99 |C_{X\bar\nu udd}^{{\tt SR,VL}-,x}|^2
+27\kappa_3^2 |C_{X\bar\nu udd}^{{\tt SR,VL}+,x}|^2
+4.1\kappa_3^2 |C_{X\bar\nu ddu}^{{\tt SR,VL},x}|^2
\nn\\
&- 199 \Re \big( C_{X\bar\nu udd}^{{\tt SR,VL}-,x} 
C_{X\bar\nu ddu}^{{\tt VL,SL},x*} \big)
+ 29 \kappa_3 \Re\big( (C_{X\bar\nu ddu}^{{\tt VL,SL},x} 
- C_{X\bar\nu udd}^{{\tt SR,VL}-,x})
C_{X\bar\nu udd}^{{\tt SR,VL}+,x*} \big)
\nn\\
&- 21 \kappa_3^2 \Re\big( C_{X\bar\nu udd}^{{\tt SR,VL}+,x} 
C_{X\bar\nu ddu}^{{\tt SR,VL},x*} \big)
+5.1\kappa_3\Re\big((C_{X\bar\nu udd}^{{\tt SR,VL}-,x}
-C_{X\bar\nu ddu}^{{\tt VL,SL},x})
C_{X\bar\nu ddu}^{{\tt SR,VL},x*}\big),
\\%
{\Gamma_{p\to\nu_x K^+ X}\over (0.1\rm GeV)^9} =\,&
3.3|C_{X\bar\nu sdu}^{{\tt VL,SL},x}|^2 
+3.2|C_{X\bar\nu uds}^{{\tt SR,VL}-,x}|^2 
+1.4|C_{X\bar\nu dsu}^{{\tt SR,VL}-,x}|^2
+0.9\kappa_3^2 |C_{X\bar\nu usd}^{{\tt SR,VL}+,x}|^2
\nn\\
&+0.2\kappa_3^2 \big(|C_{X\bar\nu dsu}^{{\tt SR,VL}+,x}|^2
+|C_{X\bar\nu uds}^{{\tt SR,VL}+,x}|^2\big)
+0.1\big(|C_{X\bar\nu dsu}^{{\tt VL,SL},x}|^2 
+|C_{X\bar\nu usd}^{{\tt SR,VL}-,x}|^2\big)
\nn\\
&-6.5 \Re\big(C_{X\bar\nu uds}^{{\tt SR,VL}-,x} 
C_{X\bar\nu sdu}^{{\tt VL,SL},x*} \big)
+4.1 \Re\big( (C_{X\bar\nu uds}^{{\tt SR,VL}-,x}
-C_{X\bar\nu sdu}^{{\tt VL,SL},x})
C_{X\bar\nu dsu}^{{\tt SR,VL}-,x*} \big)
\nn\\
&+1.6\kappa_3 \Re\big( (C_{X\bar\nu sdu}^{{\tt VL,SL},x}
- C_{X\bar\nu uds}^{{\tt SR,VL}-,x})
C_{X\bar\nu usd}^{{\tt SR,VL}+,x*} \big)
- 1.3\kappa_3 \Re\big(C_{X\bar\nu dsu}^{{\tt SR,VL}-,x} 
C_{X\bar\nu usd}^{{\tt SR,VL}+,x*} \big)
\nn\\
&+1.2 \Re\big( (C_{X\bar\nu sdu}^{{\tt VL,SL},x}
-C_{X\bar\nu uds}^{{\tt SR,VL}-,x})
(C_{X\bar\nu dsu}^{{\tt VL,SL},x*}
-C_{X\bar\nu usd}^{{\tt SR,VL}-,x*}) \big)
\nn\\
&+0.9\kappa_3^2 \Re\big( (C_{X\bar\nu uds}^{{\tt SR,VL}+,x}
-C_{X\bar\nu dsu}^{{\tt SR,VL}+,x})
C_{X\bar\nu usd}^{{\tt SR,VL}+,x*} \big)
+ 0.7\kappa_3 \Re\big( C_{X\bar\nu dsu}^{{\tt SR,VL}-,x} 
C_{X\bar\nu dsu}^{{\tt SR,VL}+,x*} \big)
\nn\\
&+0.8\kappa_3 \Re\big( (C_{X\bar\nu sdu}^{{\tt VL,SL},x}
-C_{X\bar\nu uds}^{{\tt SR,VL}-,x})
C_{X\bar\nu uds}^{{\tt SR,VL}+,x*} \big)
-0.7\kappa_3 \Re\big( C_{X\bar\nu dsu}^{{\tt SR,VL}-,x} 
C_{X\bar\nu uds}^{{\tt SR,VL}+,x*} \big)
\nn\\
&+0.8\kappa_3 \Re\big( (C_{X\bar\nu uds}^{{\tt SR,VL}-,x}
-C_{X\bar\nu sdu}^{{\tt VL,SL},x})
C_{X\bar\nu dsu}^{{\tt SR,VL}+,x*} \big)
-0.5\kappa_3^2 \Re \big(C_{X\bar\nu dsu}^{{\tt SR,VL}+,x} 
C_{X\bar\nu uds}^{{\tt SR,VL}+,x*} \big)
\nn\\
&+0.7 \Re\big( (C_{X\bar\nu usd}^{{\tt SR,VL}-,x}
-C_{X\bar\nu dsu}^{{\tt VL,SL},x})
C_{X\bar\nu dsu}^{{\tt SR,VL}-,x*} \big)
-0.3 \Re\big(C_{X\bar\nu usd}^{{\tt SR,VL}-,x} 
C_{X\bar\nu dsu}^{{\tt VL,SL},x*} \big)
\nn\\
&+0.1\kappa_3 \Re\big( (C_{X\bar\nu dsu}^{{\tt VL,SL},x}
-C_{X\bar\nu usd}^{{\tt SR,VL}-,x})
C_{X\bar\nu usd}^{{\tt SR,VL}+,x*} \big)
\nn\\
&+0.07\kappa_3 \Re\big( (C_{X\bar\nu uds}^{{\tt SR,VL}+,x}
-C_{X\bar\nu dsu}^{{\tt SR,VL}+,x})
(C_{X\bar\nu dsu}^{{\tt VL,SL},x*}
-C_{X\bar\nu usd}^{{\tt SR,VL}-,x*}) \big),
\\%
{\Gamma_{n\to\nu_x \pi^0 X} \over (0.1\rm GeV)^9} =\,&
52 |C_{X\bar\nu ddu}^{{\tt VL,SL},x}|^2 
+51 |C_{X\bar\nu udd}^{{\tt SR,VL}-,x}|^2
+14\kappa_3^2 |C_{X\bar\nu ddu}^{{\tt SR,VL},x}|^2
+1.1\kappa_3^2 |C_{X\bar\nu udd}^{{\tt SR,VL}+,x}|^2
\nn\\
&-103 \Re\big(C_{X\bar\nu udd}^{{\tt SR,VL}-,x} 
C_{X\bar\nu ddu}^{{\tt VL,SL},x*} \big)
+15\kappa_3 \Re\big( (C_{X\bar\nu udd}^{{\tt SR,VL}-,x}
-C_{X\bar\nu ddu}^{{\tt VL,SL},x})
C_{X\bar\nu ddu}^{{\tt SR,VL},x*} \big)
\\
& 
+9.7\kappa_3 \Re\big( (C_{X\bar\nu udd}^{{\tt SR,VL}-,x}
-C_{X\bar\nu ddu}^{{\tt VL,SL},x})
C_{X\bar\nu udd}^{{\tt SR,VL}+,x*} \big)
+6.4\kappa_3^2 \Re\big(C_{X\bar\nu udd}^{{\tt SR,VL}+,x} 
C_{X\bar\nu ddu}^{{\tt SR,VL},x*} \big),
\nn
\\%
{\Gamma_{n\to\nu_x \eta X} \over (0.1\rm GeV)^9} =\,&
1.7 |C_{X\bar\nu ddu}^{{\tt VL,SL},x}|^2 
+0.06 |C_{X\bar\nu udd}^{{\tt SR,VL}-,x}|^2
+0.02 \kappa_3^2 \big(|C_{X\bar\nu ddu}^{{\tt SR,VL},x}|^2
+|C_{X\bar\nu udd}^{{\tt SR,VL}+,x}|^2 \big)
\nn\\
&+0.6 \Re\big(C_{X\bar\nu udd}^{{\tt SR,VL}-,x} 
C_{X\bar\nu ddu}^{{\tt VL,SL},x*} \big)
+0.2\kappa_3 \Re\big( (
C_{X\bar\nu udd}^{{\tt SR,VL}+,x}
-C_{X\bar\nu ddu}^{{\tt SR,VL},x})
C_{X\bar\nu ddu}^{{\tt VL,SL},x*} \big)
\\
&+0.05\kappa_3 \Re\big( (C_{X\bar\nu udd}^{{\tt SR,VL}+,x}
-C_{X\bar\nu ddu}^{{\tt SR,VL},x})
C_{X\bar\nu udd}^{{\tt SR,VL}-,x*} \big)
-0.04 \kappa_3^2 \Re\big(C_{X\bar\nu udd}^{{\tt SR,VL}+,x} 
C_{X\bar\nu ddu}^{{\tt  SR,VL},x*} \big)
\nn,
\\%
{\Gamma_{n \to \nu_x K^0 X} \over (0.1\rm GeV)^9} =\,&
3.2 |C_{X\bar\nu sdu}^{{\tt VL,SL},x}|^2 
+3.1 |C_{X\bar\nu uds}^{{\tt SR,VL}-,x}|^2 
+2.1 |C_{X\bar\nu dsu}^{{\tt VL,SL},x}|^2
+1.3 |C_{X\bar\nu usd}^{{\tt SR,VL}-,x}|^2
\nn\\
&+0.9\kappa_3^2 |C_{X\bar\nu dsu}^{{\tt SR,VL}+,x}|^2
+0.2\kappa_3^2 \big(|C_{X\bar\nu usd}^{{\tt SR,VL}+,x}|^2
+|C_{X\bar\nu uds}^{{\tt SR,VL}+,x}|^2 \big)
+0.1 |C_{X\bar\nu dsu}^{{\tt SR,VL}-,x}|^2
\nn\\
&-6.3 \Re\big(C_{X\bar\nu uds}^{{\tt SR,VL}-,x} 
C_{X\bar\nu sdu}^{{\tt VL,SL},x*} \big)
+5.1 \Re\big( (C_{X\bar\nu sdu}^{{\tt VL,SL},x}
-C_{X\bar\nu uds}^{{\tt SR,VL}-,x})
C_{X\bar\nu dsu}^{{\tt VL,SL},x*} \big)
\nn\\
&+3.9 \Re\big( (C_{X\bar\nu sdu}^{{\tt VL,SL},x}
-C_{X\bar\nu uds}^{{\tt SR,VL}-,x})
C_{X\bar\nu usd}^{{\tt SR,VL}-,x*} \big)
+3.3 \Re\big(C_{X\bar\nu usd}^{{\tt SR,VL}-,x} 
C_{X\bar\nu dsu}^{{\tt VL,SL},x*} \big)
\nn\\
&+1.6\kappa_3 \Re\big( (C_{X\bar\nu uds}^{{\tt SR,VL}-,x}
- C_{X\bar\nu sdu}^{{\tt VL,SL},x})
C_{X\bar\nu dsu}^{{\tt SR,VL}+,x*} \big)
-1.4 \kappa_3 \Re\big(C_{X\bar\nu dsu}^{{\tt SR,VL}+,x} 
C_{X\bar\nu dsu}^{{\tt VL,SL},x*} \big)
\nn\\
&-1.3\kappa_3 \Re\big(C_{X\bar\nu dsu}^{{\tt SR,VL}+,x} 
C_{X\bar\nu usd}^{{\tt SR,VL}-,x*} \big)
+1.2 \Re\big( (C_{X\bar\nu sdu}^{{\tt VL,SL},x}
-C_{X\bar\nu uds}^{{\tt SR,VL}-,x})
C_{X\bar\nu dsu}^{{\tt SR,VL}-,x*} \big)
\nn\\
&+0.9\Re\big(C_{X\bar\nu dsu}^{{\tt SR,VL}-,x} 
C_{X\bar\nu dsu}^{{\tt VL,SL},x*} \big)
+0.9\kappa_3^2 \Re\big( (C_{X\bar\nu uds}^{{\tt SR,VL}+,x}
-C_{X\bar\nu usd}^{{\tt SR,VL}+,x})
C_{X\bar\nu dsu}^{{\tt SR,VL}+,x*} \big)
\nn\\
&+0.8 \kappa_3\Re\big( (C_{X\bar\nu usd}^{{\tt SR,VL}+,x}
-C_{X\bar\nu uds}^{{\tt SR,VL}+,x})
(C_{X\bar\nu sdu}^{{\tt VL,SL},x*}
-C_{X\bar\nu uds}^{{\tt SR,VL}-,x*}) \big)
\nn\\
&+0.8\kappa_3 \Re\big(C_{X\bar\nu usd}^{{\tt SR,VL}+,x} 
C_{X\bar\nu dsu}^{{\tt VL,SL},x*} \big)
-0.7\kappa_3 \Re\big(C_{X\bar\nu uds}^{{\tt SR,VL}+,x} 
C_{X\bar\nu dsu}^{{\tt VL,SL},x*} \big)
\nn\\
&+0.7 \Re\big(C_{X\bar\nu dsu}^{{\tt SR,VL}-,x} 
C_{X\bar\nu usd}^{{\tt SR,VL}-,x*} \big)
+0.7\kappa_3 \Re\big(C_{X\bar\nu usd}^{{\tt SR,VL}-,x} 
C_{X\bar\nu usd}^{{\tt SR,VL}+,x*} \big)
\nn\\ 
&- 0.6\kappa_3 \Re\big( C_{X\bar\nu usd}^{{\tt SR,VL}-,x} 
C_{X\bar\nu uds}^{{\tt SR,VL}+,x*} \big)
- 0.4\kappa_3^2 \Re\big( C_{X\bar\nu uds}^{{\tt SR,VL}+,x} 
C_{X\bar\nu usd}^{{\tt SR,VL}+,x*} \big)
\\
&- 0.1\kappa_3\Re\big(C_{X\bar\nu dsu}^{{\tt SR,VL}-,x} 
C_{X\bar\nu dsu}^{{\tt SR,VL}+,x*} \big)
+0.07\kappa_3\Re\big( (C_{X\bar\nu usd}^{{\tt SR,VL}+,x}
-C_{X\bar\nu uds}^{{\tt SR,VL}+,x} )
C_{X\bar\nu dsu}^{{\tt SR,VL}-,x*} \big),
\nn\\%
{\Gamma_{n \to e^- \pi^+ X} \over (0.1\rm GeV)^9} =\,&
10 \kappa_3^2\big(|C_{X\bar\ell ddd}^{{\tt SL,VR},e}|^2 
+|C_{X\bar\ell ddd}^{{\tt SR,VL},e}|^2 \big) 
+0.05\kappa_3^2\Re\big(C_{X\bar\ell ddd}^{{\tt SL,VR},e}
C_{X\bar\ell ddd}^{{\tt SR,VL},e*} \big)\;,
\\%
{\Gamma_{n \to \mu^- \pi^+ X} \over (0.1\rm GeV)^9} =\,&
9.7\kappa_3^2\big(|C_{X\bar\ell ddd}^{{\tt SL,VR},\mu}|^2 
+|C_{X\bar\ell ddd}^{{\tt SR,VL},\mu}|^2 \big) 
+8.5\kappa_3^2\Re\big( C_{X\bar\ell ddd}^{{\tt SL,VR},\mu}
C_{X\bar\ell ddd}^{{\tt SR,VL},\mu*} \big)\;,
\\%
{\Gamma_{n \to e^- K^+ X} \over (0.1\rm GeV)^9} =\,&
2.3 |C_{X\bar\ell dsd}^{{\tt SL,VR}-,e}|^2 
+ 1.3 |C_{X\bar\ell dds}^{{\tt VL,SL},e}|^2
+ 0.2 \kappa_3^2 \big( |C_{X\bar\ell dds}^{{\tt SL,VR},e}|^2 
+ |C_{X\bar\ell dsd}^{{\tt SL,VR}+,e}|^2\big)
\nn\\
&- 3.4 \Re\big( C_{X\bar\ell dsd}^{{\tt SR,VL}-,e} 
C_{X\bar\ell dds}^{{\tt VL,SL},e*}  \big)
- 0.7 \kappa_3 \Re\big( (C_{X\bar\ell dds}^{{\tt SL,VR},e}
+ C_{X\bar\ell dsd}^{{\tt SL,VR}+,e} )
C_{X\bar\ell dsd}^{{\tt SL,VR}-,e*}  \big)
\nn\\
&+ 0.6 \kappa_3 \Re\big( 
(C_{X\bar\ell dds}^{{\tt SL,VR},e}
+ C_{X\bar\ell dsd}^{{\tt SL,VR}+,e} )
C_{X\bar\ell dds}^{{\tt VR,SR},e*} \big)
+ 0.4 \kappa_3^2 \Re\big(
C_{X\bar\ell dsd}^{{\tt SL,VR}+,e} 
C_{X\bar\ell dds}^{{\tt  SL,VR},e*} \big)
\nn\\
&- 0.004 \kappa_3 \Re\big(
(C_{X\bar\ell dds}^{{\tt SL,VR},e}
+ C_{X\bar\ell dsd}^{{\tt SL,VR}+,e})
C_{X\bar\ell dsd}^{{\tt SR,VL}-,e*} \big)
+ 0.003 \kappa_3 \Re\big( 
C_{X\bar\ell dds}^{{\tt SL,VR},e}
C_{X\bar\ell dds}^{{\tt VL,SL},e*} \big)
\nn\\
&+0.003\kappa_3 \Re\big( 
C_{X\bar\ell dsd}^{{\tt SL,VR}+,e}
C_{X\bar\ell dds}^{{\tt VL,SL},e*} \big)
+ 0.001 \kappa_3^2 \Re\big(
C_{X\bar\ell dsd}^{{\tt SL,VR}+,e}
C_{X\bar\ell dds}^{{\tt SR,VL},e*} \big)
\nn\\
&+ 5\cdot10^{-4}\kappa_3^2\Re\big(
C_{X\bar\ell dds}^{{\tt SL,VR},e} 
C_{X\bar\ell dds}^{{\tt SR,VL},e*} 
+C_{X\bar\ell dsd}^{{\tt SL,VR}+,e}
C_{X\bar\ell dsd}^{{\tt SR,VL}+,e*} \big)
+ \tL \leftrightarrow \tR, 
\\%
{\Gamma_{n\to\mu^- K^+ X} \over (0.1\rm GeV)^9} =\,&
1.5 |C_{X\bar\ell dsd}^{{\tt SL,VR}-,\mu}|^2 
+0.9 |C_{X\bar\ell dds}^{{\tt VL,SL},\mu}|^2
+0.2\kappa_3^2 |C_{X\bar\ell dds}^{{\tt SL,VR},\mu}|^2 
+0.1\kappa_3^2 |C_{X\bar\ell dsd}^{{\tt SL,VR}+,\mu}|^2
\nn\\
&- 2.4\Re\big( C_{X\bar\ell dsd}^{{\tt SR,VL}-,\mu} 
C_{X\bar\ell dds}^{{\tt VL,SL},\mu*}  \big)
-0.5\kappa_3 \Re\big( (C_{X\bar\ell dds}^{{\tt SL,VR},\mu}
+C_{X\bar\ell dsd}^{{\tt SL,VR}+,\mu} )
C_{X\bar\ell dsd}^{{\tt SL,VR}-,\mu*} \big)
\nn\\
&- 0.5\kappa_3 \Re\big(C_{X\bar\ell dds}^{{\tt SL,VR},\mu}
C_{X\bar\ell dsd}^{{\tt SR,VL}-,\mu*} \big)
+0.5\kappa_3 \Re\big(C_{X\bar\ell dds}^{{\tt SL,VR},\mu}
C_{X\bar\ell dds}^{{\tt VR,SR},\mu*} \big)
\nn\\
&+0.4\kappa_3 \Re\big((C_{X\bar\ell dds}^{{\tt VL,SL},\mu} 
-C_{X\bar\ell dsd}^{{\tt SL,VR}-,\mu})
C_{X\bar\ell dsd}^{{\tt SR,VL}+,\mu*} \big)
+0.3\kappa_3^2\Re\big(C_{X\bar\ell dsd}^{{\tt SL,VR}+,\mu} 
C_{X\bar\ell dds}^{{\tt  SL,VR},\mu*} \big)
\nn\\
&+0.3\kappa_3 \Re\big((C_{X\bar\ell dds}^{{\tt SL,VR},\mu}
+C_{X\bar\ell dsd}^{{\tt SL,VR}+,\mu})
C_{X\bar\ell dds}^{{\tt VL,SL},\mu*} \big)
+0.2\kappa_3^2\Re\big(C_{X\bar\ell dsd}^{{\tt SL,VR}+,\mu}
C_{X\bar\ell dds}^{{\tt SR,VL},\mu*} \big)
\nn\\
&+0.1\kappa_3^2\Re\big(C_{X\bar\ell dds}^{{\tt SL,VR},\mu}
C_{X\bar\ell dds}^{{\tt SR,VL},\mu*}
+C_{X\bar\ell dsd}^{{\tt SL,VR}+,\mu}
C_{X\bar\ell dsd}^{{\tt SR,VL}+,\mu*} \big)
+ \tL \leftrightarrow \tR.
\end{align}
\end{subequations}}\normalsize

We have verified our above results by comparing them with the corresponding expressions for nucleon decays into an ALP in the $a$LEFT framework given in~\cite{Fan:2025xhi}.
By replacing the dark photon field factor (together with an additional mass factor) $m_X X_\mu$ with $\partial_\mu a$, the $X$LEFT operators in case A convert to the corresponding $a$LEFT operators used in~\cite{Fan:2025xhi}. 
Consequently, the nucleon decay amplitudes in the two EFTs are related as 
${\cal M}_X^{\rm(A)}=m_X \epsilon_{X,\mu}^* {\cal A}^\mu$ and ${\cal M}_a = k_\mu {\cal A}^\mu$, where ${\cal A}^\mu$ is their common part and $k_\mu$ is the dark photon or ALP momentum. 
In the massless limit for both the dark photon and the ALP, the spin-averaged squared matrix elements for the two cases become identical. 
This follows from the fact that $\sum_{\rm pol.}m_X^2 \epsilon_{X,\mu}^* \epsilon_{X,\nu} = m_X^2(-g_{\mu\nu} + k_\mu k_\nu/m_X^2)\to k_\mu k_\nu$ as $m_X\to 0$.  Thus, the decay widths coincide up to notational differences. 

\subsection{Case B}
\label{app:Gamma_exp_B}

In this subsection, we summarize the width results in the case where the dark photon is parametrized by an antisymmetric field strength tensor $X_{\mu\nu}$.
These results also apply when replacing the dark photon with the SM photon. 
For the two-body decay processes, we obtain
{\small
\begin{subequations}	
\label{eq:Ga_N2lXB}
\begin{align}
{\Gamma_{p\to e^+ X} \over 10^{-8}(\rm GeV)^9} =\,&    
1100 |\tilde{C}_{X\ell uud}^{{\tt TL,SL},e}|^2
+1000 |\tilde{C}_{X\ell udu}^{{\tt VR,VL}-,e}|^2
+10^{-5}\kappa_3^2\big( 
|\tilde{C}_{X\ell udu}^{{\tt VR,VL}+,e}|^2
+ |\tilde{C}_{X\ell uud}^{{\tt VR,VL},e}|^2\big)
\nn\\
&-2100 \Im\big(
\tilde{C}_{X\ell uud}^{{\tt TL,SL},e}
\tilde{C}_{X\ell udu}^{{\tt VR,VL}-,e*}\big)
-2\cdot10^{-5}\kappa_3^2\Re\big(
\tilde{C}_{X\ell udu}^{{\tt VR,VL}+,e}
\tilde{C}_{X\ell uud}^{{\tt VR,VL},e*}\big)
\nn
\\
&+0.2\kappa_3\Re\big(
(\tilde{C}_{X\ell uud}^{{\tt VL,VR},e}
-\tilde{C}_{X\ell udu}^{{\tt VL,VR}+,e})
\tilde{C}_{X\ell udu}^{{\tt VR,VL}-,e*}\big)
\nn\\
&+0.2\kappa_3 \Im\big(
(\tilde{C}_{X\ell uud}^{{\tt VL,VR},e}
-\tilde{C}_{X\ell udu}^{{\tt VL,VR}+,e})
\tilde{C}_{X\ell uud}^{{\tt TL,SL},e*}\big)
+ \tL \leftrightarrow \tR,
\\%
{\Gamma_{p\to \mu^+ X} \over 10^{-8}(\rm GeV)^9} =\,&    
1000 |\tilde{C}_{X\ell uud}^{{\tt TL,SL},\mu}|^2
+1000 |\tilde{C}_{X\ell udu}^{{\tt VR,VL}-,\mu}|^2
+0.5\kappa_3^2\big(|\tilde{C}_{X\ell udu}^{{\tt VR,VL}+,\mu}|^2
+|\tilde{C}_{X\ell uud}^{{\tt VR,VL},\mu}|^2\big)
\nn\\
&-2000 \Im\big(\tilde{C}_{X\ell uud}^{{\tt TL,SL},\mu}\tilde{C}_{X\ell udu}^{{\tt VR,VL}-,\mu*}\big)
-1.0\kappa_3^2\Re\big(\tilde{C}_{X\ell udu}^{{\tt VR,VL}+,\mu}\tilde{C}_{X\ell uud}^{{\tt VR,VL},\mu*}\big)
\nn
\\
&+44\kappa_3\Re\big(
(\tilde{C}_{X\ell uud}^{{\tt VL,VR},\mu}
-\tilde{C}_{X\ell udu}^{{\tt VL,VR}+,\mu})
\tilde{C}_{X\ell udu}^{{\tt VR,VL}-,\mu*}\big)
\nn\\
&+44\kappa_3 \Im\big(
(\tilde{C}_{X\ell uud}^{{\tt VL,VR},\mu}
-\tilde{C}_{X\ell udu}^{{\tt VL,VR}+,\mu})
\tilde{C}_{X\ell uud}^{{\tt TL,SL},\mu*}\big)
+ \tL \leftrightarrow \tR,
\\%
{\Gamma_{n\to \nu_x X} \over 10^{-8}(\rm GeV)^9}  =\,&    
1100 |\tilde{C}_{X\bar\nu ddu}^{{\tt TR,SR},x}|^2
+1000 |\tilde{C}_{X\bar\nu udd}^{{\tt VL,VR}-,x}|^2
+2100 \Im\big(\tilde{C}_{X\bar\nu ddu}^{{\tt TR,SR},x}\tilde{C}_{X\bar\nu udd}^{{\tt VL,VR}-,x*}\big).
\end{align}
\end{subequations}}\normalsize
For three-body processes:
{\small
\begin{subequations}
\label{eq:Ga_N2lXMB}
\begin{align}
{\Gamma_{p \to e^+ \pi^0 X} \over 10^{-8}(\rm GeV)^9} =\,&
41 |\tilde{C}_{X\ell uud}^{{\tt TL,SL},e}|^2+
41 |\tilde{C}_{X\ell udu}^{{\tt VR,VL}-,e}|^2
+1.2 \kappa_3^2 |\tilde{C}_{X\ell uud}^{{\tt VR,VL},e}|^2
+0.1\kappa_3^2  |\tilde{C}_{X\ell udu}^{{\tt VR,VL}+,e}|^2
\nn\\
&+0.002\kappa_4^2 |\tilde{C}_{X\ell uud}^{{\tt SL,TL},e}|^2
+82 \Im\big(\tilde{C}_{X\ell udu}^{{\tt VR,VL}-,e}\tilde{C}_{X\ell uud}^{{\tt TL,SL},e*}\big)
+0.8\kappa_3^2 \Re\big(\tilde{C}_{X\ell uud}^{{\tt VL,VR},e}\tilde{C}_{X\ell udu}^{{\tt VL,VR}+,e*}\big)
\nn\\
&+0.04\kappa_3 \Im\big(\tilde{C}_{X\ell uud}^{{\tt VL,VR},e}\tilde{C}_{X\ell uud}^{{\tt TL,SL},e*}\big)
+0.04\kappa_3 \Re\big(\tilde{C}_{X\ell udu}^{{\tt VR,VL}-,e}\tilde{C}_{X\ell uud}^{{\tt VL,VR},e*}\big)
\nn\\
& +0.007\kappa_3 \Im\big(\tilde{C}_{X\ell udu}^{{\tt VL,VR}+,e}\tilde{C}_{X\ell uud}^{{\tt TL,SL},e*}\big)
+0.007\kappa_3 \Re\big(\tilde{C}_{X\ell udu}^{{\tt VR,VL}-,e}\tilde{C}_{X\ell udu}^{{\tt VL,VR}+,e*}\big)
\\
&-3\cdot10^{-4} \kappa_3\kappa_4\Re\big(
\tilde{C}_{X\ell uud}^{{\tt VL,VR},e}
\tilde{C}_{X\ell uud}^{{\tt SL,TL},e*}\big)
-1\cdot10^{-4} \kappa_3\kappa_4\Re\big(
\tilde{C}_{X\ell udu}^{{\tt VL,VR}+,e}
\tilde{C}_{X\ell uud}^{{\tt SL,TL},e*}\big)
+ \tL \leftrightarrow \tR,
\nn
\\%
{\Gamma_{p \to \mu^+ \pi^0 X} \over 10^{-8}(\rm GeV)^9} =\,&
36 |\tilde{C}_{X\ell uud}^{{\tt TL,SL},\mu}|^2
+35 |\tilde{C}_{X\ell udu}^{{\tt VR,VL}-,\mu}|^2
+1.2\kappa_3^2 |\tilde{C}_{X\ell uud}^{{\tt VR,VL},\mu}|^2
+0.1\kappa_3^2 |\tilde{C}_{X\ell udu}^{{\tt VR,VL}+,\mu}|^2
\nn\\
&+0.002\kappa_4^2 |\tilde{C}_{X\ell uud}^{{\tt SL,TL},\mu}|^2
+71 \Im\big(\tilde{C}_{X\ell udu}^{{\tt VR,VL}-,\mu}\tilde{C}_{X\ell uud}^{{\tt TL,SL},\mu*}\big)
\nn\\
&+6.3\kappa_3 \Im\big(\tilde{C}_{X\ell uud}^{{\tt VL,VR},\mu}\tilde{C}_{X\ell uud}^{{\tt TL,SL},\mu*}\big)
+6.2\kappa_3 \Re\big(\tilde{C}_{X\ell udu}^{{\tt VR,VL}-,\mu}\tilde{C}_{X\ell uud}^{{\tt VL,VR},\mu*}\big)
\nn\\
&+1.0\kappa_3 \Im\big(\tilde{C}_{X\ell udu}^{{\tt VL,VR}+,\mu}\tilde{C}_{X\ell uud}^{{\tt TL,SL},\mu*}\big)
+1.0\kappa_3 \Re\big(\tilde{C}_{X\ell udu}^{{\tt VR,VL}-,\mu}\tilde{C}_{X\ell udu}^{{\tt VL,VR}+,\mu*}\big)
\nn\\
&+0.7\kappa_3^2 \Re\big(\tilde{C}_{X\ell uud}^{{\tt VL,VR},\mu}\tilde{C}_{X\ell udu}^{{\tt VL,VR}+,\mu*}\big)
-0.04 \kappa_3\kappa_4\Re\big(\tilde{C}_{X\ell uud}^{{\tt VL,VR},\mu}\tilde{C}_{X\ell uud}^{{\tt SL,TL},\mu*}\big)
\nn\\
&-0.01 \kappa_3\kappa_4\Re\big(\tilde{C}_{X\ell udu}^{{\tt VL,VR}+,\mu}\tilde{C}_{X\ell uud}^{{\tt SL,TL},\mu*}\big)
+ \tL \leftrightarrow \tR,
\\%
{\Gamma_{p \to e^+ \eta X} \over 10^{-8}(\rm GeV)^9} =\,&
1.3 |\tilde{C}_{X\ell uud}^{{\tt TL,SL},e}|^2
+0.05 |\tilde{C}_{X\ell udu}^{{\tt VR,VL}-,e}|^2
+0.002 \kappa_3^2 \big(|\tilde{C}_{X\ell udu}^{{\tt VR,VL}+,e}|^2
+|\tilde{C}_{X\ell uud}^{{\tt VR,VL},e}|^2\big)
\nn\\
&-0.5 \Im\big(\tilde{C}_{X\ell udu}^{{\tt VR,VL}-,e}\tilde{C}_{X\ell uud}^{{\tt TL,SL},e*}\big)
-0.005\kappa_3^2 \Re\big(\tilde{C}_{X\ell uud}^{{\tt VL,VR},e}\tilde{C}_{X\ell udu}^{{\tt VL,VR}+,e*}\big)\nn\\
&+5\cdot10^{-4}\kappa_3 \Im\big((\tilde{C}_{X\ell uud}^{{\tt VL,VR},e}-\tilde{C}_{X\ell udu}^{{\tt VL,VR}+,e})\tilde{C}_{X\ell uud}^{{\tt TL,SL},e*}\big)
\nn\\
&+8\cdot10^{-5}\kappa_3 \Re\big((\tilde{C}_{X\ell udu}^{{\tt VL,VR}+,e}-\tilde{C}_{X\ell uud}^{{\tt VL,VR},e})\tilde{C}_{X\ell udu}^{{\tt VR,VL}-,e*}\big)
+ \tL \leftrightarrow \tR,
\\%
{\Gamma_{p \to \mu^+ \eta X} \over 10^{-8}(\rm GeV)^9} =\,&
0.8 |\tilde{C}_{X\ell uud}^{{\tt TL,SL},\mu}|^2
+0.03 |\tilde{C}_{X\ell udu}^{{\tt VR,VL}-,\mu}|^2
+0.002\kappa_3^2\big(
|\tilde{C}_{X\ell udu}^{{\tt VR,VL}+,\mu}|^2
+|\tilde{C}_{X\ell uud}^{{\tt VR,VL},\mu}|^2\big)
\nn\\
&-0.3 \Im\big(\tilde{C}_{X\ell udu}^{{\tt VR,VL}-,\mu}\tilde{C}_{X\ell uud}^{{\tt TL,SL},\mu*}\big)
+0.05\kappa_3 \Im\big((\tilde{C}_{X\ell uud}^{{\tt VL,VR},\mu}-\tilde{C}_{X\ell udu}^{{\tt VL,VR}+,\mu})\tilde{C}_{X\ell uud}^{{\tt TL,SL},\mu*}\big)\nn\\
&
+0.008\kappa_3 \Re\big((\tilde{C}_{X\ell udu}^{{\tt VL,VR}+,\mu}-\tilde{C}_{X\ell uud}^{{\tt VL,VR},\mu})\tilde{C}_{X\ell udu}^{{\tt VR,VL}-,\mu*}\big)
\nn\\
&-0.004\kappa_3^2 \Re\big(\tilde{C}_{X\ell udu}^{{\tt VL,VR}+,\mu}\tilde{C}_{X\ell uud}^{{\tt VL,VR},\mu*}\big)
+ \tL \leftrightarrow \tR,
\\%
{\Gamma_{p \to e^+ K^0 X} \over 10^{-8}(\rm GeV)^9} =\,&
1.7 |\tilde{C}_{X\ell usu}^{{\tt VR,VL}-,e}|^2
+1.0 |\tilde{C}_{X\ell uus}^{{\tt TL,SL},e}|^2
+0.03\kappa_3^2\big(|\tilde{C}_{X\ell usu}^{{\tt VR,VL}+,e}|^2
+|\tilde{C}_{X\ell uus}^{{\tt VR,VL},e}|^2\big)
\nn\\
&+4\cdot10^{-5}\kappa_4^2|\tilde{C}_{X\ell uus}^{{\tt SL,TL},e}|^2
-2.6 \Im\big(\tilde{C}_{X\ell usu}^{{\tt VR,VL}-,e}\tilde{C}_{X\ell uus}^{{\tt TL,SL},e*}\big)
+0.05\kappa_3^2 \Re\big(\tilde{C}_{X\ell uus}^{{\tt VL,VR},e}\tilde{C}_{X\ell usu}^{{\tt VL,VR}+,e*}\big)\nn\\
&+0.002\kappa_3 \Re\big(\tilde{C}_{X\ell usu}^{{\tt VR,VL}-,e}\tilde{C}_{X\ell uus}^{{\tt VL,VR},e*}\big)
+0.002\kappa_3 \Re\big(\tilde{C}_{X\ell usu}^{{\tt VR,VL}-,e}\tilde{C}_{X\ell usu}^{{\tt VL,VR}+,e*}\big)
\nn\\
&-0.001\kappa_3 \Im\big(\tilde{C}_{X\ell uus}^{{\tt VL,VR},e}\tilde{C}_{X\ell uus}^{{\tt TL,SL},e*}\big)
-0.001\kappa_3 \Im\big(\tilde{C}_{X\ell usu}^{{\tt VL,VR}+,e}\tilde{C}_{X\ell uus}^{{\tt TL,SL},e*}\big)
\nn\\
&-1\cdot10^{-5} \kappa_3\kappa_4\Re\big((\tilde{C}_{X\ell usu}^{{\tt VL,VR}+,e}+\tilde{C}_{X\ell uus}^{{\tt VL,VR},e})\tilde{C}_{X\ell uus}^{{\tt SL,TL},e*}\big)
+ \tL \leftrightarrow \tR,
\\%
{\Gamma_{p \to \mu^+ K^0 X} \over 10^{-8}(\rm GeV)^9} =\,&
1.1  |\tilde{C}_{X\ell usu}^{{\tt VR,VL}-,\mu}|^2
+0.7  |\tilde{C}_{X\ell uus}^{{\tt TL,SL},\mu}|^2
+0.02\kappa_3^2 |\tilde{C}_{X\ell uus}^{{\tt VR,VL},\mu}|^2
+0.02\kappa_3^2 |\tilde{C}_{X\ell usu}^{{\tt VR,VL}+,\mu}|^2
\nn\\
&+3\cdot10^{-5}\kappa_4^2 |\tilde{C}_{X\ell uus}^{{\tt SL,TL},\mu}|^2
-1.8 \Im\big(\tilde{C}_{X\ell usu}^{{\tt VR,VL}-,\mu}\tilde{C}_{X\ell uus}^{{\tt TL,SL},\mu*}\big)\nn\\
&+0.2\kappa_3 \Re\big(\tilde{C}_{X\ell usu}^{{\tt VR,VL}-,\mu}\tilde{C}_{X\ell uus}^{{\tt VL,VR},\mu*}\big)
+0.2\kappa_3 \Re\big(\tilde{C}_{X\ell usu}^{{\tt VR,VL}-,\mu}\tilde{C}_{X\ell usu}^{{\tt VL,VR}+,\mu*}\big)
\nn\\
&-0.1\kappa_3 \Im\big(\tilde{C}_{X\ell uus}^{{\tt VL,VR},\mu}\tilde{C}_{X\ell uus}^{{\tt TL,SL},\mu*}\big)
-0.1\kappa_3 \Im\big(\tilde{C}_{X\ell usu}^{{\tt VL,VR}+,\mu}\tilde{C}_{X\ell uus}^{{\tt TL,SL},\mu*}\big)
\nn\\
&+0.04\kappa_3^2 \Re\big(\tilde{C}_{X\ell uus}^{{\tt VL,VR},\mu}\tilde{C}_{X\ell usu}^{{\tt VL,VR}+,\mu*}\big)
\nn\\
&-8\cdot10^{-4} \kappa_3\kappa_4\Re\big((\tilde{C}_{X\ell usu}^{{\tt VL,VR}+,\mu}+\tilde{C}_{X\ell uus}^{{\tt VL,VR},\mu})\tilde{C}_{X\ell uus}^{{\tt SL,TL},\mu*}\big)
+ \tL \leftrightarrow \tR,
\\%
{\Gamma_{n \to e^+ \pi^- X} \over 10^{-8}(\rm GeV)^9} =\,&
82 |\tilde{C}_{X\ell uud}^{{\tt TL,SL},e}|^2
+80  |\tilde{C}_{X\ell udu}^{{\tt VR,VL}-,e}|^2
+2.4\kappa_3^2  |\tilde{C}_{X\ell udu}^{{\tt VR,VL}+,e}|^2
+0.3 \kappa_3^2 |\tilde{C}_{X\ell uud}^{{\tt VR,VL},e}|^2
\nn\\
&+0.001\kappa_4^2 |\tilde{C}_{X\ell uud}^{{\tt SL,TL},e}|^2
+160 \Im\big(\tilde{C}_{X\ell udu}^{{\tt VR,VL}-,e}\tilde{C}_{X\ell uud}^{{\tt TL,SL},e*}\big)
-1.6\kappa_3^2 \Re\big(\tilde{C}_{X\ell uud}^{{\tt VL,VR},e}\tilde{C}_{X\ell udu}^{{\tt VL,VR}+,e*}\big)
\nn\\
&-0.08\kappa_3 \Im\big(\tilde{C}_{X\ell udu}^{{\tt VL,VR}+,e}\tilde{C}_{X\ell uud}^{{\tt TL,SL},e*}\big)
-0.08\kappa_3 \Re\big(\tilde{C}_{X\ell udu}^{{\tt VR,VL}-,e}\tilde{C}_{X\ell udu}^{{\tt VL,VR}+,e*}\big)
\nn\\
&+0.03\kappa_3 \Im\big(\tilde{C}_{X\ell uud}^{{\tt VL,VR},e}\tilde{C}_{X\ell uud}^{{\tt TL,SL},e*}\big)
+0.03\kappa_3 \Re\big(\tilde{C}_{X\ell udu}^{{\tt VR,VL}-,e}\tilde{C}_{X\ell uud}^{{\tt VL,VR},e*}\big)
\\
&-3\cdot10^{-4} \kappa_3\kappa_4\Re\big(\tilde{C}_{X\ell udu}^{{\tt VL,VR}+,e}\tilde{C}_{X\ell uud}^{{\tt SL,TL},e*}\big)
+1\cdot10^{-4} \kappa_3\kappa_4\Re\big(\tilde{C}_{X\ell uud}^{{\tt VL,VR},e}\tilde{C}_{X\ell uud}^{{\tt SL,TL},e*}\big)
+ \tL \leftrightarrow \tR,
\nn
\\%
{\Gamma_{n \to \mu^+ \pi^- X} \over 10^{-8}(\rm GeV)^9} =\,&
71 |\tilde{C}_{X\ell uud}^{{\tt TL,SL},\mu}|^2
+70 |\tilde{C}_{X\ell udu}^{{\tt VR,VL}-,\mu}|^2
+2.4\kappa_3^2  |\tilde{C}_{X\ell udu}^{{\tt VR,VL}+,\mu}|^2
+0.3 \kappa_3^2 |\tilde{C}_{X\ell uud}^{{\tt VR,VL},\mu}|^2
\nn\\
&+0.001\kappa_4^2 |\tilde{C}_{X\ell uud}^{{\tt SL,TL},\mu}|^2
+140 \Im\big(\tilde{C}_{X\ell udu}^{{\tt VR,VL}-,\mu}\tilde{C}_{X\ell uud}^{{\tt TL,SL},\mu*}\big)
-13\kappa_3 \Im\big(\tilde{C}_{X\ell udu}^{{\tt VL,VR}+,\mu}\tilde{C}_{X\ell uud}^{{\tt TL,SL},\mu*}\big)
\nn\\
&-12\kappa_3 \Re\big(\tilde{C}_{X\ell udu}^{{\tt VR,VL}-,\mu}\tilde{C}_{X\ell udu}^{{\tt VL,VR}+,\mu*}\big)
+5.2\kappa_3 \Im\big(\tilde{C}_{X\ell uud}^{{\tt VL,VR},\mu}\tilde{C}_{X\ell uud}^{{\tt TL,SL},\mu*}\big)
\nn\\
&+5.2\kappa_3 \Re\big(\tilde{C}_{X\ell udu}^{{\tt VR,VL}-,\mu}\tilde{C}_{X\ell uud}^{{\tt VL,VR},\mu*}\big)
-1.7\kappa_3^2 \Re\big(\tilde{C}_{X\ell uud}^{{\tt VL,VR},\mu}\tilde{C}_{X\ell udu}^{{\tt VL,VR}+,\mu*}\big)
\\
&-0.04 \kappa_3\kappa_4\Re\big(\tilde{C}_{X\ell udu}^{{\tt VL,VR}+,\mu}\tilde{C}_{X\ell uud}^{{\tt SL,TL},\mu*}\big)
+0.01 \kappa_3\kappa_4\Re\big(\tilde{C}_{X\ell uud}^{{\tt VL,VR},\mu}\tilde{C}_{X\ell uud}^{{\tt SL,TL},\mu*}\big)
+ \tL \leftrightarrow \tR,
\nn
\\%
{\Gamma_{p \to \nu _x \pi^+ X} \over 10^{-8}(\rm GeV)^9} =\,&
80|\tilde{C}_{X\bar\nu ddu}^{{\tt TR,SR},x}|^2+
79 |\tilde{C}_{X\bar\nu udd}^{{\tt VL,VR}-,x}|^2 +2.4\kappa_3^2 |\tilde{C}_{X\bar\nu udd}^{{\tt VL,VR}+,x}|^2+0.3\kappa_3^2 |\tilde{C}_{X\bar\nu ddu}^{{\tt VL,VR},x}|^2
\\
&+10^{-3}\kappa_4^2|\tilde{C}_{X\bar\nu udd}^{{\tt SR,TR},x}|^2
-160 \Im\big(\tilde{C}_{X\bar\nu udd}^{{\tt VL,VR}-,x}\tilde{C}_{X\bar\nu ddu}^{{\tt TR,SR},x*}\big)
-1.6\kappa_3^2\Re\big(\tilde{C}_{X\bar\nu udd}^{{\tt VL,VR}+,x}\tilde{C}_{X\bar\nu ddu}^{{\tt VL,VR},x*}\big),
\nn
\\%
{\Gamma_{p \to \nu_x K^+ X} \over 10^{-8}(\rm GeV)^9} =\,&
2.6 |\tilde{C}_{X\bar\nu uds}^{{\tt VL,VR}-,x}|^2
+2.6|\tilde{C}_{X\bar\nu sdu}^{{\tt TR,SR},x}|^2
+1.1 |\tilde{C}_{X\bar\nu dsu}^{{\tt VL,VR}-,x}|^2 
+0.1 |\tilde{C}_{X\bar\nu usd}^{{\tt VL,VR}-,x}|^2
\nn\\
&+0.1|\tilde{C}_{X\bar\nu dsu}^{{\tt TR,SR},x}|^2
+0.1\kappa_3^2|\tilde{C}_{X\bar\nu usd}^{{\tt VL,VR}+,x}|^2
+0.03\kappa_3^2(|\tilde{C}_{X\bar\nu uds}^{{\tt VL,VR}+,x}|^2
+|\tilde{C}_{X\bar\nu dsu}^{{\tt VL,VR}+,x}|^2)
\nn\\ 
&+1\cdot10^{-5}\kappa_4^2|\tilde{C}_{X\bar\nu uds}^{{\tt SR,TR},x}|^2
+5.2 \Im\big(\tilde{C}_{X\bar\nu sdu}^{{\tt TR,SR},x}\tilde{C}_{X\bar\nu uds}^{{\tt VL,VR}-,x*}\big)
\nn\\
&+3.3 \Im\big(\tilde{C}_{X\bar\nu sdu}^{{\tt TR,SR},x}\tilde{C}_{X\bar\nu dsu}^{{\tt VL,VR}-,x*}\big)
+3.3 \Re\big(\tilde{C}_{X\bar\nu uds}^{{\tt VL,VR}-,x}\tilde{C}_{X\bar\nu dsu}^{{\tt VL,VR}-,x*}\big)
\nn\\
&+1.0 \Re\big(\tilde{C}_{X\bar\nu uds}^{{\tt VL,VR}-,x}\tilde{C}_{X\bar\nu usd}^{{\tt VL,VR}-,x*}\big)
+1.0 \Im\big(\tilde{C}_{X\bar\nu dsu}^{{\tt TR,SR},x}\tilde{C}_{X\bar\nu uds}^{{\tt VL,VR}-,x*}\big)
\nn\\
&+1.0 \Im\big(\tilde{C}_{X\bar\nu sdu}^{{\tt TR,SR},x}\tilde{C}_{X\bar\nu usd}^{{\tt VL,VR}-,x*}\big)
+1.0\Re\big(\tilde{C}_{X\bar\nu sdu}^{{\tt TR,SR},x}\tilde{C}_{X\bar\nu dsu}^{{\tt TR,SR},x*}\big)
\nn\\
&+0.6 \Re\big(\tilde{C}_{X\bar\nu usd}^{{\tt VL,VR}-,x}\tilde{C}_{X\bar\nu dsu}^{{\tt VL,VR}-,x*}\big)
+0.6 \Im\big(\tilde{C}_{X\bar\nu dsu}^{{\tt TR,SR},x}\tilde{C}_{X\bar\nu dsu}^{{\tt VL,VR}-,x*}\big)
\nn\\
&+0.2 \Im\big(\tilde{C}_{X\bar\nu dsu}^{{\tt TR,SR},x}\tilde{C}_{X\bar\nu usd}^{{\tt VL,VR}-,x*}\big)
+0.1\kappa_3^2\Re\big((\tilde{C}_{X\bar\nu uds}^{{\tt VL,VR}+,x}-\tilde{C}_{X\bar\nu dsu}^{{\tt VL,VR}+,x})\tilde{C}_{X\bar\nu usd}^{{\tt VL,VR}+,x*}\big)
\nn\\   
&-0.06\kappa_3^2\Re\big(\tilde{C}_{X\bar\nu dsu}^{{\tt VL,VR}+,x}\tilde{C}_{X\bar\nu uds}^{{\tt VL,VR}+,x*}\big),
\\%
{\Gamma_{n \to \nu_x \pi^0 X} \over 10^{-8}(\rm GeV)^9} =\,&	
42|\tilde{C}_{X\bar\nu ddu}^{{\tt TR,SR},x}|^2 +
41 |\tilde{C}_{X\bar\nu udd}^{{\tt VL,VR}-,x}|^2+1.2\kappa_3^2 |\tilde{C}_{X\bar\nu ddu}^{{\tt VL,VR},x}|^2 
+0.14\kappa_3^2 |\tilde{C}_{X\bar\nu udd}^{{\tt VL,VR}+,x}|^2
\\
&+0.002\kappa_4^2|\tilde{C}_{X\bar\nu udd}^{{\tt SR,TR},x}|^2
-83 \Im\big(\tilde{C}_{X\bar\nu udd}^{{\tt VL,VR}-,x}\tilde{C}_{X\bar\nu ddu}^{{\tt TR,SR},x*}\big)
+0.8\kappa_3^2 \Re\big(\tilde{C}_{X\bar\nu ddu}^{{\tt VL,VR},x}\tilde{C}_{X\bar\nu udd}^{{\tt VL,VR}+,x*}\big),
\nn
\\%
{\Gamma_{n \to \nu_x \eta X} \over 10^{-8}(\rm GeV)^9} =\,&
1.3|\tilde{C}_{X\bar\nu ddu}^{{\tt TR,SR},x}|^2 +0.05 |\tilde{C}_{X\bar\nu udd}^{{\tt VL,VR}-,x}|^2 +0.003\kappa_3^2 (|\tilde{C}_{X\bar\nu udd}^{{\tt VL,VR}+,x}|^2+|\tilde{C}_{X\bar\nu ddu}^{{\tt VL,VR},x}|^2) \nn\\
&+0.5 \Im\big(\tilde{C}_{X\bar\nu udd}^{{\tt VL,VR}-,x}\tilde{C}_{X\bar\nu ddu}^{{\tt TR,SR},x*}\big)-0.005\kappa_3^2 \Re\big(\tilde{C}_{X\bar\nu udd}^{{\tt VL,VR}+,x}\tilde{C}_{X\bar\nu ddu}^{{\tt VL,VR},x*}\big),
\\%
{\Gamma_{n \to \nu_x K^0 X} \over 10^{-8}(\rm GeV)^9} =\,&
2.5|\tilde{C}_{X\bar\nu sdu}^{{\tt TR,SR},x}|^2
+2.5 |\tilde{C}_{X\bar\nu uds}^{{\tt VL,VR}-,x}|^2
+1.7|\tilde{C}_{X\bar\nu dsu}^{{\tt TR,SR},x}|^2
+1.1 |\tilde{C}_{X\bar\nu usd}^{{\tt VL,VR}-,x}|^2
\nn\\
&+0.1 |\tilde{C}_{X\bar\nu dsu}^{{\tt VL,VR}-,x}|^2
+0.1\kappa_3^2|\tilde{C}_{X\bar\nu dsu}^{{\tt VL,VR}+,x}|^2
+0.03\kappa_3^2(|\tilde{C}_{X\bar\nu uds}^{{\tt VL,VR}+,x}|^2+|\tilde{C}_{X\bar\nu usd}^{{\tt VL,VR}+,x}|^2)
\nn\\
&+1\cdot10^{-5}\kappa_4^2|\tilde{C}_{X\bar\nu uds}^{{\tt SR,TR},x}|^2
-5.0 \Im\big(\tilde{C}_{X\bar\nu uds}^{{\tt VL,VR}-,x}\tilde{C}_{X\bar\nu sdu}^{{\tt TR,SR},x*}\big)
\nn\\
&+4.1\Re\big(\tilde{C}_{X\bar\nu sdu}^{{\tt TR,SR},x}\tilde{C}_{X\bar\nu dsu}^{{\tt TR,SR},x*}\big)
-4.1 \Im\big(\tilde{C}_{X\bar\nu uds}^{{\tt VL,VR}-,x}\tilde{C}_{X\bar\nu dsu}^{{\tt TR,SR},x*}\big)
\nn\\
&+3.2 \Im\big(\tilde{C}_{X\bar\nu usd}^{{\tt VL,VR}-,x}\tilde{C}_{X\bar\nu sdu}^{{\tt TR,SR},x*}\big)
-3.2 \Re\big(\tilde{C}_{X\bar\nu uds}^{{\tt VL,VR}-,x}\tilde{C}_{X\bar\nu usd}^{{\tt VL,VR}-,x*}\big)
\nn\\
&+2.6 \Im\big(\tilde{C}_{X\bar\nu usd}^{{\tt VL,VR}-,x}\tilde{C}_{X\bar\nu dsu}^{{\tt TR,SR},x*}\big)
+1.0 \Im\big(\tilde{C}_{X\bar\nu dsu}^{{\tt VL,VR}-,x}\tilde{C}_{X\bar\nu sdu}^{{\tt TR,SR},x*}\big)
\nn\\
&-1.0 \Re\big(\tilde{C}_{X\bar\nu uds}^{{\tt VL,VR}-,x}\tilde{C}_{X\bar\nu dsu}^{{\tt VL,VR}-,x*}\big)
+0.8 \Im\big(\tilde{C}_{X\bar\nu dsu}^{{\tt VL,VR}-,x}\tilde{C}_{X\bar\nu dsu}^{{\tt TR,SR},x*}\big)
\nn\\
&+0.6 \Re\big(\tilde{C}_{X\bar\nu usd}^{{\tt VL,VR}-,x}\tilde{C}_{X\bar\nu dsu}^{{\tt VL,VR}-,x*}\big)
+0.1\kappa_3^2\Re\big((\tilde{C}_{X\bar\nu uds}^{{\tt VL,VR}+,x}-\tilde{C}_{X\bar\nu usd}^{{\tt VL,VR}+,x})\tilde{C}_{X\bar\nu dsu}^{{\tt VL,VR}+,x*}\big)
\nn\\
&-0.05\kappa_3^2\Re\big(\tilde{C}_{X\bar\nu usd}^{{\tt VL,VR}+,x}\tilde{C}_{X\bar\nu uds}^{{\tt VL,VR}+,x*}\big),
\\%
{\Gamma_{n\to e^-\pi^+ X} \over 10^{-8}(\rm GeV)^9} =\,&
1.1\kappa_3^2 |\tilde{C}_{X\bar\ell ddd}^{{\tt VR,VL},e}|^2
+0.01 \kappa_4^2|\tilde{C}_{X\bar\ell ddd}^{{\tt SL,TL},e}|^2
-6\cdot10^{-4} \kappa_3\kappa_4\Re\big(\tilde{C}_{X\bar\ell ddd}^{{\tt VL,VR},e}\tilde{C}_{X\bar\ell ddd}^{{\tt SL,TL},e*}\big)
\nn\\
&+ \tL \leftrightarrow \tR,
\\%
{\Gamma_{n\to\mu^-\pi^+ X} \over 10^{-8}(\rm GeV)^9} =\,&
1.0\kappa_3^2 |\tilde{C}_{X\bar\ell ddd}^{{\tt VR,VL},\mu}|^2
+0.01 \kappa_4^2 |\tilde{C}_{X\bar\ell ddd}^{{\tt SL,TL},\mu}|^2
-0.08 \kappa_3\kappa_4\Re\big(\tilde{C}_{X\bar\ell ddd}^{{\tt VL,VR},\mu}\tilde{C}_{X\bar\ell ddd}^{{\tt SL,TL},\mu*}\big)
\nn\\
&+ \tL \leftrightarrow \tR,
\\%
{\Gamma_{n \to e^- K^+ X} \over 10^{-8}(\rm GeV)^9}=\,&
1.8 |\tilde{C}_{X\bar\ell dsd}^{{\tt VR,VL}-,e}|^2
+1.1 |\tilde{C}_{X\bar\ell dds}^{{\tt TL,SL},e}|^2
+0.03\kappa_3^2|\tilde{C}_{X\bar\ell dds}^{{\tt VR,VL},e}|^2
+0.03\kappa_3^2|\tilde{C}_{X\bar\ell dsd}^{{\tt VR,VL}+,e}|^2
\nn\\
&+4\cdot10^{-5}\kappa_4^2 |\tilde{C}_{X\bar\ell dds}^{{\tt SL,TL},e}|^2
-2.7 \Im\big(\tilde{C}_{X\bar\ell dsd}^{{\tt VR,VL}-,e}\tilde{C}_{X\bar\ell dds}^{{\tt TL,SL},e*}\big)
+0.06\kappa_3^2 \Re\big(\tilde{C}_{X\bar\ell dsd}^{{\tt VL,VR}+,e}\tilde{C}_{X\bar\ell dds}^{{\tt VL,VR},e*}\big)
\nn\\
&+0.002\kappa_3 \Re\big(\tilde{C}_{X\bar\ell dsd}^{{\tt VR,VL}-,e}\tilde{C}_{X\bar\ell dds}^{{\tt VL,VR},e*}\big)
+0.002\kappa_3 \Re\big(\tilde{C}_{X\bar\ell dsd}^{{\tt VR,VL}-,e}\tilde{C}_{X\bar\ell dsd}^{{\tt VL,VR}+,e*}\big)
\nn\\
&-0.001\kappa_3 \Im\big(\tilde{C}_{X\bar\ell dds}^{{\tt VL,VR},e}\tilde{C}_{X\bar\ell dds}^{{\tt TL,SL},e*}\big)
-0.001\kappa_3 \Im\big(\tilde{C}_{X\bar\ell dsd}^{{\tt VL,VR}+,e}\tilde{C}_{X\bar\ell dds}^{{\tt TL,SL},e*}\big)
\nn\\
&-1\cdot10^{-5} \kappa_3\kappa_4\Re\big((\tilde{C}_{X\bar\ell dsd}^{{\tt VL,VR}+,e}+\tilde{C}_{X\bar\ell dds}^{{\tt VL,VR},e})\tilde{C}_{X\bar\ell dds}^{{\tt SL,TL},e*}\big)
+ \tL \leftrightarrow \tR,
\\%
{\Gamma_{n \to \mu^- K^+ X} \over 10^{-8}(\rm GeV)^9} =\,&
1.2 |\tilde{C}_{X\bar\ell dsd}^{{\tt VR,VL}-,\mu}|^2
+0.7 |\tilde{C}_{X\bar\ell dds}^{{\tt TL,SL},\mu}|^2
+0.02\kappa_3^2 |\tilde{C}_{X\bar\ell dds}^{{\tt VR,VL},\mu}|^2
+0.02\kappa_3^2 |\tilde{C}_{X\bar\ell dsd}^{{\tt VR,VL}+,\mu}|^2
\nn\\
&+4\cdot10^{-5}\kappa_4^2|\tilde{C}_{X\bar\ell dds}^{{\tt SL,TL},\mu}|^2
-1.9 \Im\big(\tilde{C}_{X\bar\ell dsd}^{{\tt VR,VL}-,\mu}\tilde{C}_{X\bar\ell dds}^{{\tt TL,SL},\mu*}\big)
\nn\\
&+0.2\kappa_3 \Re\big(\tilde{C}_{X\bar\ell dsd}^{{\tt VR,VL}-,\mu}\tilde{C}_{X\bar\ell dds}^{{\tt VL,VR},\mu*}\big)
+0.2\kappa_3 \Re\big(\tilde{C}_{X\bar\ell dsd}^{{\tt VR,VL}-,\mu}\tilde{C}_{X\bar\ell dsd}^{{\tt VL,VR}+,\mu*}\big)
\nn\\
&-0.1\kappa_3 \Im\big(\tilde{C}_{X\bar\ell dds}^{{\tt VL,VR},\mu}\tilde{C}_{X\bar\ell dds}^{{\tt TL,SL},\mu*}\big)
-0.1\kappa_3 \Im\big(\tilde{C}_{X\bar\ell dsd}^{{\tt VL,VR}+,\mu}\tilde{C}_{X\bar\ell dds}^{{\tt TL,SL},\mu*}\big)
\nn\\
&+0.04\kappa_3^2 \Re\big(\tilde{C}_{X\bar\ell dds}^{{\tt VL,VR},\mu}\tilde{C}_{X\bar\ell dsd}^{{\tt VL,VR}+,\mu*}\big)
\nn\\
&-9\cdot10^{-4} \kappa_3\kappa_4\Re\big((\tilde{C}_{X\bar\ell dsd}^{{\tt VL,VR}+,\mu}+\tilde{C}_{X\bar\ell dds}^{{\tt VL,VR},\mu})\tilde{C}_{X\bar\ell dds}^{{\tt SL,TL},\mu*}\big)
+ \tL \leftrightarrow \tR.
\end{align}
\end{subequations} }

\bibliography{references}
\bibliographystyle{JHEP}

\end{document}